\documentclass[a4paper,12pt,twoside,openright]{iitbthesis}
\usepackage{epsfig}
\usepackage{float}
\usepackage[T1]{fontenc}
\usepackage{ae,aecompl}
\usepackage{lipsum}
\usepackage[explicit]{titlesec}
\usepackage{graphics}
\usepackage{graphicx}
\usepackage{booktabs}
\usepackage{enumerate}
\usepackage{arabtex}
\usepackage{amsmath}
\usepackage{amssymb}
\usepackage{algorithm}
\usepackage{algorithmic}
\usepackage{tabularx}
\usepackage{longtable}
\usepackage{aabrf}
\usepackage{subfigure}
\usepackage{fancyhdr}
\usepackage{textcomp}
\usepackage[nottoc]{tocbibind}
\usepackage[square,comma,numbers,sort,compress]{natbib}
\usepackage[colorlinks]{hyperref}
\usepackage{multirow}
\usepackage{rotating}
\hypersetup{
    colorlinks=true,       % false: boxed links; true: colored links
    linkcolor=blue,          % color of internal links
    citecolor=magenta,        % color of links to bibliography
    urlcolor=magenta
}
\usepackage{mathptmx}
\fontfamily{cmr}\selectfont
\renewcommand{\bibname}{References}

\newcommand{\ncom}{\newcommand}

\ncom{\beqn}{\begin{eqnarray*}} \ncom{\eeqn}{\end{eqnarray*}}
\ncom{\beq}{\begin{eqnarray}} \ncom{\eeq}{\end{eqnarray}}
\ncom{\be}{\begin{equation}} \ncom{\ee}{\end{equation}} 
\ncom{\ben}{\begin{equation*}} \ncom{\een}{\end{equation*}} 
\ncom{\nn}{\nonumber}
\titleformat{\chapter}[display]{\bfseries\centering}{\huge Chapter \thechapter}{1em}{\Huge #1}
\usepackage[titletoc]{appendix}
\usepackage[nottoc]{tocbibind}
\begin{document}
%\pagestyle{fancy}
%\setlength{\headheight}{15pt}
%\fancyhf{}
%\fancyhead[LE,RO]{\thepage}
%\fancyhead[RE,LO]{\textit{\nouppercase{\leftmark}}}
%\fancyhead[LO]{\textit{\nouppercase{\rightmark}}}
%\fancyhead[RE,LO]{\textit{\nouppercase{\thechapter}}}
%\fancyhead[LO]{\textit{\nouppercase{\thechapter}}}
\pagestyle{fancy}
\setlength{\headheight}{15pt}
%%%%%%%%%%%%%%%%%%%%%%%%%%%%%%%%%%
\newcommand{\chapquote}[3]{\begin{quotation} \textit{#1} \end{quotation} \begin{flushright} - #2, \textit{#3}\end{flushright} }
\renewcommand{\headrulewidth}{.1pt}
\fancyhead[LE,RO]{{\thepage}}
\fancyhead[RE]{\nouppercase{{\leftmark}}}
\fancyhead[LO]{\nouppercase{{\rightmark}}} \fancyfoot{}
\def\baselinestretch{1.1}
\pagenumbering{roman}
\date{\today}
\clubpenalty=10000 \widowpenalty=10000

%=====================================================================
% Include the prelude for Title page, abstract, table of contents, etc
% You need to modify it to contain your details
% prelude.tex
%   - titlepage
%   - dedication (optional)
%   - approval sheet
%   - course certificate
%   - table of contents, list of tables and list of figures
%   - nomenclature
%   - abstract
%============================================================================

\clearpage\pagenumbering{roman}  % This makes the page numbers Roman (i, ii, etc)

% TITLE PAGE
%   - define \title{} \author{} \date{}
\title{Generation and effects of primordial magnetic fields during inflation}
\author{Ashu Kushwaha}
\date{October 6, 2023}

%  - Roll number, required for title page, approval sheet, and
%    certificate of course work 
\rollnum{184120007} 

%   - The default degree is ``Doctor of Philosophy''
%     (unless the document style msthesis is specified
%      and then the default degree is ``Master of Science'')
%     Degree can be changed using the command \iitbdegree{}
\iitbdegree{Doctor of Philosophy}

%   - The default report type is preliminary report.
%      * for a PhD thesis, specify \thesis
\thesis
%      * for a M.Tech./M.Phil./M.Des./M.S. dissertation, specify \dissertation
%\dissertation
%      * for a DIIT/B.Tech./M.Sc.project report, specify \project
%\project
%      * for any other type, use  \reporttype{}
%\reporttype{Pre-Synopsis}

%   - The default department is ``Unknown Department''
%     The department can be changed using the command \department{}
\department{DEPARTMENT OF PHYSICS}

%    - Set the guide's name
\setguide{Prof. S. Shankaranarayanan}
%    - Set the coguide's name (if you have one)
%\setcoguide{Prof. FirstName LastName}
%    - Set external guide (if you have one)
%\setexguide{Prof External Guide}

%   - once the above are defined, use \maketitle to generate the titlepage
\maketitle

%--------------------------------------------------------------------%
% DEDICATION
%   Dedications, if any, must be first page after title page.
%\begin{dedication}
%
%%I dedicate this thesis to\\
%% xxxxx xxxxx
%\end{dedication}

%--------------------------------------------------------------------%
% APPROVAL SHEET
%   - for final thesis, you need Approval Sheet. So, uncomment the
%     \makeapproval command.
%     it should come after dedication, if dedication is
%     present. Otherwise it is the first page after title page.
\makeapproval
%--------------------------------------------------------------------%
% CERTIFICATE OF COURSE WORK
%   - for final thesis, a course certificate is required.
%   - specify the  PhD joining date for the certificate.
%     Contact you department office or academic office if you do not
%     know it.
%\joiningdate{Date, Month, Year}
%\begin{coursecertificate}
%%  - command to add a course
%%     it accepts 3 arguments:  Course ID, Course Name, Course Credits.
%\addcourse{ME xxx}{Course Name}{6}
%\addcourse{ME xxx}{Course Name}{6}
%\addcourse{ME xxx}{Course Name}{6}
%\addcourse{ME xxx}{Course Name}{6}
%\addcourse{ME xxx}{Course Name}{6}
%\addcourse{MES802}{PhD Seminar - \emph{Title}}{4}
%\addppcourse{HS 699}{Communication and Presentation Skills}{PP}
%\end{coursecertificate}

%--------------------------------------------------------------------%
% COPYRIGHT PAGE
%   - To include a copyright page use \copyrightpage
\copyrightpage %Also includes declaration 
%-----------------------------------------------------
\begin{acknowledgements}

I am grateful for the opportunity to pursue the Ph.D. program at the Department of Physics, IIT Bombay. Throughout my research journey and the completion of this thesis, I have received support from various individuals, and I would like to acknowledge their contributions. First and foremost, I express my sincere gratitude to my supervisor, Prof. S. Shankaranarayanan (Shanki), for his exceptional guidance and mentorship during the entire duration of my doctoral research. His expertise and insightful advice have been invaluable to me. I would also like to extend my thanks to my collaborators, Dr. Sunil Malik, Dr. Debottam Nandi, and Dr. Abhishek Naskar. Their assistance and contributions during our research collaboration were instrumental in enhancing my understanding and achieving our research goals. I thank my RPC (research progress committee) members, Prof. Vikram Rentala and Prof. Urjit A. Yajnik, for their valuable inputs and critical comments.

I am indebted to my fellow group members, Dr. Archana Sangwan, Dr. Joseph P J, Dr. Avijit Chowdhury, Dr. Karthik Rajeev, Susmita, Mahesh, Semin, and Toushif. The numerous discussions we had during group meetings and their valuable comments and critiques regarding my presentations, and research have significantly influenced the quality of my work. I am also grateful to my colleagues, especially Semin and Koustav for their unwavering support throughout these years.

I extend my heartfelt appreciation to Prof. Ram Ramaswamy from IIT Delhi for his support, motivation, and guidance, which have played a pivotal role in my academic journey. I would also like to thank Dr. Poonam Mehta from JNU for her valuable advice and assistance at various stages of my MSc and Ph.D. Amidst the pressures of pursuing a Ph.D., friends have been a source of solace and joy. I am grateful to Sachin, Naba Prakash, and Bikash for being by my side and making these years enjoyable.

Last but certainly not least, I express my deepest gratitude to my family, particularly my parents, for their unwavering support, sacrifices, and unconditional love. Their constant efforts have been the cornerstone of my achievements. I would also like to thank all my childhood and college friends for their unwavering support. None of this would have been possible without them.

\end{acknowledgements}
%--------------------------------------------------------------------%
%ABSTRACT
\begin{abstract}
  
%Due to technological advancements and the availability of vast amounts of data from various observations, we are currently in an era of precision cosmology. Although precision cosmology enhances our understanding of the Universe, it also challenges established theoretical ideas, giving rise to intriguing new questions. To seek answers to these questions, new ideas are introduced, often requiring going beyond the standard model of cosmology. 

Observational evidence indicates the existence of a coherent micro-Gauss magnetic field spanning galactic scales, as well as an extremely strong magnetic field near highly gravitating compact objects. Unfortunately, we do not have any compelling theoretical model that can explain the presence of these large-scale magnetic fields and the stability of very strong magnetic fields near these compact objects. To explain the large-scale magnetic field, the well-accepted paradigm suggests that the dynamo mechanism amplifies the tiny \emph{seed} magnetic fields. However, the standard electrodynamics and early Universe cosmology can not explain the origin of the seed magnetic field. Since the standard electrodynamics action is conformally flat, the magnetic field decays rapidly due to the Universe's expansion. Therefore, one has to go beyond the standard model of cosmology to construct a model to generate the primordial seed magnetic fields in the early Universe. Various theoretical models have been proposed to generate the primordial magnetic field during the inflationary era. However, most models introduce the coupling between the scalar and the electromagnetic fields, suffering from strong coupling and backreaction problems. 
In this thesis, we aim to investigate the mechanisms underlying the genesis of magnetic fields in the Universe and explore their potential in addressing the matter-antimatter asymmetry.
%e focus on unraveling the mechanisms responsible for the genesis of magnetic fields in the Universe and how these mechanisms can aid in addressing the matter-antimatter asymmetry. 
We construct consistent models which generate the helical magnetic field consistent with observations. We also develop an effective field theory approach to magnetogenesis, where the choice of EFT parameters describes the magnetogenesis scenario in the early Universe, and different choices of parameters correspond to different models. Our EFT explicitly shows that generating primordial magnetic fields requires two necessary conditions — \emph{conformal invariance breaking} and \emph{causal propagation}. %Hence, conformal invariance breaking of the electromagnetic fields is only necessary, not sufficient.  
Furthermore, we establish that the presence of a strong magnetic field near compact objects, such as neutron stars, aids in understanding the phenomenon of \emph{Fast radio bursts} through the conversion of gravitational waves to electromagnetic waves in the background of a strong transverse magnetic field.

\newpage
  %Laser ablation...
\end{abstract}

%---------------------------------

\begin{publications}
\small
%\newpage
%\begin{center}
%{\LARGE {\textbf{List of Papers}}}\\[0.05 cm]
%\noindent\rule{16cm}{1.5pt}
%\justify
%\end{center}
 
 \section*{Journal publications}
 
\begin{enumerate}
\item[6.] \textbf{Ashu Kushwaha}, Sunil Malik, and S. Shankaranarayanan ; \textit{Gertsenshtein-Zel$'$dovich effect explains the origin of Fast Radio Bursts}, e-print: \href{https://arxiv.org/abs/2202.00032}{2202.00032} (under review)

\item[5.]  \textbf{Ashu Kushwaha}, Sunil Malik, and S. Shankaranarayanan; \textit{Fast Radio Bursts signal high-frequency gravitational waves}, Appeared online \href{https://doi.org/10.1142/S0218271823420105}{International Journal of Modern Physics D; 2342010 }.~~~{\it Received Honorable Mention in
\href{http://www.gravityresearchfoundation.org/announcements.html}{2023 Essay Competition of the Gravity Research Foundation, USA}}

\item[4.] \textbf{Ashu Kushwaha}, Abhishek Naskar, Debottam Nandi, and S. Shankaranarayanan ; \textit{Effective field theory of magnetogenesis identify necessary and sufficient conditions}, \href{https://iopscience.iop.org/article/10.1088/1475-7516/2023/01/045}{JCAP 01 (2023) 045} e-print: \href{https://arxiv.org/abs/2207.05162}{2207.05162} 

\item[3.] \textbf{Ashu Kushwaha} and S. Shankaranarayanan ; \textit{Helical magnetic fields from Riemann coupling lead to baryogenesis}, \href{https://journals.aps.org/prd/abstract/10.1103/PhysRevD.104.063502}{Phys. Rev. D 104, 063502 (2021)}, e-print: \href{https://arxiv.org/abs/2103.05339}{2103.05339} 

\item[2.] \textbf{Ashu Kushwaha} and S. Shankaranarayanan ; \textit{Helical magnetic fields from Riemann coupling}, \href{https://journals.aps.org/prd/abstract/10.1103/PhysRevD.102.103528}{Phys. Rev. D 102, 103528 (2020)}, e-print : \href{https://arxiv.org/abs/2008.10825}{2008.10825} 

\item[1.] \textbf{Ashu Kushwaha} and S. Shankaranarayanan ; \textit{Galileon scalar electrodynamics}, \href{https://journals.aps.org/prd/abstract/10.1103/PhysRevD.101.065008}{Phys. Rev. D 101, 065008 (2020)}, e-print : \href{https://arxiv.org/abs/1912.01393}{1912.01393} 

\end{enumerate}

\section*{Conference proceedings}

\begin{enumerate}

\item[2.] \textbf{Ashu Kushwaha} and S. Shankaranarayanan ; \textit{Helical magnetic fields lead to baryogenesis} in The Sixteenth Marcel Grossmann Meeting, \href{https://www.worldscientific.com/doi/10.1142/9789811269776_0216}{pp. 2692-2699 (2023)}%{\textbf{Ashu Kushwaha} and S. Shankaranarayanan ; \textit{Helical magnetic fields lead to baryogenesis} in Sixteenth Marcel Grossmann Meeting Conference (MG-16). (accepted for publication)}% \textbf{(Virtual)} }
\item[1.] \textbf{Ashu Kushwaha} and S. Shankaranarayanan. \textit{Helical magnetic fields from riemann coupling.} In Bedangadas Mohanty, Sanjay Kumar Swain, Ranbir Singh, and Varchaswi K. S. Kashyap, editors, \href{https://link.springer.com/chapter/10.1007/978-981-19-2354-8_116}{Proceedings of the XXIV DAE-BRNS High Energy Physics Symposium, Jatni, India, pages 639– 643, Singapore, 2022. Springer Nature Singapore.} 

\end{enumerate}

\end{publications}
%--------------------------------------------------------------------%
% CONTENTS, TABLES, FIGURES
\tableofcontents
\listoftables
\listoffigures

%--------------------------------------------------------------------%
% NOMENCLATURE
%\begin{nomenclature}
%\input{./Nomenclature/Nomenclature.tex}
%\end{nomenclature}

\cleardoublepage\pagenumbering{arabic} % Make the page numbers Arabic (1, 2, etc)

%=====================================================================
% Include the technical part of the report
%\pagestyle{fancy}
%\fancyhf{}
%\fancyhead[LE,RO]{\thepage}
%\fancyhead[RE,LO]{\textit{\nouppercase{\leftmark}}}
%\fancyhead[LO]{\textit{\nouppercase{\rightmark}}}

%\pagestyle{fancy}
\chapter{Introduction}
\label{ch:intro}

Cosmology is the study of the Universe at large scales, ranging from about one kpc (the size of a typical galaxy) to a few Mpc (the order of typical inter-galactic distances in the clusters of galaxies) or up to thousands of Mpc. 
The {mathematical} foundation of modern cosmology was laid in 1915 when Albert Einstein developed his \emph{theory of gravitation} known as ``General Relativity" (GR), a framework that has been successful in explaining various phenomena starting from the time difference in GPS to the largest scales in the Universe. 
Over the past few decades, cosmology has undergone remarkable advancements as theoretical predictions have been rigorously confronted against observations. This ongoing interactivity between theory and observational data has led to numerous groundbreaking discoveries and has compelled us to reassess prevailing theoretical paradigms. One illustrative example of this is the confirmation of the Universe's expansion, initially predicted by the Friedmann and Lemaître, which was subsequently confirmed by Edwin Hubble's observation in 1929~\cite{Book-Weinberg-cosmology,Book-Dodelson.Schmidt_2020_edition}. Furthermore, the accidental discovery of the Cosmic Microwave Background (CMB) radiation by Penzias and Wilson in 1964, a phenomenon initially predicted by Gamow, Alpher, and Hermann~\cite{Book-Weinberg-cosmology,Book-Dodelson.Schmidt_2020_edition}, played a pivotal role in discrediting the steady-state theory and supporting the competing theory of hot Big Bang cosmology~\cite{Book-Weinberg-cosmology,Book-Dodelson.Schmidt_2020_edition}.
This chapter will first discuss the critical milestones achieved through observational endeavors in cosmology. We will also explore the intricate theoretical foundations underpinning modern cosmological models. Furthermore, we will investigate the persisting unresolved issues and challenges that confront the astrophysicists and cosmologists today, discuss a few of them in detail, and seek potential avenues for their resolution.
We will also highlight the unresolved issues this thesis attempts to address.

In this thesis, we use $(+,-,-,-)$ metric signature and natural units where $\hbar = c = 1/(4\pi\epsilon_0) = 1$, with reduced Planck mass $M_{\rm Pl}^2 = 1/(8 \pi G) = 2.4 \times 10^{18}~{\rm GeV}$. Greek indices refer to 4-D space-time. The various physical quantities with the over-line refers to the values evaluated for the homogeneous and isotropic FRW background. An overdot denotes a derivative with respect to the cosmic time ($t$), a prime stands for a derivative with respect to conformal time ($\eta$), and $,i$ denotes a derivative w.r.t spatial coordinates.

%\cite{Book-Gorbunov.Rubakov,Book-Kolb.Turner,Book-Mukhanov,Book-Weinberg-cosmology,Book-Dodelson.Schmidt_2020_edition,Book-Padmanabhan-III,Liddle.Lyth-Book}

%Cosmology is the study of universe at very large scales, ranging from about 1 kpc (the size of a typical galaxy) to a few Mpc (the order of typical inter-galactic distances in the clusters of galaxies) or up to thousands of Mpc which is roughly the farthest observable light source. In 1915 Einstein gave a theory of gravitation called "General relativity" which has been extremely successful in explaining the universe at solar system scales and even at large scales. In the past decades, the field seen remarkable progress by confronting theoretical predictions with observations. Many new discoveries have been made and many theoretical ideas have been challenged by observations. For example, the expansion of universe predicted by Friedmann and Lemaître was confirmed by Hubble in 1929, the prediction of the existence of cosmic microwave background by Gamow, Alpher, and Hermann was accidentally discovered by Penzias and Wilson in 1964 which later ruled out the steady state theory and favoured its competing theory --- hot Big bang cosmology. In this chapter we will discuss various observational milestones and the theoretical framework of modern cosmology. We will also discuss the open problems in cosmology and discuss few of them in detail.

\section{Observational landmarks in cosmology}
\label{sec:observ_landmarks}

%Cosmology is the study of universe at very large scales, ranging from about 1 kpc (the size of a typical galaxy) to a few Mpc (the order of typical inter-galactic distances in the clusters of galaxies) or up to thousands of Mpc which is roughly the farthest observable light source. 
Currently, we are in the era of precision cosmology where the main objective of various ongoing and proposed observational missions is to refine existing understanding while simultaneously seeking novel insights into the physics that transcend our current knowledge boundaries. Let us briefly discuss the landmark observations and their impacts on our understanding of the Universe. A concise history of important observational landmarks is given in Table~\ref{table:observational_landmarks}.
\begin{table}[h]
\small
\centering
\begin{tabular}{|l|c|c|}
\hline
\rule[-2ex]{0pt}{5ex} \textbf{Year} &  \textbf{Discovery/Observational progress} & \textbf{Implications}\\ 
\hline
\rule[-2ex]{0pt}{5ex} 1924 & Edwin Hubble, \emph{Cepheid variable stars}~\cite{Book-Weinberg-cosmology,Book-Dodelson.Schmidt_2020_edition} & distance measure beyond galaxy\\
%Cepheid variable stars in M31
\hline
\rule[-2ex]{0pt}{5ex} 1929   & Edwin Hubble, \emph{Universe's expansion}~\cite{Book-Weinberg-cosmology,Book-Dodelson.Schmidt_2020_edition} & %Forced cosmologists to 
\shortstack{ dynamical models of Universe,\\  age of Universe}  \\ 
\hline
\rule[-2ex]{0pt}{5ex} 1933 &  \shortstack{ Karl G. Jansky, \emph{Radio Waves from} \\ \emph{Outside the Solar System } }~\cite{1933-Jansky-Nature} & Birth of radio astronomy\\
\hline
\rule[-2ex]{0pt}{5ex} 1933 &  Fritz Zwicky, \emph{Evidence of Dark Matter}~\cite{1933-Zwicky-HPA} & Explains the galaxy rotation curve\\ 
\hline
\rule[-2ex]{0pt}{5ex} 1949 & 
W. A. Hiltner, \emph{Magnetic fields in galaxy}~\cite{1949-Hiltner-Sci}& New probe of Universe on large scale 
\\ 
\hline 
\rule[-2ex]{0pt}{5ex} 1964 & Penzias and Wilson, \emph{isotropic CMB radiation}~\cite{1965-Penzias.Wilson-ApJ} & Favoured hot big bang theory
\\ 
\hline 
\rule[-2ex]{0pt}{5ex} 1967 & Jocelyn Bell, \emph{Pulsar}~\cite{1968-Bell.Hewish.etal-Nature} & \shortstack{Birth of high-energy astrophysics, \\ discovery of highly compact objects}
\\ 
\hline 
\rule[-2ex]{0pt}{5ex} 1997 & \shortstack{Ron Gilliland and Mark Phillips,\\ \emph{Type 1a supernova}}~\cite{1998-SupernovaSearchTeam-ApJ} & accelerating expansion of the universe
\\ 
\hline 
\rule[-2ex]{0pt}{5ex} 2003 & \shortstack{2dF Galaxy Redshift Survey ,\\ \emph{homogeneous Universe} }~\cite{2001-2DFGRS-MNRAS} & \shortstack{first map of galaxy distribution, \\ largest possible scale of Universe}  
\\ 
\hline
\rule[-2ex]{0pt}{5ex} 2007 & Duncan Lorimer, \emph{Fast radio burst}~\cite{2007-Lorimer.etal-Science} & \shortstack{unique probe in extreme conditions\\ and of the distribution of matter in the Universe}
\\ 
\hline
\rule[-2ex]{0pt}{5ex} 2015 & \shortstack{ LIGO and Virgo collaborations, \\ \emph{discovery of Gravitational waves} }~\cite{2016-LIGOScientific} & confirmation of Einstein's prediction of GW
\\ 
\hline
\rule[-2ex]{0pt}{5ex} 2018 &  
 \shortstack{ LIGO and Virgo collaborations, \\ NS-NS merger}~\cite{2017-LIGOScientific} &  \shortstack{ constrain modified gravity theory, \\ independent measure of $H_0$}
\\ 
\hline
\end{tabular}
\caption{History of observational landmarks in astronomy and cosmology.}
\label{table:observational_landmarks}
\end{table}
%
%
%
%Numerical value of critical density $\rho_{\rm{c}} = 1.88 h^2 \times 10^{-29} \rm{g cm^{-3}}$
%The field got its first kick by the simple and elegant realisation by 
					
\begin{figure}[ht]
\centering
%\subfigure[]{%
%\label{fig:hubble-law}%
%\includegraphics[height=1.8in]{chapter1/Hubble_diagram1929.jpeg}     %}%
%\quad
%\subfigure[]{%
\label{fig:Huchra}%
\includegraphics[height=2.8in]{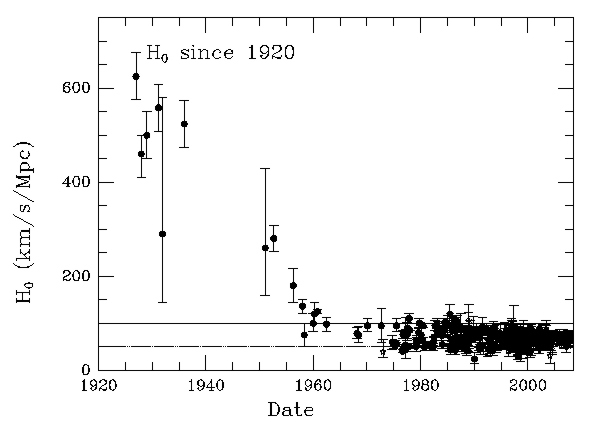}      %}%
\caption{%Left panel: 
%A modern version of the Hubble diagram from the Hubble Space Telescope Key project~\cite{Book-Dodelson.Schmidt_2020_edition}, \cite{2001-Freedman.etal-HST-ApJ}. 
Time evolution of our knoweldge of the Hubble Constant~\cite{Huchra-Hubble-constant} %(Fig source: \href{https://lweb.cfa.harvard.edu/~dfabricant/huchra/hubble/}{Huchra}) 
}
\label{fig:Hubble-law}
\end{figure}

Edwin Hubble, in 1929 found that distant galaxies are apparently receding from us~\cite{1929-Hubble-PNAS}. Also, he found a pattern in which the velocities of these distant galaxies increase with distance. This discovery of the expansion of the Universe is arguably the most crucial cosmological discovery. The expansion rate is the measure of how fast the Universe is expanding and is determined by Hubble’s law which gives the mathematical relation between the distance and relative velocity between the galaxies, in the absence of any comoving motion. Mathematically, Hubble's law is given by  
\begin{align}\label{eq:hubble_law}
v = \frac{d}{dt}(a x) = \dot{a} x = H_0 d, \qquad (v << c)
\end{align}
where 'overdot' denotes the derivative with respect to cosmic time $t$, $x$ is the comoving distance, $d$ is the physical distance between two galaxies, $c$ is the speed of light and $a(t)$ is the scale factor of the Universe. $H_0 = 100 h \, \rm{km s^{-1} Mpc^{-1}}$ is the Hubble constant which is defined with dimensionless number $h$. 
%, and the dimensionless number is estimated from current observations to be $h \simeq 0.7$. The precise value of the Hubble constant ($H_0$) has been a subject of debate, and even now, there is some controversy about its precise value at the $5\%$ level. 
 The precise value of the Hubble constant $H_0$ (or $h$) is still under debate due to conflicting results between it measurements on a large scale and the local measurements~\cite{2016-Riess.etal-ApJ}.
%First initial measurement of Hubble constant was taken by Edwin Hublble in 1929~\cite{1929-Hubble-PNAS} (shown in the figure \ref{fig:hubblediagram}), and in \textbf{[fig]} 
Fig~\ref{fig:Hubble-law} contains the evolution in the measurements of the Hubble diagram (velocity-distance relationship) over the last 100 years. 
\begin{figure}[ht]
%\centering
%\subfigure[]{%
%\label{fig:PlanckCMB}%
\includegraphics[height=2.2in]{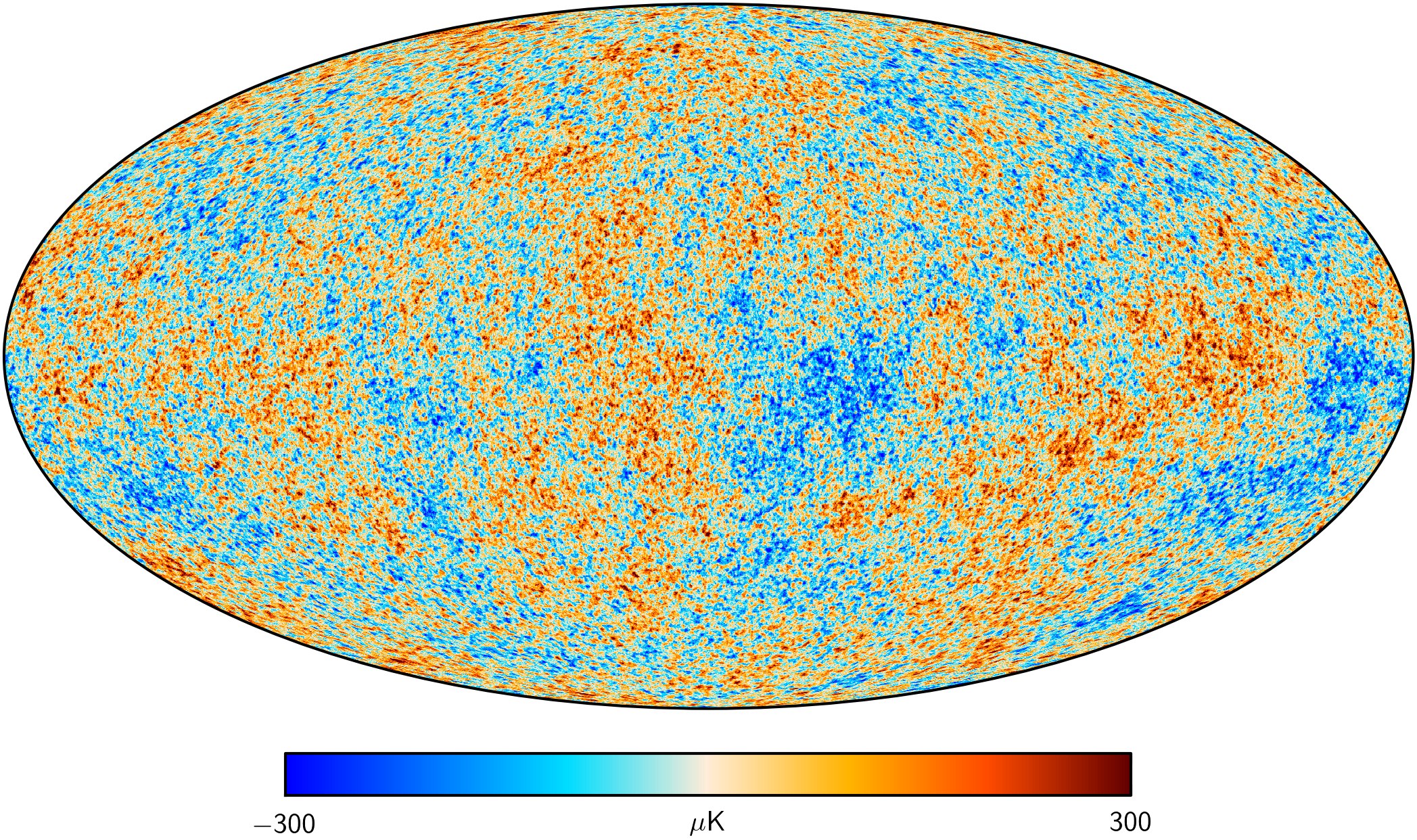}   %}%
%\qquad
%\subfigure[]{%
%\label{fig:SDSS}%
\includegraphics[height=2.4in]{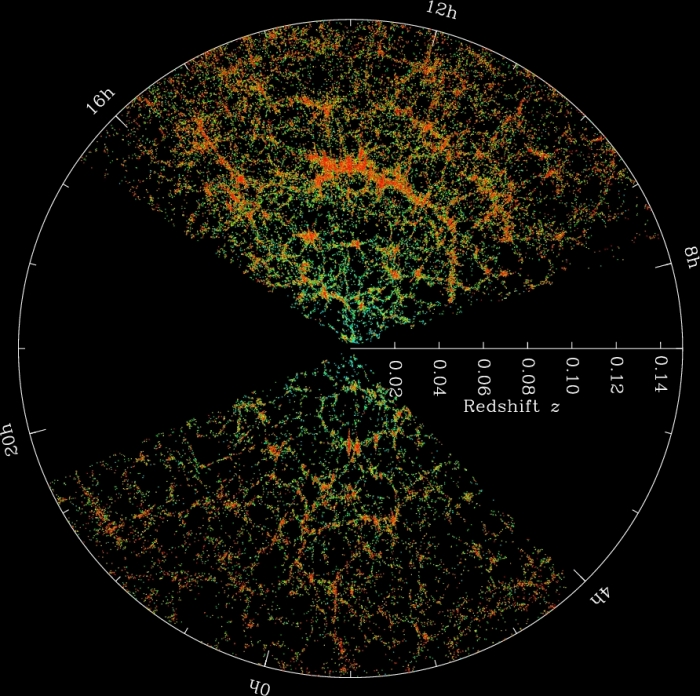}     %}%
\caption{Left plot: Planck measurement of the temperature variations in the CMB sky~\cite{Planck-2018}. Right plot: The distribution of galaxies measured in the SDSS survey~\cite{2017-SDSS_survey}}
\label{fig:Planck_CMB-SDSS}
\end{figure}

Another crucial observational landmark is the CMB radiation, which was accidentally discovered by Penzias and Wilson in 1964. It is the thermal radiation assumed to be the relic from the "Big Bang" and has a perfect black body spectrum at a temperature $2.725 \pm 0.00057 \rm{K}$~\cite{2009-Fixsen-ApJ}. In 1992, Cosmic Background Explorer (COBE) satellite discovered anisotropies in the CMB with a fractional temperature fluctuation of order $10^{-5}$, and this indicates that Universe was not completely smooth~\cite{1992-COBE-MNRAS,1992-COBE-ApJL}. However, as shown in the left Fig.~\ref{fig:Planck_CMB-SDSS}, these observations imply that, to a great extend, the Universe was perfectly isotropic and homogeneous at the epoch of recombination (at redshift $z \simeq 1100$). The observations of the temperature fluctuations in the CMB have played a crucial role in establishing the standard model of cosmology. With the advancement of precision measurements, currently it is widely accepted that these primordial fluctuations in CMB are originated from the quantum fluctuations during a period of exponential expansion known as \emph{inflation} in the early Universe~\cite{Book-Gorbunov.Rubakov,Book-Kolb.Turner,Book-Mukhanov,Book-Weinberg-cosmology,Book-Dodelson.Schmidt_2020_edition,Book-Padmanabhan-III,Liddle.Lyth-Book}.  [Sec. \ref{sec:inflation} contains a rapid review of inflation and the mechanism that generate the primordial perturbations.]
%Why the Universe is so smooth poses a problem known as the horizon problem, which we will discuss in later sections.
%\textbf{Detecting CMB photons is like taking photograph of the Universe at $z\sim 1100$. Say more coherently, and connect with the later part---Horizon problem etc.}
%

In 2003, \emph{2dF Galaxy Redshift Survey} (Two-Degree-Field Galaxy Redshift Survey) determined for the first time the galaxy distribution in the Universe up to a redshift $\sim 0.2$. Later, Sloan Digital Sky Survey improved this galaxy distribution map to include more galaxies. Fig~\ref{fig:Planck_CMB-SDSS} (right plot) shows the position of close to millions of galaxies in a slice through the volume mapped by Sloan Digital Sky Survey (SDSS)~\cite{2017-SDSS_survey}. This indicates that the distribution of galaxies in the local Universe is not homogeneous, and the Universe has a structure on large scales.
\section{ Friedmann-Robertson-Walker Universe}
The standard model of cosmology is based on the assumption that the Universe is homogeneous and isotropic at large scales, known as \emph{cosmological principle}~\cite{2006-Ellis-arXiv,Book-Gorbunov.Rubakov,Book-Kolb.Turner,Book-Mukhanov,Book-Weinberg-cosmology,Book-Dodelson.Schmidt_2020_edition,Book-Padmanabhan-III,Liddle.Lyth-Book}. 
By isotropy, we mean that there are no preferred directions in the Universe, so it looks statistically the same irrespective of the observer's direction. In contrast, homogeneity means that all locations are equivalent~\cite{1973-Collins.Hawking-ApJ}. And by large scales, we mean the length scales larger than $60-70 h^{-1} \rm{Mpc}$ (where $h=0.7$ in $\rm{\Lambda CDM}$)~\cite{2005-Yadav.Seshadri.etal-MNRAS}. %, which the galaxy distribution has tested in the Sloan Digital Sky Survey Data~\cite{2005-Yadav.Seshadri.etal-MNRAS}. 
Note that isotropy at every point automatically enforces homogeneity. 

Our current understanding of the evolution of the Universe is based on the Friedmann Lemaître Robertson Walker (FRW) cosmological model. It is successful in explaining the observed Universe and is referred to as the \emph{standard model of cosmology}. It assumes that Einstein's General Relativity is the correct description of gravity. The FRW space-time corresponding to homogeneous and isotropic spacetime is~\cite{Book-Gorbunov.Rubakov,Book-Kolb.Turner,Book-Mukhanov,Book-Weinberg-cosmology,Book-Dodelson.Schmidt_2020_edition,Book-Padmanabhan-III,Liddle.Lyth-Book}
\begin{align}\label{eq:full-frw-metric}
ds^2 = dt^2 - a^2(t) \left(  \frac{dr^2}{1-K r^2} + r^2 d\theta^2 + r^2 \sin^2\theta d\phi \right) \, ,
\end{align}
where $K$ refers to the 3-space curvature. $K =0$ corresponds to the flat Universe and $K=-1, K= +1$ is for the negative curved (open Universe) and positively curved (closed Universe) constant-time 3D hypersurface (spatial slices), respectively\footnote{Note that three values of constant 3-curvature ($K = -1,0,+1$) are the consequence of homogeneous and isotropic 3-spaces. A 3-space that is homogeneous and isotropic is maximally symmetric.}. The spatial coordinates ($r,\theta,\phi$) are the comoving coordinates, and time coordinate $t$ is the proper time measured by the comoving observer. In order to obtain the dynamics of the Universe, we use the Einstein equation
\begin{align}\label{eq:Einstein_Eq}
G_{\mu\nu} \equiv R_{\mu\nu} - \frac{1}{2} g_{\mu\nu} R = 8\pi G T_{\mu\nu}
\end{align}
where $R_{\mu\nu}$ is Ricci tensor, $R = g^{\mu\nu} R_{\mu\nu}$ is Ricci scalar, and $T_{\mu\nu}$ is the energy-momentum tensor of all the matter fields (matter, radiation, etc.). Einstein equation tells us how the spacetime curvature (determined by Einstein tensor $G_{\mu\nu}$) is affected by the matter content of the Universe (determined by $T_{\mu\nu}$). [In General Relativity the term "matter" is used for anything which is not gravitational field~\cite{Book-Mukhanov}.]
On large scales, matter can be approximated as a perfect fluid characterized by three quantities---\emph{pressure} ($p$), \emph{energy density} ($\rho$) and \emph{four-velocity} ($u^{\mu}$). Furthermore, to be consistent with the symmetries of the spatial metric, $T_{\mu\nu}$ must be diagonal, and by isotropy, the spatial components must be equal. Since the space-time is expanding, $\rho$ and $p$ are functions of time only. The expression which satisfies these criteria is the energy-momentum tensor for the perfect fluid given by 
%\footnote{One can show that the requirement of isotropy and homogeneity enforces the energy-momentum tensor to be "perfect fluid".}
%
\begin{align}
T^{\mu}_{\nu} = (\rho + p) u^{\mu}u_{\nu} - p \delta^{\mu}_{\nu}
\end{align}
where $\delta^{\mu}_{\nu} = g_{\nu\alpha}g^{\mu\alpha}$ is the 4-D Kronecker delta function which is 1 for indices $\mu = \nu$ and zero otherwise, and the 4-velocity $u^{\mu} = (1,0,0,0)$ which satisfies the condition $u^{\mu}u_{\mu} = 1$. We assume the fluid to be barotropic which means the pressure only depends on the energy density, $p = p(\rho)$. The equation of state depends on the properties of the matter and gives the relation between the pressure and energy density, which are related by the equation of state parameter $w$ as $p = w \rho$. For non-relativistic particles, there is no pressure i.e., $p=0$, which implies $w = 0$, also referred to as ``dust". For ultra-relativistic matter, for example, radiation (like Maxwell theory of electromagnetism), the energy-momentum tensor is traceless (i.e., $T^{\mu}_{\mu} = \rho - 3 p = 0$), which determines the equation of state as $p = \rho/3$ with $w = 1/3$. Moreover, for the cosmological constant, the equation of state is given by $p = -\rho$ with $w = -1$~\cite{Book-Gorbunov.Rubakov,Book-Kolb.Turner,Book-Mukhanov,Book-Weinberg-cosmology,Book-Dodelson.Schmidt_2020_edition,Book-Padmanabhan-III,Liddle.Lyth-Book}. 

The conservation of the energy-momentum tensor gives the continuity equation
\begin{align}\label{eq:continuity}
\nabla_{\mu} T^{\mu\nu} = 0 \quad \implies \quad \dot{\rho} = -3 H (\rho + p)
\end{align}
where $H = \dot{a}/a$ is the Hubble parameter.
The Friedmann equations~\cite{1922-Friedman}, which gives the evolution of the scale factor $a(t)$ are give by:
\begin{subequations}
\begin{equation}\label{eq:friedmann-1}
  H^2 + \frac{K}{a^2} = \frac{8\pi G}{3} \rho
\end{equation}    
\begin{equation}\label{eq:friedmann-2}
  \dot{H} + H^2 = -\frac{4\pi G}{3} (\rho + 3p).
\end{equation}
\end{subequations}
%
 %work with flat FRW spacetime. 
It is important to note that the second Friedmann equation~\eqref{eq:friedmann-2} can be derived by using Eq.\eqref{eq:continuity} and Eq.\eqref{eq:friedmann-1}. 
%This is because of the fact that the Bianchi identity tells us that Einstein equation and continuity equation should not be independent.
%
From first Friedmann equation \eqref{eq:friedmann-1}, we obtain:
\begin{equation}\label{eq:density-parameter-Omega}
  \Omega (t) - 1 = \frac{K}{a^2 H^2}
\end{equation} 
where $\Omega(t) = \frac{\rho (t)}{\rho_c (t)}$ is the dimensionless density parameter and $\rho_c (t) = \frac{3H^2(t)}{8\pi G}$ is the critical density.
%It is important to note that $\rho_c$ and $\Omega$ are not constant in time. 
The current value of critical density is $\rho_{c,0} = 1.879 \times 10^{-29} h^2 \rm{g \, cm^{-3}}$~\cite{Book-Dodelson.Schmidt_2020_edition}. Since current observations favour the Universe to be flat~\cite{2020-Efstathiou.Gratton-MNRAS}; we will set $K = 0$.
%
%%
%Assuming the barotropic fluids for which the equation of state can be parametrized by 
%\begin{equation}
% p = w \rho \qquad \text{where} \qquad
%    \begin{cases}
%       = 0,  & \text{for dust} \\
%      = \frac{1}{3}, & \text{for radiation}\\
%      = - 1, & \text{for cosmological constant}
%    \end{cases}       
%\end{equation}
%

Time evolution of the energy density can be obtained by using Eq.\eqref{eq:friedmann-1} and Eq.\eqref{eq:continuity} with the equation of state, which gives
\begin{equation}\label{eq:rho-for-w}
 \rho \propto a^{-3(1+w)} \quad \propto \quad
    \begin{cases}
        a^{-3},  & \text{for dust} \\
      a^{-4}, & \text{for radiation}\\
      a^0, & \text{for cosmological constant}
    \end{cases}       
\end{equation}
Therefore, the energy density during radiation domination era redshifts with scale factor as $\rho_{\rm r} \propto a^{-4}$ where the dilution includes the redshifting of the energy i.e., $E \propto a^{-1}$ as well. For matter-dominated era $\rho_{\rm m} \propto a^{-3}$ where the dilution of energy density simply reflects the expansion of the volume. Using the fact that $\rho_{\rm r}/\rho_{\rm m} \propto 1/a$ it is easy to see that the Universe would be radiation dominated in the past when the scale factor $a(t)$ was smaller, and it was matter-dominated in the most recent past. 
In the flat FRW Universe, we can solve for the scale factor as
\begin{equation}\label{eq:a-for-diff-cases}
 a(t) =
    \begin{cases}
       \propto t^{\frac{2}{3(1+w)}}, & \text{for } w \neq 1\\
      \propto e^{Ht}, & \text{for } w =  -1
    \end{cases}       
\end{equation}
We can see from the above expression that the scale factor for the negative equation of state parameter (positive cosmological constant) leads to the exponential expansion, we will discuss this scenario in the next section.
\section{Theory of cosmological inflation}\label{sec:inflation}
Inflation is an extraordinarily successful theoretical paradigm explaining why our Universe looks as it is observed. Inflation provides a causal mechanism to explain the large scale homogeneity of the Universe and provides a mechanism to generate seed perturbations that lead to observable imprints in CMB and generation of large-scale structure~\cite{1980-Starobinsky-PLB,1981-Guth-PRD,1981-Sato-MNRAS,1982-Linde-PLB,1982-Linde_2-PLB,1982-Albrecht.Steinhardt-PRL}. %\textcolor{blue}{It assumes that an infinitesimally small patch in the early Universe underwent a period of exponential expansion and becoming the currently observed Universe.} 
Originally, the inflation was proposed in the 1980s to solve the problems of hot big bang cosmology~\cite{1980-Starobinsky-PLB,1981-Guth-PRD}.
In this sense, it is always better to begin the discussion with the challenges faced by the hot big bang cosmology.
%Therefore, it is always better to start the discussion with the challenges faced by the hot big bang cosmology. 

In this section, we will first discuss two important problems of the hot big bang and then discuss how inflation solves the problems. We will also give a brief discussion on the generation of quantum fluctuations during inflation and cosmological perturbations.
\subsubsection{Flatness problem} 
From the first Friedmann equation~\eqref{eq:density-parameter-Omega}, we have 
\begin{equation}\label{eq:Omega-K}
  \Omega_K (t) \equiv \Omega (t) - 1 = \frac{K}{a^2 H^2}.
\end{equation} 
As we know, today, the value $\Omega_K (t)$ is of order unity (but less than one)\footnote{The observed Type Ia supernova redshift–distance relation and measurements of the ages of the oldest stars are consistent with a vanishing spatial curvature parameter i.e., $\Omega_K (t)=0$~\cite{Book-Weinberg-cosmology}}. Now, extrapolating backward in time and assuming that Einstein equations are valid until the Planck era (at time $t_{\rm Pl} \sim 10^{-43} \rm{s}$), we can calculate the value of $\Omega_K (t)$ at any epoch in the past. Let us estimate the following~\cite{2002-Riotto-InflationLectNotes,Book-Weinberg-cosmology}:
\begin{align}\label{eq:Omega-k-prob}
\frac{| \Omega (t) - 1 |_{t = t_{\rm Pl}}}{|\Omega (t) - 1|_{t = t_0}} %\approx \left( \frac{a_{\rm Pl}}{a_0} \right)^2
\approx \left( \frac{T_0}{T_{\rm Pl}} \right)^2 \approx \mathcal{O}(10^{-64})
\end{align}
where $T_{\rm Pl} = 10^{19} \rm{GeV}$ and $T_0 = 10^{-13} \rm{GeV}$ are the temperatures of the Universe at Planck time ($T_{\rm Pl}$) and today, respectively. This shows that to obtain the current value of $(\Omega_0 - 1)$, its value at the very early time should be extremely fine-tuned (close to zero but non-zero). This fine-tuning can not be explained within the framework of hot big bang cosmology. This is called the flatness problem.
\subsubsection{Horizon problem}
In cosmology, we can define different horizons (see, for instance, Ref.~\cite{Book-Mukhanov}):
\begin{enumerate}
\item[(I)] \emph{Particle horizon} If the universe has a finite age, then light travels only a finite distance in that time and the volume of space from which we can receive information at a given moment of time is limited. The boundary of this volume is called the particle horizon.
%If the Universe has a finite age, then the particle horizon is the boundary of the sphere of the radius of the maximum distance from which an observer can receive the signal. 
%
The particle horizon at any time $t$ is given by~\cite{2002-Riotto-InflationLectNotes}
\begin{align}\label{eq:particle_horizon}
d_p (t) = a(t) \int_{0}^{t} \frac{dt^{\prime}}{a (t^{\prime})} = a(t) \int_{0}^{a} d\ln a^{\prime}\frac{1}{a^{\prime} H(a^{\prime})} = \int_{0}^{r} \frac{dr^{\prime}}{\sqrt{1 - K {r^{\prime}}^2}} ,
\end{align}
where $t_i = 0$ is the time of begining of the Universe and $(a H)^{-1}$ is the comoving Hubble radius. The particle horizon scale is set by kinematical considerations.
\item[(II)] \emph{Event horizon} The event horizon is the complement of the particle horizon, this encloses the set of all the points from which signal sent at time $t$ will never be received by a future observer.
\item[(III)] \emph{Hubble horizon} It is the distance traveled by particles in the course of one expansion time. Hubble horizon (or curvature scale) is a dynamical scale that characterizes the expansion of the Universe.
\end{enumerate}
Among the above three kinds, the Hubble horizon and particle horizon are two different ways of determining the causal contact between the particles. For example, two particles separated by a distance larger than the physical particle horizon $d_p(t)$ have never communicated with each other. However, on the contrary, if they are separated by a distance greater than Hubble radius ($H^{-1}$) only means that they cannot communicate at a given time $t$. 
%
%\textcolor{red}{As we have mentioned in the section (\ref{sec:observ_landmarks}) that CMB gives the snapshot of the Universe at the redshift $z\sim 1100$ (the spatial hypersurface at this time is referred to as last-scattering surface).}
%

The horizon problem is related to \emph{causality}. To understand this, we compare the observable size of the Universe to the causal length scale at a given time. Let us consider the size of the Universe corresponding to our present Hubble radius ($H_0^{-1}$) at the last-scattering surface, which can be estimated as
\begin{align}\label{eq:horizon_problem-1}
\lambda (t_{\rm ls}) = d_p (t_0) \left( \frac{a(t_{\rm ls})}{a_0} \right) = d_p (t_0) \left( \frac{T_0}{T_{\rm ls}} \right)
\end{align}
where $T_0 = 2.73 \rm{K} = 10^{-13} \rm{GeV}$ and $T_{\rm ls} = 3000 \rm{K}$ are the temperatures of the Universe at present time $t_0$ and at the time of last-scattering surface $t_{\rm ls}$, respectively. In the matter-dominated era ($w = 0$), Hubble parameter changes as $H \propto T^{3/2}$~~\cite{2002-Riotto-InflationLectNotes,Book-Weinberg-cosmology}, therefore at last-scattering surface, we have
\begin{align}\label{eq:horizon_problem-2}
H_{\rm ls}^{-1} = H_0^{-1} \left( \frac{T_{\rm ls}}{T_0} \right)^{-3/2}  .
\end{align}
Let us compare the volume corresponding to both scales
\begin{align}\label{eq:horizon_problem-3}
\frac{\lambda^3 (t_{\rm ls})}{H_{\rm ls}^{-1}} = \left( \frac{T_{\rm ls}}{T_0} \right)^{-3/2} \approx 10^6  .
\end{align}
\begin{figure}[!ht]
\centering
\includegraphics[width=1\textwidth]{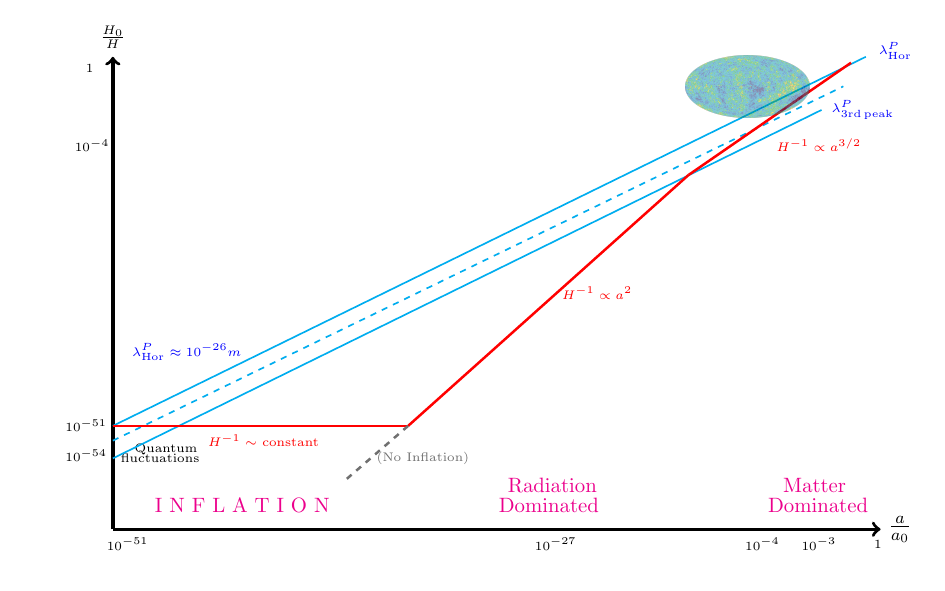} 
\caption{The evolution of horizon and perturbations in standard cosmology with
inflationary beginning and without inflation (gray dashed line). The vertical axis represents physical length scales and the horizontal axis represents the scale factor in units of current values of horizon $H_0^{-1}$ and scale factor $a_0$. The Blue lines show the physical scales which lead to CMB. }
\label{fig:inflation-figure}
\end{figure}
This shows that there were $10^6$ causally disconnected regions where the energy density was smoothly distributed with $\delta \rho/\rho \sim 10^{-4}$. Therefore, two photons that were causally disconnected (separated by a distance greater than $d_p(t_{\rm ls})$) the probability of them having the same temperature is minuscule. 
However, the data from several CMB experiments like COBE, WMAP and PLANCK reveal that the Universe was almost smooth at that time~\cite{1992-COBE-MNRAS,1992-COBE-ApJL,Planck-2018,2003-WMAP,Book-Weinberg-cosmology,Book-Dodelson.Schmidt_2020_edition} 
and all photons coming from all the directions in the sky have approximately the same temperature ($\Delta T/T\sim 10^{-5}$) (see Fig. \ref{fig:Planck_CMB-SDSS}).

Fig.(\ref{fig:inflation-figure}) depicts the Horizon problem, the red lines show the Hubble horizons and the blue lines are any given physical length (which scales $\propto a)$). Supposing the distance between both parallel solid blue lines denotes the distance between two photons we detect today. Since they are outside the Hubble radius (i.e., greater than $H^{-1}$) at recombination time, they could not have time to communicate. Thus, standard hot big bang cosmology can not provide any causal mechanism to explain the smooth Universe at the last scattering surface.

Cosmological inflation provides a causal mechanism to explain the homogeneity and isotropy of the University at the last scattering surface. Interestingly, the same mechanism can solve both --- flatness and horizon --- problems.
First, let us illustrate how a period of accelerated expansion solves the flatness problem. As we discuss in the next section, during inflation, Hubble parameter $H$ is approximately 
 constant. Setting $H \sim {\rm constant}$ in Eq.~\eqref{eq:Omega-K}, we have $(\Omega - 1) \propto a^{-2}$. Assuming that at the beginning of inflation $(\Omega_{I} - 1) \sim 1$, leads to $(\Omega_{\rm end} - 1) \sim e^{-2N}$ at the end of inflation, where $N = \ln (a_{\rm end}/a_I)$ is the number of e-foldings. 
From Eq.~\eqref{eq:Omega-k-prob}, we see that in order to have $(\Omega_0 - 1)$ of order unity today, we need $(\Omega - 1) \sim 10^{-60}$ at the beginning of radiation dominated era (i.e., end of inflation assuming instantaneous reheating). This can be achieved by taking $N = 70$. This implies that if the exponential expansion in the early Universe lasted for $70-$e foldings, then inflation solves the flatness problem.

Now, let us turn our attention to the horizon problem. During inflation, physical length scales $\lambda$ grow much faster (exponentially) than the Hubble radius $H^{-1}$, which remains approximately constant. As we can see from Fig.\ref{fig:inflation-figure} (gray dashed line), in standard hot big bang cosmology (i.e., without inflation), the scales cross the horizon only once. However, in inflationary cosmology, all scales begin sub-horizon, cross the horizon during inflation (become super-horizon) and re-enter the horizon during post-inflationary epoch. 
%\textcolor{red}{Then the physical scale $\lambda$, which are inside the horizon today (i.e., $\lambda < H^{-1}$), were outside for some time in the past, and for some period during inflation, they were again inside the Hubble radius $H^{-1}$ and 
Therefore, in inflationary cosmology, the two photons corresponding to two physical length scales (in Fig.\ref{fig:inflation-figure}) have had enough time to communicate with each other in the past. Hence, the CMB photons from causally disconnected regions in the sky have approximately the same temperature. Note that larger modes exit the Hubble radius early during inflation and re-enter the Hubble radius late in the radiation or matter-dominated era i.e. the modes follow \emph{first out, last in}!
\subsection{Inflationary mechanism}
\label{sec:inflation-mechanism}
Inflation is a stage of accelerated expansion of the Universe when the gravity acts as a repulsive force. To quantify how one can achieve the repulsive nature of gravity (i.e., accelerate the matter rather than accreting more matter into gravitating body), we consider the second Friedmann Eq.~\eqref{eq:friedmann-2} which tells us that for $\ddot{a} > 0$, we need $(\rho + 3p) < 0$ which corresponds to violation of the strong energy condition~\cite{Book-Mukhanov,Book-Weinberg-cosmology}. Furthermore, we can see that any matter field with $p = -\rho$ can give accelerated expansion as shown in Eq.\eqref{eq:a-for-diff-cases}. The condition for inflation is given by
\begin{align}
\ddot{a} > 0  \Longleftrightarrow \frac{d}{dt} \left( \frac{1}{a H} \right) < 0 \Longleftrightarrow -\frac{\dot{H}}{H^2} < 1
\end{align}
where second expression tells that the comoving Hubble radius $\mathcal{H} \equiv (a H)^{-1} $ decreases during inflation and third expression tells that the time variation of the Hubble parameter should be small. It is interesting to note that these conditions can be easily satisfied by a \emph{scalar field} which is also referred to as ``inflaton"~\cite{2002-Riotto-InflationLectNotes,Book-Gorbunov.Rubakov,Book-Kolb.Turner,Book-Mukhanov,Book-Weinberg-cosmology,Book-Dodelson.Schmidt_2020_edition,Book-Padmanabhan-III,Liddle.Lyth-Book}.
Consider the following action:
\begin{align}\label{eq:Full_action}
S = - \frac{M_{\rm Pl}^2}{2} \int d^4 x \sqrt{-g} R + \int d^4x \sqrt{-g} \left[ \frac{1}{2} \partial_{\mu}\varphi \partial^{\mu}\varphi - V(\varphi)  \right] .
\end{align}
where the first integral is the standard Einstein-Hilbert action which gives the LHS (gravity part) of the Einstein Equation~\eqref{eq:Einstein_Eq} and the second integral ($S_{\varphi}$) corresponds to the action of the scalar field (inflaton) which gives the RHS (matter part) of the Einstein Equation~\eqref{eq:Einstein_Eq}.
The energy-momentum tensor of the scalar field is given by:
\begin{align}
T_{\mu\nu} = -\frac{1}{\sqrt{-g}} \frac{ \delta S_{\varphi} }{\delta g^{\mu\nu}} = \partial_{\mu}\varphi \partial_{\nu}\varphi - g_{\mu\nu} \left( \partial_{\alpha}\varphi \partial^{\alpha}\varphi - V(\varphi)  \right) \, .
\end{align}
The energy density and pressure for a homogeneous scalar field in FRW background is
\begin{align}\label{eq:energy-pressure_for_phi}
\rho = \frac{1}{2} \dot{\overline{\varphi}}^2 + V(\overline{\varphi}), \qquad p =  \frac{1}{2} \dot{\overline{\varphi}}^2 - V(\overline{\varphi}) \, .
\end{align}
Note that, due to the symmetry of the FRW metric, the (classical) scalar field is homogeneous and hence the terms $(\partial_i \varphi)^2$ vanish.
From the equation of state parameter 
\begin{align}\label{eq:w_for_phi}
w = \frac{p}{\rho} = \frac{\frac{1}{2} \dot{\overline{\varphi}}^2 + V(\overline{\varphi})}{ \frac{1}{2} \dot{\overline{\varphi}}^2 - V(\overline{\varphi})} \, ,
\end{align}
we see that, if the kinetic energy of the scalar field is negligible compared to the potential energy, we have $w \simeq -1$ and Universe can have exponential expansion~\eqref{eq:a-for-diff-cases}. Therefore, for inflation to take place, the following condition must be satisfied
\begin{equation}
\label{eq:inflation-condition}
\dot{\overline{\varphi}}^2  << V(\overline{\varphi}) 
\end{equation}
When inflaton dominates the energy density of the Universe, the Friedmann equation~\eqref{eq:friedmann-1} reduces to
\begin{align}\label{eq:FriedmannEq-inflaton}
3 M_{\rm Pl}^2 H^2 = \frac{1}{2} \dot{\overline{\varphi}}^2 + V(\overline{\varphi}) \simeq V(\overline{\varphi}) \, .
\end{align}
%
%where we assume that inflaton dominates the energy density of the Universe and 
The equation of motion for the inflaton field can be derived by varying the action~\eqref{eq:Full_action} with respect to $\varphi$, which for the FRW metric~\eqref{eq:full-frw-metric} is given by
\begin{align}\label{eq:KG-eq}
\ddot{\overline{\varphi}} + 3 H \dot{\overline{\varphi}} + V_{\overline{\varphi}} = 0
\end{align}
where $V_{\overline{\varphi}} = \left.\partial_{\phi}V\right|_{\overline{\varphi}}$. Note that in the above equation, the term $3 H \dot{\overline{\varphi}}$ is the friction term which arises due to the expansion of the Universe.
Using \eqref{eq:inflation-condition} and \eqref{eq:KG-eq}, we obtain the slow-roll equation,
\begin{align}\label{eq:slow-roll-eq}
3 H \dot{\overline{\varphi}} \simeq - V_{\overline{\varphi}}.
\end{align}
From Eq.\eqref{eq:FriedmannEq-inflaton} and Eq.\eqref{eq:slow-roll-eq} and the condition that friction term dominates, $\ddot{\overline{\varphi}} << 3 H \dot{\overline{\varphi}}$, we obtain the slow-roll conditions as
\begin{align}\label{eq:sr-epsilon}
\epsilon = - \frac{\dot{H}}{H^2} = \frac{M_{\rm Pl}^2}{2} \left( \frac{V_{\overline{\varphi}}}{V} \right)^2 << 1
\end{align}
Similarly, the second slow-roll parameter can be obtained by using $\ddot{\overline{\varphi}} << |3 H \dot{\overline{\varphi}}|$ and Eq.\eqref{eq:slow-roll-eq}, which gives
\begin{align}\label{eq:sr-eta}
\eta % \equiv - \frac{\ddot{\varphi}}{H \dot{\varphi}} 
= M_{\rm Pl}^2  
\frac{V_{\overline{\varphi}\overline{\varphi}}}{V}  << 1 .
\end{align}
Slow-roll approximations are valid as long as both slow-roll parameters are small, $\epsilon,|\eta| << 1$. Note that $\eta$ need not be small for inflation to take place. Inflation occurs when $\epsilon <1$ regardless of value of $\eta$\cite{2009-Kinney-TASI}\footnote{Here, second slow-roll parameter $\eta$ should not be confused with conformal time, we use this notation to be consistent with the literature~\cite{2009-Kinney-TASI}.}.
\subsection{Cosmological perturbation theory: Gauge issues and Gauge-invariant quantities}
\label{subsec:perturbations}
Since Einstein's equations are non-linear, it is not possible to solve the equations for an arbitrary background. However, one can make progress by doing classical perturbation theory about a background metric. This procedure has been successfully applied during inflation and leads to one of the important predictions of inflation --- quantum fluctuations leading to the primordial perturbations~~\cite{1992-Mukhanov.etal-Phy.Rep.,2002-Riotto-InflationLectNotes,2009-Kinney-TASI,Book-Gorbunov.Rubakov,Book-Kolb.Turner,Book-Mukhanov,Book-Weinberg-cosmology, Book-Dodelson.Schmidt_2020_edition,Book-Padmanabhan-III,Liddle.Lyth-Book}. These primordial ``seed" perturbations grew over time, became classical and then collapsed due to gravitational instability to form the structures in the Universe we see today~\cite{Book-2007-Mukhanov.Winitzki-CUP,Book-Parker.Toms-CUP,1992-Mukhanov.etal-Phy.Rep.,2009-Kinney-TASI}. 
%
%\textcolor{red}{As discussed modify the text until Eq. 1.30.}

To understand the properties of the cosmological perturbations, let us expand the metric about the background FRW metric $g^{(0)}_{\mu\nu}$ given by~\cite{1992-Mukhanov.etal-Phy.Rep.,2009-Kinney-TASI}:
\begin{align}\label{eq:metpert}
g_{\mu\nu}(x^{\alpha}) = g^{(0)}_{\mu\nu}(t) + \delta g_{\mu\nu}(x^{\alpha})
\end{align}
where the background metric $g^{(0)}_{\mu\nu}$ depends only on time and the line element is given by:
\begin{align}\label{eq:FRW_metric-conformal}
    ds^2 = g^{(0)}_{\mu\nu} dx^{\mu}dx^{\nu} = dt^2 - a^2(t) d\mathbf{x}^2 =  a^2(\eta) \left( d\eta^2 - d\mathbf{x}^2 \right) ,
\end{align}
where $\eta = \int dt/a(t)$ is the conformal time, Hubble parameter in conformal time is 
\[\mathcal{H} \equiv \frac{a^{\prime}(\eta)}{a(\eta)} = H\, a(\eta)
\] 
and prime denotes the derivative with respect to $\eta$. As we will see, it is more convenient to use conformal time to study cosmological perturbations during inflation. 

The metric fluctuations $\delta g_{\mu\nu}$ depend on space and time. $\delta g_{\mu\nu}$, like the metric ($g_{\mu\nu}$) is symmetric, has 10 degrees of freedom. However, not all 10 components are independent. This is because the metric $g_{\mu\nu}$ is invariant under the coordinate transformation, $x^{\mu} \rightarrow x^{\mu} + \xi^{\mu}$ where $\xi^{\mu}$ is infinitesimal parameter~\cite{1992-Mukhanov.etal-Phys.Rept.}. Although the background metric ($g^{(0)}_{\mu\nu}$) is chosen based on physical arguments, this is not the case for the perturbed metric  
($\delta g_{\mu\nu}$). Hence, the metric split into background $g^{(0)}_{\mu\nu}$ and the perturbed 
$\delta g_{\mu\nu}$ part is not invariant under coordinate or gauge transformations. This is known as \emph{gauge problem}. There are two ways to remove the gauge redundancy: First is to choose a particular gauge (coordinate system)~\cite{1992-Mukhanov.etal-Phys.Rept.}. Second, is to construct gauge-invariant variables that are invariant under gauge transformations and can be compared to the observable quantities~\cite{1992-Mukhanov.etal-Phys.Rept.}.

At linear order, the metric perturbations ($\delta g_{\mu\nu}$) can be categorized into three distinct types --- \emph{scalar, vector} and \emph{tensor} perturbations.
This distinction is based on the transformation of the individual metric components with respect to the constant time 3D hypersurface of the background FRW spacetime. Thus the above metric perturbation can be written as:
\begin{equation}
\delta g_{\mu\nu} = \delta g_{\mu\nu}^{\rm (S)} +   \delta g_{\mu\nu}^{\rm (V)} + \delta g_{\mu\nu}^{\rm (T)}  
\end{equation}
The scalar metric perturbations can be written as
\begin{align}\label{eq:scalarpert_matrix}
\delta g_{\mu\nu}^{(S)} 
= a^2(\eta) \begin{pmatrix}
2\phi & -\partial_i B \\
-\partial_i B & 2 (\psi \delta_{ij} - \partial_{i}\partial_{j} E)   
\end{pmatrix}
\end{align}
where $\phi,B,E,\psi$ are four scalar degrees of freedom\footnote{Here we follow the notation of Ref.~\cite{1992-Mukhanov.etal-Phy.Rep.}, therefore $E,B$ should not be confused with electric and magnetic fields used in later chapters.}. Note that due to the gauge redundancy, we can choose $B = E = 0$, and the remaining two degrees of freedom correspond to the physical observables. This is referred to as \emph{longitudinal gauge}~\cite{1992-Mukhanov.etal-Phys.Rept.}. The \emph{scalar perturbations} couple of the density and pressure perturbations of the matter sector leading to density perturbations. Thus, the scalar perturbations are responsible for the temperature fluctuations of the CMB and 
the large-scale structures in the early Universe. 

The vector metric perturbations are:
\begin{align}\label{eq:vectorpert_matrix}
\delta g_{\mu\nu}^{\rm (V)} 
= a^2(\eta) \begin{pmatrix}
0 & -S_i \\
-S_i &  \partial_{i}F_j + \partial_{j} F_i 
\end{pmatrix} 
\end{align}
where $S_i$ and $F_i$ are two divergence-less vectors i.e, $\partial_i S^i = 0 = \partial_i F^i$, which give total 4 degrees of freedom. In this case also, one can remove the redundant degrees of freedom by choosing a gauge where $F_i = 0$~\cite{2004-Battefeld.Brandenberger-PRD}. The vector perturbations lead to vorticity and decay very fast in the expanding Universe; therefore, they are not considered as interesting to explore the Universe. However, they can grow in contracting Universe~\cite{2004-Battefeld.Brandenberger-PRD}. Also single-scalar field inflation can not seed rotational perturbations and hence, at linear order, can not generate vector perturbations. In Chapter~(\ref{ch:helical-PMF,ch:EFTmagnetogenesis}), we will discuss about the vector perturbations during inflation.

Lastly, the tensor metric fluctuations have two degrees of freedom which correspond to the two polarization states of the gravitational waves:
\begin{align}\label{eq:tensorpert_matrix}
\delta g_{\mu\nu}^{\rm (T)}  
= -a^2(\eta) \begin{pmatrix}
0 & 0 \\
0 &  h_{ij} 
\end{pmatrix} 
\end{align}
where $h_{ij}$ is traceless and divergenceless tensor $h_i^i = \partial_i h^{ij} = 0$. \emph{Tensor perturbations} do not couple at linear order to matter fluctuations. We will not discuss the evolution of metric perturbations in this thesis.

The matter perturbations also suffer from gauge problems. To see this, let us split the scalar field as background and perturbation\footnote{It is important to note that inflaton field $\varphi$ evolving on the potential $V(\varphi)$ described in Eq.~\eqref{eq:KG-eq} will not evolve completely classically but it will also be subjected to small quantum fluctuations about the classical trajectory which will in general be inhomogeneous.}: 
\begin{equation}
\varphi (x,\eta) = \overline{\varphi}(\eta) + \delta \varphi (\eta,\mathbf{x})  
\end{equation}
where $\overline{\varphi}(\eta)$ is the background inflaton field whose evolution is determined by Eq.\eqref{eq:KG-eq} and $\delta \varphi (\eta,\mathbf{x})$ is the fluctuations about the background field. The perturbed scalar field ($\delta \varphi (\eta,\mathbf{x})$), like the four scalar-metric perturbation $(\phi, B, \psi, E)$, suffer from the gauge problem. In other words, the value of $\delta{\varphi}$ depends on the 
choice of gauge or coordinate system. In principle, we can choose a gauge (constant scalar field gauge or comoving gauge) where $\delta \varphi$ is zero everywhere. 

However, we can construct gauge invariant variables by combining metric and matter perturbations. The gauge-invariant variables by definition are invariant under coordinate transformations and can be related to the observable quantities. In the context of scalar-field driven inflation, we can construct 
two useful gauge invariant variables that can be compared with CMB observations. These are:
\begin{equation}
Q \equiv \delta{\varphi} +  \frac{\dot{\overline{\varphi}}}{H} \psi 
\, ; \, 
\zeta = \psi + \frac{H}{\dot{\overline{\varphi}}} \delta \varphi \, .
\end{equation}
$Q$ is commonly referred to as Mukhanov-Sasaki variable~\cite{1984-Kodama.Sasaki-PTP,1992-Mukhanov.etal-Phys.Rept.,2002-Riotto-InflationLectNotes,2008-Malik.Wands-PhysRept,2009-Baumann-TASI,2010-Langlois-LectNotesPhys} while $\zeta$ is
the curvature perturbation~\cite{2002-Riotto-InflationLectNotes,2008-Malik.Wands-PhysRept,2009-Baumann-TASI,2010-Langlois-LectNotesPhys}. Note that in spatially flat gauge where $\psi = 0$, inflaton perturbation is represented by $Q$.
%
%\textcolor{red}{At very high energies, we expect that the matter will be described in terms of fields, for example, scalar fields.}
%
To study the evolution of true scalar degree of freedom, we consider the action that governs its dynamics. Note that the linearized equations of motion for the full system (gravity $+$ matter) described by action~\eqref{eq:Full_action} can be obtained from the expansion of the action~\eqref{eq:Full_action} at second order in perturbations, which in terms of the \emph{Mukhanov-Sasaki variable} $Q$ is given by (in cosmic time)~\cite{2010-Langlois-LectNotesPhys}: 
\begin{align}\label{eq:full_action-perturb}
\delta S^{(2)} = \frac{1}{2}\int dt d^3 x \, a^3(t) \left[ \dot{Q}^2 - \frac{1}{a^2(t)}\partial_i Q \partial^i Q -\mathcal{M} Q^2 \right]
\end{align}
where the effective mass term is 
\begin{align}\label{eq:effective_mass}
    \mathcal{M} = V_{\varphi\varphi} - \frac{1}{a^3} \frac{d}{dt} \left( \frac{a^3}{H}\dot{\overline{\varphi}}^2\right) .
\end{align}
In the conformal time and canonical variable $v = a Q$, we can write the action~\eqref{eq:full_action-perturb} in the canonical form:
\begin{align}\label{eq:full_action-perturb_canonical}
\delta S^{(2)} = \frac{1}{2}\int d\eta d^3 x\left[ {v^{\prime}}^2 - (\partial_i v)^2 +\frac{z^{\prime\prime}}{z} v^2 \right]
\end{align}
where $z = a\frac{\overline{\varphi}^{\prime}}{\mathcal{H}}$. It is interesting to note that the above action corresponds to the "an effective scalar field" with a varying mass. Thus, it is possible to quantize the effective scalar field using the standard field theory techniques and compare with observations. The quantization of the effective field provides a unique normalization constant~\cite{1992-Mukhanov.etal-Phy.Rep.}.

In the comoving gauge where $\delta\varphi=0$, we can obtain the relation between the canonical variable and comoving curvature perturbation:
\begin{align}\label{eq:Q-zeta-v_relation}
    v = a Q = a\frac{\overline{\varphi}^{\prime}}{\mathcal{H}} \zeta = z \zeta .
\end{align}
Varying the action~\eqref{eq:full_action-perturb_canonical} with respect to $v$ gives equation of motion, referred to as the \emph{Mukhanov-Sasaki equation}~\cite{1984-Kodama.Sasaki-PTP,1992-Mukhanov.etal-Phys.Rept.,2002-Riotto-InflationLectNotes,2008-Malik.Wands-PhysRept,2009-Baumann-TASI,2010-Langlois-LectNotesPhys}:
\begin{align}\label{eq:mukhanov-sasaki-eq}
    v^{\prime\prime}_{\textbf{k}} + \left( k^2 - \frac{z^{\prime\prime}}{z}\right)v_{\textbf{k}} = 0
\end{align}
where we have defined the Fourier modes as
\begin{align}\label{eq:fourier-v}
v_{\mathbf{k}} (\eta) \equiv \int d^3 \mathbf{x} \,  v(\eta,\mathbf{x}) e^{i\mathbf{k} \cdot \mathbf{x}} .
\end{align}
Since $z$ is depends on the background quantities, it is not possible to obtain the solution for arbitrary inflationary model. However, we can approximately evaluate the above equation \eqref{eq:mukhanov-sasaki-eq} in two regions. For modes with wavelengths much smaller than horizon, referred to as \emph{subhorizon modes} where $k^2 >> |z^{\prime\prime}/z | $, the solution is standard Bunch-Davis vacuum, and is given by
\begin{align}\label{eq:bunch-davis}
v_{\mathbf{k}} (\eta) = \frac{1}{\sqrt{2k}} e^{-ik\eta}.
\end{align}
We can see that fluctuations that are within the horizon oscillate exactly like in Minkowski spacetime. For the modes with wavelengths much larger than the horizon, referred to as \emph{superhorizon modes} $k^2 << |z^{\prime\prime}/z| $ the solution is given by~\cite{1984-Kodama.Sasaki-PTP,1992-Mukhanov.etal-Phys.Rept.,2002-Riotto-InflationLectNotes,2008-Malik.Wands-PhysRept,2009-Baumann-TASI,2010-Langlois-LectNotesPhys}
\begin{align}\label{eq:sup-mode1}
  v_{\mathbf{k}} (\eta) \propto z \quad  \implies \quad v_{\mathbf{k}} (\eta) = C(k) z
\end{align}
where $C(k)$ is the arbitrary constant which can be fixed by matching the solutions \eqref{eq:bunch-davis} and \eqref{eq:sup-mode1} at horizon exit, $k = a H$. During slow-roll, to simplify the computations, we use the relation $z^{\prime\prime}/z \simeq a^{\prime\prime}/a$ and $aH = -(1+\epsilon)/\eta$. Now, let us focus on Eq.\eqref{eq:sup-mode1} which tells that the comoving curvature perturbation freezes on superhorizon scales: $\zeta_{\mathbf{k}} = z^{-1}v_{\mathbf{k}} \propto constant$. After obtaining the evolution of the fluctuations, now let us quantize the scalar fluctuations described by canonical variable $v_{\mathbf{k}}(\eta)$ by promoting it to the quantum operator $\hat{v}_{\mathbf{k}}$ and it can be decomposed into mode functions $\hat{v}_k$ as:
\begin{align}
 \hat{v}_{\mathbf{k}}(\eta) =  a_{\mathbf{k}} \hat{v}_k (\eta) + a_{-\mathbf{k}}^{\dagger} \hat{v}^*_k (\eta),
\end{align}
where the mode functions depend on the absolute value of the comoving wavenumber $k$. 
%The equal-time commutation relation of $v$ and the canonically conjugate momentum $\hat{\Pi} = \hat{v}^{\prime}_k$ 
The creation and annihilation operators $a^{\dagger}_{-\mathbf{k}}$ and $a_{\mathbf{k}}$ satisfy the following equal-time commutation relations:
\begin{align}\label{eq:communation-rel}
\left[ a_{\mathbf{k}}, a_{\mathbf{k}^{\prime}}^{\dagger} \right] = (2\pi)^3\delta^3 (\mathbf{k} - \mathbf{k}^{\prime}), \quad \left[ a_{\mathbf{k}}, a_{\mathbf{k}^{\prime}} \right] = \left[ a_{\mathbf{k}}^{\dagger}, a_{\mathbf{k}^{\prime}}^{\dagger} \right] = 0 
\end{align}
and the vacuum state is defined as $a_{\mathbf{k}} |0 \rangle = 0$. As mentioned above, the above quantization provides a unique normalization~\cite{1992-Mukhanov.etal-Phy.Rep.}.
The vacuum two-point correlation function of the canonical variable $\hat{v}_{\mathbf{k}}$ gives the power spectrum as
\begin{align}
\langle 0 | \hat{v}_{\mathbf{k}} \hat{v}_{\mathbf{k}^{\prime}} | 0 \rangle \equiv P_v (k) \delta (\mathbf{k} + \mathbf{k}^{\prime})
\end{align}
where on the superhorizon scale, using Eq.\eqref{eq:sup-mode1} the power spectrum is given by 
\begin{align}\label{eq:power-v}
    P_v (k)  = \frac{(a H)^2}{2 k^3} .
\end{align}
Using Eq.\eqref{eq:Q-zeta-v_relation} and Eq.\eqref{eq:power-v}, we can calculate the power spectrum for the comoving curvature perturbation as $ P_{\zeta}(k) = P_v (k)/z^2$. In cosmology, we often use the dimensionless power spectrum which for the comoving curvature perturbation is given by~~\cite{2002-Riotto-InflationLectNotes,2008-Malik.Wands-PhysRept,2009-Baumann-TASI,2010-Langlois-LectNotesPhys}
\begin{align}\label{eq:zeta_powerspectrum}
\mathcal{P}_{\zeta}(k) = \frac{k^3}{2\pi^2} \frac{P_v (k)}{z^2}  \simeq \frac{1}{8\pi^2 M_{\rm Pl}^2}\frac{H_*^2}{\epsilon_*}.
\end{align}
where we have used $z^2 = 2 a^2 \epsilon$ for approximately de-Sitter spacetime, and the $*$ in the last term refers to the values evaluated at the horizon crossing where $k = aH$.
The power spectrum is a useful quantity to characterize the properties of the perturbation. Note that Eq.\eqref{eq:zeta_powerspectrum} tells that inflation predicts the almost scale-invariant power spectrum of the curvature perturbation. These scalar perturbations lead to density perturbations and explain the origin of the structures in the Universe.

As we have briefly mentioned that metric vector perturbations~\eqref{eq:vectorpert_matrix} decay very fast in the expanding Universe, hence are not considered as an important tools to study the early Universe. However, in the following chapters, we will discuss the generation and evolution of vector field perturbation and the associated electromagnetic fields during inflation.  %and its importance in solving the mystery of the origin of large-scale magnetic fields in the Universe.

\section{Open problems in cosmology}
\label{sec:open_problem}
Our understanding of the Universe is advancing through the meticulous analysis of observational data. These empirical observations have effectively ruled out several elegant theoretical ideas, for example, the steady-state theory. Nonetheless, certain puzzling observations lack an established and widely accepted theoretical framework, commonly referred to as ``open problems." Let us briefly list some of these open problems.

%
%Our understanding of the universe is becoming precise, thanks to observational data. These observations have ruled of few excellent theoretical ideas, for example steady state theory. However, there are still some puzzling observations for which there is no well-accepted theoretical framework. These are referred to as "open problems".
%For example, inflation is considered to be enormously successful to the solution of the hot big bang problem (see section~\ref{sec:inflation}). However, there are still more puzzling issues which need to be resolved, for example what is the nature of Dark Matter, why cosmological constant is so small, why there is more matter in the universe than antimatter, what is the origin of galactic magnetic fields, what are the sources of fast radio burst in the sky? 
%We have a convincing evidence to ask these questions due to the precise observational data. However, we do not have a consistent theoretical framework to answer these questions. Most of the proposal requires to go beyond the standard model. Let us first review a few unsolved problems and then in this thesis we will try to address them and propose some mechanism to resolve the problem.
%
\begin{enumerate}
\item \emph{What is Dark Energy?} 
Dark energy is a hypothetical form of energy that permeates all of space and is currently driving the Universe's accelerated expansion. It is distinct from both ordinary matter and dark matter as it is characterized by negative pressure, which means that it exerts a repulsive gravitational effect. This negative pressure is very counter-intuitive.
% since it implies that dark energy behaves differently from ordinary matter and radiation, which have positive pressure. 
Dark Energy accounts for about $68\%$ of the Universe's total energy density~\cite{2004-Sahni-LectNotesPhys,2003-Peebles.Ratra-RMP}. The nature of dark energy remains unknown, and it is one of the greatest mysteries in modern physics.
\item \emph{Why is cosmological constant so small?} Cosmological observations indicate that cosmological constant $\Lambda$ is positive with the magnitude $\Lambda (G\hbar/c^3) \approx 10^{-123}$. This number is 100-orders of magnitude smaller than theoretical estimates, i.e., the energy density of the vacuum~\cite{2003-Peebles.Ratra-RMP,2003-Padmanabhan-PhyRept,1989-Weinberg-RMP}.
\item \emph{What is the nature of Dark Matter?} 
Dark Matter is an invisible and elusive form of matter that does not interact with light or other forms of electromagnetic radiation. It does not emit, absorb, or reflect light, making it undetectable using traditional astronomical observations. The existence of dark matter is inferred through its gravitational effects on visible Matter and the Universe's structure. Dark Matter accounts for about $27\%$ of the Universe's total energy density~\cite{2004-Sahni-LectNotesPhys,2021-Arbey.Mahmoudi-PPNP}. The nature and the constituents of this elusive matter is still a mystery.
%One of the earliest pieces of evidence for dark matter came from studies of galaxy rotation curves. These curves describe how the orbital velocities of stars within a galaxy change as a function of their distance from the galactic center. Based on the visible matter alone, it was expected that the orbital velocities would decrease with increasing distance. However, observations showed that the velocities remained constant or even increased, suggesting the presence of additional matter that provides extra gravitational pull—dark matter.
%

\item \emph{Primordial lithium problem:} Big-bang nucleosynthesis (BBN), together with the precise WMAP estimate of cosmic baryon density, makes tight predictions for the abundances of the light elements. Deuterium and ${}^4\rm{He}$ measurements agree with theoretical predictions, but ${}^7\rm{Li}$ lie a factor $3-4$ below BBN+WMAP predictions. This $4-5\sigma$ mismatch is called the cosmic ``Lithium problem"~\cite{2008-Cyburt.Fields.etal-JCAP,2011-Fields-ARNPS}
\item \emph{What is the origin of large-scale magnetic fields?} A micro-Gauss strength magnetic field coherent on galactic scales has been observed in galaxies. Astrophysical mechanisms can not satisfactorily explain both properties, i.e., strength and coherence length. Therefore, the origin of these magnetic fields is still unclear. One of the aims of this thesis is to address this issue. Chapters~(\ref{ch:GalScalElect},\ref{ch:helical-PMF},\ref{ch:EFTmagnetogenesis}) provide novel mechanisms to generate magnetic field in the early universe, specifically during inflation.
\item \emph{Why is there more matter in the Universe than antimatter?}
The present Universe is observed to contain essentially only matter and no antimatter, except for the rare antiparticles produced by cosmic rays. From CMB and Big-bang nucleosynthesis (BBN), this asymmetry is of order $10^{-10}$. 
The theoretical mechanism to explain this asymmetry is still unknown.~\cite{1999-Trodden-RMP,1999-Riotto.Trodden-ARNPC}. One of the aims of this thesis is to address this issue. Chapter~\ref{ch:PMF_Baryo} provides a novel mechanism to explain the matter-antimatter symmetry.
\item \emph{What is the origin of fast radio bursts?}
Fast Radio Bursts (FRBs) are transient radio impulses lasting a few milliseconds, so rapid that it is not possible to precisely understand from which part of the sky they come from~\cite{2021-CHIME-FRB-arXiv,2022-Petroff.Lorimer-AAR}. What causes these extreme high-energy transient radio-bursts from distant galaxies, lasting only a few milliseconds, is poorly understood.  FRBs are interesting because they have the potential to provide a unique probe of physics in extreme conditions and of the distribution of matter in the Universe. We will discuss this in detail in chapter~\ref{ch:GZeffect}.
\end{enumerate}
%
%\textcolor{red}{The availability of precise, independent observational methods provides a shred of compelling evidence to pose these questions.} 
However, we currently lack a consistent theoretical framework to elucidate these questions. Most of the proposed solutions necessitate surpassing the constraints of the standard model of cosmology.
In this thesis, our primary objective is to comprehensively examine the last three unresolved issues listed above and introduce novel theoretical models for understanding and potentially resolving them. 
Specifically, the focus of this thesis is on unraveling the mechanisms responsible for the genesis of magnetic fields in the Universe and how these mechanisms can aid in addressing the other two unresolved issues. To gain insight into these mechanisms, a fundamental understanding of the characteristics of magnetic fields on relevant length scales is crucial. In the subsequent section, we will delve into the origin of magnetic fields and the constraints derived from observational data.
%\ref{ch:EFTmagnetogenesis}\ref{ch:GZeffect}\ref{ch:PMF_Baryo}\ref{ch:helical-PMF}

%===========S		E	C	T	I	O	N	=================
%
\section{Magnetic fields in the Universe}
\label{sec:mageticuniverse}
\begin{figure}[ht]
\centering
\includegraphics[height=3.4in]{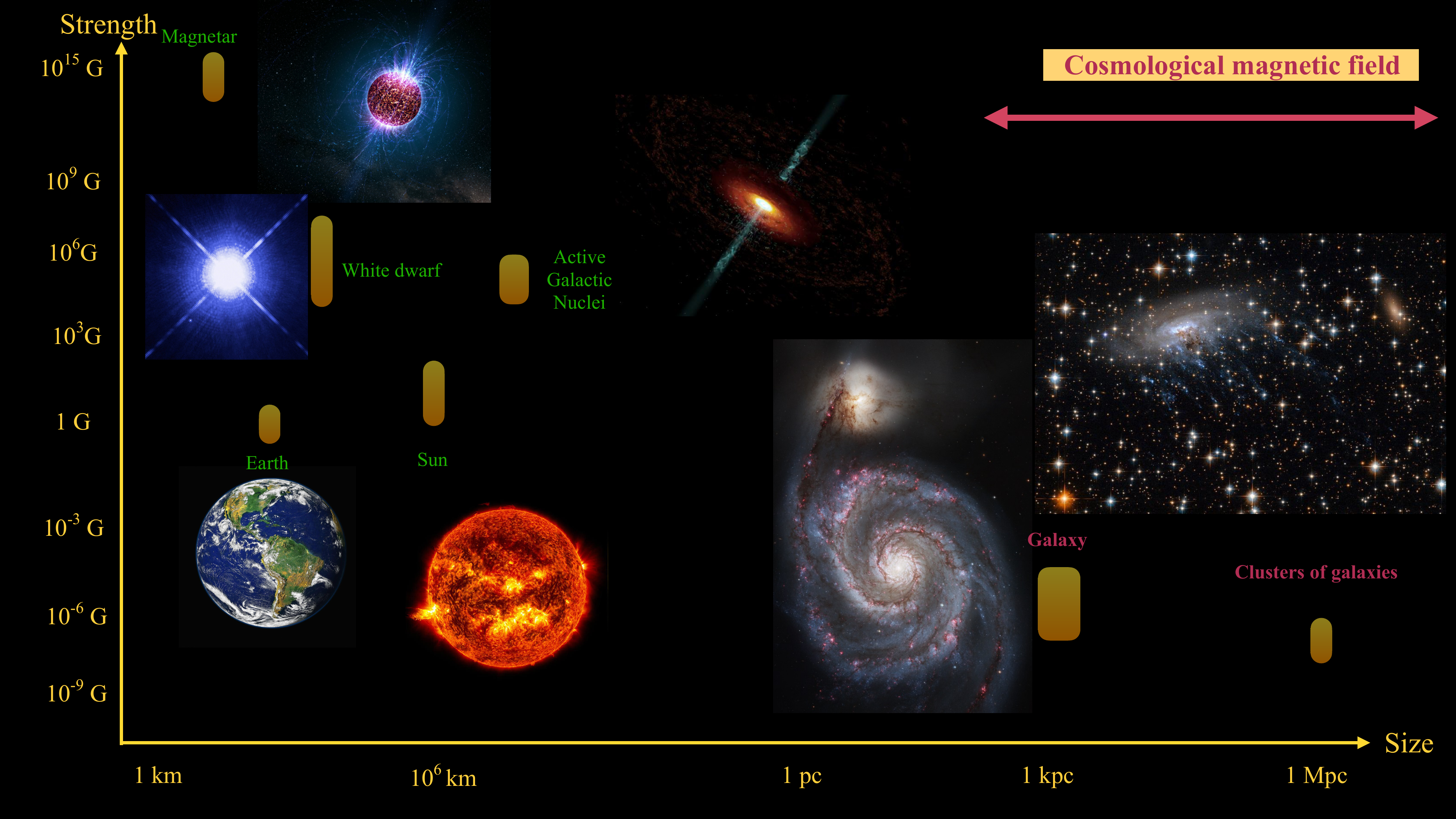}    %}%
\caption{Magnetic fields associated with some interesting astrophysical objects, vertical axis shows typical strength of the magnetic fields (height of the yellow bar) and horizontal axis tells the corresponding coherence length scales (width of the yellow bar). Note that the table is not to scale.}
%numbers are not properly scaled, they just tell the rough estimates.}
\label{fig:magneticuniverse}
\end{figure}
Magnetic fields are omnipresent in the Universe, existing across various length scales ranging from strongly magnetized magnetars to weakly magnetized regions such as Earth, Sun, galaxies, and clusters of galaxies. 
Fig.~\ref{fig:magneticuniverse} illustrates various astrophysical objects with typical magnetic field strengths and their coherence length scales. However, the most intriguing magnetic fields are those that permeate cosmological scales, spanning roughly of the order of Mpc, and are known as \emph{cosmological magnetic fields} or \emph{large-scale magnetic fields}~\cite{1994-Kronberg-Rept.Prog.Phys.,2001-Grasso.etal-PhyRep,2002-Widrow-Rev.Mod.Phys.,2008-Kulsrud.Zweibel-ReptProgPhys,2012-Widrow.etal-ScapeSciRev}. 
We have firm observational data to establish the presence of magnetic fields on large scales~\cite{1994-Kronberg-Rept.Prog.Phys.,2002-Widrow-Rev.Mod.Phys.}. For example, magnetic fields of micro-Gauss strength coherent over several kpc to Mpc scale is observed in galaxies and galaxy clusters. As discussed in the previous section, the origin of these magnetic fields is still unknown, and understanding the origin of the cosmological magnetic field is one of modern cosmology's most fascinating and challenging problems~\cite{1994-Kronberg-Rept.Prog.Phys.,2002-Widrow-Rev.Mod.Phys.,2008-Kulsrud.Zweibel-ReptProgPhys,2012-Widrow.etal-ScapeSciRev}.
% Due to high electrical conductivity of the Universe, the magnetic fields survived until today as a relic on astrophysical scales, after their generation during or after the big bang.
In order to understand these magnetic fields, it is important to ask the following questions: (i) How these regions are magnetized? (ii) how magnetic fields are amplified to the present value, (iii) what is the nature of these magnetic fields (iv) how galaxy formation and evolution is influenced by the existence of these magnetic fields, (v) how can they help to probe the physical phenomenon of associated scales~\cite{1994-Kronberg-Rept.Prog.Phys.}. To provide some answers to these questions, we need to understand their generation, evolution, and effects in the subsequent epochs. This is because the presence of magnetic fields in the Universe may have affected the physical processes. Hence looking at the effects provide a way to constrain the strength of the magnetic fields at a given epoch. In the next section, we will discuss various observational constraints on the magnetic fields at different epochs.
\subsection{Observational constraints on large scale magnetic fields}
There are several methods to detect the magnetic fields in the Universe.
%, their usefulness, and accuracy depend on the distance.
They can be broadly divided into \emph{local probes} and \emph{cosmological probes}. Local probes are limited to low redshift, for example, measurement of galactic and extra-galactic magnetic fields; and cosmological probes investigate the properties of the magnetic fields at very high redshifts, for example BBN and CMB. The main observational probes of galactic and extra-galactic magnetic fields are the Zeeman splitting of spectral lines, the intensity and polarization of synchrotron radiation from free electrons, and the Faraday rotation measurements of polarized electromagnetic radiation passing through the ionized medium~\cite{1994-Kronberg-Rept.Prog.Phys.,2002-Widrow-Rev.Mod.Phys.,2001-Grasso.etal-PhyRep}. These methods are efficient and accurate up to certain distances, for example, Faraday rotation is a very useful probe to measure the magnetic field on Mpc scales. In order to understand the origin of large-scale magnetic fields, it is crucial to determine their strength and coherence length at different redshifts or epochs. This not only reveals the evolutionary history of the magnetic fields but also constrains the values. We will discuss two probes in detail, and the constraints from various other probes are given in table~\ref{table:PMFtable}.
%Observational probes can be categorized as astrophysical, which are limited to low redshift (measurement of galactic and extra-galactic fields), and cosmological, where the field strength is estimated on very large redshifts much before the formation of structures, cosmic microwave background (CMB) and Big-Bang nucleosynthesis BBN.
%

%
\begin{table}[ht]
\centering
\begin{tabular}{|c|c|c|}
\hline 
\rule[-1ex]{0pt}{2.5ex} \textbf{Cosmological probes} & \textbf{Typical strength} & \textbf{References} \\ 
\hline 
\rule[-1ex]{0pt}{2.5ex} TeV Blazar & $ \gtrsim 3 \times 10^{-16} \rm{G}$ &  Neronov \& Vovk~\cite{2010-Neronov.Vovk-Sci}  \\ 
\hline 
\rule[-1ex]{0pt}{2.5ex} UFD & $ < 0.50 \pm 0.086 \, \rm{nG}$ & Safarzadeh \& Loeb~\cite{2019-Safarzadeh.Loeb-ApJ}   \\ 
\hline 
\rule[-1ex]{0pt}{2.5ex}  Pre-recombination & $ \leq 10^{-30} \rm{G}$ & Gopal and Sethi~\cite{2004-Gopal.Sethi-MNRAS}   \\ 
\hline 
\rule[-1ex]{0pt}{2.5ex} Reionization & $ < 2-4 \rm{nG}$ & \shortstack{Schleicher \& Miniati~\cite{2011-Schleicher.Miniati-MNRAS}, \\ 
Pandey et al~\cite{2015-Pandey.etal-MNRAS}}\\ 
\hline 
\rule[-1ex]{0pt}{2.5ex} CMB data* & $ < 1.3 - 6.4 \,\rm{nG}$  & Shaw \& Lewis~\cite{2012-Shaw.Lewis-PRD} \\ 
\hline 
%\rule[-1ex]{0pt}{2.5ex} CMB & \shortstack{$B_{\rm Mpc} < 4.4 \rm{nG}$ for non-helical, \\ $B_{\rm Mpc} < 5.6 \rm{nG}$ for helical} & Planck 2015~\cite{2015-Planck-PMF} \\ 
%\hline 
%\rule[-1ex]{0pt}{2.5ex} CMB & $\sqrt{\langle B^2 \rangle} < 0.83 \, \rm{nG}$ & Paoletti et.al (2019)  \\ 
%\hline 
\rule[-1ex]{0pt}{2.5ex} CMB Bispectrum & $ \sim 35 \,\rm{nG}$ & Seshadri \& Subramanian~\cite{2009-Seshadri.Subramanian-PRL} %\cite{2001-Seshadri.Subramanian-PRL} 
\\ 
\hline 
\rule[-1ex]{0pt}{2.5ex} Spectral distortion & $ < 30-40 \, \rm{nG}$  &  \shortstack{ Jedamzik et al~\cite{2000-Jedamzik.etal-PRL},\\
Kunze \& Komatsu \cite{2014-Kunze.Komatsu-JCAP} }  \\ 
\hline 
\rule[-1ex]{0pt}{2.5ex} Anisotropic expansion &    $ < 3.4 \, \rm{nG}$ & Barrow et al~\cite{1998-Barrow.etal-PRL}    \\ 
\hline 
\rule[-1ex]{0pt}{2.5ex} CMB & $ < 0.83 \, \rm{nG}$ & Paoletti et.al (2019)~\cite{2019-Paoletti.etal-MNRAS}  \\ 
\hline 
\rule[-1ex]{0pt}{2.5ex} CMB & $ < 4.4 \, \rm{nG}$ for non-helical & Planck 2015~\cite{2015-Planck-PMF} \\ 
\hline 
\rule[-1ex]{0pt}{2.5ex} CMB & $ < 5.6 \, \rm{nG}$ for helical & Planck 2015~\cite{2015-Planck-PMF} \\ 
\hline
%\rule[-1ex]{0pt}{2.5ex} CMB  &   $B_{\rm Mpc} < 5.6 \rm{nG}$ for helical  &   Planck 2015~\cite{2015-Planck-PMF}  \\ 
%\hline 
%\rule[-1ex]{0pt}{2.5ex}  &   &    \\ 
%\hline
%\rule[-1ex]{0pt}{2.5ex}  &   &    \\ 
%\hline
\end{tabular} 
\caption{Table showing constraints on magnetic fields from various observational probes. UFD refers to Ultra Faint Dwarf galaxies, $*$ denotes values from CMB data and including other cosmological data sets (see the given Refs).% \textbf{$B_0$ is also referred to as comoving magnetic field in some refs.}
}
\label{table:PMFtable}
\end{table}
\subsubsection{Constraints from CMB}
After decoupling at the last scattering surface ($z\sim 1100$), CMB photons free stream to the observer, observing these photons can give information about the physical conditions of the Universe. Therefore, CMB can be an important probe to detect the existence of the magnetic fields in the cosmic plasma in the early Universe.  
The presence of a sufficiently strong magnetic field in the early Universe will affect the evolution of cosmological perturbations and might induce significant additional signatures on the CMB. 
For example, due to the vector nature of the field, it has a preferred direction, which would lead to anisotropies (spatial) in the expansion of the Universe, which can further lead to anisotropies in the CMB~\cite{2008-Kahniashvili.etal.Ratra-PRD}. For a sufficiently strong magnetic field, the energy-momentum tensor perturbs the geometry of the Universe and introduces metric perturbations of three kinds---scalar, vector, and tensor~\cite{2004-Giovannini-IJMPD,2007-Barrow.etal-PhyRep,2013-Durrer.Neronov-Arxiv,2016-Subramanian-Arxiv}. These three modes affect the CMB differently, for example, scalar modes affect the shape of temperature anisotropy (TT mode) and also shifts the acoustic peaks via fast magnetosonic waves~\cite{1994-Kim.Olinto.Rosner-ApJ,2007-Kahniashvili.Ratra-PRD}. Vector modes induce $B$ mode polarization with amplitudes slightly greater than other contributions, which allows putting precise constraints~\cite{2002-Mack.Kahniashvili.etal-PRD}. The tensor modes induce gravitational wave perturbations that lead to CMB temperature and polarization anisotropies, which peak at large angular scales of the order of a degree~\cite{2001-Caprini.Durrer-PRD,2004-Caprini.Durrer.Kahniashvili-PRD}. All these effects put an upper limit on the strength of the magnetic field of the order of nano-Gauss on Mpc scales at the present epoch. Table~\ref{table:PMFtable}, shows CMB constraints on the magnetic field strength.
%where $\Omega_m$ is the ratio of matter energy density to the critical energy density, and $\delta T/T$ is the CMB temperature fluctuation with respect to the mean temperature $T$, and $z_d$ is the redshift at decoupling and $h = 0.7$. Shaw and Lewis \cite{2012-Shaw.Lewis-PRD} showed that in addition to CMB data if one includes  South Pole Telescope data (to constrain the Sunyaev-Zel’dovich effect) and Sloan Digital Sky Survey Lyman$-\alpha$ data the previous constraints ($ B < 4.1$ nG) on the lower limit of the magnetic field strength can be further lower down to $B\le 1.3$ nG
%
\subsubsection{Constraints from gamma-ray blazar observations}
Another way to probe the magnetic fields in intergalactic magnetic fields is by using \emph{blazars}, which are active galactic nuclei with relativistic jets oriented close to the line of sight with the observer, and they emit high-energy $\gamma$-rays in the TeV scale. These $\gamma$-rays, before reaching the observer, propagate through the intergalactic medium and interact with the low energy photons ($\sim 0.1 - 10 \, \rm{eV}$) from extragalactic background light. The interaction of $\gamma$-rays with low-energy photons produce electron-positron pairs. These highly-energetic charged particles interact with CMB photons, which fill up the entire Universe. Since the charged particles are energetic, these charged particles would transfer the energy to CMB photons through a process referred to as \emph{inverse Compton scattering}. Due to inverse Compton, the CMB photons can have energies in the GeV range. This process is called the \emph{electromagnetic inverse cascade}. While \emph{all} the TeV gamma rays emitted by blazars do not reach solar system, the inverse cascaded GeV gamma rays should reach us. Thus, we should observe both types of gamma rays. The HESS (High Energy Stereoscopic System) Telescope has observed the TeV scale energy emitted by blazars; however, GeV scale energy emissions have not been detected by the Fermi Gamma-ray Space Telescope for the same blazers~\cite{2010-Neronov.Vovk-Sci}. 
This leads to the following questions: What is the reason that the Fermi telescope does not detect GeV photons? Is it because these photons are not produced enough? Is there a missing physics? 

Neronov and Vovk argued that the presence of intergalactic magnetic field would bend the trajectory of the intermediate charged particles and hence reduce the GeV gamma rays flux~\cite{2010-Neronov.Vovk-Sci}.
%This inconsistency can only be explained if magnetic fields exist in these intergalactic regions or \emph{voids}.  
From the observational lower limit on the bending angle, authors derived the lower bounds on the strength of the magnetic field. This lower bound suggests the magnetic field strength $\gtrsim 3 \times 10^{-16} \rm{G}$ on Mpc scales in the voids. Note that the results derived in Ref.~\cite{2010-Neronov.Vovk-Sci} rely on the assumption that TeV sources emit gamma-rays continuously for $10^5$ years or longer. 

Cosmic voids contain very few or no galaxies, i.e., about a tenth of the average density of matter. If magnetic fields exist in the regions, they must have originated before the structure formation. Thus, Neronov and Vovk's analysis provides a tantalizing possibility that the magnetic fields on the largest possible scales might have originated in the very early Universe. 
%Lower bound of strength $B \geq 3\times 10^{-16}G$ on the magnetic field has been reported \cite{2010-Neronov.Vovk-Sci}, which is originated from the nonobservation of GeV gamma-ray emission from electromagnetic cascade initiated by tera-electron volt gamma-ray in intergalactic medium\~cite{2013-Durrer.Neronov-Arxiv}.
%\subsubsection{Difference between astrophysical and cosmological magnetic fields}

\subsubsection{Classification of magnetic fields based on helicity}
\label{subsec:classification}

As we discussed, magnetic fields are present in the Universe at almost all scales; astrophysical observations show that some galaxies might have a helical magnetic field structure~\cite{2002-Widrow-Rev.Mod.Phys.,2015-Planck-PMF}. 
Based on the helicity, primordial magnetic fields can be classified as \emph{non-helical} (with vanishing helicity density) and \emph{helical} (with non-zero helicity density). The 
%Because the conductivity of the Universe is very high in the early Universe, the magnetic flux is a conserved quantity. However, there is another interesting conserved quantity called 
\emph{magnetic helicity} is defined as the volume integral
\begin{align}\label{eq:helicitydens}
\mathcal{H}_M = \int d^3x \textbf{A} \cdot \textbf{B} ,
\end{align}
over a volume through the boundary of which no magnetic field lines cross, and $\textbf{A}$ is vector potential. Helicity is closely analogous to vorticity in fluid dynamics~\cite{2001-Grasso.etal-PhyRep,2013-Durrer.Neronov-Arxiv,2016-Subramanian-Arxiv,2020-Vachaspati-RepProgPhys}.
%Note that magnetic helicity $\mathcal{H}_M$ is zero for non-helical magnetic fields. 
In an infinitely conducting medium $\mathcal{H}_M$ is a conserved quantity. %\footnote{It is not possible to have magnetic fields which are homogeneous and carry magnetic helicity both because the magnetic helicity is determined by the condition $\langle \textbf{B} \cdot (\nabla \times \textbf{B}) \rangle$ (see chapter~\ref{ch:PMF_Baryo} for more details). Therefore, magnetic fields are often modeled as stochastic fields} 
Moreover, it is known from magnetohydrodynamics that the presence of magnetic helicity can amplify magnetic fields due to inverse cascade (transfer of power from small scale to large scale), which also increases the coherence length of the magnetic fields. It is interesting to note that \eqref{eq:helicitydens} breaks the parity symmetry (P), and consequently it affects the propagation of polarization of electromagnetic field differently. In other words, naively, the difference between helical and non-helical magnetic fields can be understood in the following way: the massless electromagnetic field has two transverse degrees of freedom, which can be thought of as two polarization, i.e., left-circular and right-circular. If both these polarization modes propagate differently, the electromagnetic field would be helical, and similar propagation would lead to the non-helical electromagnetic field.
%\textcolor{red}{The difference between helical and non-helical magnetic fields can be understood in the following way: the massless electromagnetic field has two transverse degrees of freedom, which can be thought of as two polarization, i.e., left-circular and right-circular. If both these polarization modes propagate differently, the electromagnetic field would be helical, and similar propagation would lead to the non-helical electromagnetic field.} 
%The difference between helical and non-helical magnetic fields can be understood in the following way: the massless electromagnetic field has two transverse degrees of freedom, which can be thought of as two polarization, i.e., left-circular and right-circular. If both these polarization modes propagate differently, the electromagnetic field would be helical, and similar propagation would lead to the non-helical electromagnetic field. 
%Helical magnetic fields have exciting properties which differ from the non-helical ones because the helical magnetic field changes the sign under the charge-parity (CP) symmetry, therefore detecting them provides evidence of CP violation in the early Universe~\cite{2001-Vachaspati-PRL}. 
During radiation and matter-dominated epochs, helical magnetic fields undergo an inverse cascade process that allows energy transfer from small to larger scales. Hence, the energy density and coherence length decay rate are slower for the helical magnetic fields than for the non-helical fields during these epochs~\cite{2001-Grasso.etal-PhyRep,2013-Durrer.Neronov-Arxiv,2016-Subramanian-Arxiv,2020-Vachaspati-RepProgPhys}. Furthermore, the presence of non-zero helicity would test possible modifications of Maxwell’s theory by constraining parameters describing the gauge invariance (i.e., mass
of the photon) and Lorentz invariance (i.e., existence of a preferred frame of reference)\cite{1990-Carroll.etal-PRD,2015-Planck-PMF}. CMB provides a way to detect magnetic helicity by studying the polarized power spectra. A non-zero helicity in the magnetic fields changes the amplitudes of the parity-even power spectra and induces parity-odd cross-correlations between the temperature and B-polarization anisotropies and E- and B-polarization anisotropies~\cite{2005-Kahniashvili.Ratra-PRD,2015-Planck-PMF}. Constraints on the helical magnetic fields from CMB are given in table~\ref{table:PMFtable} (last row).
\subsection{Theoretical frameworks to understand the origin of magnetic fields}
Although we have firm observational data to establish the presence of magnetic fields on large scales, the origin of magnetic fields in these regions is still unknown. %Therefore, understanding the origin of the cosmological magnetic field is one of the most fascinating and challenging problems in modern cosmology. 
It requires a consistent mechanism to generate a sufficiently strong magnetic field that could survive on large scales during its evolution from the time of generation to the present epoch. Any mechanism should explain two crucial properties of the magnetic field--- \emph{strength} and \emph{coherence length}~\cite{2001-Grasso.etal-PhyRep,2002-Widrow-Rev.Mod.Phys.,2004-Giovannini-IJMPD,2013-Durrer.Neronov-Arxiv,2016-Subramanian-Arxiv,2020-Vachaspati-RepProgPhys}. Various theoretical models have been proposed to obtain the observed field strength on large-coherence length. However, in most of these models either the generated magnetic field is found to be too small, or the large coherence length is not achieved. 
%Therefore, any magnetogenesis scenario must address these two issues --- explaining the strong magnetic field and large coherence length. 
%
%While there is a large amount of observational evidence for the presence of large-scale magnetic fields, the fundamental origin of these fields is still unclear. 

The well-accepted paradigm suggests that these large-scale magnetic fields are produced by the amplification of pre-existing tiny magnetic field--- \emph{seed} magnetic field. This \emph{seed} magnetic field has to be generated by a different mechanism that operated before the structure formation. Since this seed magnetic field could be generated from some cosmological process before structure formation, they are of primordial origin, hence often referred to as \emph{primordial magnetic fields}. Moreover, the generation mechanism is referred to as \emph{primordial magnetogenesis}~\cite{2001-Grasso.etal-PhyRep,2002-Widrow-Rev.Mod.Phys.,2004-Giovannini-IJMPD,2013-Durrer.Neronov-Arxiv,2016-Subramanian-Arxiv,2020-Vachaspati-RepProgPhys}. Since magnetogenesis could take place at any epoch before the structure formation, it is convenient to classify them in two broad categories like the cosmological models which allow us to understand the evolution history of the Universe.

Magnetogenesis mechanisms can be broadly classified as---\emph{early time} and \emph{late time} mechanisms. \emph{Early time} mechanisms provide a way of generating the primordial seed magnetic field in the early Universe, for example, during inflation and reheating, electroweak and QCD phase transitions, and at the time or before recombination~\cite{2001-Grasso.etal-PhyRep,2002-Widrow-Rev.Mod.Phys.,2004-Giovannini-IJMPD,2013-Durrer.Neronov-Arxiv,2016-Fabre.Shankaranarayanan-ApP,2016-Subramanian-Arxiv,2020-Vachaspati-RepProgPhys}. On the other hand, the \emph{late time} mechanisms provide the a way of generating the magnetic field during structure formation. 
The issue with the late time models is that they can not generate a magnetic field over a large coherence scale needed for the dynamo mechanism to amplify the seed field. This is because the coherence length of the magnetic field cannot exceed the horizon size (see section~\ref{sec:inflation}) at the time of generation~\cite{2001-Grasso.etal-PhyRep,2002-Widrow-Rev.Mod.Phys.,2004-Giovannini-IJMPD,2013-Durrer.Neronov-Arxiv,2015-Tsagas-PRD,2016-Subramanian-Arxiv,2020-Vachaspati-RepProgPhys}. In late time models, one assumes that the large scale magnetic fields are produced and maintained by dynamo mechanism, in which the seed magnetic field is amplified exponentially to the current value$ \sim B_{\rm gf} \, e^{\Gamma(t_0 - t_{\rm gf})}$, where $\Gamma $ is the amplification rate which is a model-dependent quantity, $B_{\rm gf}$ is the field strength at the time of galaxy formation ($t_{\rm gf}$) and $t_0$ refers to the present epoch. The value of maximum amplification factor between redshift $z=0$ to $z=10$, is around $10^{14}$~\cite{2001-Grasso.etal-PhyRep,2002-Widrow-Rev.Mod.Phys.,2004-Giovannini-IJMPD,2013-Durrer.Neronov-Arxiv}. Therefore, the observed value of the magnetic field at present (micro-Gauss), requires a seed magnetic field of the order of $10^{-20} G$ strength and a coherence length of 100pc, at the time of galaxy formation~\cite{2001-Grasso.etal-PhyRep,2002-Widrow-Rev.Mod.Phys.,2004-Giovannini-IJMPD,2013-Durrer.Neronov-Arxiv}. 

The electrical conductivity plays an important role in the magnetogenesis scenario, because if the conductivity is very large then electric field vanishes and magnetic field is frozen. Therefore, it becomes difficult to generate/amplify the magnetic field if conductivity is high. Because the conductivity of the universe was very high after reheating and was negligible during inflation, it is a good reason to consider the magnetogenesis during inflation or reheating.
%The conductivity of the Universe during most of the epochs after inflation was very high. 
Due to the flux freezing in the conducting medium, the magnetic flux is conserved, which implies $B\propto a^{-2}$ as the area of the region increases as the square of the scale factor. 
%Hence, it is very difficult to generate magnetic field during this epoch as it will decay very fastly.
%
%Assuming this adiabatic evolution of the magnetic field, the fact that the coherence length scale varies with the scale factor $a(t)$ as $l_B \propto a(t)$, the evolution of the magnetic field at any epoch is given by the relation
%\begin{align}\label{Bz}
%\textcolor{red}{B\left( (1+z)^{-1}l_B(z), t \right) = (1+z)^2 B(l_{B_{0}}, t_0)   }
%\end{align}
%
%where $l_B$ is the coherence scale of the magnetic field, $t_0$ refers to the current epoch, $B_0$ refers to current value of magnetic field and $z$ is the redshift. Eq. (\ref{Bz}) shows the relation between magnetic field strength at the current epoch and redshift $z$.
%
In this thesis, we will focus on the magnetogenesis during inflation.

%Using the observational data, we can obtain the magnetic energy density corresponding to the observed value at present, which is 
%%
%$$\rho_{B} = \frac{B^2}{8\pi} \sim 10^{-16}-10^{-12}\; erg\;cm^{-3} \, . $$ 
%%
%The CMB photon energy density is $\rho_{\gamma} = \sigma T^4  \sim 10^{-13} \rm{erg \, cm^{-3}}$, where $\sigma$ is Stefan-Boltzmann constant. Thus, we draw an important result: At present epoch, magnetic energy density is approximately equal to the photon energy density.
%
%======== S E C T I O N =========================
%
%
\section{Electrodynamics in curved spacetime}
\label{sec:EM-theory-curved}
In this section we discuss the properties of electromagnetic field in curved spacetime. We derive the Maxwell's equations in flat FRW spacetime. Let us begin with the action for the electromagnetic fields in generic 4D curved spacetime, given by~\cite{1975-Landau.Lifshitz-book,2016-Subramanian-Arxiv}
\begin{align}\label{eq:EM-TotalAction}
\mathcal{S}_{\rm EM} = -\frac{1}{4}  \int d^4x \, \sqrt{-g} \, F_{\mu\nu} F^{\mu\nu}  - \int d^4x \, \sqrt{-g} A_{\mu} J^{\mu} 
\end{align}
where $F_{\mu\nu} = \nabla_{\mu} A_{\nu} - \nabla_{\nu} A_{\mu} = \partial_{\mu} A_{\nu} - \partial_{\nu} A_{\mu}$ is the electromagnetic (EM) field tensor (also known as Faraday tensor), $A^{\mu}$ is the electromagnetic four vector, and $J^{\mu}$ is the four current density, $A_{\mu}$ is the electromagnetic four vector. In the action~\eqref{eq:EM-TotalAction}, the first term is the kinetic term and the second term is the interaction of the electromagnetic field with other particles.

Variation of the action (\ref{eq:EM-TotalAction}) with respect to $A_{\mu}$ gives the first pair of the Maxwell's equations

\begin{align}\label{eq:covariant-maxwellEq-1st}
\partial_{\nu} \left( \sqrt{-g} F^{\mu\nu} \right) = J^{\mu}
\end{align}
and the second pair of Maxwell's equations is given by the Bianchi identity, 
\begin{align}\label{eq:covariant-maxwellEq-2nd}
 \hspace{1cm} \nabla_{\gamma} F_{\mu\nu} + \nabla_{\nu} F_{\gamma\mu} + \nabla_{\mu} F_{\nu\gamma} = 0, \qquad \text{or} \qquad \partial_{\mu} ( \sqrt{-g}\tilde{F}^{\mu\nu} ) = 0.
\end{align}
\\
where $\tilde{F}^{\mu\nu} = \epsilon^{\mu\nu\alpha\beta} F_{\alpha\beta}/2$ is the dual electromagnetic field tensor and $\epsilon^{\mu\nu\alpha\beta} = \eta^{\mu\nu\alpha\beta}/\sqrt{-g}$ is the totally antisymmetric Levi-Civita tensor and $\eta^{\mu\nu\alpha\beta}$ is totally antisymmetric symbol which we set $\eta^{0123} = 1 = - \eta_{0123}$. Note that in flat spacetime the electric field and magnetic field can be written as time-space and space-space components of the antisymmetric electromagnetic tensor $F_{\mu\nu}$, and due to the antisymmetry, the diagonal components of $F_{\mu\nu}$ are zero and total number of independent components are 6 (three for electric field and three for magnetic field). The electromagnetic tensor $F_{\mu\nu}$ in Minkowski spacetime can be written in the matrix form ($\mu = 0,1,2,3$ labels the rows and index $\nu$ the columns) as~\cite{1975-Landau.Lifshitz-book}

\begin{align}\label{eq:EM-tensor-dd_matrix}
F_{\mu\nu}  
= \begin{pmatrix}
0 & E_x & E_y & E_z \\
-E_x & 0 & -B_z & B_y \\
-E_y & B_z & 0 & -B_x\\
-E_z & - B_y & B_x & 0  
\end{pmatrix}  , 
\qquad 
F^{\mu\nu}
= \begin{pmatrix}
0 & -E_x & -E_y & -E_z \\
E_x & 0 & -B_z & B_y \\
E_y & B_z & 0 & -B_x\\
E_z & - B_y & B_x & 0  
\end{pmatrix} 
\end{align}
It is important to note that electromagnetic four-vector potential $A^{\mu}$ gives the covariant description of the EM phenomenon. But, in order to study the observable effacts, we need to decompose it in terms of electric and magnetic fields which are intrinsically frame dependent quantities. We define the electric and magnetic field with respect to a family of the comoving observer with velocity $u^{\mu} \equiv dx^{\mu}/ds$, where $s$ is the associated proper time~\cite{2013-Durrer.Neronov-Arxiv,2015-Tsagas-PRD,2016-Subramanian-Arxiv}. The electric and magnetic fields are given by
%
%\begin{align}\label{eq:u_mu}
%u^{\mu} = \frac{dx^{\mu}}{ds} , \hspace{1cm} u_{\mu} u^{\mu} = 1
%\end{align}
%
\begin{align}\label{eq:Emu-Bmu}
E_{\mu} = F_{\mu\alpha} u^{\alpha}, \quad 
B_{\mu} = \frac{1}{2} \, \epsilon_{\mu\alpha\beta\gamma} F^{\alpha\beta} u^{\gamma} = \tilde{F}_{\mu\alpha} u^{\alpha}
\end{align}
where $u^{\mu} = (1/a(\eta),0,0,0)$ and $u_{\mu} = (a(\eta),0,0,0)$ in the conformal time.

The electric and magnetic field 4-vectors in Eq.~\eqref{eq:Emu-Bmu} are essentially 3-vector fields in the sense orthogonal to the comoving observer i.e., $E_{\mu} u^{\mu} = 0 = B_{\mu} u^{\mu}$. % hence their time components ($0^{th}$ components) are zero. 
%It is important to note that electric and magnetic four vectors defined above have purely spatial components and hence are spatial three vectors.
Inverting the relation \eqref{eq:Emu-Bmu} gives the electromagnetic field tensor and its dual as~\cite{2013-Durrer.Neronov-Arxiv,2016-Subramanian-Arxiv}
\begin{subequations}
\begin{equation}\label{eq:EMT}
 F_{\mu\nu} = u_{\mu} E_{\nu} - u_{\nu} E_{\mu} + \epsilon_{\mu\nu\alpha\beta} B^{\alpha} u^{\beta}
\end{equation}    
\begin{equation}\label{eq:EMT_dual}
\tilde{F}^{\alpha\beta} = \frac{1}{2}\epsilon^{\alpha\beta\mu\nu} F_{\mu\nu} =   \epsilon^{\alpha\beta\mu\nu} u_{\mu} E_{\nu} +(u^{\alpha} B^{\beta} - u^{\beta} B^{\alpha}  ) 
\end{equation}
\end{subequations}
Using the FRW metric~\eqref{eq:FRW_metric-conformal} and \eqref{eq:Emu-Bmu}, we obtain the electric and magnetic fields as~\cite{2013-Durrer.Neronov-Arxiv,2016-Subramanian-Arxiv}
%
%\subsection{Electrodynamics in expanding universe}
\begin{align}
E_i &= F_{i 0} u^0 = -\frac{1}{a(\eta)} {A^{\prime}}_i , \,\,\,\, E_0 = 0 
\quad \text{and} \quad
B_i = - \frac{1}{2 a(\eta)} \eta_{0 i j k} F^{j k} = \frac{1}{a(\eta)} \hat{\epsilon}_{ijk} \partial_j A_k
\end{align}
where we have used $ \hat{\epsilon}_{ijk} = - \eta_{0ijk}  $ is 3D Levi-Civita symbol with $\hat{\epsilon}_{123} = 1$ is the .
Therefore, the electric and magnetic four vectors defined in Eq.\eqref{eq:Emu-Bmu} can be written as
\begin{align}\label{eq:Emu-Bmu-2}
E_{\mu} = \left( 0, -\frac{1}{a(\eta)} {A^{\prime}}_i   \right)  = a(\eta) \, \left( 0, \textbf{E} \right) \quad \text{and} \quad
B_{\mu}  = \left(  0, \frac{1}{a(\eta)} \, \hat{\epsilon}_{ijk}  \, \partial_j A_k   \right) =  a(\eta) \, \left( 0, \textbf{B} \right) 
\end{align}
where $ \textbf{E} = -  \frac{{A^{\prime}}_i }{a^2(\eta)}$ and $ \textbf{B} = \frac{1 }{a^2(\eta)}\,  \hat{\epsilon}_{ijk}  \, \partial_j A_k $. It is important to note that $ \textbf{E} $ and $\textbf{B} $ scale as $a^{-2}(\eta)$. Now using Eq.\eqref{eq:covariant-maxwellEq-1st} and \eqref{eq:covariant-maxwellEq-2nd} with the relations \eqref{eq:Emu-Bmu-2} lead to the four Maxwell equations~\cite{2013-Durrer.Neronov-Arxiv,2016-Subramanian-Arxiv}:
%
%\begin{align}\label{eq:MW_equations}
%\nabla \cdot \mathbf{E} = a \rho_e  , \qquad - \partial_{\eta} \left( a^2 \mathbf{E} \right) +a^2  \nabla \times \mathbf{B} = a^3 \mathbf{J},
%\\
%\partial_{\eta} \left( a^2 \mathbf{B} \right) +a^2  \nabla \times \mathbf{E} = 0, \qquad  \nabla \cdot \mathbf{B} = 0
%\end{align}
%
\begin{subequations}\label{eq:MW_equations}
\begin{equation}
 \nabla \cdot \mathbf{E} = a \rho_e  , \qquad  - \partial_{\eta} \left( a^2(\eta) \mathbf{E} \right) +a^2(\eta)  \nabla \times \mathbf{B} = a^3 \mathbf{J},
\end{equation}    
\begin{equation}
 \nabla \cdot \mathbf{B} = 0, \qquad  \partial_{\eta} \left( a^2(\eta) \mathbf{B} \right) +a^2(\eta)  \nabla \times \mathbf{E} = 0. 
\end{equation}
\end{subequations}
where $\rho_e$ and $\mathbf{J}$ are the electric charge and current densities. It is important to note that $ \textbf{E} $ and $\textbf{B} $ scale as $a^{-2}(\eta)$. Thus, in terms of the rescaled quantities $ a^2(\eta) \textbf{E}, a^2(\eta) \textbf{B}$ and $a^3(\eta)  \textbf{J}$, the Maxwell equations in Eq.\eqref{eq:MW_equations} would have essentially the same form as in the Minkowski space, and the scale factor can be `scaled out'. 
This decay behaviour of electric and magnetic fields with $a(\eta)$ occurs due to the conformal invariance of standard electromagnetic action. We will discuss more about this issue in the next section.
%
%===========S		E	C	T	I	O	N	=================
%
\section{Classification of inflationary Magnetogenesis models}\label{sec:inflationary_magneto_classification}

As discussed in Sec.~(\ref{sec:inflation}), inflation provides a causal mechanism and generate growing scalar and tensor fluctuations on very large length scales. However, this is not the case for electromagnetic field. 
%in the rapidly expanding Universe, they decay very fast due to the conformal invariance of the electromagnetic action. 
To understand this, let us consider the action of free electromagnetic field in generic 4D curved spacetime,
\begin{align}\label{eq:standardEM_action}
S_{\rm EM} = -\frac{1}{4}\int d^4x\;\sqrt{-g}\;F_{\mu\nu}F^{\mu\nu} = -\frac{1}{4}\int d^4x\;\sqrt{-g}\;g^{\mu\alpha}g^{\nu\beta}\;F_{\mu\nu}F_{\alpha\beta} ,
\end{align}
%where $F_{\mu\nu} = \partial_{\mu}A_{\nu} - \partial_{\nu}A_{\mu}$ is electromagnetic field tensor and $A^{\mu}$ is electromagnetic four-vector. 
and the conformal transformation, which is defined as $\tilde{g}_{\mu\nu} = \Omega^{2}(x) g_{\mu\nu}$, where conformal factor $\Omega (x)$ is a spacetime dependent quantity\footnote{Note that in this section $\Omega(x)$ should not be confused with density parameter defined in the previous sections. Here we use the standard notation for the conformal factor, which is denoted by $\Omega(x)$~(see Ref.\cite{Book-General-Relativity-Wald})}. Under the conformal transformation, we have $\sqrt{-g} \rightarrow \sqrt{-\tilde{g}} = \Omega^4 \sqrt{-g}, g^{\mu\nu} \rightarrow \tilde{g}^{\mu\nu} = \Omega^{-2} g^{\mu\nu}$, and $F_{\mu\nu} \rightarrow \tilde{F}_{\mu\nu} = F_{\mu\nu}$. Thus, using these relations in Eq.\eqref{eq:standardEM_action} gives:
\begin{align}
S_{\rm EM} \longrightarrow \tilde{S}_{\rm EM} = -\frac{1}{4}\int d^4x \sqrt{-\tilde{g}} \tilde{g}^{\mu\alpha}\tilde{g}^{\nu\beta} \tilde{F}_{\mu\nu}\tilde{F}_{\alpha\beta}
 =-\frac{1}{4}\int d^4x \sqrt{-g} F_{\mu\nu}F^{\mu\nu} = S_{\rm EM} .
\end{align} 
Hence, electromagnetic action~\eqref{eq:standardEM_action} is invariant under the conformal transformation. %, and so is the equation of motion of the electromagnetic fields. 
It is important to note that this invariance is specific to ${\rm D} = 4$, and we can write the FRW metric as $ds^2_{\rm{FRW}} = a^2(\eta)ds^2_{\rm{Minikowski}}$ where the scale factor $a(\eta)$ can be identified as a conformal factor.
%
%if we make the transformation $dt = a(t)d\eta$ where $t$ and $\eta$ are cosmic time and conformal time respectively, i.e.
%\begin{align}\label{eq:FRW_metric}
%ds^2 = dt^2 -a^2(t)(dx^2 + dy^2 + dz^2)= a^2(\eta)(d\eta^2 - dx^2 -dy^2 -dz^2)
%\end{align}
%
This tells us that the electromagnetic field in the FRW background is identical to that in the Minkowski spacetime. Therefore, equation of motion in FRW background will be identical to those in Minkowski spacetime with suitable rescaling by $a(\eta)$, as we discussed in the previous section~(see Eq.~\eqref{eq:MW_equations}).
So, in the FRW Universe magnetic field will decay as $B \sim a^{-2}$ (due to magnetic flux conservation). 

Thus, any seed magnetic field ($B_i$) generated at the beginning of inflation will decay to a value $\sim e^{-120} B_i$ at the end of the inflation, if we assume inflation lasts for 60 e-foldings after the generation of magnetic field. So, it is impossible to get a residual magnetic field of significant strength with standard electromagnetic action~\eqref{eq:standardEM_action}. Therefore, the generation of sufficiently strong magnetic field during inflation requires breaking the conformal invariance of the electromagnetic action~\eqref{eq:standardEM_action}. As we know that geometrical quantities like Christoffel symbol $\Gamma^{\lambda}_{\mu\nu}$, Riemann tensor $R_{\mu\nu\alpha\beta}$, Ricci tensor $R_{\mu\nu}$, and Ricci scalar $R$, or higher derivative terms in the action break the conformal invariance explicitly (see appendix~\ref{CT} and the corresponding discussion in Chapter~\ref{ch:GalScalElect})~\cite{Book-General-Relativity-Carroll,Book-General-Relativity-Weinberg,Book-General-Relativity-Wald,Book-Gravitation_MTW}.
Therefore, coupling electromagnetic field with these terms break the conformal invariance of the action and consequently will change the decaying behaviour of the electromagnetic fields in the FRW spacetime as long as the modification due to conformal breaking terms is significant. 
In a 1988 seminal paper, Turner and Widrow proposed several ways to break the conformal invariance of the electromagnetic action and generation of magnetic field~\cite{1988-Turner.Widrow-PRD}. Few years later, various other exciting mechanisms have been proposed~\cite{1991-Ratra-Apj.Lett,1991-Vachaspati-PLB,1993-Dolgov-PRD}. %This has been first proposed by Turner and Widrow in a seminal paper~\cite{1988-Turner.Widrow-PRD}, later various other interesting ways are proposed~\cite{1991-Ratra-Apj.Lett,1991-Vachaspati-PLB,1993-Dolgov-PRD}. 
Various inflationary magnetogenesis models in the literature can be broadly classified 
into the following three categories:
%First, Let us briefly list various ways to break the conformal invariance of the action discussed in the literature\footnote{We would like to emphasise that this list only contains the broad class of possible inflationary magnetogenesis scenarios.}. 
%
\begin{enumerate}
\item Introducing non-minimal coupling of the electromagnetic field with the curvature by adding terms of the forms: $R_{\mu\nu\alpha\beta} F^{\mu\nu} F^{\alpha\beta}, R_{\mu\nu} F^{\mu\alpha} F^{\nu}\,_{\alpha}, R F_{\mu\nu} F^{\mu\nu}$ and $R^{\mu\nu}A_{\mu}A_{\nu}, R A_{\mu} A^{\mu}$. Note that the last two terms break the gauge invariance of the electromagnetic field due to the mass term of the photon and leading to ghost instability~\cite{2009-Himmetoglu.Contaldi.Peloso-PRD}.
\item Modification of the standard electromagnetic action with a time-dependent coupling to another matter field. 
The most common way of achieving this is by coupling to a scalar field, i. e., $f(\phi(t)) F_{\mu\nu}F^{\mu\nu}$, where $\phi(t)$ is the scalar field which depends on time~\cite{1991-Ratra-Apj.Lett}. This is referred to as Ratra model.
%\cite{2004-Bamba.Yokoyama-PRD}.
%
\item Constructing electromagnetic action containing higher derivative terms which leads to second order equations of motion without having ghost instabilities~ \cite{2017-Debottam.Shankaranarayanan-JCAP,2019-Kushwaha.Shankaranarayanan-PRD}. %In these kinds of models conformal invariance is naturally broken due to the presence of higher derivative terms. 
%, authors have shown that we can construct a model whose action is purely electromagnetic by using vector Galileon fields and complex scalar Galileon fields, respectively, which preserve the gauge invariance.
\end{enumerate}
In this thesis, we will discuss more about constructing higher derivative electromagnetic action (see chapter \ref{ch:GalScalElect}). We will also discuss the possible ways to consistently incorporate all these conformal breaking terms in a single ---\emph{effective field theory} --- framework. 
%Galileon fields are interesting as they break the conformal invariance of the action explicitly due to the presence of Christoffel connection in the action (see appendix \ref{CT}), and it is shown in refs \cite{2010-Kobayashi.etal-PRL}\cite{2014-Unnikrishnan.Shankaranarayanan-JCAP} that Galileon fields drive inflation, so we can use Galileon models during inflation to generate seed fields.
The models discussed above generate the non-helical primordial magnetic field. Recently, there has been much excitement on about the helical magnetic fields \cite{2001-Vachaspati-PRL,2018-Sharma.Subramanian.Seshadri.PRD,2011-Durrer.Hollenstein.Jain-JCAP}. %The difference between both fields can be understood in the following way: the massless electromagnetic field has two transverse degrees of freedom, which can be thought of as two polarization, i.e., left-circular and right-circular.
As we discussed, if both left-circular and right-circular polarization modes propagate differently, the electromagnetic field would be helical.
To generate the helical fields, the action must contain term that breaks the conformal invariance as well as parity symmetry. Thus any of the models listed above with parity-breaking terms could generate the helical magnetic field. 

In chapter~\ref{ch:helical-PMF}, we will discuss the generation of the helical magnetic field by introducing one such term: $\epsilon^{\mu\nu\sigma\kappa}R_{\mu\mu\alpha\beta}R^{\alpha\beta}\,_{\sigma\kappa}$ where totally antisymmetric tensor $\epsilon^{\mu\nu\sigma\kappa}$ breaks the parity symmetry. Helical magnetic fields have exciting properties which differ from the non-helical ones, for example detecting them, provides evidence of CP (charge parity) violation in the early Universe~\cite{2001-Vachaspati-PRL}. 
%During radiation and matter-dominated epochs, helical magnetic fields undergo an inverse cascade process that allows the transfer of energy from small to larger scales. Hence, the energy density and coherence length decay rate are slower for the helical magnetic fields than for the non-helical fields during these epochs~\cite{2001-Grasso.etal-PhyRep,2013-Durrer.Neronov-Arxiv,2016-Subramanian-Arxiv,2020-Vachaspati-RepProgPhys}.
%\subsection{Magnetic helicity}
Because the conductivity of the Universe is very high in the early Universe, the magnetic flux is a conserved quantity. However, there is another interesting conserved quantity called \emph{magnetic helicity}, which is defined in Eq.\eqref{eq:helicitydens}. 
%\textcolor{red}{as the volume integral
%
%\begin{align}%\label{eq:helicitydens}
%\mathcal{H}_M = \int d^3x \textbf{A} \cdot \textbf{B} ,
%\end{align}
%over a volume through the boundary of which no magnetic field lines cross, and $\textbf{A}$ is vector potential. Helicity is closely analogous to vorticity in fluid dynamics~\cite{2001-Grasso.etal-PhyRep,2013-Durrer.Neronov-Arxiv,2016-Subramanian-Arxiv,2020-Vachaspati-RepProgPhys}.}
%
As we discussed in subsection~(\ref{subsec:classification}), the magnetic helicity $\mathcal{H}_M$ is zero for non-helical magnetic fields.
%
%\textcolor{red}{Note that magnetic helicity $\mathcal{H}_M$ is zero for non-helical magnetic fields.} 
It is known from magnetohydrodynamics that the presence of magnetic helicity can lead to the amplification of magnetic fields due to inverse cascade (transfer of power from small scale to large scale). In chapter~\ref{ch:PMF_Baryo}, we will discuss that the relation between $\mathcal{H}_M$ (defined in Eq.~\eqref{eq:helicitydens}) and Chern-Simon number density~\cite{2001-Grasso.etal-PhyRep,2013-Durrer.Neronov-Arxiv,2016-Subramanian-Arxiv,2020-Vachaspati-RepProgPhys}, and show how non-zero helical magnetic field can provide a way to create matter-antimatter asymmetry in the early Universe.
\section{Motivation of the thesis}\label{sec:motivation_thesis}
%\textcolor{blue}{Still working on this section.}

As discussed in section~\ref{sec:open_problem}, there are many open problems in modern cosmology. The solution to most of them requires either extending the standard model or extreme fine-tuning of model parameters. This thesis addresses three open problems: the origin of large-scale magnetic fields, the creation of matter-antimatter asymmetry in the early Universe, and the origin of Fast radio bursts. The primary objective is to comprehensively examine these problems and propose novel mechanisms to resolve them. 
%The thesis also gives a novel picture of how we can use magnetic fields as a tool to probe the Universe. 
As we discussed, the magnetic fields are present in the Universe at all length scales, starting from small-scale but very strong magnetic fields of magnetars to large-scale weak magnetic fields in galaxies~(see sec \ref{fig:magneticuniverse}). Interestingly, they provide novel probes to the understanding properties of the Universe at relevant scales. In this thesis, we discuss in detail how
we can use these small-scale and large-scale magnetic fields to understand various physical phenomena. For example, the magnetic fields generated during inflation can explain the large-scale magnetic fields if the modes exit the horizon at around 50 e-folding, and the modes which exit the horizon near the end of inflation (around 5 e-foldings) can solve the baryogenesis problem. Also, the strong small-scale magnetic fields in neutron stars/magnetars can convert passing gravitational waves to electromagnetic waves, which helps us explain the origin of Fast radio bursts in the sky. Let us first systematically analyze all these problems, and the following chapters will provide a detailed discussion of the proposed mechanisms to resolve the problems.

In section~\ref{sec:mageticuniverse}, we have seen that observations reveal the presence of a micro-Gauss magnetic field coherent over galactic scales. Unfortunately, we do not have any compelling theoretical model that can explain the presence of these large-scale magnetic fields. The well-accepted paradigm suggests that the pre-existing tiny seed magnetic fields are amplified by the dynamo mechanism~\cite{1994-Kronberg-Rept.Prog.Phys.,2001-Grasso.etal-PhyRep,2002-Widrow-Rev.Mod.Phys.,2004-Giovannini-IJMPD,2008-Kulsrud.Zweibel-ReptProgPhys,2012-Widrow.etal-ScapeSciRev,2013-Durrer.Neronov-Arxiv,2016-Subramanian-Arxiv,2020-Vachaspati-RepProgPhys}. However, the standard electrodynamics and early Universe cosmology cannot explain the origin of the seed magnetic fields. This is because the standard electrodynamics action is conformally flat, hence the magnetic field decays rapidly due to the Universe's expansion. Therefore, one has to go beyond the standard model of cosmology to construct a model to generate the primordial seed magnetic fields in the early Universe. Various cosmological models have been proposed in the literature to generate the magnetic field during inflation~\cite{1988-Turner.Widrow-PRD,1991-Ratra-Apj.Lett,1991-Vachaspati-PLB,1993-Dolgov-PRD,2001-Grasso.etal-PhyRep,2002-Widrow-Rev.Mod.Phys.,2004-Giovannini-IJMPD,2008-Kulsrud.Zweibel-ReptProgPhys,2012-Widrow.etal-ScapeSciRev,2013-Durrer.Neronov-Arxiv,2016-Subramanian-Arxiv,2020-Vachaspati-RepProgPhys}. However, most models either generate a very weak magnetic field or suffer from two kinds of problems; a strong-coupling problem, which arises when the effective coupling constant becomes much larger so that theory cannot be treated perturbatively, and a backreaction problem, which is caused by the overproduction of electromagnetic fields. Thus, these models require fine-tuning of the coupling function. In chapters~(\ref{ch:GalScalElect}) and  (\ref{ch:helical-PMF}), we propose a novel mechanisms to generate the observed magnetic fields during inflation and discuss the implications in explaining the large-scale magnetic fields. In chapter (\textcolor{red}{\ref{ch:EFTmagnetogenesis}}), we construct a generic effective field theory framework to study magnetogenesis, which can reproduce a large class of inflationary magnetogenesis models.
%
%There is no mechanism which can produce sufficiently strong magnetic field over large coherence length without fine-tuning
%
%\subsection{Matter-antimatter asymmetry}
%\label{subsec:BAU-problem}
%

\begin{figure}[h]
%\centering
%\subfigure[]{%
\label{fig:BBN-eta}%
\includegraphics[height=3.1in]{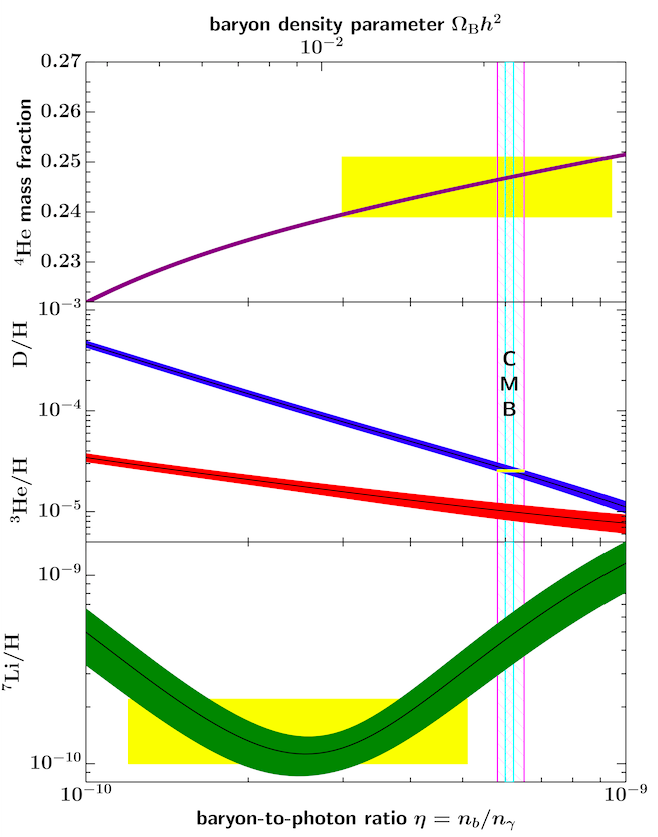}   %}%
%\qquad
%\subfigure[]{%
\label{fig:CMB_eta}%
\includegraphics[height=2.6in]{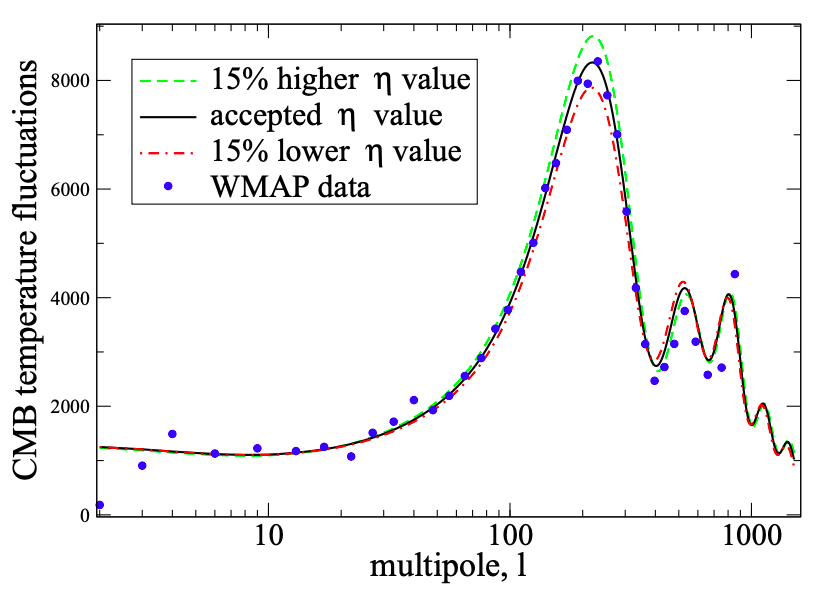}     %}%
\caption{The figure shows the constraints on the baryon to photon ration from BBN (left panel) and CMB (right panel) probes~\cite{2006-Cline-Lecture}.}
\label{fig:BAU-observe}
\end{figure}

The present Universe is observed to contain essentially matter and no antimatter, except for the rare antiparticles produced by cosmic rays. The asymmetry between baryons and antibaryons, referred to as Baryon Asymmetry of the Universe (BAU), is characterized by baryon to photon ratio $\eta_B = n_B/n_{\gamma}$ where $n_B$ is the baryon number density and $n_{\gamma}$ is the photon number density. Fig. \ref{fig:BAU-observe} shows the observational constraints from BBN and CMB on the value of $\eta_B$ to be of order $\sim 10^{-10}$~\cite{1999-Riotto.Trodden,2003-Dine.Kusenko-RevModPhy,2006-Cline-arXiv,2011-Riotto-JPCS}. Inflation has been proven to be an extremely successful paradigm for solving the hot big bang problems and explaining the origin of large-scale structures in the Universe. If that is the case, then we know any asymmetry created or existed before inflation would dilute to a vanishingly small value. 
%\textcolor{red}{Therefore, we would have a zero matter-antimatter asymmetry in the Universe, to begin with.} 
Creating this asymmetry within the standard model of particle physics and cosmology is still an unresolved issue.
%Origin of this matter-antimatter asymmetry is still an unresolved issue in particle physics and modern cosmology.

In a remarkable 1967 paper, Sakharov proposed three necessary conditions for creating the BAU, which must be satisfied by any particle physics model. These are known as \emph{Sakharov's conditions}: (i) baryon number violation, (ii) charge ($C$) and charge parity ($CP$) violation, and (iii) departure from thermal equilibrium~\cite{1967-Sakharov}. The $CP$-violating effects are not sufficiently pronounced to account for as large a BAU as we observe. As a result, there must have been additional physics beyond the standard model to produce it. This physics could have operated anywhere between the weak and GUT scales.
%
%~\cite{1999-Riotto.Trodden,2003-Dine.Kusenko-RevModPhy,2006-Cline-arXiv,2011-Riotto-JPCS}. 
%Corresponding to out-of-equilibrium conditions, the baryogenesis scenarios are divided into two categories: (a) by the universe expansion itself or (b) by fast phase transition and bubble nucleation. In particular, the latter concerns the electroweak baryogenesis schemes, while the former is typical for a GUT type baryogenesis or leptogenesis~\cite{1999-Riotto.Trodden,2003-Dine.Kusenko-RevModPhy,2006-Cline-arXiv,2011-Riotto-JPCS}.
%
At present, it is difficult to say which known mechanisms are responsible for the observed BAU. However, primordial magnetic fields in the early Universe can provide an exciting way to look at the origin of matter-antimatter asymmetry. This is because the broken symmetries in the presence of magnetic fields are similar to that of Sakharov demanded to generate baryogenesis~\cite{1996-Davidson-PLB}.
In chapter~\ref{ch:PMF_Baryo}, we will discuss how to consistently map all three Sakharov's conditions with the symmetries of the Universe in the presence of magnetic fields, and we will also identify a missing link in the earlier proposal by Davidson. This missing link is related to the helicity properties of the magnetic field. Furthermore, we show that the mechanism that leads to primordial helical magnetic fields also leads to baryogenesis at the beginning of the radiation-dominated epoch.
%the presence of background magnetic fields in the early Universe could lead to the breaking of C, CP, SO(3)
Helical magnetic fields are exciting to understand the baryon asymmetry problem because they have non zero helicity which might suggest a non zero contribution in the CP violation term. Therefore, the presence of a helical magnetic field in the early Universe with a sufficient amount of helicity density can help us to resolve the problem of the baryon asymmetry of the Universe. In the literature, it has been suggested that helical magnetic fields can produce the observed amount of baryon asymmetry through the chiral anomaly without any ingredients beyond the standard model of particle physics \cite{2016-Fujita.Kamada-PRD,2009-Semikoz.Sokoloff.Valle-PRD,2015-Anber.Sabancilar-PRD,2016-Kamada.Long-PRD,2016-Kamada.Long-2,2014-Long.Sabancilar.Vachaspati-JCAP}. 
%
%\subsection{Origin of Fast Radio Bursts}
%

Fast Radio Bursts (FRBs) are transient astrophysical phenomena in the sky, observed as intense bursts of radio waves (lasting a few milliseconds) coming from distant cosmic sources~\cite{2021-CHIME-FRB-arXiv,2022-Petroff.Lorimer-AAR}. FRBs are considered mysterious because of the poor understanding of their features. For example, the mechanism responsible for the observed peak flux, pulse width, coherent nature of radiation, and their repeating nature is still unknown, despite numerous proposed theoretical models for the progenitor of FRBs, such as neutron stars or magnetars~\cite{2021-CHIME-FRB-arXiv,2022-Petroff.Lorimer-AAR}. The high intensity of these bursts raises questions about the physical processes behind them and the astrophysical objects capable of producing such extreme events.
Moreover, they occur randomly, and their transient nature makes it difficult to predict when and where they will appear. The physical phenomenon that causes these bursts is still a mystery. In chapter~\ref{ch:GZeffect}, we discuss a mechanism known as \emph{Gertsenshtein-Zel$'$dovich effect}, which is a conversion of gravitational waves to electromagnetic waves in the presence of transverse background magnetic field. We show that this mechanism can explain certain features of FRBs, like coherent radiation, observed flux, and pulse width. 

A chapter-wise summary of the thesis is given below.

%
%\section{Organisation of thesis}
%

In chapter~\ref{ch:GalScalElect}, we discuss the higher derivative field theory model of a complex scalar Galileon action in flat space-time by demanding the following conditions: the action must satisfy the Galileon symmetry, equations of motion must be second-order, electromagnetic gauge invariance is preserved. We show that naturally, the new action has the standard scalar electrodynamics with new terms that contribute at high-energies. We also discuss the phenomenology of the model in the early Universe. Since, the model breaks the conformal invariance of the electromagnetic action due to the presence of higher derivative terms, it could be a potential candidate for the magnetogenesis. This chapter is based on the publication~\cite{2019-Kushwaha.Shankaranarayanan-PRD}.

Chapter~\ref{ch:helical-PMF} discusses the generation of primordial helical magnetic field during inflation. As we have mentioned most models in the literature introduce the coupling between the scalar field and electromagnetic field which breaks the conformal invariance of the action. However, those models suffers from strong coupling and backreaction problems, and hence require the fine-tuning. Therefore, to avoid the strong-coupling problem and fine-tuning of coupling functions, we discuss the model that couples the electromagnetic fields with the dual of Riemann tensor and due to the parity breaking term in the action, the generated fields are helical. Then we discuss the key features of the model like free from strong coupling and backreaction problem for a range of scale-factor during inflation. We show that the power-spectrum of the fields has slight red-tilt for slow-roll inflation. We also discuss the implications and differentiate the model with the earlier proposals. This chapter is based on the publication~\cite{2020-Kushwaha.Shankaranarayanan-PRD}.

In chapter~\ref{ch:PMF_Baryo}, we study the creation of matter-antimatter asymmetry in the Universe. We show that the generation of primordial helical magnetic fields from the model discussed in chapter~\ref{ch:helical-PMF} leads to baryogenesis. First, we address a very crucial issue which is related to the Sakharov's three necessary conditions for baryogenesis. We consistently map the symmetries of the Universe in the presence of primordial helical magnetic field with the Sakharov's conditions. The fact that magnetic fields are generated at all scales, we show that modes leaving the horizon during the last ten e-foldings of inflation will re-enter the horizon very early in the radiation-dominated era and lead to the creating of asymmetry between matter and antimatter. We also discuss the model's parameter space for achieving observed amount of asymmetry. This chapter is based on the publication~\cite{2021-Kushwaha.Shankaranarayanan-PRD}.

Effective field theory (EFT) provides a more systematic and generic way to look at the physics at a given energy scale by considering the effects (or operators constructed from combinations of the physical degrees of freedom of the system) only relevant to the energy scale of the problem. 
In chapter~\ref{ch:EFTmagnetogenesis}, we systematically write the \emph{effective field theory} (EFT) of magnetogenesis in the early Universe, where choice of EFT parameters describes the magnetogenesis scenario in the early Universe, and different choices of parameters correspond to different models. We explicitly show that the vector perturbations do not have temporal evolution; hence, only the gauge field is the relevant gauge-invariant variable for the EFT. We will also explicitly show that the generation of primordial magnetic fields requires two necessary conditions --- conformal invariance breaking and causal propagation. We also show that conformal invariance breaking of the electromagnetic fields is only a necessary condition, not a sufficient condition in the sense that we also need to consider the causal propagation of the gauge field fluctuations. We have confirmed this by considering a specific model of primordial magnetogenesis. We will also discuss and differentiate our EFT framework with other EFT proposals to study the magnetogenesis in the literature. This chapter is based on the publication~\cite{2022-Kushwaha.Naskar.etal-JCAP}.

In chapter~\ref{ch:GZeffect}, we discuss another interesting aspect of magnetic field in the Universe which are very strong and are associated with the compact objects such as neutron stars or magnetars. We discuss the mechanism of conversion of passing gravitational waves to electromagnetic waves in the presence of transverse magnetic field, referred to as the \emph{Gertsenshtein-Zel'dovich (GZ) effect}. We discuss how this mechanism can explain the origin of transient astrophysical phenomena like \emph{Fast radio bursts} in the sky. This chapter is based on the publications~\cite{2022-Kushwaha.etal-arXiv}.

In chapter~\ref{ch:conclusion}, we discuss the conclusions and some future directions.

%Also, we comment on the possibility of utilizing high-frequency gravitational waves to understand the mysteries of the Universe.

%Let us briefly mention the notations and conventions used in this thesis. We use $(+,-,-,-)$ metric signature and natural units where $\hbar = c = 1/(4\pi\epsilon_0) = 1$, with reduced Planck mass $M_{\rm Pl}^2 = 1/(8 \pi G) = 2.4 \times 10^{18}~{\rm GeV}$. The various physical quantities with the over-line refers to the values evaluated for the homogeneous and isotropic FRW background. A dot denotes a derivative with respect to the cosmic time ($t$), a prime stands for a derivative with respect to conformal time ($\eta$), and $,i$ denotes a derivative w.r.t spatial coordinates. We will work in the We also work in Heaviside-Lorentz units such that $\hbar = c = \epsilon_0 = \mu_0 = k_B = 1$.

%\pagestyle{fancy}
\chapter{Galileon scalar electrodynamics}
%\label{ch:intds}
\label{ch:GalScalElect}

The standard relativistic field theories describing physical phenomena contain the second-order time and spatial derivatives. The perturbative approach of these field theories is highly successful in explaining the experiments and observations. However, these field theories have ultraviolet divergences, and higher-derivative terms were introduced in an attempt to remove the ultraviolet divergence~\cite{1948-Podolsky.etal-RevModPhys,1950-Pais.Uhlenbeck-PRD,1950-Thirring-PR}. 

The higher derivative field theories are known to have Ostrogradsky instability~\cite{2015-Woodard-arXiv}. This is because Hamiltonian contains a term proportional to the momenta, thus leading to an unbounded Hamiltonian from below~\cite{1990-Simon-PRD}. The instability itself is not a concern if the energy is an integral of motion or the higher derivative field (say, $\pi$) does not interact with any other field (say, $\phi$) whose Hamiltonian is bounded from below. The interaction between $\pi$ and $\phi$ will pump out energy from $\pi$, leading to a runaway situation. Hence, these negative energy states can be traded by negative norm states (or ghosts), leading to non-unitary theories~\cite{2002-Hawking.Hertog-PRD} and, therefore, unsuitable to describe physical phenomena.

Nonetheless, higher-derivative theories have some salient features that
make them indispensable to understand the high-energy behavior of a theory~\cite{1983-Barth.Christensen-PRD}. Specifically, the divergence structure of quantum fields is expected to improve when higher-derivative terms are taken into account~\cite{1983-Barth.Christensen-PRD,2002-Hawking.Hertog-PRD}. This also leads to improved convergence of Feynman diagrams. Although Einstein's gravity is not renormalizable, conformal gravity is~\cite{1977-Stelle-PRD,1982-Fradkin.Tseytlin-NPB}.

There is a resurgence of interest in higher-derivative theories primarily from the modifications of gravity in the short and long distances~\cite{2010-Sotiriou.Faraoni-RMP,2011-Capozziello.DeLaurentis-PRep,2011-Nojiri.Odintsov-PRep,2012-Clifton.etal-PRep,2017-Nojiri.etal-PhyRept,2019-Ishak-LivingRevRel}. 
It is now known that higher derivative gravity theories like $f(R)$ are degenerate, and do not suffer from Ostrogradsky instability~\cite{2007-Woodard-Proc}. A consistent scheme has been implemented where the Ostrogradsky instability is suppressed by requiring that the equations of motion are second order~\cite{1974-Horndeski-IJTP}. Starting from the following general action:
\begin{equation}
L=L\left(g_{i j} ; g_{i j, i_{1}} ; \ldots ; g_{i j, i_{1} \cdots i_{p}} ; \phi ; \phi ; \phi,_{i_{1}} ; \ldots ; \phi_{, i_{1}} \ldots i_{q}\right) \, ,
\end{equation}
Horndeski showed that it is possible to obtain a second-order equation for $\phi$ even if the action contains higher-order derivatives of the scalar field~\cite{1974-Horndeski-IJTP}. It has recently shown that the field equations in a subclass of Horndeski models have Galilean shift symmetry in flat space, i.e., 
\begin{equation}
\label{eq:GalileanShift}
\phi \rightarrow \phi + a_{\mu}x^{\mu} + b \, ,
\end{equation}
%
%where $a_{\mu}$ and $b$ are infinitesimal four-vector (shift parameter) and a constant scalar respectively~\cite{2009-Deffayet.etal4-PRD}. Galileon scalar fields are the most general non-canonical and non-minimally coupled single-field model, which yields second-order equations.
%
where $a_{\mu}$ and $b$ are infinitesimal four-vector (shift parameter) and a constant scalar respectively~\cite{2009-Deffayet.etal4-PRD}. 
Note that under the Galileon symmetry~\eqref{eq:GalileanShift}, the action is invariant up to a constant term as $\partial_{\mu}\phi \longrightarrow \partial_{\mu}\phi + a_{\mu}$ in flat spacetime, and in 4D this symmetry constrains the equations of motion with upto second order derivative terms. The vectorial parameter $a_{\mu}$ corresponds to the shift of gradient, and this is the spacetime generalization of the Galilean symmetry $\dot{x} \longrightarrow \dot{x} + v$ of non-relativistic mechanics ($0+1$ field theory), where $v$ is constant velocity shift~\cite{2009-Nicolis.etal-PRD}.
Galileon scalar fields are the most general non-canonical and non-minimally coupled single-field model, which yields second-order equations.

These theories, while naively look like \emph{higher-derivative} theories, are healthy non-higher-derivative theories; their equations of motion are second order in time derivatives and do not suffer from the instability~\cite{2009-Deffayet.etal4-PRD}. This is achieved by the addition of structure in the Lagrangian --- usually by a subtle cancellation of higher derivative terms in the equations of motion, much like Lanczos-Lovelock theories of gravity~\cite{1971-Lovelock-JMP,1972-Lovelock-JMP,1986-Zumino-PRep,2013-Padmanabhan.Kothawala-PRep}. 
Like Galileon scalars, these theories of gravity are exceptional in that the resulting equations of motion are no more than the second order. They are also free of ghosts when expanded about the flat space-time.

Recently, a vector Galileon field was constructed in a curved space-time~\cite{2017-Debottam.Shankaranarayanan-JCAP}. It was shown that the electromagnetic action breaks conformal invariance and has the following three properties: model is described by vector potential $A_{\mu}$ and its derivatives, Gauge invariance is preserved, and equations of motion are linear in second derivatives of the vector potential. This is an essential result as earlier it was proven that such an action could not be constructed in flat space-time.

While the scalar and vector Galileon fields have been investigated, there has been \emph{no study} on the interaction between the scalar and vector Galileons \emph{preserving gauge-invariance.} (In Ref. \cite{2018-Heisenberg.etal-PRD}, Heisenberg et al have studied Scalar-Vector Galileon models that do not preserve gauge invariance.) In this chapter, we bridge this gap by constructing a higher derivative action of interacting fields (complex scalar Galileons interacting with a vector field), which do not lead to ghosts. We refer to this model as \emph{Galileon scalar electrodynamics}. We construct a complex scalar Galileon action in flat space-time by demanding the following conditions: action must satisfy the Galileon symmetry, equations of motion must be second-order, gauge invariance is preserved. Naturally, the new action has the standard scalar electrodynamics with new terms that contribute at high-energies. We apply this model in the early Universe. As we discussed in section \ref{sec:inflationary_magneto_classification}, the key requirement to generate a sufficiently strong magnetic field is the breaking of conformal invariance. The \emph{Galileon scalar electrodynamics} action satisfy this requirement, and hence this could be a potential candidate for inflationary magnetogenesis.
%In the Coulomb gauge of the Vector field, we obtain the non-zero contributions from the new terms. In the flat FLRW background, we show that the model leads to inflation and is driven by the real scalar field. 

The rest of the chapter is organized as follows: 
In Sec. \eqref{sec:ComplexGalileon}, we explicitly construct the 
Galileon scalar electrodynamics in flat space-time. We show that 
the action is invariant under the local gauge invariance in flat space-time. In Sec. \eqref{Min}, we extend the analysis to curved space-time. We show that the minimal coupling of the matter and gravity leads to higher-derivative terms in the equations of motion. We then include non-minimal terms to the action that will lead to subtle cancellation of terms and, hence, lead to second-order equations of motion. We then apply the model to the early Universe in Sec. \eqref{sec:energy_pressure} and show that the model can lead to inflation. In Sec. \eqref{sec:conc}, we conclude this chapter by briefly discussing the importance of these results. Appendix~\ref{ap:GalScED} contains the details of the calculations in this chapter. The results reported in this chapter are based on Ref.~\cite{2019-Kushwaha.Shankaranarayanan-PRD}

%In this work, we use (+,-,-,-) metric signature and natural units $\hbar = c = 1/(4\pi \epsilon_0) = 1$. We set $8\pi G = M_P^{-2}$.  
%The real Galileon scalar field is denoted by $\varphi$, complex Galileon scalar field is denoted by $\pi$ and  $\Box = \nabla_{\mu}\nabla^{\mu}$. An overdot denotes the derivative for cosmic time.

\section{Galileon Scalar Electrodynamics in flat space-time}
\label{sec:ComplexGalileon}

In this section, we obtain the higher derivative complex scalar field action coupled to the electromagnetic field in flat space-time. In the next subsection, we list the real scalar Lagrangians obtained by Nicolis et al.~\cite{2009-Nicolis-PRD} that lead to \emph{healthy} non-higher-derivative theories. In Sec. \eqref{sec:CSG}, we obtain the complex scalar Lagrangian that leads to non-higher-derivative theories. In Sec. \eqref{sec:GaugeFlat}, we couple the Galileon complex scalar with the electromagnetic field $A_{\mu}$. 

\subsection{Real scalar Galileon}
\label{sec:ScalarGalileon}

As mentioned earlier, the action of the real scalar Galileons ($\varphi$) is invariant under the \emph{Galilean transformation} \eqref{eq:GalileanShift}. In 4-D Minkowski space-time, Nicolis et al.~\cite{2009-Nicolis-PRD} have shown that the following five Lagrangians are invariant under the Galilean transformation:
\begin{subequations}
\begin{align} \label{eq:ScalarL2}
    & \mathcal{L}_1^{\prime} = \varphi \qquad 
    \mathcal{L}_2^{\prime} = -\frac{1}{2}\partial_{\mu}\varphi\partial^{\mu} \varphi\\
    \mathcal{L}_3^{\prime} &= -\frac{1}{2}\Box \varphi\partial_{\mu}\varphi\partial^{\mu} \varphi\\
    \label{eq:ScalarL4}
    \mathcal{L}_4^{\prime} &= -\frac{1}{4}[ (\Box \varphi)^2 \partial_{\mu}\varphi\partial^{\mu} \varphi -2\Box \varphi\partial_{\mu}\varphi\partial^{\mu}\partial_{\nu}\varphi\partial^{\nu}\varphi- \partial_{\mu}\partial_{\nu}\varphi \partial^{\mu}\partial^{\nu}\varphi\partial_{\alpha}\varphi\partial^{\alpha} \varphi 
    \nonumber\\ 
    &{}{}\hspace{0.5cm} +2\partial_{\mu}\varphi \partial^{\mu}\partial_{\nu}\varphi\partial^{\nu}\partial_{\alpha}\varphi
    \partial^{\alpha}\varphi ]\\
    \mathcal{L}_5^{\prime} &= -\frac{1}{5}[ (\Box \varphi)^3 \partial_{\mu}\varphi\partial^{\mu} \varphi -3(\Box \varphi)^2 \partial_{\mu}\varphi \partial^{\mu}\partial_{\nu}\varphi\partial^{\nu} \varphi - 3\Box \varphi\partial_{\mu}\partial^{\nu}\varphi\partial^{\mu}\partial_{\nu}\varphi\partial_{\alpha}\varphi \partial^{\alpha} \varphi \nonumber\\
    &{}{}\hspace{0.5cm}+ 6\Box \varphi\partial_{\mu}\varphi\partial^{\mu}\partial_{\nu}\varphi \partial^{\nu}\partial_{\alpha}\varphi\partial^{\alpha} \varphi 
+2\partial_{\mu}\partial^{\nu}\varphi\partial_{\nu}\partial^{\alpha}\varphi\partial_{\alpha}\partial^{\mu}\varphi\partial_{\lambda}\varphi \partial^{\lambda} \varphi 
\nonumber\\
&{}{}\hspace{0.5cm} 
+3\partial_{\mu}\partial^{\nu}\varphi\partial_{\nu}\partial^{\mu}\varphi\partial_{\alpha}\varphi \partial^{\alpha}\partial_{\lambda}\varphi\partial^{\lambda} \varphi -6\partial_{\mu}\varphi \partial^{\mu}\partial_{\nu}\varphi\partial^{\nu}\partial_{\alpha}\varphi\partial^{\alpha}\partial_{\lambda}\varphi\partial^{\lambda} \varphi ]
    \end{align}
    \end{subequations}
We want to emphasize that the above Lagrangians are the linear combinations of Lorentz invariant terms which are added (up to total derivative term) in such a way that the action is Galilean invariant. The equations of motion are second order. 
Also, note that the Lagrangian $\mathcal{L}_1^{\prime}$ does not have any dynamics and often referred to as \emph{tadpole} term.\footnote{We can see that under transformation~\eqref{eq:GalileanShift}, the Lagrangian changes as $\delta \mathcal{L}_1^{\prime} = a_{\mu}x^{\mu} + c = \partial_{\mu} (a_{\rho}x^{\prime}x^{\mu}/5 + c x^{\mu}/4)$, which is invariant upto total derivative term.} In fact, it is the only non-daynamical Lagrangian out of five possible Lorentz invariant Lagrangian that can be constructed in 4D which obey the Galileon symmetry in field space.
In the next subsection, we use these five Lagrangians to construct the complex scalar Galileon action.
 
\subsection{Complex scalar Galileon}
\label{sec:CSG}
    
We aim to write down the Lagrangian that satisfies the following three conditions: First, it must satisfy shift symmetry (\ref{eq:GalileanShift}) on $\pi$ and the complex conjugate ($\pi^*$). Second, it must be Lorentz invariant. Third, the Lagrangian must be real. It is easy to see that the Lagrangians with the odd number of $\pi$'s lead to complex action and, hence, are discarded. Having $\mathcal{L}_2^{\prime}$ and $\mathcal{L}_4^{\prime}$ in hand, we start with $\mathcal{L}_2^{\prime}$. The complex scalar Galileon Lagrangian for this case is straightforward:
    \begin{align}\label{L2a}
    \mathcal{L}_2^{\prime} = \frac{1}{2}\partial_{\mu}\pi \partial^{\mu}\pi \implies {\mathcal{L}_2} = \frac{1}{2}\partial_{\mu}\pi \partial^{\mu}\pi^*
    \end{align}
 where $\pi^*$ is the complex conjugate. The above action is invariant 
 under the Galilean shift symmetry (\ref{eq:GalileanShift}) on $\pi$ and $\pi^*$.  In order to avoid confusion, we have used unprimed for complex scalar Lagrangian to distinguish it from the scalar Galileon Lagrangian (primed). 
 
Next, we consider the Lorentz invariant terms of $\mathcal{L}_4^{\prime}$ (with two fields $\pi$ and $\pi^*$). By symmetrizing each term in 
the RHS of \eqref{eq:ScalarL4} and ignoring the double-counting, 
we write $\mathcal{L}_4$ as a linear combination of these invariants, i.e.,
\begin{align}\label{L4}
    \mathcal{L}_4 &= A_1\, (\, (\Box\pi)^2 \partial_{\alpha}\pi^* \partial^{\alpha}\pi^* + c.c. ) + A_2 \,\,\Box\pi^* \, \Box\pi \partial_{\alpha}\pi^* \partial^{\alpha}\pi   
    +  B_1(\, \Box \pi\partial_{\nu}\pi \partial^{\nu} \partial^{\alpha}\pi^* \partial_{\alpha}\pi^*+ c.c.) 
    \nonumber \\ 
    &{}+ B_2\,(\, \Box \pi\partial_{\nu}\pi^* \partial^{\nu} \partial^{\alpha}\pi \partial_{\alpha}\pi^* + c.c.)  + C_1\,(\, \partial_{\mu}\partial_{\nu}\pi \partial^{\mu}\partial^{\nu}\pi \partial^{\alpha}\pi^* \partial_{\alpha}\pi^* + c.c.) 
    \nonumber \\ &{} 
    + C_2\,\, \partial_{\mu}\partial_{\nu}\pi^*\partial^{\mu}\partial^{\nu}\pi \partial^{\alpha}\pi^* \partial_{\alpha}\pi 
    + D_1 \,(\partial_{\mu}\pi^*\partial^{\mu}\partial^{\nu}\pi \partial_{\nu}\partial_{\alpha}\pi \partial^{\alpha}\pi^* + c.c.) 
    \nonumber \\ &{}
    + D_2 \,\,\partial_{\mu}\pi^*\partial^{\mu}\partial^{\nu}\pi^* \partial_{\nu}\partial_{\alpha}\pi \partial^{\alpha}\pi +  D_3 \,\,\partial_{\mu}\pi^*\partial^{\mu}\partial^{\nu}\pi \partial_{\nu}\partial_{\alpha}\pi^* \partial^{\alpha}\pi
\end{align}
where $A_i, B_i, C_i, D_i, D_3 \, (i = 1, 2)$ are unknown complex constants, c.c. in the parentheses denote the complex conjugate part (hence total part contributing to real) of the corresponding Lorentz invariant term and rest of the terms are real. Hence, the above  Lagrangian is a real scalar.

However, for any arbitrary constants, the Lagrangian will not be Galilean invariant. Demanding the Galilean invariance of the Lagrangian \eqref{L4} (up to some total derivative) leads to the constraints on the constants. Appendix \eqref{Generic} contains detailed calculations 
where we show that $\mathcal{L}_4 $ depends only on 
two arbitrary coefficients $A_1$ and $A_2$ and the resultant equations of motion match with Nicolis et al.~\cite{2009-Nicolis-PRD}. In Appendix \eqref{Generic}  we also explicitly show that the equations of motion for this Lagrangian is the same for any value of $A_1$ and $A_2$. This is because the action $\mathcal{L}_4 $ is invariant under the shift symmetry \eqref{eq:GalileanShift} for any value of these arbitrary constants. For simplicity, we set $A_1 = 0$. Setting $A_1 = 0$ and $A_2 = \omega \, \lambda^{-6}$ in Eq.~(\ref{LCGG}) leads to:
\begin{align}\label{GIL}
\mathcal{L}_4 &= \frac{\omega}{2 \lambda_{\pi}^6} \left[ 2\, \Box \pi^* \,\Box \pi \,\partial_{\alpha}\pi \partial^{\alpha}\pi^* 
- 2\, \partial_{\mu}\partial_{\nu}\pi^* \partial^{\mu}\partial^{\nu}\pi \partial_{\alpha} \pi\partial^{\alpha}\pi^* \right. \nonumber \\
&+ \left. (\,\Box \pi \,\partial_{\nu}\pi^* \partial^{\nu}\partial^{\alpha}\pi \partial_{\alpha}\pi^* 
- \partial^{\mu}\partial^{\nu}\pi \partial_{\nu}\partial_{\alpha}\pi\,\partial_{\mu}\pi^* \partial^{\alpha}\pi^* +\rm{c.c.}) \right]
\end{align}
Note that $A_2$ has a dimension $[L]^{6}$, thus $\lambda$ has dimensions of inverse length and $\omega$ can take $+1$ or $-1$. 
The value of $\omega$ will be fixed in Sec. \eqref{sec:coeffiFixing} by demanding that the energy density of the field is always positive. In the next section, we will show that the generalization of the above Lagrangian in curved space-time matches with Deffayet et al. for some suitable value of the coefficient $\lambda_{\pi}$~\cite{Deffayet2009}. The Galilean invariant action corresponding to the above Lagrangian is given by
\begin{align}\label{action4}
{S}_{4} &= \frac{\omega}{2 \lambda_{\pi}^6}\int d^4x\,\,[\,2\, \Box \pi^* \,\Box \pi \,\partial_{\alpha}\pi \partial^{\alpha}\pi^* - 2\, \partial_{\mu}\partial_{\nu}\pi^* \partial^{\mu}\partial^{\nu}\pi \partial_{\alpha} \pi\partial^{\alpha}\pi^*   \nonumber\\&{}\hspace{2cm} +  (\,\Box \pi \,\partial_{\nu}\pi^* \partial^{\nu}\partial^{\alpha}\pi \partial_{\alpha}\pi^* 
- \partial^{\mu}\partial^{\nu}\pi \partial_{\nu}\partial_{\alpha}\pi\,\partial_{\mu}\pi^* \partial^{\alpha}\pi^* +\rm{c.c.}) ] 
\end{align}
Equations of motion for $\pi$ corresponding to the above action is:
\begin{align}\label{EOM}
\mathcal{E}_{4} &= \frac{\omega}{2 \lambda_{\pi}^6}\left[ -(\Box \pi^*)^2\,\Box \pi + \Box \pi\, \partial_{\mu}\partial_{\nu}\pi^* \,\partial^{\mu} \partial^{ \nu}\pi^* - 2\,\partial_{\mu}\partial_{\nu}\pi^*\, \partial^{\nu}\partial^{\alpha}\pi\,\partial^{\mu}\partial_{\alpha}\pi^* \right. 
\nonumber \\
&{}{}\qquad \left. + 2\,\Box \pi^*\, \partial_{\mu}\partial_{\nu}\pi^*\, \partial^{\mu}\partial^{\nu}\pi\,\,\right]
\end{align}
The equations of motion contain derivatives up to second order and, hence, does not lead to extra degrees of freedom. The complex scalar Galileon action in 4D Minkowski space-time is given by
\begin{align}\label{action_CSG}
{S}_{\rm{flat}} &= S_{2} + S_{\rm 4} \, ,
\end{align}
where $S_{\rm 4}$ is given by \eqref{action4} and 
\begin{equation}
\label{eq:CanonicalScalar}
S_{2} = \frac{1}{2}\int d^4x \,\,\partial_{\mu}\pi\, 
\partial^{\mu}\pi^*  \, .
\end{equation}
Thus, we have constructed a complex Galileon field action. In the low-energy limit, the Galileon term will not be significant. However, at high-energies $S_4$ plays a crucial role in the dynamics. 
We will discuss the implications of this in Sec. \eqref{sec:energy_pressure}.

\subsection{Coupling to the electromagnetic field}
\label{sec:GaugeFlat}

The action \eqref{action_CSG} is invariant under the global transformation, i.e. $\pi \to \pi e^{-i\, e \, \theta}$, where $\theta$ is a constant parameter and $e$ is the electric charge. However, the action is not invariant if the parameter $\theta$ is space-time dependent or the local $U(1)$ gauge transformation.  In order 
for the action to be invariant under the transformation: 
$\pi \rightarrow \pi e^{-ie\theta(x)}$ and $\pi^* \rightarrow \pi e^{ie\theta(x)}$, we need to replace the partial derivatives as
\begin{align}\label{eq:EMcovariant}
\partial_{\mu} \rightarrow {D}_{\mu} \equiv \partial_{\mu} + ie A_{\mu} \hspace{0.5cm} \text{and} \hspace{0.5cm}
\partial_{\mu}\partial^{\nu} \rightarrow {D}_{\mu}{D}^{\nu}= (\partial_{\mu} + ie A_{\mu} )(\partial^{\nu} + ie A^{\nu})
\end{align}
where $A_{\mu}$ is the electromagnetic field vector. Under the following gauge transformations 
\begin{align}\label{eq:GaugeTransformation}
\pi \rightarrow \pi e^{-ie\theta(x)}\,;\hspace{.5cm}\pi^* \rightarrow \pi e^{ie\theta(x)}\,; \hspace{.5cm}
A_{\mu} \rightarrow A_{\mu} + \partial_{\mu}\theta
\end{align}
the action (\ref{action_CSG}) is invariant. Appendix \eqref{GaugeAppend} contains the details of the calculation. 
The complete Galileon scalar electrodynamics action in flat space-time is given by:
\begin{align}\label{action_full}
{S}_{\rm{flat}}^{\rm G} &= \frac{1}{2}\int d^4x\,\, {D}_{\mu}\pi\, {D}^{\mu}\pi^*+\frac{\omega}{2 \lambda_{\pi}^6}\int d^4x\,\,[\,2\, {D}_{\mu}{D}^{\mu} \pi^*\, {D}_{\nu}{D}^{\nu} \pi \,{D}_{\alpha}\pi \,{D}^{\alpha}\pi^* \nonumber\\&{}\hspace{0.5cm} - 2 \,{D}_{\mu}{D}_{\nu}\pi^*\,{D}^{\mu}{D}^{\nu}\pi\, {D}_{\alpha} \pi\, {D}^{\alpha}\pi^* + ( {D}_{\mu}{D}^{\mu} \pi \,{D}_{\nu}\pi^*\, {D}^{\nu}{D}^{\alpha}\pi \,{D}_{\alpha}\pi^* \nonumber\\&{}\hspace{0.5cm}   
 - {D}^{\mu}{D}^{\nu}\pi\, {D}_{\nu}{D}_{\alpha}\pi\,{D}_{\mu}\pi^*\,{D}^{\alpha}\pi^* + {\rm c.c.}) ] -\frac{1}{4}\int d^4 x F_{\mu\nu}F^{\mu\nu}
\end{align}
where $F_{\mu\nu} = \partial_{\mu}A_{\nu} - \partial_{\nu}A_{\mu}$ is the electromagnetic field tensor. Unlike scalar Galileons, there exists a 
no-go theorem that states that, higher derivative vector Galileons cannot be constructed in flat space-time~\cite{2017-Debottam.Shankaranarayanan-JCAP}. Thus, for the Galileon scalar electrodynamics in the flat space-time, the non-linear part of the electromagnetic fields appear through the gauge coupling $D_{\mu}$. It is important to note that the action (\ref{action_full}), in addition to U(1) gauge invariance, also satisfies the Galilean symmetry (for $\pi$ and $\pi^*$), and hence, we refer to  this action as \emph{Galileon Scalar Electrodynamics}. In Sec. \eqref{sec:energy_pressure}, we show that the Galileon term leads to interesting features in inflationary dynamics.

\section{Galileon Scalar Electrodynamics in curved space-time}\label{Min}

One of the critical features of the action of the Galileon field 
\eqref{action_CSG} is that it contains second derivatives of the field. 
Assuming a minimal coupling of the matter and gravity leads to $\partial_{\mu} \to \nabla_{\mu}$. While the partial derivatives $\partial_{\mu}\partial_{\nu}$ commute, this is not the case for covariant derivatives. We need to take into account the commutation properties of the covariant derivatives. The procedure we will adopt is similar to that of Deffayat et al~\cite{2009-Deffayet.etal-PRD}. However, due to complex scalar fields, there are some differences in the final expression.

\subsection{Coupling to gravity}
\label{sec:CurvedST}

For the minimal coupling, Galilean symmetry is preserved for $\mathcal{L}_2$ defined in \eqref{L2a}. However, the 
Galilean symmetry is broken explicitly for the fourth order Lagrangian (\ref{GIL}). Assuming a minimal coupling of the complex field with gravity, the fourth order action (\ref{action4}) becomes:
\begin{align}\label{action_m}
{S}_4^{\rm{min}} &= \frac{\omega}{2 \lambda_{\pi}^6}\int d^4x\,\,\sqrt{-g}\,\,[\,\,2\, \Box \pi^*\, \Box \pi \,\nabla_{\alpha}\pi\, \nabla^{\alpha}\pi^* - 2 \,\nabla_{\mu}\nabla_{\nu}\pi^*\, \nabla^{\mu}\nabla^{\nu}\pi\, \nabla_{\alpha} \pi\, \nabla^{\alpha}\pi^*\nonumber\\&{}\hspace{0.5cm} + (\,\, \Box \pi\, \nabla_{\nu}\pi^* \, \nabla^{\nu}\nabla^{\alpha}\pi \,\nabla_{\alpha}\pi^* 
  - \nabla^{\mu}\nabla^{\nu}\pi\, \nabla_{\nu}\nabla_{\alpha}\pi\,\nabla_{\mu}\pi^*\, \nabla^{\alpha}\pi^*  + \rm{c.c.} )\,\,\,] 
\end{align}
here $\nabla_{\alpha}$ denotes the covariant derivative with respect to the metric $g_{\mu\nu}$ and $\Box = \nabla_{\mu}\nabla^{\mu}$. Varying the action (\ref{action_m}) with respect to $\pi$ yields the equation of motion of $\pi$. Using the commutation properties of covariant derivatives as given in Appendix \eqref{app:C}, we obtain the 
following equation of motion:
\begin{small}

\begin{align}
\label{eq:EOM4-Curved}
\mathcal{E}_4^{\rm{min}} = \frac{\omega}{2 \lambda_{\pi}^6}[\,-2\,\nabla_{\alpha}\pi\, \nabla^{\alpha}\pi^*\,\nabla^{\nu}\nabla^{\mu}\pi^*\,R_{\mu\nu} -\nabla_{\alpha}\pi\, \nabla^{\alpha}\pi^*\,\nabla^{\mu}\pi^*\,\nabla_{\mu}R - \nabla^{\mu}\pi^*\,\nabla^{\alpha}\pi^*\,\nabla^{\rho}\pi\,\nabla_{\rho}R_{\alpha\mu}
 \nonumber\\ - \nabla^{\mu}\pi^*\,\nabla^{\alpha}\pi^*\,\nabla_{\mu}\nabla^{\nu}\pi\, R_{\nu\alpha} +2\,\nabla^{\mu}\pi^*\,\nabla^{\alpha}\pi^*\,\nabla^{\nu}\nabla^{\rho}\pi\, R_{\rho\mu\alpha\nu} -3\,\nabla^{\alpha}\nabla^{\nu}\pi\, \nabla_{\alpha}\pi^* \,\nabla^{\mu}\pi^*\, R_{\mu\nu}\nonumber\\ -2\,\nabla^{\mu}\pi^*\,\nabla_{\alpha}\pi\,\nabla^{\nu}\nabla^{\alpha}\pi^*\,R_{\mu\nu} + \nabla^{\nu}\pi^*\,\nabla^{\alpha}\pi^*\,\Box\pi\, R_{\nu\alpha} -2\,\nabla^{\nu}\nabla^{\alpha}\pi^*\,\nabla_{\alpha}\pi^*\,\nabla^{\mu}\pi\, R_{\mu\nu}\nonumber\\ +2\,\Box\pi^*\, \nabla_{\nu}\nabla_{\alpha}\pi\,\nabla^{\nu}\nabla^{\alpha}\pi^* -(\Box\pi^*)^2\, \Box\pi - 2\,\nabla_{\mu}\nabla_{\nu}\pi^*\, \nabla^{\mu}\nabla^{\alpha}\pi^*\nabla^{\nu}\nabla_{\alpha}\pi +\Box\pi\, \nabla^{\mu}\nabla^{\nu}\pi^* \,\nabla_{\mu}\nabla_{\nu}\pi^*\,\,]
\end{align}
\end{small}
where $R_{\rho\mu\alpha\nu}, R_{\mu\nu}$ and $R$ are Riemann tensor, Ricci tensor and Ricci scalar respectively. One can immediately notice that the second and third terms in the RHS of the above
contain third order derivative terms. Thus, the minimal coupling of the matter with gravity lead to the higher-derivative equations of motion. 
As mentioned in the introduction, it is possible to cancel these higher derivative terms adding suitable terms in the action~\cite{2009-Deffayet.etal-PRD}\cite{2017-Debottam.Shankaranarayanan-JCAP}. The following non-minimal action 
\begin{align}\label{non-min}
{S}_4^{\rm{nm}}&= -\frac{\omega}{4 \lambda_{\pi}^6}\int d^4x\,\,\sqrt{-g}\,\, \nabla_{\alpha}\pi\nabla^{\alpha}\pi^* \nabla_{\mu}\pi \nabla_{\nu}\pi^*g^{\mu\nu}R \nonumber\\&{} \nonumber\\ &{}\hspace{0.5cm}  -\frac{\omega}{4 \lambda_{\pi}^6}\int d^4x\,\,\sqrt{-g}\,\, \left[\nabla_{\alpha}\pi\nabla^{\alpha}\pi \nabla_{\mu}\pi^*\nabla_{\nu}\pi^* \left( R^{\mu\nu} - \frac{1}{4}g^{\mu\nu}R \right) +\rm{c.c.} \right]
\end{align}
can remove the higher-derivative terms that appear in Eq. \eqref{eq:EOM4-Curved}. See Appendices \eqref{app:C} and \eqref{sec:A1_terms}, for details. 
If we add the above non minimal action in the action (\ref{action_m}) then it will cancel all the higher order derivative terms in the EOM. So varying the action
${S}_4^{\rm{min}} + {S}^{\rm{nm}}_4$ with respect to $\pi$ gives the equation of motion  (see appendix Eq. \ref{eq:S4eomCurved}).
%
%Hence equation of motion for the total action 
%\begin{align}\label{S'}
%{S}^{\prime}_{\rm{CGG}} = {S}_{\rm{CGG}} + {S}_{\rm{non\; min}}
%\end{align}
%will have only second order terms in the Galileon fields and metric, which is 
%\begin{align}\label{Eprime}
%\mathcal{E}^{\prime} &= \frac{A_2}{2}\,\,[\,\,2\,\Box\pi^*\, \nabla_{\nu}\nabla_{\alpha}\pi\,\nabla^{\nu}\nabla^{\alpha}\pi^* -(\,\Box\pi^*)^2\, \Box\pi +\Box\pi\, \nabla^{\mu}\nabla^{\nu}\pi^*\, \nabla_{\mu}\nabla_{\nu}\pi^* - 2\,\nabla_{\mu}\nabla_{\nu}\pi^*\, \nabla^{\mu}\nabla^{\alpha}\pi^*\,\nabla^{\nu}\nabla_{\alpha}\pi \nonumber\\ &{} \hspace{1cm} + \Box\pi^*\,\nabla_{\mu}\pi^*\,\nabla^{\mu}\pi\, R -\frac{1}{2}\Box\pi\,\nabla_{\alpha}\pi^*\,\nabla^{\alpha}\pi^* \,R + \nabla_{\alpha}\pi^*\,\nabla_{\mu}\pi^*\,\nabla^{\mu}\nabla^{\alpha}\pi \,R +2\,\Box\pi\, \nabla^{\nu}\pi^*\,\nabla^{\mu}\pi^* \,R_{\nu\mu} \nonumber\\ &{} \hspace{1cm} +\nabla_{\alpha}\pi^*\, \nabla^{\alpha}\pi^*\,\nabla^{\nu}\nabla^{\mu}\pi \,R_{\mu\nu}  - 2\,\nabla_{\alpha}\pi\, \nabla^{\alpha}\pi^*\,\nabla^{\nu}\nabla^{\mu}\pi^*\,R_{\mu\nu} - \nabla^{\mu}\pi^*\,\nabla^{\alpha}\pi^*\,\nabla_{\mu}\nabla^{\nu}\pi\, R_{\nu\alpha} \nonumber\\ &{} \hspace{1cm}+ 2\, \nabla^{\mu}\pi^*\,\nabla^{\alpha}\pi^*\,\nabla^{\nu}\nabla^{\rho}\pi\, R_{\rho\mu\alpha\nu} -3\,\nabla^{\alpha}\nabla^{\nu}\pi\, \nabla_{\alpha}\pi^*\, \nabla^{\mu}\pi^*\,R_{\mu\nu} ]
%\end{align}
%
Hence, the complete Galileon Complex scalar action in an arbitrary curved space-time is given by
\begin{align}\label{action_curved}
{S}_{\rm{Curved}} &= S_2  + S_4^{\rm{min}} + S_4^{\rm{nm}} \, .
%
%\frac{1}{2}\int d^4x\, \sqrt{-g}\,\,\nabla_{\mu}\pi\nabla^{\mu}\pi^* +\frac{A_2}{2}\int d^4x\,\,\sqrt{-g}\,\,[\,\,2\, \Box \pi^*\, \Box \pi \,\nabla_{\alpha}\pi\, \nabla^{\alpha}\pi^* - 2 \,\nabla_{\mu}\nabla_{\nu}\pi^*\, \nabla^{\mu}\nabla^{\nu}\pi\, \nabla_{\alpha} \pi\, \nabla^{\alpha}\pi^*\nonumber\\&{}\hspace{0.5cm} + (\,\, \Box \pi\, \nabla_{\nu}\pi^* \, \nabla^{\nu}\nabla^{\alpha}\pi \,\nabla_{\alpha}\pi^* 
%- \nabla^{\mu}\nabla^{\nu}\pi\, \nabla_{\nu}\nabla_{\alpha}\pi\,\nabla_{\mu}\pi^*\, \nabla^{\alpha}\pi^*  + \rm{c.c.} )\,\,\,] \nonumber\\&{}\hspace{0.5cm}
%-\frac{A_2}{4}\int d^4x\,\,\sqrt{-g}\,\, \nabla_{\alpha}\pi\nabla^{\alpha}\pi^* \nabla_{\mu}\pi \nabla_{\nu}\pi^*g^{\mu\nu}R   -\frac{A_2}{4}\int d^4x\,\,\sqrt{-g}\,\, \left[\nabla_{\alpha}\pi\nabla^{\alpha}\pi \nabla_{\mu}\pi^*\nabla_{\nu}\pi^* \left( R^{\mu\nu} - \frac{1}{4}g^{\mu\nu}R \right) + \rm{c.c.} \right]
\end{align}
This is one of the key result regarding which we would like to stress the following: First, the non-minimal coupling terms in \eqref{non-min} is different for the complex scalar as compared to the non-minimal terms that arises for the real scalar field~\cite{2009-Deffayet.etal-PRD}. Second, as expected, the non-minimal coupling terms vanish and the above action matches with flat space time action \eqref{action_CSG}. Third, in the limit of $\pi = \pi^*$, $S_4^{\rm min} $ and $S_4^{nm}$ reduce to:
\begin{subequations}
\label{eq:S4pi}
\begin{align}
\left. {S}_4^{min} \right|_{\pi = \pi^*}  &= \frac{\omega}{\lambda_{\pi}^6}\int d^4x\,\,\sqrt{-g}\,\, \left[\, (\Box\pi)^2 \,\nabla_{\alpha}\pi\, \nabla^{\alpha}\pi + \Box \pi\, \nabla_{\nu}\pi \, \nabla^{\nu}\nabla^{\alpha}\pi \,\nabla_{\alpha}\pi  \right.
\nonumber\\
\label{eq:S4minpi}
& ~~ - \left. \,\nabla_{\mu}\nabla_{\nu}\pi\, \nabla^{\mu}\nabla^{\nu}\pi\, \nabla_{\alpha} \pi\, \nabla^{\alpha}\pi 
 - \nabla^{\mu}\nabla^{\nu}\pi\, \nabla_{\nu}\nabla_{\alpha}\pi\,\nabla_{\mu}\pi\, \nabla^{\alpha}\pi\, \right] \\
\left. {S}_4^{nm} \right|_{\pi = \pi^*}  &=
- \frac{\omega}{4 \, \lambda_{\pi}^6}\int d^4x\,\,\sqrt{-g}\,\, \nabla_{\alpha}\pi\nabla^{\alpha}\pi \nabla_{\mu}\pi \nabla_{\nu}\pi g^{\mu\nu}R \nonumber\\
\label{eq:S4nmpi}
&{}   - \frac{\omega}{2 \lambda_{\pi}^6} \int d^4x\,\,\sqrt{-g}\,\,\nabla_{\alpha}\pi \nabla^{\alpha}\pi \nabla_{\mu}\pi\nabla_{\nu}\pi\,\,\left( R^{\mu\nu} - \frac{1}{4}g^{\mu\nu}R \right)
\end{align}
\end{subequations}
Although the above action looks different compared to that of  Deffayet et al \cite{2009-Deffayet.etal-PRD}, it is possible to show that the two actions are related by a boundary term. In Appendix \eqref{sec:consistencyRealGal}, we have explicitly shown that the above action for real scalar field is identical to the action used in Ref.~\cite{2009-Deffayet.etal-PRD}. Hence, the equations of motion from the above action matches with the equations of motion derived in Ref. \cite{2009-Deffayet.etal-PRD}.

\subsection{Coupling to the electromagnetic field}
\label{G}

Like in the flat space-time, the action \eqref{action_curved} is 
invariant under the global transformation, i.e. $\pi \to \pi e^{-i\, e \, \theta}$, where $\theta$ is a constant parameter and $e$ is the electric charge. In order for the action to be invariant under the 
local gauge transformation, we can replace $\nabla_{\mu} \to \mathcal{D}_{\mu} \equiv \nabla_{\mu} + ieA_{\mu}$ in action \eqref{action_curved}. Beside this, the electromagnetic field will have additional Galileon terms that vanish in the flat space-time~\cite{2017-Debottam.Shankaranarayanan-JCAP}. The complete Galileon scalar electrodynamics action in curved space time is given by
\begin{align}\label{action_full_curved}
{S}^{\rm{G}}_{Curved} &=  \frac{1}{2}\int d^4x\sqrt{-g} \mathcal{D}_{\mu}\pi \mathcal{D}^{\mu}\pi^*+\frac{\omega}{2 \lambda_{\pi}^6}\int d^4x \sqrt{-g}[\,\,2 \mathcal{D}_{\mu}\mathcal{D}^{\mu} \pi^* \mathcal{D}_{\nu}\mathcal{D}^{\nu} \pi \mathcal{D}_{\alpha}\pi \mathcal{D}^{\alpha}\pi^*\nonumber\\&{}- 2 \mathcal{D}_{\mu}\mathcal{D}_{\nu}\pi^* \mathcal{D}^{\mu}\mathcal{D}^{\nu}\pi \mathcal{D}_{\alpha} \pi \mathcal{D}^{\alpha}\pi^* + (\, \mathcal{D}_{\mu}\mathcal{D}^{\mu} \pi \mathcal{D}_{\nu}\pi^*  \mathcal{D}^{\nu}\mathcal{D}^{\alpha}\pi \mathcal{D}_{\alpha}\pi^*  
\nonumber\\&{}
 - \mathcal{D}^{\mu}\mathcal{D}^{\nu}\pi \mathcal{D}_{\nu}\mathcal{D}_{\alpha}\pi\mathcal{D}_{\mu}\pi^* \mathcal{D}^{\alpha}\pi^* + \rm{c.c.}\,\,)\, ]
 \nonumber\\&{}
 - \frac{\omega}{4 \lambda_{\pi}^6}\int d^4x\,\,\sqrt{-g}\,\, \left[\mathcal{D}_{\alpha}\pi\mathcal{D}^{\alpha}\pi \mathcal{D}_{\mu}\pi^*\mathcal{D}_{\nu}\pi^* \left( R^{\mu\nu} - \frac{1}{4}g^{\mu\nu}R \right) + \rm{c.c.}\right]
\\&{}  
  -\frac{\omega}{4 \lambda_{\pi}^6}\int d^4x\,\sqrt{-g}\, \mathcal{D}_{\alpha}\pi\mathcal{D}^{\alpha}\pi^* \mathcal{D}_{\mu}\pi \mathcal{D}_{\nu}\pi^*g^{\mu\nu}\,R -\frac{1}{4}\int d^4x \sqrt{-g} F_{\mu\nu}F^{\mu\nu} + S_{\rm{VEG}} \nonumber
\end{align}
where the last term $S_{\rm{VEG}}$ is the vector Galileon action obtained in Ref.~\cite{2017-Debottam.Shankaranarayanan-JCAP}. 
We have listed the terms in Appendix \eqref{app:D}. We will use this 
action to study the effects of the Galileon term in the early Universe.
\subsection{Fixing the value of $\omega$}
\label{sec:coeffiFixing}

$\lambda_{\pi}$ is the new coupling constant of the model and can only be fixed with observations. As mentioned in Sec. \eqref{sec:CSG}
$\omega$ can take either $+1$ or $-1$. In this subsection, we fix the value of $\omega$ by evaluating the energy density corresponding to the minimal and non-minimal terms in \eqref{action_full_curved} in the Coulomb gauge ($A^0 = 0, \partial_i A^{i} = 0$). 

To make the calculations transparent, we evaluate the energy density 
in the FRW background which includes arbitrary Lapse function $N(t)$: 
\begin{equation}
\label{eq:FRW}
ds^2 = N^2(t) \, dt^2 - a^2(t)(dx^2 + dy^2 + dz^2)
\end{equation}
where $a(t)$ is the scale factor. To satisfy the homogeneity and isotropy of the FRW background, the Galileon scalar electrodynamics must satisfy the condition $A^i = 0$. The equation of motion of $\pi$ corresponding to $S_4^{\min} + S_4^{nm}$ is given by (using Eq.(\ref{eq:S4eomCurved})):
\begin{align}\label{EOM_FRW}
2\dot{\pi}^*\,\ddot{\pi}^*\dot{\pi}\left(\frac{\dot{a}}{a}\right)^2-2(\dot{\pi}^*)^2\,\dot{\pi}\left(\frac{\dot{a}}{a}\right)^2\frac{\dot{N}}{N} +33(\dot{\pi}^*)^2\dot{\pi}\left(\frac{\dot{a}}{a}\right)^3 + (\dot{\pi}^*)^2\ddot{\pi}\left(\frac{\dot{a}}{a}\right)^2\nonumber\\ -(\dot{\pi}^*)^2\dot{\pi}\left(\frac{\dot{a}}{a}\right)^2 \frac{\dot{N}}{N}+ 8(\dot{\pi}^*)^2\dot{\pi}\left(\frac{\ddot{a}}{a}\right)\left(\frac{\dot{a}}{a}\right) -8(\dot{\pi}^*)^2\dot{\pi}\left(\frac{\dot{a}}{a}\right)^2\left(\frac{\dot{N}}{N}\right)= 0
\end{align}

where $H(t) = \dot{a}(t)/a(t)$. Varying the action $ {S}_4^{\rm{min}} + {S}^{\rm{nm}}_4$ with respect to the $g^{00} = N^{-2}(t)$ leads to:
\begin{align}
\delta {S}_4^{\rm{min}} + \delta {S}^{\rm{nm}}_4 &= - \frac{3\omega}{2 \lambda_{\pi}^6} \int d^4 x \frac{a^3}{N^2}
\left[ 79\, ({\dot{\pi}}^*)^2 \,\dot{\pi}\ddot{\pi} \, H(t) 
 + 79\, ({\dot{\pi}})^2 \,\dot{\pi}^*\ddot{\pi}^* \, H(t) \right.\nonumber\\&{}\hspace{1.5cm}\left.
 + 20 \, ({\dot{\pi}}^*)^2 \,(\dot{\pi})^2 H^2(t) 
 +23\, ({\dot{\pi}}^*)^2 \,(\dot{\pi})^2 \left(\frac{\ddot{a}}{a}\right)\right]\,\, \delta N
\end{align}

Using the definition, 
\begin{equation}
T_{0 0} = \frac{2}{\sqrt{-g}}\frac{\delta S}{\delta g^{0 0}} 
= - \frac{N^2}{a^3} \frac{\delta S}{\delta N} \, ,
\end{equation}
we get,
\begin{align}\label{energyd}
\rho =\,\,T^0_0 =  \rho = \frac{3 \omega }{2\,\lambda_{\pi}^6}\,\, \left[ \frac{79}{2}\,\partial_{0}N_1 \,H + 20\,N_1 \,H^2 + 23\,\frac{\ddot{a}}{a}\,N_1\,\,\right]
\end{align}
where $N_1 = |\dot{\pi}|^4$.  In order to get the value of $\omega$ we simplify the analysis further by setting $N(t) = 1$ and taking the limit, $\pi = \pi^*$. Thus, the energy density and equation of motion reduce to
\begin{align}
\label{energy_back}
& \rho =  \frac{3 \omega}{2 \lambda_{\pi}^6} \left[ 2\times79\, ({\dot{\pi}})^3 \,\ddot{\pi} \,H + 20 \, (\dot{\pi})^4 \,H^2+23\, (\dot{\pi})^4 \,(H^2 + \dot{H})\right] \\
\label{eom_back}
& \ddot{\pi} +\frac{41}{3}\,\dot{\pi}H + \frac{8}{3}\,\dot{\pi}\left(\frac{\dot{H}}{H}\right) = 0
\end{align}
Substituting Eq.~(\ref{eom_back}) in Eq.~(\ref{energy_back}), we get
\begin{align}
\rho = -\frac{\omega \dot{\pi}^4\,H^2}{2 \lambda_{\pi}^6}\left[ 6349 + 1195\,\frac{\dot{H}}{H^2} \right]
\end{align}
Note that $\dot{H}/H^2 = - \epsilon$ which is a slow-roll parameter. 
During inflation, when the non-linear terms contribution can not be ignored, $\epsilon < 1$, hence the quantity in the square bracket, is positive. The condition that the energy density is positive implies that $\omega = -1$. 

%the sign of coefficient $\lambda^6$ must be negative because energy density is a positive definite quantity. Hence the original coefficient $A_2$ should be negative. So we can set $\omega = -1$, the expression for energy density becomes
%\begin{align}\label{rho_final}
%\rho = \frac{ \dot{\pi}^4\,H^2}{2 \lambda_{\pi}^6}\left[ 6349 + 1195\,\frac{\dot{H}}{H^2} \right]
%\end{align}
%After fixing the sign of the coefficient at the background. We write the energy density expression in terms of $\pi$ and $\pi^*$ and some convenient quantities which will be useful for further calculations. These quantities are defined as
%\begin{align}
%\alpha_1 = (\dot{\pi}^*)^2 \dot{\pi}\,\ddot{\pi}\,a, \,\,\, \alpha_2 = (\dot{\pi})^2 \dot{\pi}^*\,\ddot{\pi}^*\,a,\,\,\, \alpha_3 = (\dot{\pi}^*)^2 \,(\dot{\pi})^2\,\dot{a}, \,\,\, \alpha_4 = (\dot{\pi}^*)^2 \,(\dot{\pi})^2\,a
%\end{align}
%using these relations we get, $ (\dot{\pi}^*)^2 \,(\dot{\pi})^2\,\ddot{a}\,a = a\,\partial_{0}\alpha_3 - 2\dot{a}\,(\alpha_1 + \alpha_2)$ and $\partial_{0}\alpha_4 = 2\,\alpha_1 + 2\,\alpha_2 + \alpha_3$, so the energy density in Eq. (\ref{energyd}) becomes
%\begin{align}\label{rho}
%\rho = \frac{3 \omega}{2 \lambda_{\pi}^6 \, a} \left[ 33\, \alpha_1\,H + 33\, \alpha_2\,H + 20\,\alpha_3 \,H+ 23\,\partial_{0}\alpha_3\right]
%\end{align}

%%%%%%55
\section{Applications to Early-Universe}
\label{sec:energy_pressure}

To extract interesting features of the Galileon model, in this section, we show that the action \eqref{action_full_curved} leads to accelerated expansion in the early Universe. We obtained the energy density of the Galileon field in \eqref{energy_back}. Using the same procedure, the 
pressure of the Galileon field is given by:
\begin{align}\label{pressure}
p &= \frac{\omega }{2\,\lambda_{\pi}^6}\,\,[\,\,17 \dot{N_1} \,H- 72\, N_1 \,H^2+ 8\, \frac{\ddot{a}}{a}\,N_1  + 3\,\ddot{N_1} \,\,]
%p  = - \frac{1}{\lambda_{\pi}^6}\,\,[\,\, 
%6 N_2 \,H + 6 N_3 \,H -3 6 N_1 \,H + 
%3\, \partial_t \left(N_2 \, a\right) + 
%3\, \partial_{t} \left(N_3 \, a \right) + 
%4\, \partial_{t} \left(N_1 \dot{a} \right)\,\,]
\end{align}
%%
%where 
%\begin{align}\label{N123}
%N_2 &= (\dot{\pi}^*)^2 \dot{\pi}\,\ddot{\pi};~~
%N_3 =(\dot{\pi})^2 \dot{\pi}^*\,\ddot{\pi}^*;
%\end{align}
%%
Taking the case that the phase variation of the complex field is almost constant, the Friedmann equation becomes:
\begin{align}\label{eq:slow-roll}
1 = \frac{\dot{\pi}^4}{6\,M_P^2\,\lambda_{\pi}^6}\left[ 6349 -1195\,\epsilon \right]
\end{align}
Inverting the above equation, we get 
\begin{align}\label{eq:epsilon}
\epsilon = \frac{6349}{1195}\left[1 - \Gamma\right] \quad 
\mbox{where} \quad \Gamma = 
\frac{6\,M_P^2\,\lambda_{\pi}^6}{6349\,\dot{\pi}^4}\,
\end{align}
In order for $\epsilon$ to be less than unity,  $\Gamma \simeq 1$. 
To identify the parameter space of $(\dot{\pi}, \lambda_{\pi})$ that can lead to $\Gamma \simeq 1$, we define $\dot{\pi} \equiv \Delta_P \,M_P^2, \lambda_{\pi} \equiv \delta_P \, M_P$ where $\Delta_P$ and $\delta_P$ are dimensionless quantities. \eqref{fig:Plot} contains 
the plot of $\log \Gamma$ as a function of $\Delta_P$ for different values of $\delta_P$. We notice the following: First, the value of $\Gamma$ is weakly dependent on $\delta_P$. In other words, for different values of $\delta_P$, the value of $\Gamma$ is almost the same. Second, for a large values of $\Delta_P$, $\Gamma$ is almost close to unity and hence, the slow-roll parameter is very small. More specifically, for $\Delta_P$ in the range $10^{-5}$ to $0.3$, the slow-roll parameter is of the order of $0.1$. Thus, for large parameter range, the model can lead to extended period of inflation. 

\begin{figure}[h]
	\centering
	\includegraphics[width=0.85\textwidth]{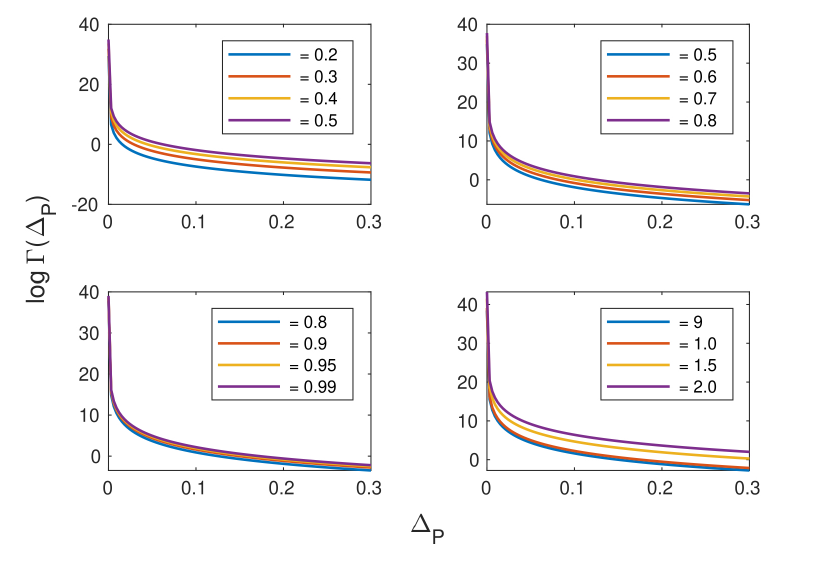}
	\caption{Plot of $\log\Gamma(\Delta_P) $ versus $\Delta_P$ for different values of $\delta_P$. Note that for a range of values of the two parameters, $\Gamma$ is also zero.}
	\label{fig:Plot}
\end{figure}

\section{Conclusions and Discussions}
\label{sec:conc}

In this chapter, we constructed Galileon scalar electrodynamics action, which preserves Galilean symmetry and local gauge invariance. Due to the complex scalar field, the number of Galilean symmetry 
invariant terms are reduced. For an earlier result, in a different context, see Ref. \cite{2012-Goon.etal-PLB}. In the flat space-time, we have explicitly shown that the equations of motion are second order. 
In curved space-time, due to the non-commutative nature of the covariant derivatives, the minimal coupling of the matter and gravity term leads to higher-derivatives in the equation of motion. We introduced non-minimal coupling terms to the Galileon field that makes the equations of motion second-order in an arbitrary curved space-time. The non-minimal coupling term is different compared that the one used in the literature for the real scalar field~\cite{2009-Deffayet.etal-PRD}. However, in the real scalar field limit, the complex scalar Galileon action is identical to the real scalar Galileon action obtained by Deffayet et al~\cite{2009-Deffayet.etal-PRD}.

As an application of the model, we considered the case when the Galileon scalar electrodynamics dominated the early Universe. We have shown that for a range of parameters, the model leads to inflation. One possible application of the model is to study the perturbations generated during the inflation dominated by the model. 
Since the model breaks conformal invariance, this can generate primordial magnetic fields~\cite{2017-Debottam.Shankaranarayanan-JCAP}. 
%We hope to report this in more detail shortly.

The model has two free parameters -- $\lambda_{\pi}$ and $\lambda_{VG}$. In the flat FRW background, the electromagnetic field vanishes and hence the scale of inflation fixes $\lambda_{\pi}$. 
The other parameter $\lambda_{VG}$ will play a critical role in the first order perturbations. In principle, the two parameters can be different.

The model presented here can be part of the scalar-vector-galileon (SVG) theories. However, as mentioned earlier, not all SVG gravity theories are gauge-invariant, while our model is gauge-invariant~\cite{2018-Heisenberg.etal-PRD}. 
The complexification of a gauge-invariant subset of the SVG theories should be identical to our model with an appropriate choice of coefficients $A_{i}, B_{i}, C_{i}, D_{i}$ (for $i=1,2$) and $D_{3}$.

\chapter{Helical magnetic field from Riemann coupling}
\label{ch:helical-PMF}

Effective field theory (EFT) now forms a standard tool in early-Universe cosmology~\cite{2017-Burgess-arXiv,2014-Agarwal.etal-JCAP,2008-Weinberg-PRD,2007-Cheung.etal-JHEP}. Effective field theories rely on the separation of the energy scales of interest for observations and the underlying physics of the early Universe near the singularity~\cite{1994-Donoghue-PRD,1996-Donoghue.etal,2003-Burgess-LivRev}. In an effective field theory approach, the usual requirement of renormalizability is too strong. Instead, it demands that a finite number of parameters describe the physics up to effects suppressed by $(E/M)^n$ where $ E $ is the energy of the particles corresponding to the quantum fields, and $ M $ is the energy scale below which the effective field theory description is valid. As the value of $n$ increases, more parameters are required to describe the Physics using EFT. 

Effective field theory description for gravity has shown that gravity and quantum mechanics can be compatible with the energies that have been experimentally probed~\cite{Donoghue:2015hwa}. EFT has been successfully applied to inflationary cosmology, especially for the single scalar field inflation model were the constants in the higher derivative terms of the effective Lagrangian take values that are powers of $M$, with coefficients roughly of order unity~\cite{2008-Weinberg-PRD,2007-Cheung.etal-JHEP}. However, these EFTs aim to obtain a generic prediction for density perturbations in a single-scalar field inflationary models~\cite{2007-Cheung.etal-JHEP,2019-Bastero-Gil.etal-arXiv}. Hence, the analyses can not be extended to other fields (like electromagnetic fields) during inflation.

Recently, the general effective field theory of gravity coupled to the Standard Model of particle physics was constructed~\cite{2019-Ruhdorfer.etal-JHEP}. The authors systematically showed that the first gravity operators appear at mass dimension $6$ in the series expansion, and these operators only couple to the standard model Bosons. They also showed that (i) no new gravity operators appear at mass dimension $7$, (ii) in mass dimension $8$ the standard model Fermions appear, and (iii) coupling between the scalar (Higgs) field and the standard model gauge Bosons appear \emph{only} at mass dimension $8$. (Note that these corrections do not include higher derivative (Galileon) terms in the electromagnetic action~\cite{2017-Debottam.Shankaranarayanan-JCAP,2019-Kushwaha.Shankaranarayanan-PRD}.)

In chapter~(\ref{ch:EFTmagnetogenesis}), we will discuss a generic framework---\emph{effective field theory} (EFT) of magnetogenesis to generate the primordial non-helical magnetic fields. We will show that our EFT construction can reproduce a large class of models proposed in the literature. In this chapter, we limit to mass dimension $6$ operators coupling to the gauge field, specifically, to the electromagnetic field. We concentrate on the coupling of the electromagnetic field with the dual Riemann tensor $\tilde{R}^{\mu \nu \rho \sigma} ( \equiv \epsilon^{\mu \nu \alpha \beta} R_{\alpha \beta}^{~~~\rho \sigma} / 2$). We show that such a term leads to the generation of a helical magnetic field during inflation~\cite{2001-Vachaspati-PRL}. 

As mentioned in Chapter~\ref{ch:intro}, magnetic fields have been observed at all scales in the Universe; however, there is no compelling model of the origin of large scale magnetic fields. Observations from Faraday rotation and synchrotron radiation show the presence of micro-Gauss strength magnetic fields in the galaxies and the clusters of galaxies~\cite{1994-Kronberg-Rept.Prog.Phys.,2001-Grasso.etal-PhyRep,2001-Grasso.etal-PhyRep,2002-Widrow-Rev.Mod.Phys.,2013-Durrer.Neronov-Arxiv,2016-Subramanian-Arxiv}. 

While the magnetic field measurements from Faraday rotation and synchrotron radiation provide upper bounds of the magnetic fields, the FERMI measurement of gamma-rays emitted by blazars provides a lower bound of the order of $10^{-15}$ G in intergalactic voids~\cite{2010-Neronov.Vovk-Sci}. However, inflation can not generate these large scale magnetic as the standard 4-D electromagnetic action is conformally invariant. To amplify the quantum fluctuations of the electromagnetic fields in the early Universe,  
one needs to break the conformal invariance of the action~\cite{1988-Turner.Widrow-PRD,2002-Widrow-Rev.Mod.Phys.,2013-Durrer.Neronov-Arxiv,2016-Subramanian-Arxiv,2018-Subramanian-arXiv}. 

While mechanisms to generate non-helical fields were proposed four decades ago~\cite{1988-Turner.Widrow-PRD,1991-Ratra-Apj.Lett,1993-Dolgov-PRD}, the generation of helical fields is recent~\cite{2001-Vachaspati-PRL,2003-Caprini.etal-PRD,2005-Campanelli-Giannotti-PRD}. 
One of the interests in primordial magnetic helicity is that it can be a direct indication of parity violation in the early Universe and may be related to the matter-antimatter asymmetry in the early Universe~\cite{1996-Davidson-PLB}. Besides, the conservation of helicity leads to an inverse cascade in the turbulent plasma era, that can move power from small to large scales~\cite{2003-Caprini.etal-PRD}. Hence, the decay rate of energy density and coherence length is slower than the non-helical fields during these epochs. It has been shown that helical magnetic fields will leave distinct signatures in CMB, such as TE- and EB-cross-correlations~\cite{2006-Kahniashvil-NAR,2018-Chowdhury.etal-JCAP}.

As mentioned above, there has been a lot of interest in generating a primordial helical field in the early Universe~\cite{2001-Vachaspati-PRL,2003-Caprini.etal-PRD,2005-Campanelli-Giannotti-PRD,2018-Sharma.Subramanian.Seshadri.PRD,2009-Caprini.Durrer.Fenu-JCAP,2009-Campanelli-IJMPD,2019-Shtanov-Ukr.PJ}. However, most models introduce non-minimal coupling of the electromagnetic fields with a (pseudo-)scalar. While this leads to the breaking of conformal invariance of the electromagnetic field, due to non-minimal coupling, extra degrees of freedom are present at all energies. More importantly, these extra degrees of freedom propagate even at the low-energy and can potentially lead to the strong coupling problem~\cite{2009-Demozzi.etal-JCAP}. It is possible to overcome the strong-coupling problem for a narrow range of coupling functions~\cite{2018-Sharma.Subramanian.Seshadri.PRD}.

In this chapter, to avoid the strong-coupling problem and restricted coupling functions, we propose a model that couples the electromagnetic fields with the Riemann tensor. To our knowledge, Riemann tensor coupling has not be discussed in the literature to generate helical fields. The model has three key features: First, it does not require the coupling of the electromagnetic field with the scalar field. Hence, there are no extra degrees of freedom and will not lead to a strong-coupling problem. Second, the conformal invariance is broken due to the coupling to the Riemann tensor. Since the curvature is large in the early Universe, the coupling term will introduce non-trivial corrections to the electromagnetic action. However, at late-times, the new term will not contribute, and the theory is identical to standard electrodynamics. {The power-spectrum of the fields has slight red-tilt for slow-roll inflation}. Third, as we show explicitly, our model is free from backreaction for a range of scale-factor during inflation. This is different from other models where a specific form of coupling function is chosen to avoid any back-reaction~\cite{2018-Sharma.Subramanian.Seshadri.PRD}. 

In Sec. \eqref{sec:Model}, we introduce the model and discuss its properties. We discuss the classical properties and define the relevant quantities. We also briefly discuss the procedure to quantize these fields in the FRW background. In Sec. \eqref{sec:Helical}, {we explicitly evaluate the helical magnetic field generation in our model and show that the power-spectrum is red-tilted.} This is different compared to the other models in the literature. We also show that the model does not have a backreaction problem for a range of scale-factor. We discuss the implications of the results in Sec. \eqref{sec:conc}. Appendix~\ref{ap:helical_pmf} contains the details of the calculations in this chapter. The results reported in this chapter are based on Ref.~\cite{2020-Kushwaha.Shankaranarayanan-PRD}
%
%%%%%%%%%%  S E C T I O N %%%%%%%%%%%%%%%%%%%%
%
%
\section{The model}
\label{sec:Model}

We consider the following action:
\begin{align}\label{eq:action}
S  = S_{\rm{Grav}} + S_{\phi} + S_{\rm{EM}} + S_{\rm CB}
\end{align}
where $ S_{\rm{Grav}}$ is the Einstein-Hilbert action
\begin{align}\label{eq:EH-action}
S_{\rm Grav} = -\frac{M_{\rm P}^2}{2}\int d^4x \sqrt{-g} \, R \, ,
\end{align}
and $ S_{\phi} $ is the action for the minimally coupled, self-interacting canonically scalar field:
\begin{align}\label{eq:inflation-action}
S_{\phi} = \int d^4x \sqrt{-g} \left[  \frac{1}{2} \partial_{\mu}\phi \partial^{\mu}\phi -  V(\phi) \right].
\end{align}
We assume that the scalar field ($\phi$) dominates the energy density in the early Universe (during inflation) and leads to $60 \, - \, 70$ e-foldings of inflation with $H \simeq 10^{14} {\rm GeV}$. $S_{\rm{EM}}, S_{\rm CB}$ refer to the standard electromagnetic (EM)  and conformal breaking part of the electromagnetic terms, respectively, and given by:
\begin{align}\label{eq:S_EM}
 S_{\rm{EM}} &= -\frac{1}{4} \int d^4x \, \sqrt{-g} \, F_{\mu\nu} F^{\mu\nu}, \hspace{0.5cm}\\  
 \label{eq:S_h}
 S_{\rm{CB}} &= - \frac{\sigma}{M^2} \,\int d^4x \, \sqrt{-g} \, R_{\rho\sigma}\,^{\alpha\beta} F_{\alpha\beta} \, \tilde{F}^{\rho\sigma} = - \frac{\sigma}{M^2} \,\int d^4x \, \sqrt{-g} \, \tilde{R}^{\mu\nu\alpha\beta} F_{\alpha\beta} \, F_{\mu\nu} \, ,
 \end{align}
where $R_{\rho\sigma}\,^{\alpha\beta}$ is the Riemann tensor and its dual is $\tilde{R}^{\mu\nu\alpha\beta} = \frac{1}{2}\epsilon^{\mu\nu\rho\sigma} R_{\rho\sigma}\,^{\alpha\beta}$, $A_{\mu}$ is the four-vector potential of the electromagnetic field, $F_{\mu\nu} = \nabla_{\mu}A_{\nu} - \nabla_{\nu}A_{\mu} $ and $\tilde{F}^{\rho\sigma} = \frac{1}{2} \epsilon^{\mu\nu\rho\sigma}F_{\mu\nu} $ is the dual of $F_{\mu\nu}$. $\epsilon^{\mu\nu\rho\sigma} = \frac{1}{\sqrt{-g}}\, \eta^{\mu\nu\rho\sigma}$ is fully antisymmetric tensor, $\eta^{\mu\nu\rho\sigma}$ is Levi-Civita symbol whose values are $\pm1$ and we set $\eta^{0123} = 1 = - \eta_{0123}$. 

The standard electromagnetic action $S_{\rm{EM}}$ is conformally invariant; however, the presence of Riemann curvature in $S_{\rm CB}$ breaks the conformal invariance. $M$ is the energy scale, which sets the scale for the breaking of conformal invariance. For our discussion below, we assume that
$10^{-3}  \leq (H/M) \leq 1$~\cite{2018-Nakonieczny-JHEP,2016-Goon.Hinterbichler-JHEP,2016-Goon-JHEP,2013-Balakin.etal-CQG}.  Due to the Riemann tensor, in FRW background, $M$ appears as a time-dependent coupling~\cite{1971-Prasanna-PLA}. To see this, let us evaluate Riemann tensor for the flat FRW background \eqref{eq:FRW}: 
\[
R_{\mu\nu}\,^{\sigma\gamma} \sim \frac{{a^{\prime}}^2}{a^4} ~\rm{or}~ \frac{a^{\prime\prime}}{a^3}
\]
Thus, the coupling function is time-dependent, i. e.,
\begin{align}\label{eq:coupl-I_for_constant_M}
\frac{1}{M_{\rm eff}} \sim \frac{1}{M} \frac{a^{\prime}}{a^2} \, .
% \propto \frac{1}{\eta^{2\beta + 4}}. 
\end{align} 
Before proceeding with the analysis, we want to highlight the following salient features of this model compared to the earlier models that introduce non-minimal scalar field coupling in action: First, our model does not require the coupling of the electromagnetic field with the scalar field. Hence, there are no extra degrees of freedom and will not lead to a strong-coupling problem. Second, the conformal invariance is broken due to the coupling to the Riemann tensor. Since the curvature is significant in the early Universe, the coupling term will introduce non-trivial corrections to the electromagnetic action. However, at late-times, $S_{\rm CB}$ will not contribute, and the model is identical to standard electrodynamics.  Third, as we show explicitly, our model is free from backreaction for a range of scale-factor during inflation. This is different from other models where a specific form of coupling function is chosen to avoid any back-reaction~\cite{2018-Sharma.Subramanian.Seshadri.PRD}. 

As mentioned earlier, we aim to generate the helical magnetic field during inflation. Hence, the scalar field's energy density dominates over the standard electromagnetic and conformal breaking term in action \eqref{eq:action}.
Since the single-scalar field inflation can not generate vector perturbations, the magnetic field generated will be due to the conformal breaking term $S_{\rm CB}$ in action.
 
The variation of the action (\ref{eq:action}) with respect to gauge field $A_{\mu}$ leads to the following equation:
\begin{align}\label{eq:eom}
\partial_{\mu} \left( \sqrt{-g} \, F^{\mu\nu} + \frac{1}{M^2}  \sqrt{-g} \,  \epsilon^{\alpha\beta\rho\sigma} R_{\rho\sigma}\,^{\mu\nu}   \, F_{\alpha\beta} +  \frac{1}{M^2} \sqrt{-g}   \, \epsilon^{\mu\nu\rho\sigma} R_{\rho\sigma}\,^{\alpha\beta} \, F_{\alpha\beta} \,  \right) = 0
\end{align}
As mentioned above, we will consider a flat  Friedman universe described by the line-element:
\begin{align}\label{eq:FRW}
ds^2 = a^2(\eta) \,(d\eta^2 - \delta_{ij} dx^i dx^j)
\end{align}
where $\eta$ is the conformal time. For $\nu = i$, Eq.~(\ref{eq:eom}) reduces to: 
\begin{align}\label{eq:eom1}
 \partial_{0} \left(F_{0 i} + \frac{2}{M^2}  \,  \eta^{0 i j k}  \frac{a^{\prime\prime} }{a^3} \,F_{j k} \,  \right) 
- \partial_{l} \left( \, F_{l i} - \frac{4}{M^2}  \,  \eta^{0 i l j}  \frac{  a^{\prime\prime} }{a^3}  F_{j0} \,\right)   = 0  
\end{align}
where we have substituted the following components of Riemann tensor: 
\[
R_{i j}\,^{k l} =  \frac{    {a^{\prime} }^2}{a^4} \, \left( \delta^l_i \,  \delta^k_j  - \delta^k_i \,  \delta^l_j  \right),  R_{0 i}\,^{0 j} =  \left( \frac{ {a^{\prime}}^2}{a^4} -\frac{a^{\prime\prime}}{a^3} \right) \, \delta^j_i \, .
\]
In the Coulomb gauge ($A^{0} = 0, \partial_iA^i = 0$), and using $\eta^{0 i j l} = \epsilon_{i j l}$ (where $\epsilon_{i j l}$ is the Levi-Civita symbol in 3D Euclidean space), the above equation motion (of $A^i$) leads to the following evolution equation:
\begin{align}\label{eq:equation_of_motion}
A_i^{\prime\prime} + \frac{4 \, \epsilon_{i j l}}{M^2} \, \left( \frac{a^{\prime\prime\prime}}{a^3} - 3\frac{a^{\prime\prime} a^{\prime} }{a^4} \right) \partial_j A_l 
- \partial_j \partial_j A_i = 0
\end{align}
Note that the above equation differs from other models in the literature by an overall factor in the term with $\epsilon_{i j l} \partial_j A_l$, especially, the third derivative of the scale factor. Thus, the model can lead to different evolution of the fluctuations in comparison to non-minimally coupled scalar field models.

In this chapter, we consider two inflationary scenarios --- power-law inflation and the slow-roll inflation --- to evaluate the power-spectrum of the electromagnetic fluctuations. For power law inflation the scale factor (in cosmic time) is given by  $a(t) = a_0 t^p$ where $p>1$ and $a_0$ is arbitrary constant. In the conformal time, the scale factor is~\cite{2004-Shankaranarayanan.Sriramkumar-PRD}:
 \begin{align}\label{eq:powerLaw}
a(\eta) =  \left( - \frac{\eta}{\eta_0} \right)^{(\beta+1)}
\end{align}
where $\eta_0$ is an arbitrary constant and denotes the scale of inflation. During inflation, $\eta \in (-\infty, 0)$.  $\beta$ and $\eta_0$ are given by:
\begin{align}\label{eq:mathcalH}
\beta = - \left( \frac{2p - 1}{p - 1} \right)  \hspace{0.5cm} \text{and} \hspace{0.5cm} \eta_0 = \left[ (p - 1)a_0^{1/p}  \right]^{-1}.
\end{align}
Note that $\beta \leq -2$ and $\beta = -2$ corresponds to the de Sitter. The Hubble parameter ($\mathcal{H} \equiv {a^{\prime}(\eta) }/{a(\eta)}$) is given by:
 \begin{align}\label{eq:H-eta_relation}
 \mathcal{H} = \frac{a^{\prime} }{a } = \frac{\beta + 1}{\eta} \implies \eta = \frac{\beta + 1}{\mathcal{H}}
 \end{align}
Slow-roll inflation is a generic inflationary paradigm that leads to an accelerated expansion independent of a particular model (or potential). In this case, we have 
\begin{equation}
\beta \approx -2-\epsilon, \quad \mathcal{H} \approx - \frac{1 + \epsilon}{\eta}   
\end{equation}
where $\epsilon $ is the slow roll parameter. 

\subsection{Physical quantities of interest}

Although, the four-vector potential $A^{\mu}$ provides the covariant
description of the electromagnetic processes, to compare with observations, we need to decompose the physical quantities in terms of the electric and magnetic fields, that are intrinsically frame-dependent. Hence, it is always useful to define a comoving observer with velocity $u^{\mu} = \left( 1/a(\eta),0, 0, 0 \right)$ satisfying $u_{\mu} u^{\mu} = 1$. The electric and magnetic field four-vector for this observer is given by projecting the EM field tensor with $u^{\mu}$ as
\begin{align}
E_{\mu} = u^{\alpha} F_{\alpha\mu}, \hspace{.5cm}
B_{\mu} = \frac{1}{2} \,u^{\gamma} \, F^{\alpha\beta} \epsilon_{\gamma\alpha\beta\mu}   = u^{\alpha}\tilde{F}_{\alpha\mu} .
\end{align}
Note that the electric and magnetic field four-vectors are both three-vector fields in a sense that they are orthogonal to the comoving observer, i.e., $E_{\mu} u^{\mu} = 0 = B_{\mu} u^{\mu}$, and we have:
\begin{align}
E_{\mu} =  a(\eta) \left( 0, \textbf{E} \right), \quad
B_{\mu}  =  a(\eta) \left( 0, \textbf{B} \right) 
\end{align}
where 
\[
 \textbf{E} =   \frac{A_i^{\prime}}{a^2(\eta)},  \quad  \textbf{B} = -\frac{1 }{a^2(\eta)}\,  \epsilon_{ijk}  \, \partial_i A_j \, . 
 \]
Electromagnetic energy densities are defined as
\begin{align}
\rho_B \equiv - \frac{1}{2} B_{\mu} B^{\mu} = \frac{1}{2}  \textbf{B} \cdot \textbf{B}, \qquad
\rho_E \equiv - \frac{1}{2} E_{\mu} E^{\mu} = \frac{1}{2}  \textbf{E} \cdot \textbf{E}.
\end{align}
and the magnetic helicity density is
\begin{align}
\rho_h \equiv - A_{\mu} B^{\mu}.
\end{align}
We will evaluate these quantities for the quantum fluctuations generated during inflation.

\subsection{Quantization in the Helicity basis}
In this section, we briefly discuss the evolution of the quantum fluctuations of the electromagnetic field in the helicity basis~~\cite{2018-Sharma.Subramanian.Seshadri.PRD}.  Decomposition of the vector potential in Fourier space, we have:
\begin{align}\label{eq:FourierT}
A^{i}(\vec{x}, \eta) =  \int \frac{d^3 k}{(2\pi)^3} \sum_{\lambda = 1,2} \varepsilon^i_{\lambda} \left[ A_{\lambda}(k,\eta) b_{\lambda}(\vec{k}) e^{ik\cdot x}  
+ A^*_{\lambda}(k,\eta)  b^{\dagger}_{\lambda}(\vec{k}) e^{- ik\cdot x} \right]
\end{align}
where $b(\textbf{k})$ and $b^{\dagger}(\textbf{k})$ are the annihilation and creation operators respectively for a given comoving mode $\textbf{k}$, and $\varepsilon_{\lambda}^i$ is the orthogonal basis vector which in right-handed coordinate system~\cite{2018-Sharma.Subramanian.Seshadri.PRD} is given by
\begin{align}\label{eq:basisVector}
\varepsilon^{\mu} = \left( \frac{1}{a}, \textbf{0} \right), \,\,\,\, \varepsilon^{\mu} = \left( 0, \frac{ \hat{\varepsilon}^i_{\lambda} }{a} \right), \,\,\,\, \varepsilon^{\mu}_3 = \left(  0, \frac{\hat{\textbf{k}}}{a} \right) \quad  \text{for} \quad \lambda = 1, 2 \, ,
\end{align}
3-vectors $\hat{\varepsilon}^i_{\lambda}$ are unit vectors orthogonal to $\hat{\textbf{k}}$ and to each other. Substituting Eq.~(\ref{eq:basisVector}) in 
Eq.~(\ref{eq:FourierT} ) and defining the new variable 
$\bar{A}_{\lambda} = a(\eta) \,  A_{\lambda}(k,\eta)$, we have:
\begin{align}\label{eq:Fdecomposition}
A_{i}(\textbf{x}, \eta) = \int \frac{d^3 k}{(2\pi)^3} \sum_{\lambda = 1,2} \,\hat{\varepsilon}_{i \lambda} \left[ \bar{A}_{\lambda} b_{\lambda}(\textbf{k}) e^{ i \textbf{k} \cdot \textbf{x} }  
+ \bar{A}^*_{\lambda}  b^{\dagger}_{\lambda}(\vec{k}) e^{- i \textbf{k} \cdot \textbf{x} } \right] \, .
\end{align}
Substituting  Eq.~\eqref{eq:Fdecomposition} in Eq.~\eqref{eq:equation_of_motion}, we get:
\begin{align}\label{eq:EOM_fourier_space}
\sum_{\lambda = 1,2}b_{\lambda} \left[  \hat{\varepsilon}_{i \lambda}  \bar{A}_{\lambda}^{\prime\prime} + \frac{4i}{M^2} \epsilon_{i j l} k_j \hat{\varepsilon}_{l \, \lambda} \bar{A}_{\lambda} \, \left( \frac{a^{\prime\prime\prime} }{a^3} - 3\frac{ a^{\prime\prime}  a^{\prime}  }{a^4} \right) +  k^2 \hat{\varepsilon}_{i \lambda} \bar{A}_{\lambda}\right] = 0 
\end{align}
where we have used $\partial_j \partial_j = -k^2$. 

Since the action \eqref{eq:action} contains parity breaking term (helicity term), it is always useful to work in the helicity basis. The helicity basis vectors $\varepsilon_+$ and $\varepsilon_-$ corresponding to $h = +1$ and $h = -1$ are defined as
\begin{align}\label{eq5:helicity_basis}
\varepsilon_{\pm} = \frac{1}{\sqrt{2}} \left(   \hat{\varepsilon}_1 \pm i  \hat{\varepsilon}_2  \right).
\end{align}
Assuming that the wave propagates in the $z-$direction, the vector potential in the helicity basis is given by:
%the case where the wave is propagating along the $\varepsilon_3$ direction, in %coulomb gauge (radiation gauge), the vector potential takes the form
%
\begin{align}\label{eq5:decomp_A_helicity}
\bar{\textbf{A}} = \bar{A}_1 \hat{\varepsilon}_1 + \bar{A}_2  \hat{\varepsilon}_2 = A_+ \varepsilon_+ + A_- \varepsilon_-
\end{align}
where $A_+$($A_-$) refer to the vector potential with positive (negative) helicity. 
The ground state in the helicity basis is defined  as
\begin{align}\label{eq:GS_helicity}
b_h(\textbf{k}) | 0 \rangle = 0 
\end{align}
and commutation relation are:
\begin{align}\label{eq:comm-b_h}
\left[ b_h(\textbf{k}), b^{\dagger}_{h^{\prime}}(\textbf{q})  \right] &= \left( 2\pi \right)^3 \, \delta^3(\textbf{k} - \textbf{q}) \, \delta_{h h^{\prime}} \\
\left[ b_h(\textbf{k}), b_{h^{\prime}}(\textbf{q})  \right] &= 0 = \left[ b^{\dagger}_h(\textbf{k}), b^{\dagger}_{h^{\prime}}(\textbf{q})  \right] \, .
\end{align}

Rewriting \eqref{eq:EOM_fourier_space} in the Helicity basis and replacing
$\epsilon_{i j l} \partial_j A_l \longrightarrow  -k \sum_{h = \pm 1} h A_h \varepsilon_{h}$, we have:
\begin{align}\label{eq:eom_helicity}
A_h^{\prime\prime} + \left[  k^2 - \frac{4kh}{M^2} \, 
\Gamma(\eta)  \right] A_h= 0 \, ,
\end{align}
where,
\begin{equation}
\label{def:Gamma}
  \Gamma(\eta) = \frac{a^{\prime\prime\prime}}{a^3} - 3\frac{a^{\prime\prime} a^{\prime} }{a^4} = \frac{1}{a^2} \left(\mathcal{H}'' - 2 \mathcal{H}^3\right) \, .  
\end{equation}
We would like to stress the following points regarding the above expression: First, unlike the scalar or tensor perturbations, the mode functions contain third-order derivatives of the scale factor. This implies that the spectrum of perturbations may be different in our model. Second, since the perturbations equations contain second-order derivatives of $\mathcal{H}$, the helicity modes will be different for inflation and bounce models~\cite{2020-Nandi-PLB}. Third, since $h$ takes two values, $A_h$ evolves differently for the two modes leading to non-zero helicity. 

The EM energy densities of the ground state with respect to the comoving observer are:
\begin{align}\label{eq:rhoB}
& \rho_B\left(\eta, k \right) \equiv -\frac{1}{2}\langle 0 | B_i B^i | 0 \rangle = \int \frac{dk}{k} \, \frac{d\rho_B}{d\rm{ln}k} = \int \frac{dk}{k} \frac{1}{\left( 2\pi \right)^2} \frac{k^5}{  a^4} \left( \,\, \left| A_+\left(\eta, k \right) \right|^2 + \left| A_- \left(\eta, k \right) \, \, \right|^2  \right) \\
\label{eq:rhoE}
& \rho_E\left(\eta, k \right) \equiv -\frac{1}{2}\langle 0 | E_i E^i | 0 \rangle = \int \frac{dk}{k} \, \frac{d\rho_E}{d\rm{ln}k} = \int \frac{dk}{k} \frac{1}{\left( 2\pi \right)^2}  \frac{k^3}{ a^4} \left( \,\,  \left| A^{\prime}_+\left(\eta, k \right) \right|^2 + \left| A^{\prime}_-\left(\eta, k \right) \right|^2 \,\, \right)
\end{align}
and the ground state helicity density as
\begin{align}\label{eq:rhoh}
\rho_h \left(\eta, k \right) \equiv -\langle 0 | A_i B^i | 0 \rangle = \int \frac{dk}{k} \, \frac{d\rho_h}{d\rm{ln}k} = \int \frac{dk}{k} \frac{1}{2\pi^2} \frac{k^4}{ a^3} \left( \,\, \left| A_+\left(\eta, k \right) \right|^2 - \left| A_-\left(\eta, k \right) \right|^2 \,\,  \right).
\end{align}
where $d\rho_{\Upsilon}/d (\ln k)$ for $\Upsilon \in \{ E,B,h\}$ is the spectral energy contained in logarithmic interval in $k-$space. Note that the helicity density is the difference between the two helicity spectrum. Hence, it is possible to maximize the magnetic helicity density, if one helicity is enhanced and the other helicity is suppressed~\cite{2013-Durrer.Neronov-Arxiv}. For most of the calculation, we will keep both the terms and evaluate the energy density for both helicity modes. 

%
%%%%%%%   S E C T I O N %%%%%%%%%%%%%%%%
%

\section{Helical magnetic field generation}
\label{sec:Helical}

In this section, we explicitly calculate the power-spectrum and energy densities for our model in power-law inflation. We obtain the power-spectrum in the slow-roll limit. 

Substituting the power-law scale factor (\ref{eq:powerLaw}) in Eq.(\ref{eq:eom_helicity}) leads to:
\begin{align}\label{eq:arbitrayBeta_eta}
{A_h^{\prime\prime} + \left[ k^2 - \frac{8kh}{M^2} 
\frac{ \beta (\beta+1) ( \beta + 2)}{\eta_0^3} 
\left( \frac{-\eta_0}{\eta} \right)^{(2 \beta+5)} \right]  \, A_h = 0 }
\end{align}
As expected, for de-sitter case ($\beta = -2$), the helicity term $(\Gamma(\eta)$) vanishes. This is consistent with the fact that the de Sitter symmetry will not be preserved in the presence of helicity terms. However, it will be non-zero for the approximately de-sitter universe i.e., $\beta = -2-\epsilon$. 

For the power-law inflation model, the scalar and tensor perturbations can be evaluated exactly. However, as can be seen, it is not possible to obtain an exact expression. To obtain the solution, we consider two regions. In Region I (sub-horizon limit), the wavelength of the mode is smaller than the Hubble radius, i. e., $H << k$. In this region, we can neglect $\Gamma(\eta)$ in Eq. \eqref{eq:arbitrayBeta_eta}.  In Region II (super-Horizon scales), the mode is outside the Hubble radius i. e., $k << H$. In this region, we can neglect $k^2$ in Eq. \eqref{eq:arbitrayBeta_eta}. The constants are fixed by matching 
$A_h$ and $A_h'$ at the transition time between regions I and II
at $\eta_*$. While evaluating the mode-functions is trivial in Region I, it is highly non-trivial in Region II. In the rest of this section, we obtain the mode functions and calculate the power-spectrum.

In Region I ($\left| - k \eta \right| \gg 1$), Eq.~(\ref{eq:arbitrayBeta_eta}) simplifies to:
\begin{align}\label{eq:sub-horizon}
A_h^{\prime\prime} + k^2 A_h \approx 0 
\end{align}
and assuming that the quantum field is in the vacuum state at asymptotic past 
(Bunch-Davies vacuum state), we have:
\begin{align}
A_h = \frac{1}{\sqrt{k}} e^{-ik\eta}
\end{align}
In Region II ($\left| - k \eta \right| << 1$),  Eq.~(\ref{eq:arbitrayBeta_eta}) becomes:
\begin{align}\label{eq:sup_mode_eq_alpha}
{A_h^{\prime\prime} + h  k \frac{\varsigma^2}{\eta^2} 
\left( \frac{-\eta_0}{ \eta} \right)^{2 \alpha }   \, A_h = 0 }
\end{align}
where  
\begin{equation}
\label{def:varsigma}
{\varsigma^2 \equiv -\frac{1}{M^2 \, \eta_0} (2\alpha - 3)(2\alpha - 1)(2\alpha + 1)\, ,
\quad \alpha = \beta + \frac{3}{2} }
\end{equation}
$\alpha$ makes the expressions look tidier! Note that $\alpha = -\frac{1}{2}$ corresponds to de-sitter and $ \alpha \leq - \frac{1}{2} $.

As mentioned above, unlike in the scalar and tensor perturbations during inflation, the above equation is not exactly solvable. To do this, we introduce a new dimensionless variable $\tau$, and is defined as:
\begin{align}\label{eq:tau_to_eta}
\tau = \left( -\frac{\eta_0}{ \eta} \right)^{\alpha} \quad \implies  \quad \eta = - \frac{\eta_0}{\tau^{\frac{1}{\alpha}}} \, .
\end{align}
Note that $\tau$ is directly proportional to $\eta$ and is a positive definite quantity ($0 < \tau < \infty$). [At the start of inflation, $\tau$ is large and vanishes 
at the end of inflation.] In terms of $\tau$, the scale factor for the power-law inflation \eqref{eq:powerLaw} is 
\[
a (\tau)= \left( \frac{1}{\tau} \right)^{1 - \frac{1}{2\alpha}} \, . 
\]
Rewriting Eq. \eqref{eq:sup_mode_eq_alpha} in terms of $\tau$  \eqref{eq:tau_to_eta}, we have:
\begin{align}\label{eq:sup_mode_alpha_tau-varsigma}
{   \alpha^2 \frac{d^2 A_h}{d\tau^2} + \frac{\alpha(\alpha+1)}{\tau} \frac{d A_h}{d \tau} + h \, k \, \varsigma^2 A_h  = 0     }
\end{align}
The above equation is a Bessel differential equation, and it has a complete solution as
\begin{subequations}
\begin{align}\label{eq:sup_mode_h+}
A_{+}(\tau,k) &= \tau^{- \frac{1}{2\alpha} } \, J_{  \frac{1}{2 \alpha}} \left( \frac{\varsigma \sqrt{k} }{\alpha} \tau \, \right)  C_1+ \tau^{- \frac{1}{2\alpha} } \, Y_{ \frac{1}{2 \alpha} }  \left( \frac{ \varsigma \sqrt{k} }{\alpha}\tau \right)  C_2
\\
\label{eq:sup_mode_h-}
A_{-}(\tau,k) &=  \tau^{- \frac{1}{2\alpha} } \, J_{  \frac{1}{2 \alpha}} \left(  -i \frac{ \varsigma \, \sqrt{k} }{\alpha} \tau  \right)  C_3+ \tau^{- \frac{1}{2\alpha} }  \, Y_{  \frac{1}{2 \alpha} } \left(  - i \, \frac{ \varsigma\, \sqrt{k} }{\alpha} \tau  \right)  C_4
\end{align}
\end{subequations}
where $C_1, C_2, C_3, C_4$ are arbitrary constants of dimension $L^{1/2}$. 
As mentioned above, for the two helicity modes, we fix the constants $C_1, C_2$ ($C_3, C_4)$ by matching $A_h$ and $A_h'$ at the transition time between between regions I and II at $k_* \sim \eta_*^{-1}$ where $*$ refers to the 
quantities evaluated at the horizon-exit.

Although the analysis can be done for any general value of $\alpha$, to keep the 
calculations tractable, we obtain the constants for $\alpha = -1$.  There are two reasons for this choice: First, in this special case, $\tau \propto \eta$ and the 
super-horizon modes can be written in terms of $\eta$ using the linear relation. 
Second, the constants $C_1, C_2, C_3, C_4$ have a weak dependence of $\alpha$ 
and, hence, finding the value for a given value of $\alpha$ will be accurate within a order. Thus, matching he solutions and the derivatives at the horizon-exit, we get:
\begin{align}\label{eq:Coefficients}
C_1 &= -e^i \,  \sqrt{ \frac{\pi \eta_0}{ 2} } \left( \frac{1}{\sqrt{\Theta}} \rm{sin}\Theta 
 + i \sqrt{ \Theta }  \,  \rm{cos} \Theta   \right), \,\,\,\,
C_2 = -i \, e^i \,  \sqrt{ \frac{\pi \eta_0}{ 2} } \left( \frac{1}{\sqrt{\Theta}} \rm{cos}\Theta 
 - i \sqrt{ \Theta }  \,  \rm{sin} \Theta   \right) \\
C_3 &= e^i \,  \sqrt{ \frac{\pi \eta_0}{ 2 } } \left( \frac{1}{\sqrt{i \Theta}} \rm{sinh}\Theta 
 +  \sqrt{ i \Theta }  \,  \rm{cosh} \Theta   \right), \,\,\,\,
C_4 = -i \, e^i \,  \sqrt{ \frac{\pi \eta_0}{ 2 } } \left( \frac{1}{\sqrt{i \Theta}} \rm{cosh} \Theta
 +  \sqrt{ i \Theta }  \,  \rm{sinh} \Theta   \right). \nonumber
\end{align}
where $\Theta = \sqrt{ \frac{15 \eta_*}{M^2 \eta_0^3} }$ is the dimensionless constant.

\begin{figure}
%\centering
%\subfigure[]{%
\label{fig:first}%
\includegraphics[height=2in]{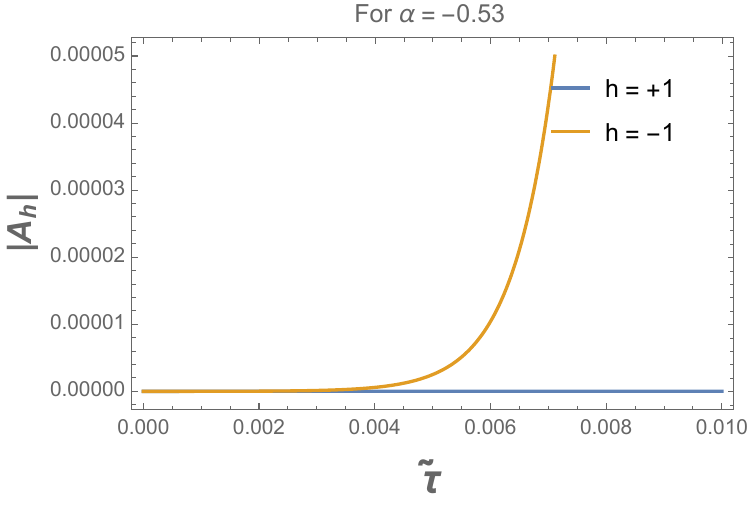}     %}%
\qquad
%\subfigure[]{%
\label{fig:second}%
\includegraphics[height=2in]{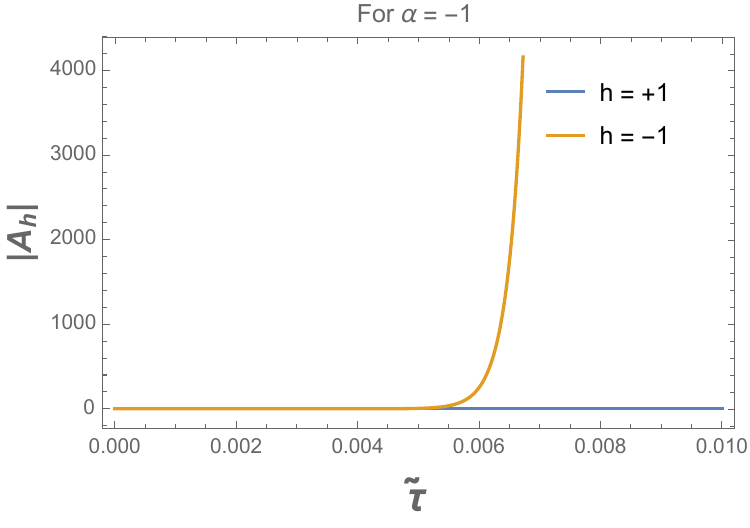}      %}%
\caption{Figure showing the behaviour of positive and negative helicity mode  $\alpha = -0.53$ (left plot) and $\alpha = -1$ (right plot). $\tilde{\tau} = 10^{-\frac{63}{2}} \,\tau $ and the vertical axis is in $\rm{GeV}^{-1/2}$. }
\label{fig:helicitymode}
\end{figure}

To obtain the dominating helicity mode during inflation, we need to obtain the 
values of the coefficients $C_1, C_2, C_3, C_4$. To obtain these values, we 
take: $\mathcal{H} \sim {\eta_0}^{-1} \sim 10^{14} \rm{GeV} \sim {10^{52}} \rm{Mpc}^{-1}$, and $M \sim 10^{17} \rm{GeV}$~\cite{2004-Shankaranarayanan.Sriramkumar-PRD}. This gives $\Theta \sim 10^{-3} $ which is small value. Approximating trigonometric functions in Eq.~(\ref{eq:Coefficients}), we obtain
\begin{align}\label{eq:Coeff-value}
\left| C_1  \right| \approx \left| C_3  \right| \approx 10^{-17/2} \rm{GeV}^{-\frac{1}{2}}, \hspace{0.5 cm} \text{and} \,\, \,\,\,
\left| C_2  \right| \approx \left| C_4  \right| \approx 10^{-11/2} \rm{GeV}^{-\frac{1}{2}}.
\end{align}
Using these values in Eqs.~(\ref{eq:sup_mode_h+}, \ref{eq:sup_mode_h-}), 
we have plotted the two modes for $\alpha = -0.53$ and $\alpha = -1$ in 
\ref{fig:helicitymode}. The plots show that the positive helicity modes are growing compared to the negative helicity modes. Specifically, from Fig.\ref{fig:helicitymode} (right plot), we see that negative helicity mode is decaying. (For $\alpha = -1$ we have $\tau \propto -\eta$ which means negative mode is decaying from $-\infty$ to zero in conformal time). Hence, we can set $\left|  A_{-}(\tau,k) \right| = 0$. The helicity density \eqref{eq:rhoh} is the difference between the two helicity spectrum, and maximum helicity is achieved if one helicity is enhanced compared to other.  In our case, the negative helicity mode is negligible and, has been set to zero.

Using the series expansion of the Bessel functions (see appendix \ref{app:Series-exp} for details), in the leading order, positive helicity mode (\ref{eq:sup_mode_h+}), takes the following form: 
%
%\begin{subequations}
\begin{align}\label{eq:A+Series}
A_+(\tau,k) &= C \, k^{\frac{1}{4\alpha}} 
  - C_2 \frac{\mathcal{F}^{-1} }{\pi} 
  \Gamma \left( \frac{1}{2\alpha}  \right) \,k^{-\frac{1}{4\alpha}} \tau^{ - \frac{1}{\alpha} }  
  %
%  \label{eq:A-Series}
%A_-(\tau,k) &=\tilde{C} \, k^{\frac{1}{4\alpha}} 
%-C_4 \frac{\tilde{\mathcal{F}}^{-1} }{\pi} 
%  \Gamma \left( \frac{1}{2\alpha}  \right) \, k^{-\frac{1}{4\alpha}} \tau^{ -\frac{1}%{\alpha} } .
\end{align}
%\end{subequations}
%
where $C, \mathcal{F}$ are constants defined in Appendix \eqref{app:Series-exp}. The value of these constants are evaluated using (\ref{eq:Coeff-value}). Note that $\varsigma \approx 10^{-10} \rm{GeV}^{-1/2}$ \, the floor function $F(\tau) = 1$. 
Fig.~(\ref{fig:Plot}) contains the plot of Floor function for the range of parameters used in evaluated  (\ref{eq:Coeff-value}). Also, $\left| \mathcal{F} \right|  \sim 10^{-\frac{5}{\alpha}} \,\, \rm{GeV}^{-1/4\alpha}, \text{and} \,
\left| C \right| \sim 10^{-\frac{5}{\alpha} - \frac{11}{2}}$.
%
%
%
%%%%%%%%%   S U B - S E C T I O N %%%%%%%%%%%%%%%%%
%
%
\subsection{Energy densities, power spectrum, and backreaction}

Substituting Eq.~(\ref{eq:A+Series}) in the relation (\ref{eq:rhoB}), 
the spectral magnetic energy density is given by:
\begin{align}\label{eq:spetralB1}
\frac{d\rho_B}{d\rm{ln}k} =  \frac{1}{\left( 2\pi \right)^2} \frac{k^5}{  a^4(\eta(\tau))}
 \left(  \,\,\,\,   \left|   C \right|^2 \, k^{\frac{1}{2\alpha}} 
  + \left| C_2 \frac{\mathcal{F}^{-1} }{\pi} 
  \Gamma \left( \frac{1}{2\alpha}  \right)  \right|^2   \,
  \left( k \,   \tau^4 \right)^{ - \frac{1}{2\alpha} }     \right).  
\end{align}
{Using  the expression: $\tau_* =   \left(  -\frac{2 \, \eta_0 k_*}{2\alpha - 1}  \right)^{\alpha}$ at the exit of inflation, spectral energy density at horizon exit is given by:}
\begin{align}\label{eq:powerSpetrum}
 \left. \frac{d\rho_B}{d\rm{ln}k}  \right|_{k_* \sim \mathcal{H} } =  \frac{(-\eta_0)^{4\alpha - 2 }}{\left( 2\pi \right)^2 }   \left[ \, \left| C \right|^2 \,  k_*^{3 + 4\alpha +\frac{1}{2\alpha}}  +  \left| C_2 \frac{\mathcal{F}^{-1} }{\pi}  \Gamma \left( \frac{1}{2\alpha}  \right) \right|^2 \,
    \frac{ (2\alpha - 1)^2 }{ 4 \eta_0^2} \, k_*^{1 + 4\alpha - \frac{1}{2\alpha}}     \right]        
\end{align}  
 Let us understand the properties of the spectral energy density obtained above: First, 
it has two branches. The first branch (setting $C_2 = 0$) has
scale-invariant spectrum  for $\alpha = -\frac{1}{2}, -\frac{1}{4}$.  Similarly, 
the second branch (setting $C = 0$) has scale invariant spectrum  for $\alpha = -\frac{1}{2}, \frac{1}{4}$. Note that the physically allowed values of $\alpha \leq -1/2$. Hence, $\alpha = \pm 1/4$ is ruled out. Thus, the two branches has scale-invariant power-specrum for exact de Sitter  ($\alpha = -\frac{1}{2}$). Second, 
for slow-roll type of inflation $\alpha = -\frac{1}{2} - \epsilon$, the two branches scale differently ---  $k_*^{-2\epsilon}$ (first branch) and $k_*^{-6\epsilon}$ (second branch). Since $\epsilon$ is positive, this implies that our model produces more power on the large scales. Thus, our model predicts a red spectrum for the helical modes for slow roll inflation. In the next subsection, we compare the results of our model with other models. 

Thus, the magnetic energy density can be obtained as: 
\begin{align}\label{eq:spetralB}
 \rho_B =  \int_{ \frac{\mathcal{H}}{100}}^{\mathcal{H}} \frac{dk}{k} \frac{d\rho_B}{d\rm{ln}k} =  \frac{1}{\left( 2\pi \right)^2 a^4(\eta(\tau)) }   \left[  \left| C\right|^2 \,  \frac{\mathcal{H}^{5 +\frac{1}{2\alpha}} }{ 5 +\frac{1}{2\alpha}  }  +  \left| C_2 \frac{\mathcal{F}^{-1} }{\pi}  \Gamma \left( \frac{1}{2\alpha}  \right) \right|^2 \,
     \tau^{ - \frac{2}{\alpha} }  \frac{\mathcal{H}^{5 - \frac{1}{2\alpha}} }{ 5 -\frac{1}{2\alpha}  }     \right]   
\end{align}
where we have used $\mathcal{H}^{5 + \frac{1}{2\alpha}} - \left(\frac{\mathcal{H}}{100}\right)^{5 + \frac{1}{2\alpha}} 
%= \mathcal{H}^{5 + \frac{1}{2\alpha}} \left( 1 - \frac{1}{100^{5 + \frac{1}{2\alpha}}} %\right) 
\approx \mathcal{H}^{5 + \frac{1}{2\alpha}} $.
 Total magnetic energy density at the exit of inflation can be obtained as   
\begin{align}\label{eq:powerSpetrumB}
\left. \rho_B \right|_{k_* \sim \mathcal{H} } =  \frac{(-\eta_0)^{4\alpha - 2 }}{\left( 2\pi \right)^2 }   \left[ \, \left| C \right|^2 \, \frac{ k_*^{3 + 4\alpha +\frac{1}{2\alpha}} }{ 5 +\frac{1}{2\alpha}  } +  \left| C_2 \frac{\mathcal{F}^{-1} }{\pi}  \Gamma \left( \frac{1}{2\alpha}  \right) \right|^2 \,
    \frac{ (2\alpha - 1)^2 }{ 4 \eta_0^2} \, \frac{ k_*^{1 + 4\alpha - \frac{1}{2\alpha}} }{ 5 -\frac{1}{2\alpha}  }     \right]    
\end{align} 
Using the fact that at super-horizon scales, we can approximate $\partial_{\eta} \sim \mathcal{H}$ (see, for instance, Ref.~\cite{2011-Durrer.Hollenstein.Jain-JCAP}), the total electric energy density at horizon exit is given by: 
\begin{align}\label{eq:powerSpetrumE}
\left. \rho_E \right|_{k_* \sim \mathcal{H} } &=  \frac{(-\eta_0)^{4\alpha - 2 }}{\left( 2\pi \right)^2 }   \left[ \,  \left| C\right|^2 \,  \frac{ k_*^{3 + 4\alpha +\frac{1}{2\alpha}} }{ 3 +\frac{1}{2\alpha}  } + \left| C_2 \frac{\mathcal{F}^{-1} }{\pi}  \Gamma \left( \frac{1}{2\alpha}  \right) \right|^2 \,
    \frac{ (2\alpha - 1)^2 }{ 4 \eta_0^2} \, \frac{ k_*^{1 + 4\alpha - \frac{1}{2\alpha}} }{ 3 -\frac{1}{2\alpha}  }     \right].
\end{align}
Thus, using the above two expressions, the total energy density at horizon exit is given by 
 \begin{align}\label{eq:totalenergy-EB}
\left. ( \rho_B + \rho_E) \right|_{k_* \sim \mathcal{H} } &= \rho_{T}^{(1)} + \rho_{T}^{(2)} 
\end{align}
where
\begin{align}
 \rho_{T}^{(1)} &= \frac{(-\eta_0)^{4\alpha - 2 }}{\left( 2\pi \right)^2 } \, \left| C \right|^2 \frac{ 4\alpha (8\alpha + 1) }{ (10\alpha + 1)(6\alpha + 1)  } k_*^{3 + \alpha +\frac{1}{2\alpha}} \\
 \rho_{T}^{(2)} & =   \frac{(-\eta_0)^{4\alpha - 4 }}{\left( 2\pi \right)^2 }   \left| C_2 \right|^2  \, \, \left| \frac{\mathcal{F}^{-1} }{\pi}  \Gamma \left( \frac{1}{2\alpha}  \right) \right|^2  \frac{ 4\alpha (8\alpha - 1) (2\alpha - 1)^2  }{(10\alpha - 1)(6\alpha - 1)  }  k_*^{1 + 4\alpha - \frac{1}{2\alpha}}  
\end{align}
Since the power-spectrum is red-titled, the power in the long wavelengths is more than in the short wavelengths. Thus, there is a possibility that these helical modes can backreact on the metric. Since the effect is cumulative, all fluctuation modes contribute to the change in the background geometry. Consequently, the backreaction effect can be large, even if the amplitude of the fluctuation spectrum is small. To identify whether these modes lead to backreaction on the metric, we define $R$, which is the ratio of the total energy density of the  fluctuations and 
background energy density during inflation~\cite{2020-Talebian.etal-arXiv}: 
\begin{align}
R = \frac{ \left. ( \rho_B + \rho_E) \right|_{k_* \sim \mathcal{H} } }{6 M_P^2H^2} \, .
\end{align}
Using $M_P = 10^{19}\rm{GeV} $ and $H = 10^{15}~\rm{GeV} $, the background energy density during inflation ($M_P^2H^2$) is $10^{68}~\rm{GeV}^4$. The table below contains estimates of the total energy density at the horizon exit for different values of $\alpha$. 
\begin{table}[h]
\centering
\begin{tabular}{|c|l|l|l|c| }
	\hline
	$\alpha$  &  $\rho^{(1)}_{T}$ (in $\rm{GeV}^4$)  & $\rho^{(2)}_{T}$ (in $\rm{GeV}^4$)& Total (in $\rm{GeV}^4$) &~~~$R$~~~~ \\ \hline
 $-\frac{1}{2} - \epsilon$ & $ \sim 10^{64} $  &  $ \sim 10^{52}$ & $ \sim 10^{64} $  & $\sim 10^{-4}$  \\ \hline
 $-\frac{3}{4} $ & $  \sim 10^{62}$ &  $\sim 10^{54}$ & $ \sim 10^{62} $ 
 & $\sim  10^{-6}$  \\ \hline
 $-1 $ & $  \sim 10^{61}$ &  $\sim 10^{55}$ & $ \sim 10^{61} $  & $\sim  10^{-7}$ 
 \\ \hline
$-3 $ & $  \sim 10^{59}$ &  $\sim 10^{57}$  & $ \sim 10^{59} $ & $\sim 10^{-9}$  
\\ \hline
\end{tabular}
\caption{The total energy density at the exit of inflation for different values of $\alpha$. To estimate  $\rho^{(1)}_{T}$ and $\rho^{(2)}_{T}$, we 
take: $\mathcal{H} \sim {\eta_0}^{-1} \sim 10^{14}~\rm{GeV} \sim {10^{52}}~ \rm{Mpc}^{-1}$, and $M \sim 10^{17}~\rm{GeV}$.}
\end{table}
We see from the above table that for varied values of $\alpha$, $R << 1$, implying that the backreaction of the helical modes on the background metric during inflation is negligible. The ratio $R$ is maximum ($10^{-4}$) for slow-roll 
inflation. Thus, while our model produces helical magnetic fields with more power at large length-scales, the backreaction of these on the metric is negligible, and these modes do not stop inflation. 

Although spectral helicity density can not be directly measured, for completeness, 
we give the expression for the spectral helicity density:
\begin{align}\label{eq:helicityPowerSpectrum}
 \frac{d\rho_h}{d\rm{ln}k} &=  \frac{1}{ 2\pi^2} \frac{k^4}{  a^3(\eta(\tau))}
\left(  \,\,\,\,   \left|   C(\tau) \right|^2 \, k^{\frac{1}{2\alpha}} 
  + \left| C_2 \frac{\mathcal{F}(\tau)^{-1} }{\pi} 
  \Gamma \left( \frac{1}{2\alpha}  \right)  \right|^2   \,
  \left( k \,   \tau^4 \right)^{ - \frac{1}{2\alpha} }  \right).
\end{align}

\subsection{Comparison of our model with scalar-field ($f^2(\phi)\, F\tilde{F}$) coupling models}

Often in the literature, the breaking of conformal invariance
of the electromagnetic action is through the non-minimal coupling of the 
electromagnetic field ($f^2(\phi)\, F\tilde{F}$) with scalar field (possibly inflaton). For a suitable choice of the coupling parameters, it has been shown that a sufficient amount of large-scale magnetic fields can be generated~\cite{2001-Vachaspati-PRL,2003-Caprini.etal-PRD,2005-Campanelli-Giannotti-PRD,2018-Sharma.Subramanian.Seshadri.PRD,2009-Caprini.Durrer.Fenu-JCAP,2009-Campanelli-IJMPD,2019-Shtanov-Ukr.PJ}. 

In contrast, our model does not rely on the fine-tuning of the extra coupling parameter to the electromagnetic field and depends on the background quantities through the Riemann tensor. Due to this, the mode functions \eqref{eq:eom_helicity} contain higher derivatives of ${\cal H}$ compared to scalar field coupled models. In Appendix \ref{app:slow-roll-case}, we use a naive approximation of our model for power-law and slow-roll inflation, which effectively ignores higher-derivatives of ${\cal H}$. Under this approximation, we show that the model leads to a blue-tilt spectrum. {Thus, the presence of a higher-derivative of ${\cal H}$ leads to the red-tilt.} This is an important difference between our model compared to scalar-field coupled models. 

To further understand this, we define the overall coupling function in our model by a dimensionless coupling function:
\begin{align}
I = \frac{ R_{\mu\nu}\,^{\sigma\gamma} }{ M^2 } \qquad 
\end{align}
In the flat FRW line-element, Riemann tensor $R_{\mu\nu}\,^{\sigma\gamma} \sim \frac{{a^{\prime}}^2}{a^4}$ and $ \sim \frac{a^{\prime\prime}}{a^3}$. Let us now compare our model with the two specific forms of scalar-field coupled models~ \cite{2011-Durrer.Hollenstein.Jain-JCAP,2018-Sharma.Subramanian.Seshadri.PRD}.

In our model, for power-law inflation, the coupling function $I$ is
\begin{align}\label{eq:coupl-I-constant_M}
 I \sim \frac{ {a^{\prime}}^2 }{M^2 a^4}~~\left(\mbox{or}~\frac{ {a^{\prime\prime}} }{M^2 a^3}\right)
 \propto \frac{1}{\eta^{2\beta + 4}} \propto \eta^{\delta} \qquad (  \delta > 0  )    
\end{align} 
As mentioned above, for the scalar field coupled models, many authors have used different forms of $f^2(\phi)$: 
\begin{equation}
    f_1(\phi) \propto  e^{\frac{\phi}{m_P}}~;~
f_2(\phi) \propto a^2     
\end{equation}
Note that $f_1$ was used in Ref.  \cite{2011-Durrer.Hollenstein.Jain-JCAP} while 
$f_2$ was used in Ref. \cite{2018-Sharma.Subramanian.Seshadri.PRD} . In both the cases, the coupling function is of the form:
\begin{equation}
\label{eq:coupl-f-constant_M}
  f(\phi(\eta)) \propto \frac{1}{\eta^\sigma} \qquad ( \sigma > 0 )      
  \end{equation}
We want to make the following remarks regarding the two coupling forms  \eqref{eq:coupl-I-constant_M} and \eqref{eq:coupl-f-constant_M}:  First, the functional form of the coupling function in our case is different compared to the scalar field coupling models. Since $\eta$ is large at early times, the Riemann coupling term contributes significantly at early times, and hence, the modes that leave the horizon at early times will have large helicity. In the scalar field coupled models, the modes generated at early times will not have significant helicity modes. In contrast, the modes generated close to the end of inflation will have significant helicity. This provides a qualitative understanding of the red-tilt power-spectrum in our case. Second, since most of the helical modes are generated at early times, unlike in Ref.  \cite{2018-Sharma.Subramanian.Seshadri.PRD}, the generated helical fields are not sensitive to the reheating dynamics. Thus, 
the helical modes generated evolve similar to the scalar and tensor perturbations generated during inflation. Today's observable scales (in the CMB and LSS) span roughly three orders in the comoving wavenumber $k$. The largest observable wavelength $\lambda_{\rm max}$, associated with the wavenumber $k_{\rm max}$, corresponds to the horizon radius. For a model with 60 e-foldings of inflation,  the observable cosmological wavelengths exit the Hubble radius around 30 e-foldings before the end of inflation. This will also apply to the helical modes generated in our model. Third, the total energy density of the helical fields in our model is larger compared to the scalar-field coupled models. Specifically, the energy density of the helical fields in our model is at least of an order of magnitude larger than for the coupling function $f_2(\phi)$ in Ref. \cite{2018-Sharma.Subramanian.Seshadri.PRD}. However, as shown in the previous subsection, our model is free from the backreaction problem for a range of scale-factor during inflation.

\subsection{Estimating the strength of the helical magnetic fields}

As mentioned above, the model generates helical fields around $30$ e-folding before the end of inflation. To estimate the current value of the helical fields, we assume instantaneous reheating, and the Universe becomes radiation dominated after inflation.  
Due to flux conservation, the magnetic energy density will decay as ${1}/{a^4}$, i. e.:
\[
\rho_{B}(0) = \rho_B^{(f)} \left(  \frac{a_f}{a_0} \right)^4
\]
where $a_0$ is the present day scale-factor, $\rho_B^{(f)}$ and $a_f$ refer to the magnetic energy density and the scale-factor at the end of inflation, respectively. Using the entropy conservation i.e., $g \, T^3 \, a^3 = constant$ where $g$ refers to the effective relativistic degrees of freedom and $T$ is the temperature of the relativistic fluid, we get $ {a_0}/{a_f} \approx 10^{30} \left( {H_f}/{10^{-5}  M_{\rm{Pl} } }  \right)^{1/2}$~\cite{2016-Subramanian-Arxiv}. 

Using the fact that the relevant modes exited Hubble radius around 30 e-foldings of inflation, with energy density $\rho_B \approx 10^{64} \rm{GeV}^4$, the primordial helical fields at ${\rm GPc}$ scales is:
\begin{align}
B_0 \approx 10^{-20} \rm{G} 
\end{align}
where we have used $1 G = 1.95 \times 10^{-20} \rm{GeV}^2 $ and $H_f=10^{-5} M_{\rm{Pl}}$ is the Hubble parameter during inflation. Our model predicts the following primordial helical fields that re-entered the horizon at two different epochs:
\[
\left. B \right|_{50~{\rm MPc}} \sim 10^{-18}~G~(z \sim 20)~; 
~ \left. B \right|_{1~{\rm MPc}} \sim 10^{-14}~G~(z \sim 1000)\, .
\]
Thus, the model generates sufficient primordial helical magnetic fields at all observable scales.

\section{Conclusions and Discussions}
\label{sec:conc}

We have proposed a viable scenario for the generation of
helical magnetic fields during inflation, which does not require coupling to the scalar field. The generation of the helical fields is
due to the coupling of the electromagnetic fields with the Riemann tensor. To our knowledge, Riemann tensor coupling has not been discussed in the literature to generate helical fields.

The model has many key features: First, it does not require the coupling of the electromagnetic field with the scalar field. Hence, there are no extra degrees of freedom and will not lead to a strong-coupling problem. Second, the conformal invariance is broken due to the coupling to the Riemann tensor. Since the curvature is large in the early Universe, the coupling term will introduce non-trivial corrections to the electromagnetic action. However, at late-times, the new term will not contribute, and the theory is identical to standard electrodynamics.  
{ Third, the power spectrum of the helical fields generated has a slight red-tilt for slow-roll inflation.} This is different compared to the scalar field coupled models where the power-spectrum has a blue-tilt. We have also identified the reason for this difference. 
Fourth, our model is free from backreaction for a range of scale-factor during inflation. This is different from other models where a specific form of coupling function is chosen to avoid any back-reaction~\cite{2018-Sharma.Subramanian.Seshadri.PRD}. 

In this chapter, we did not discuss the generation of non-helical magnetic fields. The generation of the non-helical magnetic field with Riemann coupling has been discussed in the seminal paper by Turner and Widrow~\cite{1988-Turner.Widrow-PRD}. An analysis including parity preserving term can be done straightforwardly, and we can obtain total energy density of the non-helical ($\rho_{B}$) and helical energy density ($\rho_H$). As shown recently, the two energy densities must satisfy the realizability condition~\cite{2014-Kahniashvili.etal-PRD}, i. e.,  $\rho_H \leq 2 \, \xi_M \rho_{B}$, where $\xi_M$ is the magnetic correlation length. Assuming that the non-helical and helical power spectra are a power-law:
\[
 P_{B} = A_{B} k^{n_{B}}, \quad P_{H} = A_{H} k^{n_{H}}   \, ,    
\]
{for maximal helicity, it was shown that the helical magnetic fields must have red-tilt.} More specifically, for the WMAP nine-year data, using the cross power-spectrum between the temperature and B-mode polarization they set $95\%$ confidence level upper limit on the helicity amplitude to be $10 \rm{nG}^2 \, \rm{Gpc} $ for the helical spectral index $n_H = -1.9$ and for a cosmological magnetic field with effective field strength of $3~\rm{nG}$ and $n_B = -2.9$. PLANCK 2015 data placed constraints on the strength for causally generated magnetic fields with spectral index $n_B = 2$ and fields with almost scale-invariant spectrum with $n_B = -2.9$ are $B_{1\rm{Mpc}} < .011~\rm{nG}$ and  $B_{1\rm{Mpc}} < 0.9~\rm{nG}$ at $95\%$ confidence level~\cite{2015-Planck-PMF}. Thus, the PLANCK 2015 data also prefers $n_H$ to be negative. With improved B-mode polarization measurements, helicity modes can be better constrained and put our model's prediction to test with the CMB data. We hope to address this soon. 

The perturbations equations \eqref{eq:equation_of_motion} contain second-order derivatives of $\mathcal{H}$. Since, ${\cal H}$, and ${\cal H}''$ are different for inflation and bounce models~\cite{2020-Nandi-PLB}, the helicity modes may provide signatures to distinguish the two paradigms. %This is currently under investigation.

In this chapter, we have focussed on the modes which exit the horizon around 40-50 e-foldings and shown that they can potentially explain the large scale magnetic fields observed in galaxies. In the next chapter~\ref{ch:PMF_Baryo}, we will discuss how the same mechanism can help us to understand the baryogenesis problem. We will show that focusing on the modes which leave the horizon at around 5-10 e-foldings during inflation can lead to the creation of matter-antimatter asymmetry at the beginning of radiation dominated era.

\chapter{Helical magnetic field from Riemann coupling lead to baryogenesis}
\label{ch:PMF_Baryo}

%\section{Introduction}
%
Understanding the physical processes in the very early Universe is a crucial ingredient for deciphering the physics at energies that we cannot currently probe in terrestrial experiments. While most observables have been washed away by the thermal bath of the pre-recombination era and do not have observational consequences, \emph{three observables} provide crucial information of the physics at high-energies. These are the spectrum of energy density fluctuations~\cite{Book-Kolb.Turner,Book-Mukhanov,Book-Padmanabhan-III,Book-Gorbunov.Rubakov}, excess of baryons over antibaryons (baryon asymmetry)~\cite{1999-Riotto.Trodden,2003-Dine.Kusenko-RevModPhy,2006-Cline-arXiv,2011-Riotto-JPCS,2007-Yoshimura-JPSJ,2015-Cui-ModPhyLetA,2020-Garbrecht-PPNP}, and coherent large-scale magnetic fields~\cite{2001-Grasso.etal-PhyRep,2002-Widrow-Rev.Mod.Phys.,2013-Durrer.Neronov-Arxiv,2016-Subramanian-Arxiv,2004-Giovannini-IJMPD,2020-Vachaspati-RepProgPhys}. 

As mentioned in Chapter~\ref{ch:intro}, the inflationary paradigm provides an attractive mechanism to generate the primordial density perturbations that lead to anisotropies in the cosmic microwave background (CMB) and the formation of large-scale structures~\cite{Book-Kolb.Turner,Book-Mukhanov,Book-Padmanabhan-III,Book-Gorbunov.Rubakov}. During inflation, the early Universe underwent an accelerated expansion, stretching quantum fluctuations to super-horizon scale density perturbations. Besides providing a causal mechanism to density perturbations, inflation also solves the standard cosmological model's long-standing puzzles, such as the horizon, flatness, and monopole problems.

As mentioned in Chapter~\ref{ch:intro}, the predictions of inflation are in good agreement with the present-day observations of CMB anisotropies and polarization~\cite{2018-Planck}. However, within the standard electrodynamics, inflation cannot provide a mechanism to generate large-scale B fields. This is because in 4-dimensions electromagnetic field is conformally invariant. Since FRW models are conformally flat, the electromagnetic field vacuum in FRW is the same as the Minkowski space-time. Hence, the standard electromagnetic fields generate negligible magnetic fields. More importantly, even if the baryon asymmetry or cosmological magnetic fields existed before the epoch of inflation, these would have been diluted by a factor of $e^{-3 N}$, where $N$ is the number of e-foldings of inflation~\cite{2016-Fujita.Kamada-PRD,2019-Domcke.etal-JCAP,2014-Long.Sabancilar.Vachaspati-JCAP}. 

As mentioned in Sections~(\ref{sec:open_problem},\ref{sec:motivation_thesis}), the present Universe is observed to contain essentially only matter and no antimatter, except for the rare antiparticles produced by cosmic rays. The asymmetry between baryons and antibaryons, referred to as Baryon Asymmetry of the Universe (BAU), can be expressed as~\cite{PartDataGroup,2018-Planck}
\begin{equation}
\label{def:etaObs}
\eta_B =\frac{n_{b}-n_{\bar{b}}}{n_{\gamma}}=\left\{\begin{array}{r}
{[5.8-6.6] \times 10^{-10}}~~\text{(from BBN)} \\
(6.09 \pm 0.06) \times 10^{-10}~~\text{(from CMB)}
\end{array}\right.
\end{equation}
where $n_{b}, n_{\bar{b}}, n_{\gamma}$ refer to the density of baryons, antibaryons and photons, respectively. 
Magnetic fields permeate the Universe. Coherent magnetic fields in spiral galaxies and clusters of galaxies have a magnitude of the order of $\mu$Gauss~\cite{2001-Grasso.etal-PhyRep,2013-Durrer.Neronov-Arxiv,2016-Subramanian-Arxiv,2002-Widrow-Rev.Mod.Phys.}. There is also indirect evidence of a lower limit of order $10^{-16}~$G for the magnetic field contained in the voids between galaxies and clusters of galaxies~\cite{2010-Neronov.Vovk-Sci}.  

As mentioned in Sections~(\ref{sec:open_problem},\ref{sec:motivation_thesis}), the origin of primordial magnetic fields and baryon asymmetry of the Universe are still unresolved issues and require physics beyond the standard models of cosmology and particle physics. This leads to the following questions: As the Universe cooled, from the early Universe to today, what were the processes responsible for generating baryon asymmetry and large-scale magnetic fields? 
Are these processes cosmological or particle physics or both? 
Since both require physics beyond the standard model, there is a \emph{tantalizing possibility} that the same new physics can solve both. In this chapter, we consider such a possibility and show that the mechanism that leads to primordial helical magnetic fields also leads to baryogenesis at the beginning of the radiation-dominated epoch. Interestingly, our mechanism \emph{also requires} stretching of the primordial helical magnetic fields to super-horizon scales 
during inflation --- the same mechanism that leads to primordial density perturbations.

Before we discuss the model itself, it is necessary to understand the key ingredients to generate baryon-asymmetry and magnetic fields and why the same new physics can potentially solve both these problems~\cite{1967-Sakharov,1996-Davidson-PLB}. In 1967, Sakharov listed three necessary conditions for creating the BAU~\cite{1967-Sakharov,1999-Riotto.Trodden}: (1) baryon number violation, (2) charge ($C$) and charge parity ($CP$) violation, and (3) departure from thermal equilibrium. All three of the Sakharov conditions are satisfied in the Standard Model; however, the electroweak phase transition is not sufficiently strong in the first order~\cite{1999-Riotto.Trodden,2003-Dine.Kusenko-RevModPhy,2006-Cline-arXiv,2011-Riotto-JPCS}. The CP-violating effects are not sufficiently pronounced to account for as large a BAU as we observe. As a result, there must have been additional physics beyond the standard model to produce it. This physics could have been operating anywhere between the weak scale and the GUT scale. Corresponding to out-of-equilibrium conditions, the baryogenesis scenarios are divided into two categories: (a) by the universe expansion itself or (b) by fast phase transition and bubble nucleation. In particular, the latter concerns the electroweak baryogenesis schemes, while the former is typical for a GUT
type baryogenesis or leptogenesis~\cite{1999-Riotto.Trodden,2003-Dine.Kusenko-RevModPhy,2006-Cline-arXiv,2011-Riotto-JPCS}.
%At present, it is not possible to say which of the known mechanisms is responsible for the observed BAU. 

More than two decades ago, Davidson pointed out an interesting relation between the primordial magnetic field and Sakharov's conditions~\cite{1996-Davidson-PLB}. She argued that the presence of background magnetic fields in the early Universe could lead to the breaking of $C, CP, SO(3)$ symmetries and thermal equilibrium. Specifically, she argued that the presence of the magnetic fields leads to the following three conditions: (1) There should be some moderately out-of-thermal-equilibrium dynamics because in equilibrium, the photon distribution is thermal, and there are no particle currents to sustain a "long-range" field, (2) Since $\textbf{B}$ is odd under $C$ and $CP$, the presence of magnetic field will lead to $CP$ violation, (3) Since the magnetic field is a vector quantity, it chooses a particular direction hence breaks the isotropy (rotational invariance). Thus, Davidson provided a possible link between the presence of magnetic fields to the conditions required for baryogenesis~\cite{1996-Davidson-PLB}.

Davidson's conditions are necessary \emph{but not} sufficient. One key missing ingredient, as shown in this chapter, is the requirement of \emph{primordial helical magnetic fields} (details in Sec. \ref{sec:Baryo-magnetic}). Primordial helical magnetic fields are generated by the terms that break conformal invariance and parity symmetry~\cite{2001-Vachaspati-PRL,2003-Caprini.etal-PRD,2005-Campanelli-Giannotti-PRD,2018-Sharma.Subramanian.Seshadri.PRD,2009-Caprini.Durrer.Fenu-JCAP,2009-Campanelli-IJMPD,2019-Shtanov-Ukr.PJ,2020-Kushwaha.Shankaranarayanan-PRD}.
If we could measure them, primordial helical magnetic fields provide evidence of CP violation in the early Universe. Interestingly, the presence of primordial helical fields leads to non-zero Chern-Simons number~\cite{2016-Fujita.Kamada-PRD,2015-Anber.Sabancilar-PRD,2016-Kamada.Long-PRD} and, eventually, the change in the Fermion number. 

In chapter~\ref{ch:helical-PMF}, we discussed a simple model of inflationary magnetogenesis that couples the electromagnetic fields with the Riemann tensor~\cite{2020-Kushwaha.Shankaranarayanan-PRD}. We have seen that the model leads to a primordial helical magnetic field where one helical mode is enhanced while the other mode is suppressed. The model has two key advantages over other models~\cite{2005-Campanelli-Giannotti-PRD,2018-Sharma.Subramanian.Seshadri.PRD,2009-Caprini.Durrer.Fenu-JCAP,2009-Campanelli-IJMPD,2019-Shtanov-Ukr.PJ}: First, it does not require the coupling of the electromagnetic field with any scalar field. Hence, unlike Ratra model~\cite{1991-Ratra-Apj.Lett,2019-Shakeri.etal-PRD,2009-Demozzi.etal-JCAP}, there is no strong-coupling problem caused by the extra degrees of freedom. 
Second, the model is free from backreaction for generic slow-roll inflation models~\cite{2020-Kushwaha.Shankaranarayanan-PRD}. In Ref.~\cite{2018-Sharma.Subramanian.Seshadri.PRD}, authors have shown the strong-coupling problem in Ratra model~\cite{1991-Ratra-Apj.Lett} can be avoided by choosing a particular coupling function. In chapter~\ref{ch:helical-PMF}, we used the general effective field theory of gravity coupled to the Standard Model of particle physics framework to obtain leading order gravity terms that couple to the standard model Bosons~\cite{2019-Ruhdorfer.etal-JHEP}. As we have done in the previous chapter, we limit to mass dimension 6-operators coupling to the gauge field Lagrangian, specifically, to the electromagnetic field.

\begin{figure}[!ht]
\centering
\includegraphics[width=1\textwidth]{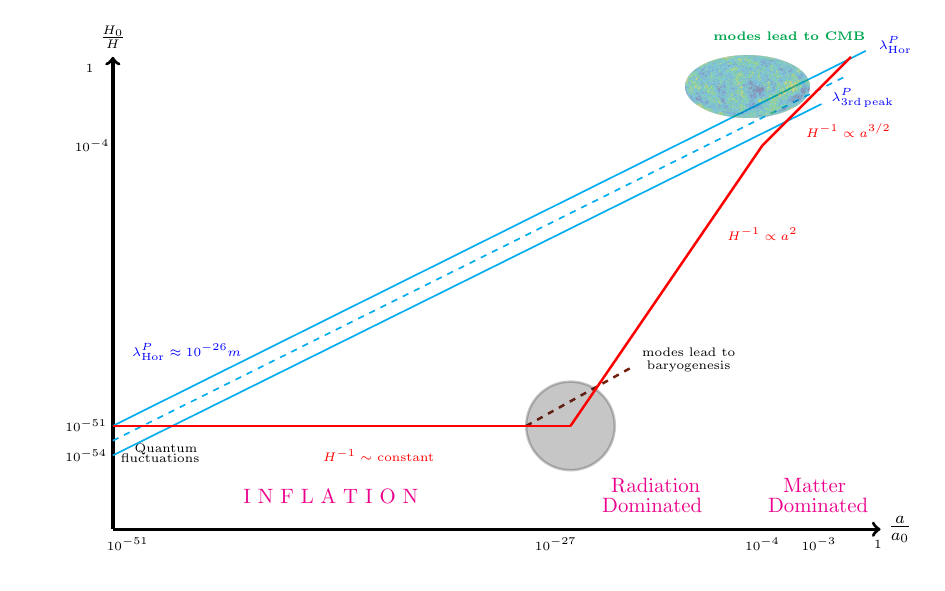} 
\caption{Modes exiting the horizon around last 10 e-foldings (gray shaded circle) will re-enter the Universe early and lead to baryogenesis.}
\label{fig:inflation-figure-BAU}
\end{figure}

Like in Chapter 3, in this chapter, we limit to mass dimension 6-operators coupling to the gauge field, specifically, to the electromagnetic field.
We show that the generation of primordial helical magnetic fields in the previous chapter~\ref{ch:helical-PMF} leads to baryogenesis. Since the model produces helical fields over large length scales, we show that the Chern-Simons (CS) number density is non-zero (details in Sec. \ref{sec:Baryo-magnetic}). Considering that the model generates primordial helical modes at all length scales, we focus on the last ten e-foldings of inflation. As we have discussed in chapter~\ref{ch:intro}, the modes that leave the Hubble radius during the last 10 e-foldings of inflation will reenter the Universe just after the reheating (see Fig.~\ref{fig:inflation-figure-BAU}); these primordial helical modes will lead to baryogenesis just at the beginning of the radiation-dominated epoch.  Furthermore, we show that the BAU is independent of inflation models and depends \emph{only on} the energy scale at the exit of inflation and reheating temperature.

In Sec. (\ref{sec:Baryo-magnetic}), we discuss the relation between primordial helical magnetic fields and baryogenesis, in particular, the chiral anomaly in the presence of the magnetic field, and obtain the expression for Chern-Simon number density. In Sec. (\ref{sec:Model}), we discuss the generation of primordial helical modes and show that primordial helical modes lead to a non-zero CS number density. Then we evaluate the baryon asymmetry parameter in Sec.(\ref{sec:baryon_asymm}). Sec. \eqref{sec:conc} contains the implications of the results. Appendix~\ref{apdetails:PMF_baryo} contains the details of the calculations in this chapter. The results reported in this chapter are based on Ref.~\cite{2021-Kushwaha.Shankaranarayanan-PRD}.

%%%%%%%%%%  S E C T I O N %%%%%%%%%%%%%%%%%%%%
%
\section{Conditions on baryogenesis in the presence of primordial magnetic field}
\label{sec:Baryo-magnetic}

As we mentioned in the introduction, Davidson's conditions are necessary but not sufficient. One key missing ingredient is the requirement of \emph{primordial helical magnetic fields}. In this section, we briefly discuss this. 

In the very early Universe, just after the exit of inflation, the energy scale of the Universe was close to $10^{14}~{\rm GeV}$. All particles, including Fermions, are highly relativistic and can be treated as massless. 
Although the massless Dirac equation is invariant under chiral transformations in the classical theory, the chiral symmetry is broken due to quantum mechanical effects in the presence of the external electromagnetic fields. This phenomenon, known as the quantum axial anomaly, affects the transport properties of the chiral medium, leading to experimentally accessible signatures such as the chiral magnetic effect~\cite{2008-Fukushima.etal-PRD} and the 
chiral separation effect~\cite{2005-Metlitski.Zhitnitsky-PRD}.

In the early Universe, the generation of the non-zero primordial helical magnetic fields leads to a chiral anomaly resulting from the imbalance between left and right-handed fermions. 
In  the presence of an electromagnetic field in curved space-time, the chiral anomaly is given by the following equation~\cite{2009-Parker.Toms-Book,2014-Barrie.Kobakhidze-JHEP}:
\begin{align}\label{eq:chiralAnomaly}
 \nabla_{\mu}J_A^{\mu}  = -\frac{1}{384 \pi^2} \epsilon^{\mu\nu\rho\sigma}  R_{\mu\nu\alpha\beta} R^{\alpha\beta}\,_{\rho\sigma} + \frac{e^2}{16 \pi^2} \epsilon^{\mu\nu\alpha\beta} F_{\mu\nu} F_{\alpha\beta}  
\end{align}
where $J^{\mu}_A$ is the chiral current, $R_{\rho\sigma}\,^{\alpha\beta}$ is the Riemann tensor and  $A_{\mu}$ is the four-vector potential of the electromagnetic field, $F_{\mu\nu} = \nabla_{\mu}A_{\nu} - \nabla_{\nu}A_{\mu} $. $\epsilon^{\mu\nu\rho\sigma} = \frac{1}{\sqrt{-g}}\, \eta^{\mu\nu\rho\sigma}$ is a fully antisymmetric tensor, $\eta^{\mu\nu\rho\sigma}$ is Levi-Civita symbol whose values are $\pm1$ and we set $\eta^{0123} = 1 = - \eta_{0123}$. It is easy to see from the above equation that the
anomaly contribution from the electromagnetic field and the gravity act independently and, for most parts, can be treated independently.

In the case of flat FRW background in conformal time ($\eta$):
\begin{align}\label{eq:FRW}
ds^2 = a^2(\eta) \,(d\eta^2 - \delta_{ij} dx^i dx^j)
\end{align}
the contribution of the first term in the RHS of Eq.~(\ref{eq:chiralAnomaly}) vanishes, i. e.,
\begin{align}
\epsilon^{\mu\nu\rho\sigma} R_{\mu\nu\alpha\beta} R^{\alpha\beta}\,_{\rho\sigma} = 0 \, .
\end{align}
It can be shown that even at the first-order, the gravitational contribution vanishes, and the non-zero contribution arises only at second order~\cite{2006-Alexander.Peskin.Jabbari-PRL}. Due to the presence of the antisymmetric tensor, the gravitational fluctuations lead to gravitational birefringence and can lead to net chiral current. 
%In this work, we consider the contributions from the first-order and, hence, we will not consider the anomaly from the background electromagnetic fields generated during inflation.

In the flat FRW background, the second term in the RHS of Eq.(\ref{eq:chiralAnomaly}) is given by:
\begin{align}
\frac{e^2}{16 \pi^2} \epsilon^{\mu\nu\alpha\beta} F_{\mu\nu} F_{\alpha\beta} = \frac{e^2}{4 a^4} \epsilon_{ijk} \partial_j A_k \, \partial_0 A_i  \, .
%=  \frac{4}{a^4} \textbf{E} \cdot\textbf{B}
\end{align}
In the presence of the magnetic field, this term is non-zero and hence leads to 
a net chiral current. 
Thus, if we consider only up to the first-order in perturbations, 
\emph{only} the second term in the RHS of  Eq.~(\ref{eq:chiralAnomaly})  contributes and the chiral anomaly equation reduces to:
\begin{align}\label{eq:chiralAnomaly_FF}
\partial_{\mu}\left(  \sqrt{-g} J_A^{\mu} \right) =  \frac{e^2}{16 \pi^2} \eta^{\mu\nu\alpha\beta} F_{\mu\nu} F_{\alpha\beta}  \, ,
\end{align}
where we have used 
\[
\nabla_{\mu}J^{\mu}_A = \frac{1}{\sqrt{-g}} \partial_{\mu} \left( \sqrt{-g} J^{\mu}_A \right), \quad \epsilon^{\mu\nu\alpha\beta} = \frac{1}{\sqrt{-g}} \eta^{\mu\nu\alpha\beta} \, .
\]

Note that during inflation, LHS in Eq. (\ref{eq:chiralAnomaly_FF}) is zero, 
and due to the exponential expansion, standard model particles are diluted. However, if we can generate non-zero primordial helical fields during inflation, then 
these non-zero primordial helical fields can lead to chiral current at the radiation-dominated epoch (or during reheating when the standard model particles are created). To see this, we rewrite Eq.~(\ref{eq:chiralAnomaly_FF})
using $\eta^{\mu\nu\alpha\beta} F_{\mu\nu} F_{\alpha\beta} = 4 \partial_{\mu} \left( \eta^{\mu\nu\alpha\beta} A_{\nu} \partial_{\alpha} A_{\beta}  \right)$, i.e.,
\begin{align}\label{eq:chiral_topo_current}
 \partial_{\mu} \left( \sqrt{-g} J^{\mu}_A \right)  =  \frac{e^2}{4 \pi^2} \partial_{\mu} \left(  \eta^{\mu\nu\alpha\beta} A_{\nu} \partial_{\alpha} A_{\beta}  \right) =  \frac{e^2}{4 \pi^2} \partial_{\mu} \left( \sqrt{-g} K^{\mu} \right) 
\end{align}
where 
\[
K^{\mu} =  \frac{\eta^{\mu\nu\alpha\beta} }{\sqrt{-g} } A_{\nu} \partial_{\alpha} A_{\beta} 
\]
is the topological current. For FRW background, the components are given by
\begin{align}
K^0 =  a^{-4}(\eta) \, \epsilon_{ijk} A_{i} \partial_j A_k  \qquad \text{and} \qquad K^i = a^{-4}(\eta) \, \epsilon_{ijk} A_{j} \partial_0 A_k.
\end{align}
Solving Eq.~(\ref{eq:chiral_topo_current}), we get, 
\[
J^{\mu}_A   =  \frac{e^2}{4 \pi^2} K^{\mu}  \, .
\]
Thus, the net baryon number density, $n_B = n_b - n_{\bar{b}} = a(\eta) \langle 0 | J^0_A | 0 \rangle$ is related to Chern-Simon number density $n_{CS} = \langle 0 | K^0 | 0 \rangle$ as~\cite{2014-Barrie.Kobakhidze-JHEP},
\begin{align}\label{eq:n_B-n_CS-definition}
n_B \equiv \frac{e^2}{4\pi^2} a(\eta) n_{CS}.
\end{align}
Note that $n_{CS} = 0$ at the start of inflation, and due to the absence of standard model particles $n_B = 0$ during inflation. Using the expression for $K^0$, we can write the Chern-Simon number density as
\begin{align}\label{eq:n_cs-relation}
n_{CS} = \frac{1}{a^4} \epsilon_{i j k} \langle 0 | A_i \, \partial_j A_k | 0 \rangle = \frac{1}{a^4}\int_{\mu}^{\Lambda} \frac{dk}{k} \frac{k^4}{2\pi^2} \left(  | A_+ |^2 - |A_-|^2  \right) \, ,
\end{align}
where $\Lambda$, and $\mu$ set the possible energy range (or epoch) during which baryon asymmetry is generated after inflation, and $A_{\pm}$ refer to the positive and negative helicity modes of the electromagnetic field. The above expression is key in illuminating a useful relation between primordial helical magnetic fields generated during inflation and baryogenesis: 
First, we see that the contribution to $n_{CS}$ is from all the modes that reenter the horizon at the beginning of the radiation-dominated epoch. Thus, the value of $n_{CS}$ depends on the upper cut-off $\Lambda$. 
Second, the expression corresponds to the total Chern-Simons number density generated from the modes 
in the energy range $[\mu, \Lambda]$ ---  when these helical modes re-enter during the radiation-dominated epoch. 
The helicity modes $A_+ $ and $ A_-$ are generated during inflation, and ${a^{-4}(\eta)}$ is the dilution due to the expansion of the Universe during this epoch. 
Finally, $n_{CS}$ vanishes if the primordial magnetic fields are non-helical, i. e. $|A_+ | = |A_-|$. Hence, as mentioned at the beginning of this section, the generation of non-helical magnetic fields will not lead to baryogenesis. Thus, the key missing ingredient of Davidson's argument is the requirement of primordial helical magnetic fields. 

In the following two sections, we explicitly evaluate the Chern-Simons number for our model and show that it is not sensitive to inflationary and reheating dynamics. 
\section{The model and the primordial helical fields}
\label{sec:Model}
Like in chapter~\ref{ch:helical-PMF}, we consider the following action \cite{2020-Kushwaha.Shankaranarayanan-PRD} :
\begin{align}\label{eq:action}
S  = S_{\rm{Grav}} + S_{\phi} + S_{\rm{EM}} + S_{\rm CB}
\end{align}
where $ S_{\rm{Grav}}$ is the Einstein-Hilbert action
\begin{align}\label{eq:EH-action}
S_{\rm Grav} = -\frac{M_{\rm P}^2}{2}\int d^4x \sqrt{-g} \, R \, ,
\end{align}
and $ S_{\phi} $ is the action for the minimally coupled, self-interacting canonical scalar field:
\begin{align}\label{eq:inflation-action}
S_{\phi} = \int d^4x \sqrt{-g} \left[  \frac{1}{2} \partial_{\mu}\phi \partial^{\mu}\phi -  V(\phi) \right].
\end{align}
$S_{\rm{EM}}, S_{\rm CB}$ refer to the standard electromagnetic (EM)  and conformal breaking part of the electromagnetic terms, respectively, which are given by:
\begin{align}\label{eq:S_EM}
 S_{\rm{EM}} &= -\frac{1}{4} \int d^4x \, \sqrt{-g} \, F_{\mu\nu} F^{\mu\nu}, \hspace{0.5cm}\\  
 \label{eq:S_h}
 S_{\rm{CB}} &= - \frac{1}{M^2} \,\int d^4x \, \sqrt{-g} \, R_{\rho\sigma}\,^{\alpha\beta} F_{\alpha\beta} \, \tilde{F}^{\rho\sigma} = - \frac{1}{M^2} \,\int d^4x \, \sqrt{-g} \, \tilde{R}^{\mu\nu\alpha\beta} F_{\alpha\beta} \, F_{\mu\nu} \, ,
 \end{align}
where $\tilde{R}^{\mu\nu\alpha\beta} = \frac{1}{2}\epsilon^{\mu\nu\rho\sigma} R_{\rho\sigma}\,^{\alpha\beta}$ is the dual of Riemann tensor and $\tilde{F}^{\rho\sigma} = \frac{1}{2} \epsilon^{\mu\nu\rho\sigma}F_{\mu\nu} $ is the dual of $F_{\mu\nu}$.  The standard electromagnetic action $S_{\rm{EM}}$ is conformally invariant; however, the presence of Riemann curvature in $S_{\rm CB}$ breaks the conformal invariance. $M$ is the energy scale, which sets the scale for the breaking of conformal invariance. Note that the signs of $S_{\rm{EM}}$ and $S_{\rm{CB}} $ are chosen with respect to the positive electromagnetic energy density.

In Ref. \cite{2019-Ruhdorfer.etal-JHEP}, the authors systematically showed that the first gravity operators appear at mass dimension 6 in the series expansion of the coupling between gravity and the standard model of particle physics. These operators only couple to the standard model Bosons. They also showed that (i) no new gravity operators appear at mass dimension 7, (ii) in mass dimension 8, the standard model Fermions
appear, and (iii) coupling between the scalar (Higgs) field and the standard model gauge Bosons appear only at mass dimension 8. Since mass dimension 8 operators are highly suppressed, like in Ref. \cite{2020-Kushwaha.Shankaranarayanan-PRD}, we limit ourselves to mass dimension 6  operators. Due to Riemann coupling, $M$ appears as a time-dependent coupling in the FRW background i.e., $1/M_{\rm eff} \sim H/M$.
At the current epoch where 
$H_0 \approx 10^{-42} \rm{GeV}$ and assuming the parameter $M \approx 10^{17} \rm{GeV}$, we obtain ${H_0}/{M} \sim 10^{-59}$. Therefore, the coupling (Riemann tensor) is tiny and the non-minimal coupling term in the electromagnetic action will have significant contribution only in the early universe.
We also would like to point that the coupling term  
($S_{\rm CB}$) is tiny near the Schwarzschild radius of a solar mass black-hole  
(for details, see appendix \ref{app:blackhole}).

We assume that the scalar field ($\phi$) dominates the energy density in the during inflation and leads to $60 \, - \, 70$ e-foldings of inflation with $H_{\rm Inf} \sim 10^{14} {\rm GeV}$. Specifically, we consider power-law inflation in which the scale factor (in conformal time) is~\cite{2004-Shankaranarayanan.Sriramkumar-PRD}:
 \begin{align}\label{eq:powerLaw}
a(\eta) =  \left( - \frac{\eta}{\eta_0} \right)^{(\beta+1)}
\end{align}
where, the constant $\eta_0$ denotes the scale of inflation and $\beta \leq -2$. $\beta = -2$ corresponds to exact de Sitter.  During inflation, $\eta \in (-\infty, 0)$. For slow-roll inflation $ \beta \approx -2-\epsilon$ and $ \mathcal{H} \equiv a^{\prime}/{a} \approx - (1 + \epsilon)/{\eta}$, where 
$\mathcal{H}$ is the Hubble parameter in conformal time and 
$\epsilon $ is the slow roll parameter. For our discussion below, we also assume that $10^{-3}  \leq (H_{\rm Inf}/M) \leq 1$~\cite{2018-Nakonieczny-JHEP,2016-Goon.Hinterbichler-JHEP,2016-Goon-JHEP,2013-Balakin.etal-CQG}.

Equation of motion of the gauge field can be obtained by varying the action (\ref{eq:action}) with respect to $A^{\mu}$. In the Coulomb gauge ($A^{0} = 0, \partial_iA^i = 0$), we have:
\begin{align}\label{eq:equation_of_motion}
A_i^{\prime\prime} + \frac{4 \, \epsilon_{i j l}}{M^2} \, \left( \frac{a^{\prime\prime\prime}}{a^3} - 3\frac{a^{\prime\prime} a^{\prime} }{a^4} \right) \partial_j A_l 
- \partial_j \partial_j A_i = 0
\end{align}
where $\epsilon_{i j l}$ is the Levi-Civita symbol in the 3-D Euclidean space. The above equation is different from other models in the literature and leads to distinct evolution of the magnetic field fluctuations in comparison to non-minimally coupled scalar field models~\cite{2020-Kushwaha.Shankaranarayanan-PRD}. In the helicity basis, the above equation reduces to (see appendix \ref{app:helicity_basis}):
\begin{align}\label{eq:eom_helicity}
A_h^{\prime\prime} + \left[  k^2 - \frac{4kh}{M^2} \, 
\left( \frac{a^{\prime\prime\prime}}{a^3} - 3\frac{a^{\prime\prime} a^{\prime} }{a^4} \right)  \right] A_h= 0 \, .
\end{align}
For the two helicity states ($h = \pm$), the above expression leads to two different evolution equations [cf. Eqs.~(\ref{eq:sup_mode_h+}, \ref{eq:sup_mode_h-})]. From Eq. \eqref{eq:n_cs-relation} we see that to obtain appreciable value of Chern-Simons number ($n_{CS}$), the difference between the two helicity states should be non-zero, and it is maximum 
if one helicity mode is enhanced compared to other. 

In the previous chapter~\ref{ch:helical-PMF}, we showed that for a range of parameters of interest, negative helicity mode decays while the positive helicity mode is enhanced~\cite{2020-Kushwaha.Shankaranarayanan-PRD}. Hence, negative helicity mode ($A_-$) will have negligible contribution and can be set to zero, i. e., 
$|A_-| = 0$. Using the series expansion of the Bessel functions, in the leading order, the positive helicity mode takes the following form
(\ref{eq:sup_mode_h+}): 
%
%\begin{subequations}
\begin{align}\label{eq:A+Series}
A_+(\tau,k) &= C \, k^{\frac{1}{4\alpha}} 
  - C_2 \frac{\mathcal{F}^{-1} }{\pi} 
  \Gamma \left( \frac{1}{2\alpha}  \right) \,k^{-\frac{1}{4\alpha}} \tau^{ - \frac{1}{\alpha} }  
  %
%  \label{eq:A-Series}
%A_-(\tau,k) &=\tilde{C} \, k^{\frac{1}{4\alpha}} 
%-C_4 \frac{\tilde{\mathcal{F}}^{-1} }{\pi} 
%  \Gamma \left( \frac{1}{2\alpha}  \right) \, k^{-\frac{1}{4\alpha}} \tau^{ -\frac{1}%{\alpha} } .
\end{align}
where,
\begin{align}\label{eq:C_F}
| C | \approx \varsigma^{-1} |C_2| \approx 
\frac{M^{3/2} \eta_0}{\sqrt[4]{\eta_{end} 10^{45} GeV^3}},~~  
\mathcal{F} \approx |\varsigma|^{-1} \approx \sqrt{M^2 \eta_0}, ~~
%
%\varsigma \approx \sqrt{ \frac{1}{M^2 \eta_0} } ,~~
%
\alpha = -\frac{1}{2} -\epsilon
\end{align}
For details, see Appendix \eqref{app:Helical}. 

Our model generates primordial magnetic fields through the non-minimal coupling of the electromagnetic field. The model requires inflation. Inflation generates density perturbations at all scales and provides a causal mechanism to generate the structure formation. Similarly, our model generates magnetic fields at all length scales, including the current Horizon radius~\cite{2001-Grasso.etal-PhyRep,2013-Durrer.Neronov-Arxiv,2016-Subramanian-Arxiv,2002-Widrow-Rev.Mod.Phys.,2004-Giovannini-IJMPD}. This has to be contrasted from the models where the magnetic field is generated during recombination. In these models, the coherence scale of the generated fields cannot exceed the size of the horizon radius at that time.

In Appendix \ref{app:Helical}, we have plotted the power spectrum of the present-day helical magnetic field ($B_0$) as a function of $k$. Assuming $M = 10^{17} GeV$, our model predicts the  primordial helical magnetic fields of strength $10^{-20} \rm{G}$ on Gpc  scales at the current epoch. From \ref{fig:PowerSpectrum} we can see that our model predicts the present-day helical magnetic field of strength $10^{-15} G$ on Mpc scales. The primordial fields generated from our model are within the upper bounds on the strength of the seed magnetic fields needed to explain the current galactic magnetic fields~\cite{2010-Kahniashvili.Ratra.etal-PRD}. These primordial fields are amplified by the dynamo mechanism and can lead to the observed magnetic fields; hence our model requires the dynamo mechanism.

%
%
%%============ S E C T I O N =====================
%
\section{Baryon Asymmetry of the Universe}
\label{sec:baryon_asymm}
In this section, we compute the baryon asymmetry parameter due to the primordial helical magnetic fields. Specifically, we compute it for the maximum helicity modes --- one mode is enhanced compared to the other. Substituting Eq~(\ref{eq:A+Series}) in Eq.~(\ref{eq:n_cs-relation}), we obtain
\begin{align}\label{eq:n_CS-integral}
n_{CS} = \frac{1}{2\pi^2 \, a^4(\eta)}\int^{\Lambda}_{\mu} dk  \left( \left| C \right|^2 \, k^{3 + \frac{1}{2\alpha}} 
  + \left| C_2 \frac{\mathcal{F}^{-1} }{\pi} 
  \Gamma \left( \frac{1}{2\alpha}  \right) \right|^2\,k^{3 -\frac{1}{2\alpha}} \tau^{ - \frac{2}{\alpha} }    \right).
\end{align}
Integrating the above expression, we get
\begin{align}\label{eq:n_CS}
n_{CS} =  \frac{1}{2\pi^2 \, a^4(\eta)}\left[  \left.  \left| C \right|^2 \,  \frac{ k^{4 + \frac{1}{2\alpha}} }{ 4 + \frac{1}{2\alpha}} \right|^{\Lambda}_{\mu}
  + \left. \left| C_2 \frac{\mathcal{F}^{-1} }{\pi} 
  \Gamma \left( \frac{1}{2\alpha}  \right) \right|^2 \,\frac{ k^{4 -  \frac{1}{2\alpha}} }{ 4 - \frac{1}{2\alpha}} \tau^{ - \frac{2}{\alpha} } \right|^{\Lambda}_{\mu} \,\,    \right].
\end{align}
We want to make the following remarks regarding the above expression: First, the BAU is generated similarly to the inflationary mechanism of the generation of density perturbation. During inflation, the primordial helical magnetic field fluctuations are stretched exponentially and exit the horizon. The modes that reenter during the radiation-dominated epoch are responsible for the generation of baryon asymmetry.  Second, the generation of baryon asymmetry does not strongly depend on the reheating dynamics since only the modes that 
reenter the Hubble radius during the radiation-dominated epoch are relevant.  

Assuming a de-Sitter (or approximately de-Sitter) Universe, from Eq. \eqref{eq:C_F}, 
we have $\tau^{-\frac{2}{\alpha}} = a^{-2}(\eta)$. Substituting this in the 
Eq. (\ref{eq:n_CS}), we see that the 
the second term in the RHS decays faster compared to the first term by $a^{-2}(\eta)$. Hence, we can neglect the second term. Substituting the 
resulting form of $n_{CS}$ in Eq.~(\ref{eq:n_B-n_CS-definition}) leads to:
\begin{align}\label{eq:n_B}
n_{B} = \frac{e^2}{4\pi^2} \frac{1}{2\pi^2 \, a^3(\eta)}  \left.  \left| C \right|^2 \,  \frac{ k^{4 + \frac{1}{2\alpha}} }{ 4 + \frac{1}{2\alpha}} \right|^{\Lambda}_{\mu}.
\end{align}
To obtain the ranges of $\Lambda$ and $\mu$,  we need to know the modes exited during inflation. For the density perturbations, the largest scales observed in the CMB are produced around 40 - 60 e-foldings before the end of inflation~\cite{2006-Bassett.Tsujikawa.Wands-RevModPhys}.
This is because the adiabatic quantum fluctuations responsible for the density perturbations reenter the Hubble radius around $z \sim 1500$. Hence, in Ref. \cite{2020-Kushwaha.Shankaranarayanan-PRD}, the current authors only looked at primordial helical fields generated around 40 - 60 e-foldings before the end of inflation. However, in this case, we will concentrate on the primordial helical fields that renter the horizon very early (at the beginning of the radiation-dominated epoch) to generate the required BAU. This means that the modes that left the horizon around the last 5 to 10 e-foldings of inflation are only relevant. Since these modes have already left the Hubble radius during inflation, the reheating dynamics do not alter these primordial helical modes. Hence, the model is insensitive to the reheating dynamics.  

Our focus now shifts to explicitly evaluating BAU for our model.  First step is to evaluate the 
dilution factor $a^{-3}$ in Eq. \eqref{eq:n_B}. To do this, we define 
$a_{\Lambda}$ (and $a_{\mu}$) as the scale factor at the time when the maximal helicity mode with energy $\Lambda$ (and $\mu$) left the Hubble radius during inflation. Assuming an instant reheating, and following the calculations given in Appendix \eqref{app:Calculations}, we have 
$a_{\mu} = 10^6 a_{\Lambda}$. Taking into account that these modes 
exited the Hubble radius during inflation in the last 5 e-foldings, the 
the dilution factor [prefactor in Eq. \eqref{eq:n_B}]  
becomes $a^{-3} \sim 10^{-24}$. 

The second step is to obtain the constant $C$. 
As discussed in previous section, for slow-roll inflation, $|C|$ is given by 
Eq. \eqref{eq:C_F}. Thus, Eq. \eqref{eq:n_B} reduces to:
\begin{align}\label{eq:n_B-Lambda}
n_{B} \approx  \quad \frac{10^{-24} \cdot |C|^2 \cdot e^2}{24\pi^4}  \left( \Lambda^3 - \mu^3 \right) \, .
\end{align}
Third step is to compare the theoretically derived quantity ($n_B$) 
with observations Eq.~\eqref{def:etaObs}. However, $n_{\gamma}$ is not constant in the early Universe (since the photon chemical potential is zero) and is approximately constant only after the last scattering surface. 
Since entropy density per comoving volume is conserved, the quantity $n_B/s$ is better suited for theoretical calculations~\cite{Book-Kolb.Turner}.
Assuming that there was no significant entropy production after reheating phase, entropy density in the radiation-dominated epoch is: 
\begin{equation}
s \simeq  \frac{2\pi^2}{45} g \,  T^3_{\rm{RH}} \, ,
\label{def:entdens}
\end{equation}
where $T_{\rm{RH}}$ is the reheating temperature and the effective relativistic degrees of freedom $g \sim 100$ at reheating. From Eqs. (\ref{eq:n_B-Lambda}, \ref{def:entdens}), we can define the following dimensionless BAU parameter: 
\begin{align}\label{eq:baryon_Asym}
\eta_B = \frac{n_B}{s} \approx 10^{-24} 
\frac{|C|^2 \cdot e^2}{24\pi^4 }  \left( \Lambda^3 - \mu^3 \right)
\frac{45}{2\pi^2 g T_{RH}^3}  \approx 10^{-29} |C|^2
\frac{\Lambda^3 }{  T^3_{\rm{RH}}  }
\end{align}
where in the last expression we have neglected $\mu^3$ i.e., $ \Lambda^3 - \mu^3 \approx \Lambda^3$. Appendix \eqref{app:Calculations} contains
plots for different values of $\Lambda$ and $\mu$. From these plots, we infer that the results do not strongly depend on the exact value of $\mu$. 

Finally, substituting the value of $|C|^2$ (from Eq. \eqref{eq:C_F} and 
using the values in Appendix \ref{app:Helical}) in Eq.~\eqref{eq:baryon_Asym}, we obtain:
\begin{align}\label{eq:baryon_Asym-M}
\eta_B  \approx  \frac{ 10^{-29} \cdot \eta_0^2 }{ \sqrt{\eta_{end} \cdot 10^{45} GeV^3} } \frac{M^3 \Lambda^3 }{  T^3_{\rm{RH}}  }  \approx 
10^{-2} \left( \frac{M}{M_P} \right)^3  
\left( \frac{\Lambda}{T_{ \rm{RH}}} \right)^3 
\end{align}
This is one of the crucial expressions in this chapter regarding which we would like to stress the following: First, the BAU parameter depends on three quantities --- $M$ (the conformal invariance breaking scale), $T_{\rm{RH}}$ (reheating temperature scale) and $\Lambda$ (the largest helical mode that catalyses
baryogenesis). 
Second, the BAU parameter is inversely proportional to the reheating temperature. This behavior is different from the results of Ref. \cite{2006-Alexander.Peskin.Jabbari-PRL,2014-Barrie.Kobakhidze-JHEP,2014-Long.Sabancilar.Vachaspati-JCAP,2016-Fujita.Kamada-PRD}. In some of these models, BAU is linearly dependent on the reheating temperature. The difference in the relationship is because the detailed reheating dynamics is not required, only the information about the entropy production is required in our model. In other models, the exact detailed reheating dynamics is required, which is avoided in our approach. 
Third, the BAU parameter is linearly proportional to $M$ and $\Lambda$. For smaller $M$, the contribution of the conformal breaking term \eqref{eq:S_h} will be much larger, and hence, more primordial helical fields are produced during inflation. However, for the same reheating temperature, $\Lambda$ has to be larger to produce the same amount of BAU. 
Fourth, to get a better understanding of the dependence of BAU on various parameters, we use the following parametrization:
\begin{align}\label{eq:parametrization}
\eta_B = n \times 10^{-10}, \quad M = m \times 10^{14} GeV,  
\quad \Lambda = \delta \times 10^{12} GeV, \quad 
T_{RH} = \gamma \times 10^{12} GeV 
\end{align}
where $n, m, \delta, \gamma$ are dimensionless parameters. The maximum reheating 
corresponds to the inflation scale~\cite{2006-Bassett.Tsujikawa.Wands-RevModPhys}. With supersymmetry, the 
requirement that not too many gravitinos are produced 
after inflation provides a stringent constraint on the reheating temperature,
$T_{\rm RH} \sim 10^{10} - 10^{11}~$GeV~\cite{1984-Ellis.Kim.Nanopoulos-PLB,1999-Benakli.Davidson-PRD}. Hence, we consider the range of $\gamma$ 
to be  $\{ 10^{-2}, 1000  \} $. Since the value of $M$ should be between the GUT and Planck scale, we consider the range of $m$ to be $\{ 1, 1000  \}$.  
We assume that the modes that reenter during radiation epoch is around $10^{12}~$GeV. Hence, we consider the range of $\delta$ to be $\{ 1, 100  \} $. Using the above parametrization in Eq.~(\ref{eq:baryon_Asym-M}), we get:
\begin{align}\label{eq:baryon_Asym-Parameter}
\frac{m^3 \times \delta^3 }{\gamma^3}  \approx  n \, 10^7 . 
\end{align}
\begin{figure}
	\centering
	\includegraphics[height=2in]{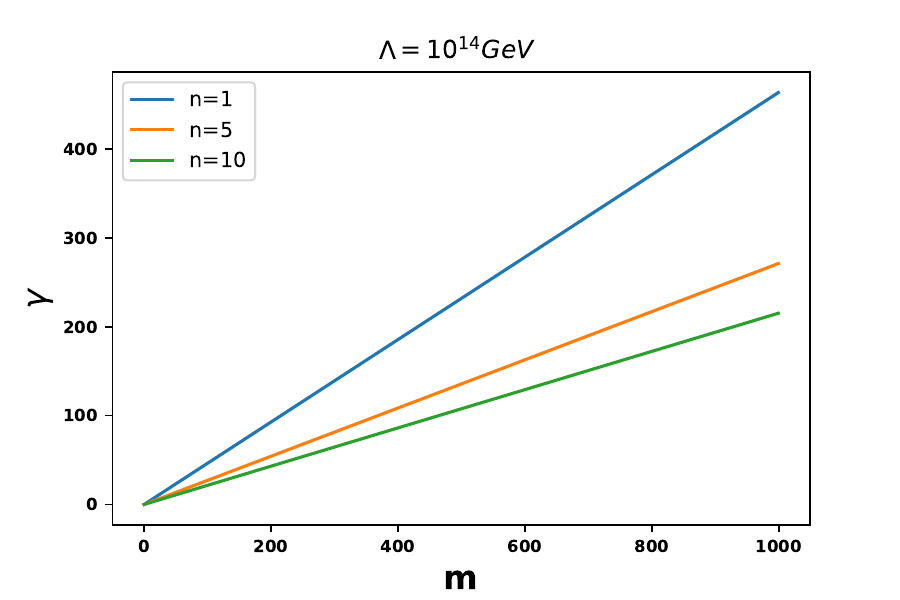}
	\caption{Plot of the rescaled reheating temperature $T_{RH}$  with the rescaled conformal symmetry breaking  parameter $M$, for different values of $n$. Here, we have set $\Lambda = 10^{14} GeV, \mu = 10^{10} GeV$.}
	\label{fig:Plot}
\end{figure}

\begin{figure}[!hbt]
%\centering
%\subfigure[]{%
%\label{fig:first}%
\includegraphics[height=2in]{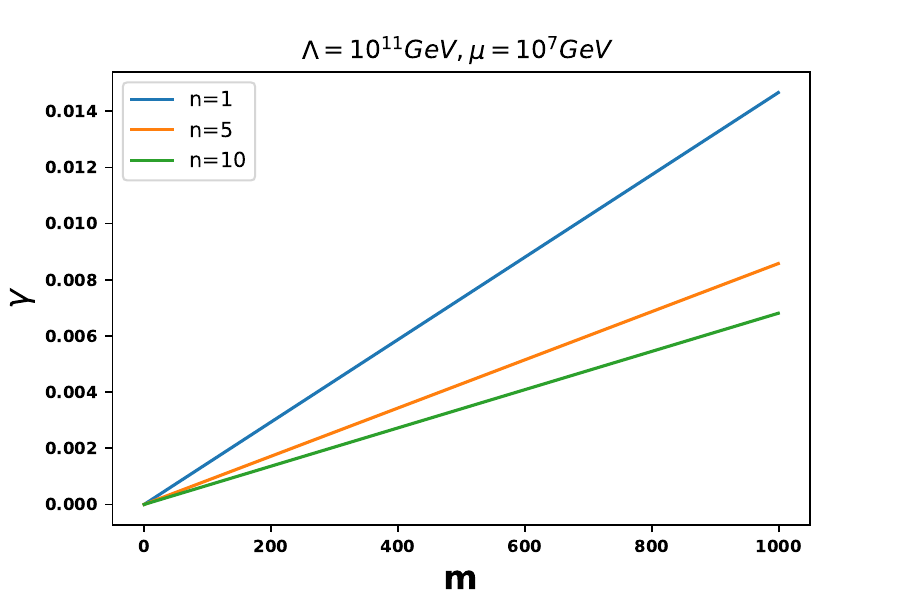}%
\quad
%\subfigure[]{%
%\label{fig:third}%
\includegraphics[height=2in]{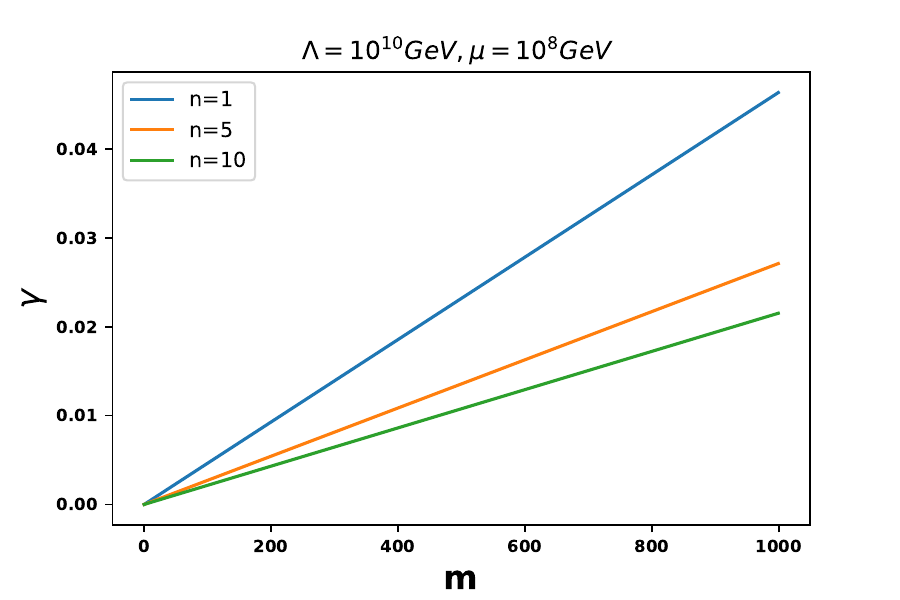}%
\quad
%\subfigure[]{%
%\label{fig:third}%
\includegraphics[height=2in]{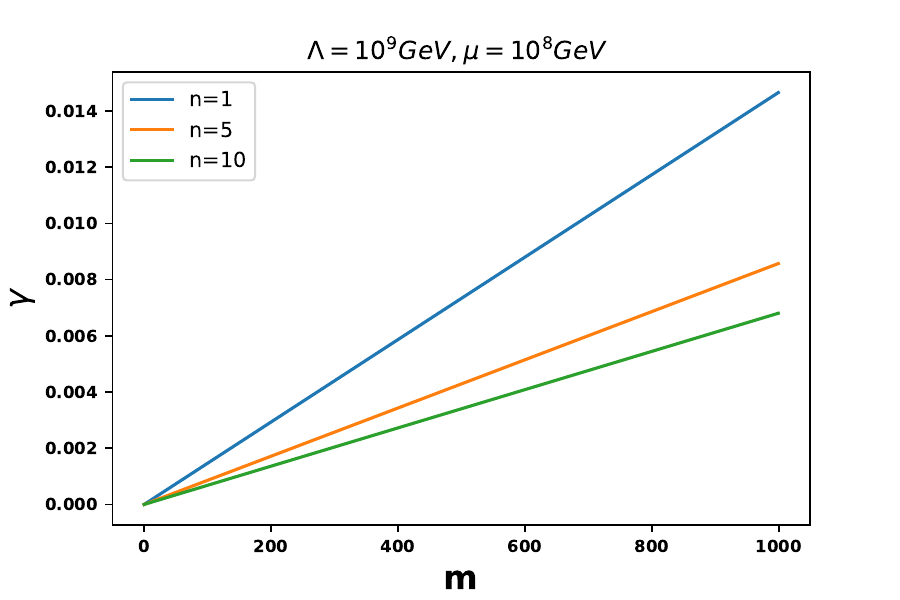}%
\quad
%\subfigure[]{%
%\label{fig:fourth}%
\includegraphics[height=2in]{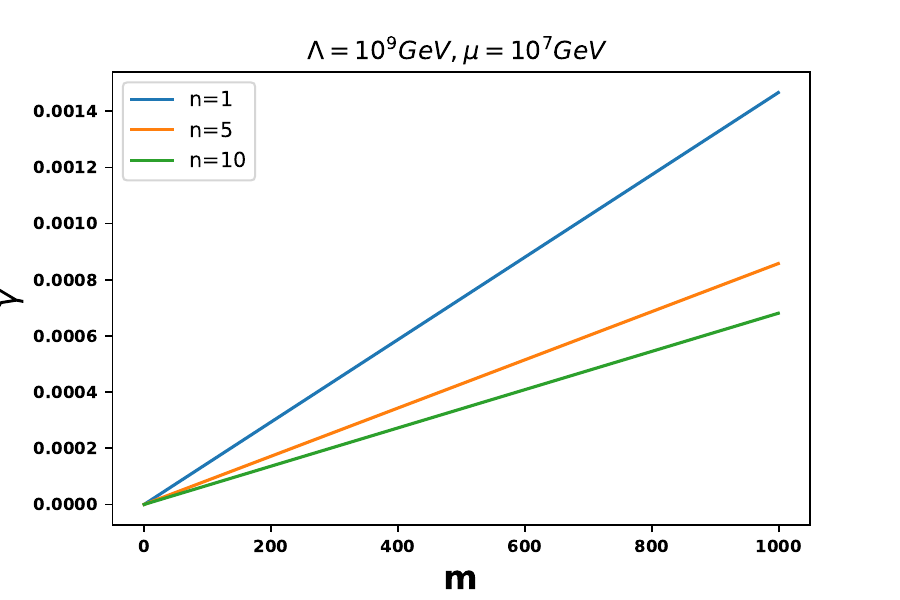}%
\caption{Plots showing the behaviour of reheating temperature $T_{RH}$ (vertical axis) with parameter $M$ (horizontal axis), for lower energy scales 
of $\Lambda$ and $\mu$.}
\label{fig:reheatingGeneric2}
\end{figure}

\ref{fig:Plot} and \ref{fig:reheatingGeneric2} contain the plots of $\gamma$ versus $m$ for different values of $n$ and fixed $\delta$. In Appendix \eqref{app:Calculations} we have plotted the same for other values of $\delta$. From these plots, we deduce the following: First, for a range of values of 
$\gamma,$ $\delta$, and $m$, 
BAU can have values between $10^{-10}$ to $10^{-9}$. Thus, the model 
can lead to the observed amount of baryon asymmetry of the Universe consistent with the Planck data~\cite{2018-Planck}. Second, the model does not depend on the nature of the reheating dynamics. As can be seen from the plots, for a range of values of $m, \delta$, the model can lead to BAU for a range of reheating temperatures. This has to be contrasted with other models in the literature~\cite{2014-Long.Sabancilar.Vachaspati-JCAP,2016-Fujita.Kamada-PRD} which requires detailed knowledge of the reheating phase of the Universe. Third, the unknown parameter in the model is $M$. In Ref.~\cite{2020-Kushwaha.Shankaranarayanan-PRD}, we showed that for the model to be consistent with the lower limit of $10^{-16}$ Gauss magnetic fields in the voids~\cite{2010-Neronov.Vovk-Sci}, then $M \sim 10^{17} GeV$. The 
current analysis shows that $M \sim 10^{17} GeV$ is consistent with baryogenesis. Thus, the model is \emph{tantalizingly close} to solving baryogenesis and magnetogenesis using the same causal mechanism that solves the origin of density perturbations.

\section{Conclusions and Discussions}
\label{sec:conc}

In this chapter, we have proposed a viable baryogenesis scenario in the early Universe that does not require any extension to the Standard Model of particle physics. The crucial ingredient is the generation of primordial helical magnetic fields
due to Riemann coupling. The advantage of the primordial helical fields is that the non-zero helicity suggests a non-zero contribution in the CP violation term. 
An interesting feature of our model is the stretching of the primordial helical magnetic fields to super-horizon scales during inflation --- the same mechanism that leads to primordial density perturbations. While the helical modes generated around 40 - 60 e-foldings before the end of inflation lead to the observed large-scale magnetic fields, the helical modes that renter the horizon very early (at the beginning of the  radiation-dominated epoch) lead to the baryon asymmetry.  Thus, our mechanism provides \emph{possible testable evidence} for the entire inflationary epoch. 

More than two decades ago, Davidson pointed out an interesting relation between the primordial magnetic field and Sakharov's conditions~\cite{1996-Davidson-PLB}. In this chapter, we have explicitly shown that Davidson's conditions are necessary \emph{but not} sufficient. The key missing ingredient is the requirement of \emph{primordial helical magnetic fields}. While the helical and non-helical fields break the isotropy and lead to CP violation, only the modes with maximal helicity contribute significantly to the Chern-Simon number density. We have shown that the BAU parameter predicted by our model is independent of any specific inflation model and reheating dynamics; however, it depends on the scale at which inflation ends and reheating temperature. 
 
The BAU parameter \eqref{eq:baryon_Asym-M} obtained in our model is inversely proportional to reheating temperature. Assuming the exit of inflation at $10^{14}$~GeV, for the observed amount of baryon asymmetry $\eta_B \sim 10^{-10}$, we obtained  that the reheating temperature should be in the range 
$10^{12} - 10^{14}$~GeV, which is consistent with the constraints on the reheating temperature \cite{2006-Bassett.Tsujikawa.Wands-RevModPhys,1984-Ellis.Kim.Nanopoulos-PLB,1999-Benakli.Davidson-PRD}. 
This means that our model \emph{does not} prefer a very low-energy reheating temperature~\cite{1999-Benakli.Davidson-PRD}. 

In the literature, various mechanisms have been discussed to solve the BAU problem using the primordial helical magnetic fields~\cite{2014-Barrie.Kobakhidze-JHEP,2014-Long.Sabancilar.Vachaspati-JCAP,2016-Fujita.Kamada-PRD,2015-Anber.Sabancilar-PRD,2019-Domcke.etal-JCAP}. 
In Ref.~\cite{2016-Fujita.Kamada-PRD}, the authors obtained the required BAU by assuming the presence of helical magnetic fields of present-day strength $10^{-14} G < B_0 < 10^{-12} G $ and coherence length $1 \rm{pc} < \lambda < 1 \rm{Mpc}$, and taking into account of the MHD effects. In Ref.~\cite{2014-Long.Sabancilar.Vachaspati-JCAP}, authors studied the generation of a primordial magnetic field in conjunction with the BAU generation through leptogenesis; however, the predicted value of the present-day coherence length of such magnetic fields is very small$\sim 10$~pc.

In Refs.~\cite{2015-Anber.Sabancilar-PRD,2019-Domcke.etal-JCAP}, the authors consider pseudoscalar inflation (axion inflation) model with a dimension five couplings. In these models, the authors assumed the scale of the baryogenesis to be electroweak scale, and they obtained the required BAU assuming the scale of inflation to be $10^{10} \rm{GeV}$ --- $10^{12} \rm{GeV}$~\cite{2019-Domcke.etal-JCAP,2015-Anber.Sabancilar-PRD}.  In Ref. ~\cite{2014-Barrie.Kobakhidze-JHEP}, the authors considered the extension of the Standard Model with anomalous gauge symmetry. They obtained the 
required BAU for $H_{Inf} \sim 10^{14} \rm{GeV}$ and reheating temperate at $10^{16}\rm{GeV}$. In Ref.~\cite{1999-Brustein.Oaknin-PRL}, the authors argued that to generate the observed baryon asymmetry, some asymmetry in the initial conditions of either $\bf{B}$ or scalar field $\phi$ is required, which can be induced from temperature-dependent potential or asymmetry in quantum fluctuations. Our model is robust to inflationary/reheating dynamics and uses the same success of inflationary perturbations to generate BAU. Thus, our model is \emph{tantalizingly close} to solving baryogenesis and magnetogenesis using the same causal mechanism that solves the origin of density perturbations.

In this chapter, we did not consider the gravity contribution to the chiral anomaly equation. In Ref.~\cite{2006-Alexander.Peskin.Jabbari-PRL}, the authors considered the phenomenon of gravitational birefringence to show that the gravitational fluctuations generated during inflation can give the Universe's observed amount of baryon asymmetry. However, as we showed in. 
Sec.~\eqref{sec:Baryo-magnetic}, $R\tilde{R}$ contributes only in the second-order, and hence we have ignored it in this analysis. It may be interesting to look at the second-order corrections and analyze the parameter constraints.

In this chapter, we have used the general effective field theory of gravity coupled to the Standard Model of particle physics framework to obtain leading order gravity terms that couple to the standard model Bosons~\cite{2019-Ruhdorfer.etal-JHEP}.  We have considered only the mass dimension 6-operators coupling to the gauge field Lagrangian, specifically, to the electromagnetic field. 
The coupling to the Fermions arises at the mass dimension 8. Thus, coupling of Fermion-anti-Fermion with $U(1)$ field will play a role only at this order. While these are expected to be suppressed compared to mass-dimension 6 operators, they are relevant at Planck scale. We plan to look at the effects of mass dimension 8 operators on the baryogenesis.

In this chapter, we focused on the electromagnetic fields and the effects of the helical fields on baryogenesis. It will be interesting to extend the analysis to 
Gluons and study the effects on the asymmetry generated in quarks and the Baryons. It is particularly important, and we will discuss this in detail in chapter~\ref{ch:conclusion}. 
%a study on this is currently in progress to acquire more stringent constraints on the parameters $M$ and $T_{\rm RH}$~\cite{2021-Sharma.Ashu.Shanki}. 

%\pagestyle{fancy}
\chapter{Effective field theory of magnetogenesis}%Effective field theory of magnetogenesis identify necessary and sufficient conditions
\label{ch:EFTmagnetogenesis}

As discussed in Section~\ref{sec:mageticuniverse}, various observations have confirmed the existence of magnetic fields in the Universe~\cite{1994-Kronberg-Rept.Prog.Phys.,2001-Grasso.etal-PhyRep,2010-Neronov.Vovk-Sci}. In galaxies and galaxy clusters, the typical magnetic field strength is found to be on the order of micro-Gauss with a coherence length of kpc to Mpc~\cite{1994-Kronberg-Rept.Prog.Phys.,2001-Grasso.etal-PhyRep,2002-Widrow-Rev.Mod.Phys.,2004-Giovannini-IJMPD,2013-Durrer.Neronov-Arxiv,2016-Subramanian-Arxiv}. Existing data on the magnetic fields in these regions cannot directly constrain the properties and origin of cosmic-scale magnetic fields. Therefore, it is unknown whether their origin is astrophysical or primordial. However, magnetic field measurements from Faraday rotation and Synchrotron radiation provide an upper bound for magnetic fields. In contrast, FERMI measurements of gamma-rays emitted by Blazars provide a lower bound of the order of $10^{-15}~{\rm G}$ in the intergalactic voids~\cite{2010-Neronov.Vovk-Sci}. 

As mentioned in Chapter~\ref{ch:intro}, according to the widely accepted paradigm, magnetic fields in these regions are produced by the dynamo amplification of the weak primordial magnetic field~\cite{2016-Subramanian-Arxiv}. There are various mechanisms for producing the primordial magnetic field, but in most of them, either the produced magnetic field is too weak to be amplified via dynamo or the coherence length is too short to be sustained due to Universe expansion~\cite{2016-Fabre.Shankaranarayanan-ApP}. Inflation provides a causal mechanism to generate the magnetic field over a large scale~\cite{1988-Turner.Widrow-PRD}. 
However, one of the pre-requisites for generating the primordial magnetic field during inflation is breaking the conformal invariance of the 4-D electromagnetic action. 

As discussed in previous chapters~(\ref{ch:GalScalElect},\ref{ch:helical-PMF},\ref{ch:PMF_Baryo}), Several models have been proposed to break the conformal invariance of the action without breaking the gauge invariance. Broadly they can be classified into two categories --- coupling electromagnetic fields with other matter (scalar) fields and higher-derivative terms in the electromagnetic action leading to the non-minimal coupling of the electromagnetic field with curvature~\cite{1988-Turner.Widrow-PRD,1991-Ratra-Apj.Lett,1993-Dolgov-PRD,2020-Talebian.etal-arXiv,2020-Bamba.Odintsov.etal-JCAP,2021-Giovannini-JCAP,2014-Basak-Shanki-JCAP,2017-Debottam.Shankaranarayanan-JCAP,2019-Kushwaha.Shankaranarayanan-PRD,2020-Kushwaha.Shankaranarayanan-PRD}. Due to simplicity, the first class of models, especially, scalar field coupled models are extensively studied~\cite{1991-Ratra-Apj.Lett,2020-Talebian.etal-arXiv}. However, these models suffer from strong coupling and back-reaction problems that necessitate parameter tweaking. 
For the possible resolution of these issues, see Refs.~\cite{2009-Demozzi.etal-JCAP,2017-Sharma.etal-PRD,2021-Nandi-JCAP,2021-Tripathy.etal-arXiv}. 

The second class of models is more natural as higher derivative terms are expected to arise when quantum gravitational effects are taken into account~\cite{1981-DeWitt-PRL,1996-Padmanabhan-PRL}. Demanding that the theory be Lorentz invariant in flat space-time, the field action (in Fourier space) can only be a function of 
$k^2 (\equiv k_{\mu} k^{\mu})$~\cite{1983-Barth.Christensen-PRD,1990-Simon-PRD,2002-Hawking.Hertog-PRD}. Besides, the divergence
structure of quantum field theory is expected to vastly improve when
the quantum gravitational effects are taken into account~\cite{1983-Barth.Christensen-PRD}. For instance, higher-derivative electromagnetic theory by Podolsky-Schwed~\cite{1948-Podolsky.etal-RevModPhys} removes the divergence of the Coloumb potential. One problem with higher derivative theories is the appearance of negative energy states. Although they can be traded by negative norm states (or ghosts), they normally lead to non-unitary theories. However,  vector Galileons do not have ghosts~\cite{2017-Debottam.Shankaranarayanan-JCAP,2019-Kushwaha.Shankaranarayanan-PRD}.

Ideally, we require a fundamental theory of quantum gravity to obtain a generic magnetic field power spectrum generated in the early Universe. However, since we do not have such a consistent model of quantum gravity yet, we aim to obtain an \emph{effective field theory} (EFT) description of primordial magnetogenesis during inflation (based on expansion about the Hubble parameter $(H)$ and its derivatives). In this chapter, we obtain a generic magnetic field power spectrum from a low-energy effective field theory of magnetogenesis. While broken conformal invariance is a common requirement for primordial magnetogenesis, for the first time, we show that causal propagation is also a necessary condition.

While EFT of inflation has been systematically analyzed following Ref.~\cite{2007-Cheung.etal-JHEP}, there is no such systematic analysis for magnetogenesis. See, for instance, Refs.~\cite{2021-Giovannini-PLB,2021-Maity.etal-JCAP}. 
However, two key differences exist between the EFT of magnetogenesis and inflation.  
First, the EFT of inflation is a model-independent framework for studying scalar (and tensor) perturbations. In this setup, 
time-translation invariance needs to be broken as inflation ends at a finite time. In the case of EFT of magnetogenesis, the conformal invariance of the gauge fields also needs to be broken in the cosmological background besides breaking time diffeomorphism. 
Second, in EFT of inflation, one makes a specific gauge choice (unitary gauge) 
where the scalar-field (inflaton) fluctuations are zero, add gravitational operators to the Lagrangian that preserves spatial-diffeomorphism and breaks the time diffeomorphism. As the system breaks time diffeomorphism, one can write down a Lagrangian for the Goldstone Boson associated with the broken symmetry with the Stuckleberg trick. Now, this Goldstone Boson $\pi$ is related to the gauge-invariant quantity the curvature perturbation $\zeta$ as $\zeta = - H \pi$ with $H$ being the Hubble parameter. So by analyzing the dynamics of $\pi$, one can analyze the scalar mode of perturbation through the gauge-invariant quantity $\zeta$ produced during inflation. 
In the case of magnetogenesis, we do not have to make any specific gauge choice where the perturbed gauge fields vanish. This is because, unlike the scalar perturbations, 
the vector modes of the gravitational operators do not 
dynamically affect the gauge field. Hence, we do not need to construct any gauge-invariant variable out of the gauge field and the vector modes of perturbation, and for the EFT expansion, we need only to consider the gauge field $A_{\mu}$. In Appendix \eqref{sec:Decoupling}, we show this explicitly for general relativity.%; to our knowledge, this has not been shown earlier in the literature. 

In this chapter, we systematically write the EFT of magnetogenesis in the early Universe in terms of the Hubble parameter $H(t)$ and its derivatives. More specifically, we expand the Lagrangian in the powers of the cut-off scale $\Lambda$ and consistently analyze the conditions for generating a primordial magnetic field without any other assumptions. 
These terms containing $\Lambda$ break the conformal invariance of the gauge field and satisfy one of the key criteria. Our approach is different from the approaches in Ref.~\cite{2021-Maity.etal-JCAP, 2021-Giovannini-PLB}. In Ref.~\cite{2021-Giovannini-PLB} the effective action terms only consist of four derivatives associated with functions of background inflaton. This expansion leads to different susceptibilities for 
electric and magnetic fields, but the results are produced with a particular parametrization of these susceptibilities. In Ref.~\cite{2021-Maity.etal-JCAP}, the EFT Lagrangian is written in second-order with all possible contractions of electromagnetic tensor $F_{\mu\nu}$ with itself associated with time-dependent analytic functions $f_i(\eta)$. The functions $f_i(\eta)$ is chosen to be proportional to either
$\left({a(\eta)}/{a_f(\eta_f)}\right)^2$ or higher powers where $\eta_i$ and $\eta_f$ are (conformal) time at beginning and end of inflation. 
More specifically, in both these cases, the authors did not
include terms that can be proportional to $({H}/{\Lambda})$ where in our expansion scheme, it naturally arises. 

The EFT method we use to study the physics of magnetogenesis in model-insensitive. We explicitly show that the conformal invariance breaking is \emph{only} a necessary condition, not a sufficient condition.  We show that we can have large amplifications even for super-luminal fluctuations. To avoid EFTs with superluminal fluctuations, we need another physical condition --- the modes should be sub-luminal. As explained in Ref.~\cite{2006-Adams.etal-JHEP}, local quantum field theories contain a Lorentz-invariant concept of causality and satisfy the typical S-matrix axioms. This should also be satisfied by the expansion scalar functions (cf. Eq. \eqref{EFT:L}) describing effective field theory of magnetogenesis. By construction, these functions are arbitrary, one need to impose additional conditions for a well-defined relativistic field theory. For a recent discussion in the context of black holes, see Ref.~\cite{2022-Serra.etal-JHEP}.
%We then show that the speed of propagation of the perturbed gauge fields need not be less than $1$. 
Thus, there are two necessary conditions for the generation of primordial magnetic fields --- conformal invariance breaking and causal propagation. We confirm this by considering a specific model of primordial magnetogenesis. In this chapter, we will only look at the generation of non helical magnetic field, the modification terms for the helical magnetic fields, for example as discussed in chapters~(\ref{ch:helical-PMF},\ref{ch:PMF_Baryo}) can be obtained in straightforward manner.

The rest of the chapter is organized as follows: 
In section \ref{sec:EFT} we write down the EFT action of magnetogenesis and compare it with models proposed in the literature.  
Section \ref{sec:EFT-Spectrum} obtains the magnetic power spectrum for the EFT in slow-roll inflation and identifies the loophole in the magnetogenesis model building.  
In section \ref{sec:vector_galileon_example}, we take a specific example and show that magnetic field amplification is possible with the cost of super-luminal propagation. 
Appendix~\ref{apdetail:EFTmagnetogenesis} contains the details of the calculations in this chapter. The results reported in this chapter are based on Ref.~\cite{2022-Kushwaha.Naskar.etal-JCAP}.

\section{EFT action of magnetogenesis}
\label{sec:EFT}

Like any effective field theory, the EFT of magnetogenesis includes two components: Symmetries and degrees of freedom~\cite{2007-Burgess-ARNPS,2020-Penco-Arxiv}. The vector perturbations 
do not influence the dynamical evolution of the gauge field. See Appendix \ref{sec:Decoupling} for details. Hence, the vector perturbations and gauge fields are decoupled, and the only relevant degree of freedom is the gauge field. Having identified the relevant degrees of freedom, our next step is to write down the action. 

In principle, effective action can have infinite terms. Therefore, we need to identify (broken) symmetries that describe the physics to expand the action about the expansion parameter.  Because of the conformal invariance of the standard electromagnetic action 
   \begin{align}
   \label{eq:SEM}
        S_{\rm EM} = -\frac{1}{4} \int d^4x \sqrt{-g} F_{\mu\nu} F^{\mu\nu} 
    \end{align}
it is impossible to produce a detectable amount of magnetic field from this setup\footnote{Note that $F_{\mu\nu} \tilde{F}^{\mu\nu}$ is a total derivative term in FRW background and does not contribute to the dynamics.}. Splitting the gauge-field in the background and perturbations \eqref{eq:VectPert}, we have:
\begin{equation}
    A_{\mu} = \bar{A}_{\mu} + \delta A_{\mu} = \delta A_{\mu}.
\end{equation}
Due to the background symmetry, we have considered the background field $\bar{A}_{\mu} =0$ and if we use the well-known choice of Coulomb gauge  (see Appendix \ref{sec:Decoupling} for details), 
\begin{equation}
\delta A_0 = 0, \partial_i \delta A^{i}=0 \, ,
\label{def:CoulombG}
\end{equation}
the Lagrangian for the fluctuation field $\delta A_{\mu}$ from \eqref{eq:SEM} can be written as,
\begin{align}
        S_{\rm EM} = \int d^4x \left[  (\delta A_i^{\prime})^2 - (\partial_i \delta A_j)^2 \right]
    \end{align}

This Lagrangian for the fluctuation is also conformally invariant and thus, the generation of primordial magnetic fields in the early Universe requires conformal invariance breaking of the electromagnetic action~\cite{1988-Turner.Widrow-PRD,1991-Ratra-Apj.Lett,1993-Dolgov-PRD}. In other words, the terms in the effective action must break the conformal invariance. Here, we demand that the effective action satisfies the following symmetries: 
\begin{enumerate}
\item \textbf{Local Lorentz invariance}: In the Minkowski limit, we demand that the effective action is Lorentz invariant. In other words, in the limit, $a(\eta) \to {\rm constant} (H(t) \to 0)$, the EFT action of magnetogenesis reduces to the standard electromagnetic action.
%second integral in the RHS of Eq.~\eqref{vec-L}.

\item \textbf{Gauge invariance}: As mentioned in detail in Ref.~\cite{2001-Jackson.Okun-RMP}, an implicit assumption behind the formalism of the gauge invariance is that the field equations must have unique solutions. We impose the validity of this condition in the early Universe. Thus, the EFT action \emph{do not} contain terms proportional to $A_{\mu} A^{\mu}$. Note that from now on, we will use $A_i$ instead of $\delta A_i$ to denote the gauge field fluctuation.
\end{enumerate}

While the above symmetry requirements will restrict the form of the EFT action of magnetogenesis, we need to identify the expansion parameters of the action. Although the two expansion parameters are well-defined in the time-independent system, it is not straightforward for time-dependent systems like in the early Universe~\cite{2017-Burgess-arXiv}.
In general, it is not possible to construct an EFT without energy conservation, as EFTs divide states based on energy. However, if the time-evolution of the degrees of freedom $(A_{i})$ is adiabatic --- they vary sufficiently slowly compared to the UV scales of interest $\Lambda$ --- we can treat the Hamiltonian obtained from the EFT action to be approximately conserved Hamiltonian with an approximate time-dependent low/high energy split~\cite{2015-Donoghue.Holstein-JPG,2017-Burgess-arXiv}. 

In the case of slow-roll inflation, the energy scale during inflation $H$ is approximately constant. Let us define energy scale $\Lambda$ that represents a cutoff scale for which the effects of high-scale physics are described by non-renormalizable operators, which can be thought of as originating from integrating out all particles with mass $m > \Lambda$~\cite{2003-Burgess-LivRev,2020-Penco-Arxiv}. Thus, at these energies, the scale dependence between the characteristic energy scale $H$ and the cutoff scale $\Lambda$ 
is given by the expansion parameter $H/\Lambda$. Like in general EFT, we fix the level of precision and sort by $H/\Lambda$ at a given order all terms that contribute to the action.
Note that we cannot use standard perturbation
theory to quantize, and the problem of non-renormalizability becomes an actual problem at energies close to $\Lambda$. We assume $\Lambda$ to be at least \emph{one order} higher than $H$ and less than $M_{\rm Pl}$.

Thus, the second-order EFT action of magnetogenesis, based on expansion about the Hubble parameter $(H)$ and its derivatives, is 
\begin{equation}
    \mathcal{S}_{\rm EFT} = \int d^4 x \left[f_1(H,a,\Lambda) ({A}_i^{\prime})^2 - f_2(H,a,\Lambda) 
    (\partial^j A_i  \, 
    \partial_j A_i )\right]
    \label{EFT:L}
\end{equation}
where, the \emph{expansion scalar functions} --- $f_1(H,a,\Lambda)$ and $f_2(H,a,\Lambda)$ --- depend of Hubble parameter $H$, scale factor $a(\eta)$, and cutoff scale $(\Lambda)$. This is a crucial expression regarding which we want to discuss the following: 
First, as explained in Ref.~\cite{2006-Adams.etal-JHEP}, local quantum field theories contain a Lorentz-invariant concept of causality and satisfy the typical S-matrix axioms. This should also be satisfied by the functions $f_1$ and $f_2$ describing effective field theory of magnetogenesis. By construction, these functions are arbitrary, one need to impose additional conditions for a well-defined relativistic field theory. For a recent discussion in the context of black holes, see Ref.~\cite{2022-Serra.etal-JHEP}.
Second, we have chosen the Coulomb gauge condition for the EM fields and hence resulting in only two physical degrees of freedom\footnote{The Coulomb gauge allows us to evaluate the observables like magnetic and electric power-spectrum efficiently.}. In principle, we can have separate expansion scalar functions in front of $A_1'^2, A_2'^2, A_3'^2$ but this will lead to different propagation speeds for the three components. (See Appendix \ref{general-propagation} for details). In this chapter, we focus on action \eqref{EFT:L}. 

Also, note that one can add parity breaking term  {$\epsilon^{ijk} A_i^{\prime} \partial_j  A_k$} in the EFT action (\ref{EFT:L}) with additional scalar expansion function $f_3$. In the limit of $\Lambda \to \infty $ and $a(\eta) \to {\rm constant}$ (or $H(t) \to 0$), $f_3 \to~\rm{constant}$. Since the parity symmetry only determines the nature of primordial magnetic field, we will not consider the parity breaking term in the action.
Third, in the limit of $\Lambda \to \infty$,
$f_1 = f_2 \simeq {\rm constant}$. In other words, the above expression will reduce to Eq. \eqref{eq:SEM}.
%second integral in the RHS of Eq.~(\ref{vec-L}). 
Also, in the limit of $a(\eta) \to {\rm constant}$ or $H(t) \to 0$), $f_1 = f_2 \simeq {\rm constant}$ and reduce to Eq. \eqref{eq:SEM}.
%second integral in the RHS of Eq.~(\ref{vec-L}).
%
Fourth, $f_1(H,a,\Lambda)$ and $f_2(H,a,\Lambda)$ capture all possible interactions of electromagnetic field that leads to breaking of conformal invariance.

Fifth, since $A_{\mu}$ has mass dimension $1$, both $f_1(H,a,\Lambda)$ and $f_2(H,a,\Lambda)$ have mass dimension zero~\cite{1992-Polchinski-arXiv}. 
Formally, the effective action can be written as~\cite{1992-Polchinski-arXiv}:
\begin{equation*}
    S \sim \int d^4 x ~ \mathcal{O}_{p,q} \sim \left(\frac{E}{\Lambda}\right)^{p +q - 4},
\end{equation*}
where, the operator $\mathcal{O}$ is made up of $p$ fields and $q$ derivatives. Since, all the observable quantities in cosmology are related to the Hubble parameter $H(t)$ and it determines the energy scale of the epoch. To compare the inflationary scale with EFT scale, a broad class of magnetogenesis models can be reproduced from the EFT that is a sum of series in $H/\Lambda$, time-derivatives of $H$ --- ($H'/\Lambda^2$), ($H''/\Lambda^3$), $\cdots$ --- and their products, for instance, $H H'/\Lambda^3$, $H H''/\Lambda^4$, $\cdots$. Since, $p$ and $q$ are integers, the effective field theory can be expanded only as a series in terms of the Hubble parameter ($\mathcal{H}$) in conformal 
time\footnote{Higher derivatives of $\mathcal{H}$ are also present, however, not shown in the expansion.}:
{\small
\begin{align}\label{eft:A}
\begin{split}
\!\!\!\!\!\!\! f_1(H, a, \Lambda) & = 
\sum_{n=0}^{\infty} s_n \frac{1}{a^n} \left(\frac{\mathcal{H}}{\Lambda}\right)^n
    + \sum_{m=1}^{\infty} b_m \frac{1}{a^{2m}}
    \left(\frac{\mathcal{H}^{\prime}}{\Lambda^2}\right)^m + \sum_{m,n=1}^{\infty} v_{n,m} \frac{1}{a^{m+2n}}
    \left(\frac{\mathcal{H}}{\Lambda}\right)^m \left(\frac{\mathcal{H}^{\prime}}{\Lambda^2}\right)^n 
\\
\!\!\!\!\!\!\!  f_2(H, a, \Lambda) &= 
    \sum_{n=0}^{\infty} d_n \frac{1}{a^n} \left(\frac{\mathcal{H}}{\Lambda}\right)^n
    + \sum_{m=1}^{\infty} e_m \frac{1}{a^{2m}}
    \left(\frac{\mathcal{H}^{\prime}}{\Lambda^2}\right)^m + \sum_{m,n=1}^{\infty} w_{n,m} \frac{1}{a^{m+2n}} \left(\frac{\mathcal{H}}{\Lambda}\right)^m \left(\frac{\mathcal{H}^{\prime}}{\Lambda^2}\right)^n 
\end{split}
\end{align}
}

where $s_n, b_m, d_n, e_n, v_{n,m}, w_{n,m}$, are the unknown real (postive or negative) parameters and can be fixed for a particular magnetogenesis model. $s_0$ and
$d_0$ correspond to the values in standard electrodynamics satisfying the local Lorenz invariance. Note that $H' = (\mathcal{H^{\prime}} - \mathcal{H}^2)/a$, hence the expansion in either of the two variables are equivalent. In the above expression, we have not included series in higher derivatives of $\mathcal{H}$ like $(\mathcal{H}''/\Lambda^3)$. In principle, these terms should also be included in the EFT and appear in higher-order gravitational coupling. As mentioned above, since the coefficients $s_n, b_m, d_n, e_n, v_{n,m}, w_{n,m}$ are unknown, one need to impose additional conditions, like causality, for a well-defined relativistic field theory~\cite{2006-Adams.etal-JHEP,2022-Serra.etal-JHEP}.

Sixth, in the literature, 
the odd powers of ${\mathcal{H}}/{\Lambda}$ are not 
included and the first-order correction is taken to be ${\mathcal{H}^2}/{\Lambda^2}$ or ${\mathcal{H}^{\prime}}/{\Lambda^2}$. However, we have included odd powers in
the expansion parameter to keep the analysis general. As mentioned above, the above expansion is valid only for $H < \Lambda$.

Lastly, Table (I) identifies the early Universe magnetogenesis models and the corresponding EFT parameters. The list is not exhaustive but gives a good representation of the various magnetogenesis models discussed in the context of inflation. Thus, we see that the EFT action \eqref{EFT:L} can reproduce most of the known magnetogenesis models. For most models, it is sufficient to consider up to the second order in the expansion parameter. Appendix \ref{app:coup-details} 
contains detailed calculations that provide a one-to-one mapping between the magnetogenesis model 
and EFT parameters.

\begin{table}[!htb]
\centering
\label{table:comparison}
\begin{tabular}{|c|c|}
\hline
{\bf Magnetogenesis models} & {\bf Non-zero EFT parameters} \\ \hline
Class of Ratra Model: $f(\phi) F_{\mu\nu}F^{\mu\nu}$  \cite{1991-Ratra-Apj.Lett} & 
$s_n, d_n$ (depending on $f(\phi)$) \\ \hline
Higgs Starobinsky Inflation  \cite{2022-Durrer.etal-arXiv} & 
$s_n, d_n$ \\ \hline
Vector Galileon Model~\cite{2017-Debottam.Shankaranarayanan-JCAP}        &      $s_2, e_1$          \\ \hline
Gravitational Coupling: $RF_{\mu\nu}F^{\mu\nu}$ \cite{1988-Turner.Widrow-PRD}&  $s_2, b_1, d_2, e_1$  %$= -3$
\\ \hline
Gravitational Coupling: $R_{\mu \nu}F^{\mu\alpha}F^{\nu}_{\alpha}$ \cite{1988-Turner.Widrow-PRD} & 
$s_2, b_1, d_2, e_1$
%$s_2=-\frac{1}{2}, b_1 =-1$ 
%and $d_2 =-1, e_1=-\frac{1}{2}$
\\ \hline 
Gravitational Coupling: $R_{\mu\nu\alpha\beta} F^{\mu\nu}F^{\alpha\beta}$
\cite{1988-Turner.Widrow-PRD} & 
$b_1, d_1$
%$b_1=d_1=-1$
\\ \hline 
Higher order Gravitational Coupling:  &  
$s_6, b_3, v_{4,1}, v_{2,2}$ \\
%$s_6 = -6, b_3 = -6, v_{4,1} = -18, v_{2,2} = -18$; \\
 $R^3 F_{\mu\nu}F^{\mu\nu}$ \cite{2022-Bertolami.etal-arXiv} & 
$d_6, e_3, w_{4,1}, w_{2,2}$ 
% $d_6 = -6, e_3 = -6, w_{4,1} = -18, w_{2,2} = -18$
\\ \hline 
\end{tabular}
\caption{One-to-one mapping between the magnetogenesis model and the EFT parameters. See Appendix \ref{app:coup-details} for details.}
\end{table}
Before we proceed with the rest of the analysis, we compare the above EFT action \eqref{EFT:L} with the ones recently proposed in the literature~\cite{2021-Giovannini-PLB,2021-Maity.etal-JCAP}. In Ref.~\cite{2021-Giovannini-PLB} the effective action terms only consist of four derivatives associated with functions of background inflaton. This expansion leads to different susceptibilities for 
electric and magnetic fields, but the results are produced with a particular parametrization of these susceptibilities. In Ref.~\cite{2021-Maity.etal-JCAP}, the EFT Lagrangian is written in second-order with all possible contractions of electromagnetic tensor $F_{\mu\nu}$ with itself associated with time-dependent analytic functions $f_i(\eta)$. The functions $f_i(\eta)$ are chosen to be proportional to either
$\left({a(\eta)}/{a_f(\eta_f)}\right)^2$ or higher-order where $\eta_i$ and $\eta_f$ are time at beginning and end of inflation. 
More specifically, in both these cases, the authors did not
include terms that can be proportional to $({H}/{\Lambda})$ where in our expansion scheme, it naturally arises. 
%Also, these authors do not identify the true degrees of freedom for magnetogenesis.

\section{Generic magnetic field power-spectrum from EFT action}
\label{sec:EFT-Spectrum}

In the previous section, we constructed EFT action \eqref{EFT:L} of magnetogenesis based on symmetries and degrees of freedom. We also constructed a form of the
expansion scalar functions ($f_1(H, a, \Lambda)$ and $f_2(H, a, \Lambda)$) and showed that the EFT parameters in this generic form indeed correspond to the various magnetogenesis models. To make the computation of the power-spectrum tractable and to highlight the importance of speed of perturbations, we truncate the series \eqref{eft:A} up to second order. However, the truncation of the series to compute the power spectrum has no bearing on the EFT expansion \eqref{eft:A}.
%\textcolor{red}{We also noted that for most models, it is sufficient to consider up to the second-order in the expansion parameter.}
%In this section, we derive the equation of motion (EOM) of the gauge field by truncating the expansion scalar functions to $\Lambda^{-2}$. We then obtain the power spectrum in the slow-roll inflation scenario. 

\subsection{Equation of motion from EFT action}

Truncating the expansion scalar functions ($f_1(H, a, \Lambda)$ and $f_2(H, a, \Lambda)$) in Eq.~\eqref{eft:A} to $\Lambda^{-2}$ order, we have,
\begin{equation}
\begin{split}
&   f_1(H, a, \Lambda) \simeq 
s_0 + \frac{s_1}{a(\eta)} \left(\frac{\mathcal{H}}{\Lambda}\right) +  
\frac{s_2}{a^2(\eta)}\left(\frac{\mathcal{H}}{\Lambda}\right)^2
+ \frac{b_1}{a^2(\eta)} \left(\frac{\mathcal{H}^{\prime}}{\Lambda^2}\right),  \\
&  f_2(H, a, \Lambda) \simeq d_0 
+ \frac{d_1}{a(\eta)}\left(\frac{\mathcal{H}}{\Lambda}\right) 
+ \frac{d_2}{a^2(\eta)} \left(\frac{\mathcal{H}}{\Lambda}\right)^2
+  \frac{e_1}{a^2(\eta)} \left( \frac{\mathcal{H}^{\prime}}{\Lambda^2}\right).
\end{split}\label{eft:truncated}
\end{equation}
As mentioned earlier, in the limit of $\Lambda \to \infty$, the expansion should reduce to Eq. \eqref{eq:SEM}.
%coefficients in the second integral in the RHS of Eq. \eqref{vec-L}. 
Hence, we have $s_0 = d_0 = 1/2$. (See Appendix \eqref{app:rhoEFT} for more details.)

To obtain the equation of motion corresponding to the action \eqref{EFT:L}, we first need to rewrite the action in canonical form. To do that, we define  
$A_i = \mathcal{A}_i/Z$ in the effective action \eqref{EFT:L} and we have: 
\begin{equation}
\label{EFT:L2}
 \mathcal{S}_{\rm EFT} = \int d^4 x \left[ (\mathcal{A}_i^{\prime})^2 + \frac{Z''}{Z}
    \mathcal{A}_i^2 - \frac{f_2}{f_1}(\partial_j \mathcal{A}_i)^2 \right].
\end{equation}
where $Z = f_1^{1/2}$. Using Eq. \eqref{eft:truncated}, we have: 
\begin{multline}
\frac{Z^{\prime \prime}}{Z} = 
\frac{s_1}{a(\eta)}
\left(\frac{\mathcal{H}}{\Lambda}\right) 
\left[\frac{\mathcal{H}^2}{2} - 3  \mathcal{H}^{\prime} 
+ \frac{1}{2} \frac{\mathcal{H}^{\prime\prime}}{\mathcal{H}} \right] 
%%%
+ \frac{1}{a^2(\eta)}
\left(\frac{\mathcal{H}}{\Lambda}\right)^2 
\left[ \left(\frac{s_1^2}{4}+ 2 s_2 \right) \mathcal{H}^2  \right. \\ 
  \left. 
+ \left(s_1^2-5s_2+2b_1\right) \mathcal{H}^{\prime} 
+ \left( \frac{s_1^2}{4}+s_2-b_1 \right) 
\left(\frac{\mathcal{H}^\prime}{\mathcal{H}}\right)^2 
- \left( \frac{s_1^2}{2}+s_2+2 b_1 \right) 
\frac{\mathcal{H}^{\prime\prime}}{\mathcal{H}} 
+ b_1 \frac{\mathcal{H}^{\prime\prime\prime}}{\mathcal{H}^2}
\right] 
\end{multline}
Note that we have only kept terms up to $1/\Lambda^2$ and ignored higher-order $\Lambda$ contributions. Like in non-canonical scalar fields, $f_2/f_1$ in the effective action \eqref{EFT:L2} can be identified as the adiabatic sound speed:
\begin{eqnarray}
\label{def:CA}
c_A^2 &=& 
\frac{1+ \frac{d_1}{a(\eta)} \left(\frac{\mathcal{H}}{\Lambda}\right) 
+ \frac{d_2}{a^2(\eta)} \left(\frac{\mathcal{H}}{\Lambda}\right)^2
+ \frac{e_1}{a^2(\eta)} \left(\frac{\mathcal{H}^{\prime}}{\Lambda^2}\right)}
{1+ \frac{s_1}{a(\eta)} 
\left(\frac{\mathcal{H}}{\Lambda}\right) 
+ \frac{s_2}{a^2(\eta)} 
\left(\frac{\mathcal{H}}{\Lambda}\right)^2 
+ \frac{b_1}{a^2(\eta)} \left(\frac{\mathcal{H}^{\prime}}{\Lambda^2}\right)} \\
\label{def:CA2}
&\simeq& 1 + \frac{d_1-s_1}{a(\eta)} \left(\frac{\mathcal{H}}{\Lambda} \right)
+ \frac{1}{a^2} \left(\frac{\mathcal{H}}{\Lambda}\right)^2 \left[ 
s_1^2-s_2-s_1 d_1+d_2 + (e_1-b_1) (1 - \epsilon_1)
\right]
\end{eqnarray}
where in arriving at the above expression, we have assumed that $H/\Lambda$ is small and higher-order 
terms are negligible. This is the second key result of this chapter regarding which we want to stress the following points: First, as mentioned above, since the coefficients $s_1, s_2, d_1, d_2, b_1, e_1$ are unknown, one need to impose additional conditions, like causality, for a well-defined relativistic field theory~\cite{2006-Adams.etal-JHEP,2022-Serra.etal-JHEP}. More specifically, considering both $d_1$ and $s_1$ to be positive, the above expression implies that $c_A^2 > 1$ if $d_1 > s_1$, irrespective of the value of $H/\Lambda$. Thus, such models violate standard causality condition~\cite{2007-Ellis.etal-GRG}. Earlier,  effective field theories have been rejected based on super-luminal fluctuations as such
propagation generally leads to a global breakdown of causality~\cite{2006-Adams.etal-JHEP}. Second, models with $s_1 <0$ and $d_1 > 0$, will always lead to superluminal modes.
Third, when $d_1 = s_1$ and assuming all EFT parameters are positive, the causality condition implies $s_2 > d_2 + (e_1-b_1) (1 - \epsilon_1)$  during the entire inflationary epoch [$\epsilon_1$ is the first-order slow-roll parameter defined in Eq. \eqref{def:Slowroll}]. Lastly, in general, $c_A$ is a function of time. Since $H/\Lambda$ is small, one can assume that $c_A$ has a weak time dependence. Note that, by construction, the EFT action \eqref{EFT:L} is locally Lorenz invariant, and the causality condition imposes restrictions on the EFT parameters.

Fourth, in Appendix \eqref{app:rhoEFT} we have obtained the energy density \eqref{eq:EFT-rho} corresponding to the EFT action \eqref{EFT:L}. We infer the following from the energy-density \eqref{eq:EFT-rho}: $\rho_{\rm mixing}$ decays faster than $\rho_{\rm E}$ and $\rho_{\rm B}$. Hence, we can ignore $\rho_{\rm mixing}$ contribution in evaluating the energy density of the EFT. Take the extreme scenario where $\rho_{\rm E}$ and $\rho_{\rm B}$ contribute equally, imposing the condition that the energy density is always positive provides a condition that $s_1 > d_1$ (assuming $s_1, d_1$ are positive). This is consistent with the causality condition we obtained earlier. We show that all these features are satisfied for the specific Galileon vector model in Sec. \eqref{sec:vector_galileon_example}.

Lastly, the equation of motion corresponding to the action \eqref{EFT:L2}, in the Fourier domain $(k)$, is 
\begin{equation}
\label{eq:GFieldEOM}
\mathcal{A}_k ^{''} + \left[c_A^2 k^2 - \frac{Z^{''}}{Z} \right] \mathcal{A}_k = 0 \, ,
\end{equation}
where, $k = |{\bf k}|$ and ${\bf k}$ is the comoving wave vector. For brevity, we have defined $\mathcal{A}_k = \mathcal{A}_i^{(k)}$. In the rest of this section, we now compute the magnetic field power spectrum during inflation. 

\subsection{Generic magnetic power spectrum during inflation}

In this section, we quantize the effective gauge field given by the action \eqref{EFT:L2}, and obtain the general expression for the primordial magnetic field  (PMF) power spectra.  On quantization, the gauge field $\mathcal{A}$ can be expressed 
as follows:
\begin{eqnarray}\label{eq:fourierd}
\mathcal{\hat A}_i (\eta, {\bf x}) =  \int \frac{d^3 {\bf
     k}}{(2\pi)^{3/2}} \sum_{\lambda=1}^{2} \epsilon_{\lambda i}({\bf
   k}) \Big[\hat{b}_{\bf k}^\lambda \mathcal{A}_{\bf k} e^{i{\bf k.x}} +
   \hat{b}_{\bf k}^{\lambda\dagger} \mathcal{A}_{\bf k}^*(\eta) e^{- i{\bf k.x}}\Big]
 \, ,
 \end{eqnarray}
where $\lambda$ corresponds to two orthonormal transverse
polarizations, $\epsilon_{\lambda i}$ are the polarization
vectors and the creation (${\hat a}_{\bf k}$) and the annihilation (${\hat a}_{\bf k}^{\dag}$) operators obey the usual commutation relations. Like scalar and tensor perturbations~\cite{1992-Mukhanov.etal-Phy.Rep.},  the power spectrum as well as the statistical properties of the gauge field is characterized by the Wightman function of the gauge field. The power spectrum (the two-point correlation in Fourier space) is~\cite{2013-Durrer.Neronov-Arxiv}: 
\begin{equation}
\left\langle B_{i}^{*}(\mathbf{k}) B_{j}\left(\mathbf{k}^{\prime}\right)\right\rangle=(2 \pi)^{3} \delta^{3}\left(\mathbf{k}-\mathbf{k}^{\prime}\right) P_{i j} \mathcal{P}_{B}(k) \, ,
\end{equation}
where $P_{i j}$ is a projector onto the transverse plane and is given by
\begin{equation}
P_{i j}=\delta_{i j}-\frac{\mathbf{k}_{i} \mathbf{k}_{j}}{k^{2}},~~ P_{i j} P_{j k}=P_{i k}, ~~ P_{i j} \mathbf{k}^{j}=0 \, ,
\end{equation}
and $\mathcal{P}_{B}(|\mathbf{k}|)$ is the gauge field power spectrum. Since $B$ is statistically homogeneous and isotropic, the correlation depends only on the distance $|\mathbf{x}-\mathbf{y}|$.  
Using the decomposition (\ref{eq:fourierd}), the
PMF spectrum per logarithmic interval can then be written in terms of the modes $\mathcal{A}_{k}$ as
\begin{equation}
{\cal P}_{B}(k) 
= \frac{k^5}{2 \pi^2 a^4}\left\vert
\frac{\mathcal{A}_k}{Z}\right\vert^2
\label{eq:pMS}
\end{equation}
and the expression on the right-hand side is to be evaluated when 
the physical wavelength $(k/a)^{-1}$ of the mode corresponding to the comoving wavenumber ${\bf k}$ equals the \emph{effective sound horizon} of the electromagnetic fluctuations $c_A\, H^{-1}$~\cite{1999-Garriga.Mukhanov-PLB}.  The above condition translates to $(-k \eta) c_A = 1$ corresponding electromagnetic fluctuations exiting the sound horizon during inflation. This sound horizon is not the same sound horizon of the scalar perturbations during inflation~\cite{1999-Garriga.Mukhanov-PLB,2011-Hu-PRD}.
The normalization constant takes into account both modes of polarization~\cite{2016-Subramanian-Arxiv}. In the rest of this subsection, we obtain the PMF power spectrum for the slow-roll inflation. Appendix \ref{app:power-law} contains the results for power-law and de Sitter inflation.

To obtain the solution to Eq.~\eqref{eq:GFieldEOM} in slow-roll inflation, we introduce a new set of variables~\cite{2003-Martin.Schwarz-PRD,2004-Shankaranarayanan.Sriramkumar-PRD}:
\begin{equation}\label{new-var2}
x = \ln{\frac{aH}{k c_A}},~~
\mathcal{A}_k = e^{-\frac{x}{2}} (1-\epsilon_1)^{-\frac{1}{2}} u_k \, ,
\end{equation}
where $\epsilon_1$ is the first-order slow-roll parameter defined as:
\begin{equation}
\label{def:Slowroll}
    \epsilon_1 = -\frac{\dot{H}}{H^2} = 1 - \frac{\mathcal{H}^{\prime}}{\mathcal{H}^2}
\end{equation}
Note that the above variables are well-defined up to the exit of inflation $(\epsilon_1 = 1)$. To solve the differential equation ~\eqref{eq:GFieldEOM}, we need to evaluate $Z$ in terms of the slow-roll parameters.  
Rewriting Eq.~\eqref{def:Slowroll}, we have~\cite{2001-Schwarz.etal-PLB},
\begin{equation}
\label{def:eta-epsilon}
\eta = - \frac{1}{(1 - \epsilon_1) \mathcal{H}} - \int \frac{2 \epsilon_1 \, 
(\epsilon_1 - \epsilon_2)}{(1 - \epsilon_1)^3} \, d\left(\frac{1}{\mathcal{H}}\right) \, ,
\end{equation}
where $\epsilon_2$ is the second slow-roll parameter \eqref{def:Slowrollpara}. Note that the second term in the above expression can be ignored when $\mathcal{H}$ is approximately constant and/or $\epsilon_1 \simeq \epsilon_2$. Under this condition, we have:
\begin{eqnarray} \label{slow-r}
aH \simeq -\frac{1}{\eta (1 - \epsilon_1)} ~~ \Longrightarrow~~x \simeq \ln{\frac{1}{c_A k \eta (\epsilon_1 - 1)}} \, .
\end{eqnarray}
We would like to note the following points: We do not assume $\epsilon_1 << 1$. $\eta$ is negative during inflation, hence $x$ is a well-defined. Using Eqs.~\eqref{new-var2} and \eqref{slow-r}, we have: 
\begin{eqnarray}
    \mathcal{H} = e^x k c_A \, , &~~~~& 
    \mathcal{H}^{\prime} = e^{2x}k^2 (1-\epsilon_1) c_A^2 \, ,\\
    \mathcal{H}^{\prime\prime} = 2 e^{3x}k^3 (1-\epsilon_1)^2 c_A^3\, , &~~~~& \mathcal{H}^{\prime\prime\prime} = 6 e^{4x}k^4(1-\epsilon_1)^3 c_A^4.
\end{eqnarray}

In the new variables, the adiabatic sound speed ($c_A$) and $Z''/Z$ become:
\begin{eqnarray}
\label{eq:cAslowroll}
c_A^2 &\simeq& 1 + (d_1-s_1) \frac{H}{\Lambda} + \left[s_1^2-s_2-s_1d_1+d_2+(e_1-b_1)
(1-\epsilon_1)\right] \left(\frac{H}{\Lambda}
\right)^2 \\
%%%
\frac{Z''}{Z} &\simeq&
 \frac{e^{2x} k^2}{4} \left\{2 s_1 \epsilon_1(\epsilon-1)
\frac{H}{\Lambda} \right. \nonumber \\
\label{eq:Zslowroll}
& + & \left.
\left[a_1^2 (2+2\epsilon_1-3 \epsilon_1^2) -a_2 (4-7\epsilon_1+\epsilon^2)-b_1 \epsilon_1
 (1-4\epsilon_1+3\epsilon_1^2)\right] \left(\frac{H}{\Lambda} \right)^2\right\} \, .
\end{eqnarray}
where we have truncated the series up to $(H/\Lambda)^2$ and we have not imposed any slow-roll approximation. Substituting the above expressions (\ref{new-var2}, \ref{eq:cAslowroll}, \ref{eq:Zslowroll}) in Eq. \eqref{eq:GFieldEOM} leads to: 
\begin{equation}
\label{eq:FinVecEq}
\frac{d^2 u_k}{dx^2} + \left[ q_1^2 e^{-2x} 
- q_2^2 \right] u_k =0 \, .
\end{equation}
where, 
{\small
\begin{eqnarray}
q_1 &=& (1-\epsilon_1)^{-1} \\
%%%%    
q_2 &=& \frac{1}{2}+ \left\{\frac{s_1
\epsilon_1 (1 - 2\epsilon_1)}{2}\frac{H}{\Lambda} - \left[\frac{s_1^2}{4}(2+2\epsilon_1-3 \epsilon_1^2) -s_2 (4-7\epsilon_1+\epsilon_1^2)-b_1 \epsilon_1(1-4\epsilon_1 - 3\epsilon_1^2)\right]\left(\frac{H}{\Lambda} \right)^2\right\} 
\nonumber \\
\label{q2-sr}
& & \times (1-\epsilon_1)^{-2} - \frac{s_1^2 \epsilon_1^2}{8} 
\frac{(1-2 \epsilon_1)^2}{(1-\epsilon_1)^{4}} 
\left(\frac{H}{\Lambda} \right)^2 
\end{eqnarray}
}
We want to note that we have not assumed $\epsilon_1 << 1$. Assuming $\epsilon_1$ is approximately constant, the solution to the above differential equation \eqref{eq:FinVecEq} is Hankel functions~\cite{abramowitz+stegun}
\begin{equation}
u_k(x) = \alpha \, H_{q_2}^{(1)}(e^{-x}q_1) 
+ \beta \, H_{q_2}^{(2)}(e^{-x}q_1) \, .
\end{equation}
Using the relation \eqref{new-var2}, the mode functions of the gauge field $\mathcal{A}_k$ are:
%
\iffalse
\begin{eqnarray}
    \mathcal{A} &=& (1-\epsilon_1)^{-\frac{1}{2}}
    \frac{(-q_1 k\eta)^{\frac{1}{2}}}{(1+\epsilon_1)^{\frac{1}{2}}}
    \left\{\alpha H_{q_2}^{(1)}(e^{-x}q_1) + \beta H_{q_2}^{(2)}(e^{-x}q_1)\right\}\\
  \mathcal{A}  &=& (-q_1k\eta)^{\frac{1}{2}}  
    \left\{\alpha H_{q_2}^{(1)}(e^{-x}q_1) + \beta H_{q_2}^{(2)}(e^{-x}q_1)\right\} ~~~
    \text{(first order in slow-roll)}.
\end{eqnarray}
\fi
%
\begin{equation}
\label{eq:EFT-Modefunction}
\mathcal{A}_k  = (-c_A k\eta)^{\frac{1}{2}}  
\left\{\alpha H_{q_2}^{(1)}(e^{-x}q_1) + \beta H_{q_2}^{(2)}(e^{-x}q_1)\right\}.
\end{equation}

In the sub-horizon limit ($c_A k |\eta| \gg 1$), we assume that the modes satisfy the Bunch-Davies vacuum, i. e., the mode approaches the Minkowski space behavior in the asymptotic past:
\begin{align}
\label{eq:BunchDaviesEFT}
\lim_{k\eta \to -\infty}  \mathcal{A}_k (\eta) =  \frac{1}{\sqrt{2c_A k}}e^{-ic_A k\eta}  
\, .
\end{align}
This leads to:
\begin{equation}
    \alpha = 0;~~~~ \beta = \sqrt{\frac{\pi}{4 c_A k}} \, .
\end{equation}
Substituting the mode function 
\eqref{eq:EFT-Modefunction}, 
with the Bunch-Davies vacuum initial condition,  in Eq.~\eqref{eq:pMS}, the 
power spectrum for the super-horizon modes ($c_A k |\eta| << 1$) is:
\begin{equation}
     \mathcal{P}_B=\frac{k^5}{a^4} \frac{c_A^2}{f_2}\frac{\eta}{8\pi^3}(\Gamma(q_2))^2
    \left(\frac{- c_A k\eta }{2}\right)^{-2 q_2}
\end{equation}
Substituting Eq. \eqref{slow-r}, 
we have:
\begin{equation}\label{PS-sr}
\mathcal{P}_B = \frac{H^4 (1-\epsilon_1)^4}{8 \pi^3} 
\frac{\Gamma^2(q_2)}{c_A^3 f_2}
(c_A k \eta)^{5-2 q_2} \, .
\end{equation}
By expanding $c_A$ and $f_2$ up to the second-order in $(\mathcal{H}/\Lambda)$, we have:
\begin{multline}\label{PS-srFin}
\mathcal{P}_B = H^4 (1-\epsilon_1)^4 
\Gamma^2(q_2) 
(c_A k \eta)^{5-2 q_2} \times \\
\left\{1+\frac{3 s_1-5d_1}{2}\frac{H}{\Lambda}+\frac{1}{2}[3 s_2-6s_1 d_1 -5d_2 +8d_1^2 +(3b_1-5e_1)(1-\epsilon)]\left(\frac{H}{\Lambda}\right)^2\right\} \, .
\end{multline}

This is the third key result of this chapter, regarding which we want to discuss the following points: 
First, the analysis is valid for all values of $\epsilon_1$, assuming that the contribution of the second term in Eq.~\eqref{def:eta-epsilon} can be ignored. While this is true during most of the inflation, the above expression may not be valid at the end of inflation.
Second, in Appendix \ref{app:power-law}, we have derived the power spectrum for de Sitter and Power-law inflation. While the power spectrum for the de Sitter inflation is exact, we have used the WKB approximation for the power-law case. Setting 
the $\epsilon_1 = 0$ in the above power spectrum \eqref{PS-sr} matches with the magnetic power spectrum \eqref{de-sitter-ps} during de Sitter. 
Third, from Eq.~(\ref{PS-srFin}), we see that the amplification in the magnetic power spectrum --- evaluated at the horizon crossing $(-k \eta c_A = 1$) ---  is possible whenever $3 s_1 - 5 d_1 > 0$.
Specifically, we see for sub-luminal ($d_1 < s_1$) or super-luminal ($s_1 < 0$) modes, the leading order correction term in the power-spectrum ${3 s_1-5d_1} > 0$.
In other words, we can have large amplifications even for super-luminal fluctuations. To avoid EFTs with superluminal fluctuations~\cite{2006-Adams.etal-JHEP}, we need another physical condition --- the modes should be sub-luminal.
Thus, our analysis clearly shows that magnetogenesis models must satisfy two necessary conditions --- conformal symmetry breaking and causal propagation. In the next section, we consider a specific model and show that this is the case.

\begin{figure}[ht]
\centering
%\subfigure[]{%
\label{b1-eps1}%
\includegraphics[height=2.85in]{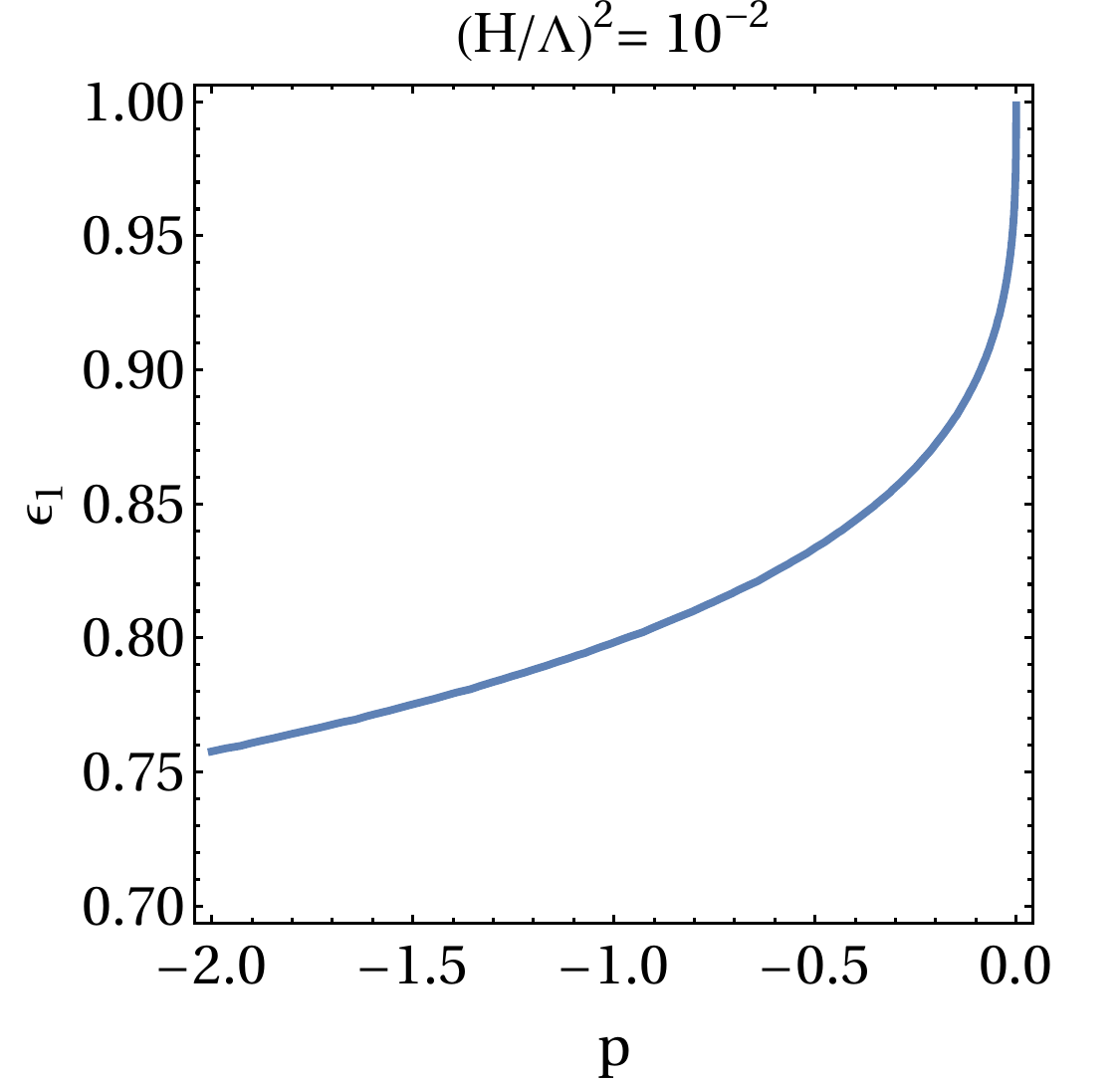}%}%
\quad
%\hspace{1.0cm}
%\subfigure[]{%
\label{b1-eps2}%
\includegraphics[height=2.85in]{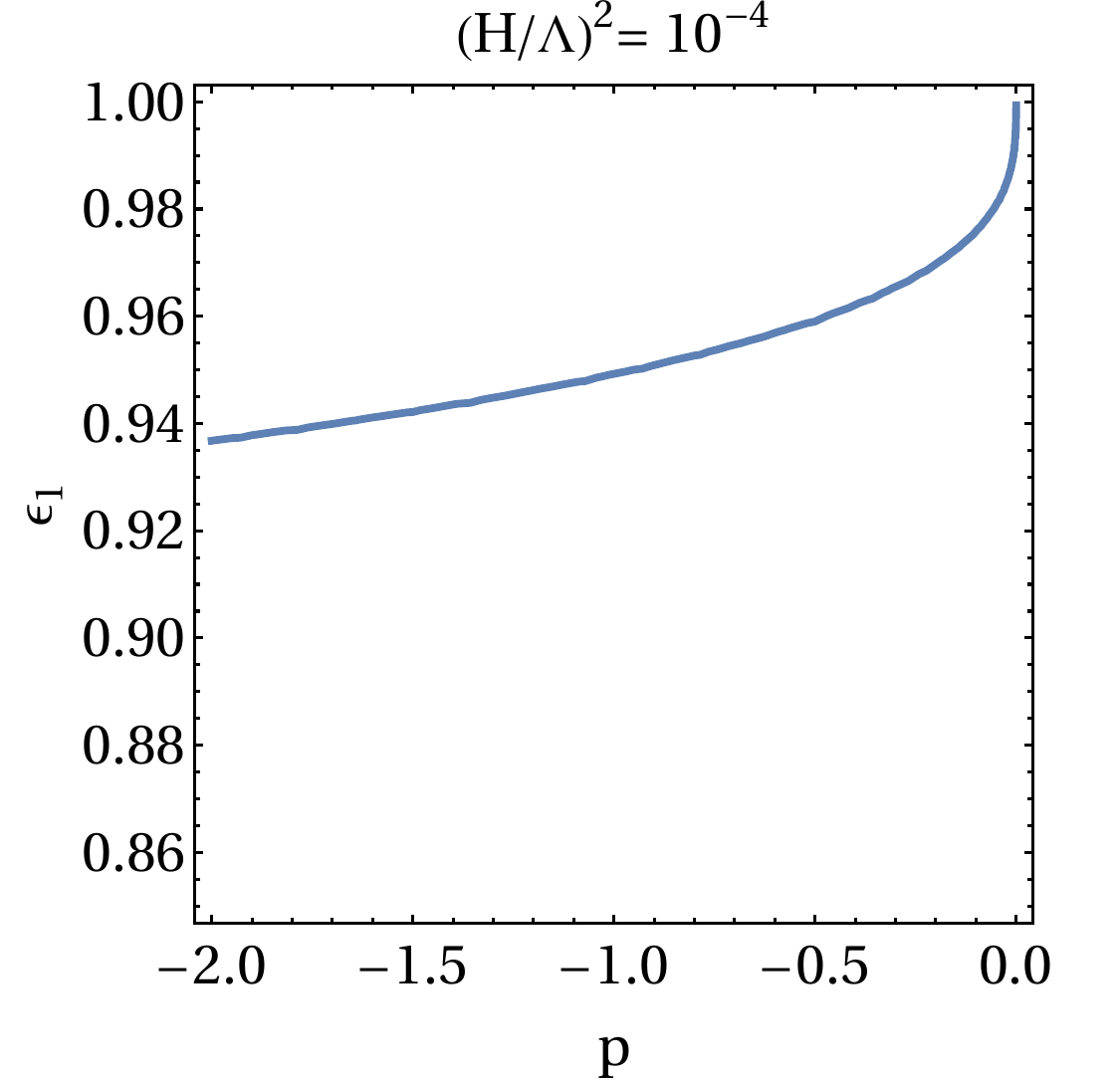}%}%
\caption{The allowed curves for two different values of $(H/\Lambda)^2$, for which we can get a scale invariant power spectrum for the magnetic field. The parameter $p$ is
$\mathcal{O}(1)$ and this curve is only valid at the end of inflation as
we have considered the value of the first slow roll parameter 
$\epsilon_1 \rightarrow 1$. In the left panel we have set $(H/\Lambda)^2 = 10^{-2}$ and in the right panel we have set
$(H/\Lambda)^2 = 10^{-4}$.}
\label{fig:b1-eps}
\end{figure}

Lastly, the above expression provides the following condition under which the magnetic field power spectrum \eqref{PS-sr} will be scale-invariant:
\begin{equation}\label{index-sr}
5-2 q_2 =0 \, .
\end{equation}
From Eq.~\eqref{q2-sr} we see that parameters $s_1$ and $s_2$ are associated with the $(H/\Lambda)$ expansion terms, while the parameter $b_1$ is associated with $\left(\mathcal{H}^{\prime}/a \Lambda\right)^2$.
While the terms containing the parameters $s_1$ and $s_2$ do not have time-dependent factors, the term containing $b_1$  has time-dependent factors. This can be seen by rewriting $\left(\mathcal{H}^{\prime}/a \Lambda\right)^2$ as $(1-\epsilon_1) \left(H/\Lambda \right)^2$. If we demand that the prefactors in the expansion are $\mathcal{O}(1)$, then this implies that $s_1, s_2$ 
and $((1-\epsilon_1) b_1)$ are all order unity during the entire inflationary epoch. In order for this to be satisfied, in general, $b_1$ can have a large value even with $\left(H/\Lambda\right)^2$ suppression. Hence, in order 
to get a scale-invariant power spectrum, the above condition reduces to:
\begin{equation}
\label{index-sr2}
4 - \frac{p \epsilon_1 (1-4 \epsilon_1 - 3 \epsilon_1^2)}{(1-\epsilon_1)^3} \left(\frac{H}{\Lambda}\right)^2 = 0 \,  
\quad {\rm where} \quad  p = b_1 (1 - \epsilon_1) \, .
\end{equation}
Note that the above redefinition ensures that $p$ is $\mathcal{O}(1)$ close to the end of inflation.  \ref{fig:b1-eps} contains the allowed values of $\epsilon_1$ for different values of $p$ for a fixed $({H}/{\Lambda})^2$. From the plot we see that for the scale invariant power spectrum in slow-roll scenario requires $b_1<0$.

\section{Example for EFT Magnetogenesis: Vector Galileon model}
\label{sec:vector_galileon_example}
In the earlier section, using EFT, we showed that conformal invariance breaking is not a sufficient condition for generating the primordial magnetic field during inflation. In this section, we take a specific model and show that the generation of primordial magnetic fields require two necessary conditions --- conformal invariance breaking and causal propogation. 

In Ref.~\cite{2017-Debottam.Shankaranarayanan-JCAP}, two of the current authors have constructed a consistent vector Galileon model by demanding the following conditions in action: theory be described by vector potential $A_{\mu}$ and its derivatives, $U(1)$ gauge invariance should be preserved, and the equations of motion must be second order. The exciting feature of the model is that it only contains the derivative coupling --- higher derivative terms. Hence, it explicitly breaks the conformal invariance of the action, and therefore it may lead to the generation of magnetic fields in the early Universe.
Hence, this model has a clear advantage over other models, particularly the scalar field coupled models~\cite{1991-Ratra-Apj.Lett,2017-Sharma.etal-PRD}. Moreover, due to the absence of coupling between scalar field and electromagnetic field, the strong coupling problem is avoided naturally; hence, the model does not require fine-tuning.

In this section, we consider the total action, i.e., standard electromagnetic action with modification due to vector Galileon, and show that the conformal invariance breaking is the necessary condition to amplify the magnetic field generated during inflation, but it is not sufficient. More specifically, we show that the amplification of order ($\sim 10^7$) comes with a price that speed of sound $c_s>1$, without leading to the back-reaction problem. For $c_s<1$, the model generates tiny magnetic fields on large scales. Thus, despite having Lorentz-invariant Lagrangians, such theories admit super-luminal fluctuations can not be low-energy effective field theories of magnetogenesis.

\subsection{Generation of primordial magnetic fields}
\label{sec:PMF_generation}
This subsection will discuss the phenomenological consequences of the vector Galileon model in the early universe. The vector Galileon action is:
\begin{equation}
\label{def:VECaction}
 \mathcal{S}_{\rm{tot}} =  S_{\rm EM} + 
 \mathcal{S}_{\rm{VEC}} 
\end{equation}
where $S_{\rm EM}$ is the standard electromagnetic action given in Eq.~\eqref{eq:SEM} and $\mathcal{S}_{\rm{VEC}}$ is the vector Galileon part. In the flat 
FRW background (\ref{def:FRW}), with $N(\eta) = a(\eta)$, we have\footnote{ Note that in obtaining Eq.(\ref{eq:EMaction-FRW}) and Eq.(\ref{eq:TheModelFRW}), we have set $N(\eta) = a(\eta)$ in 
Eq.~(\ref{appeq:EMaction-FRW}) and 
Eq.~(\ref{appeq:TheModelFRW-simplify}), respectively} :
\begin{align}\label{eq:EMaction-FRW}
S_{EM} = \frac{1}{2} \int d^4x \, \left[  {A_i^{\prime}}^2 - (\partial_i A_j)^2  \right] \, ,
\end{align}

\begin{figure}
\centering
%\subfigure[]{%
\label{fig:csgt1}%
\includegraphics[height=2in]{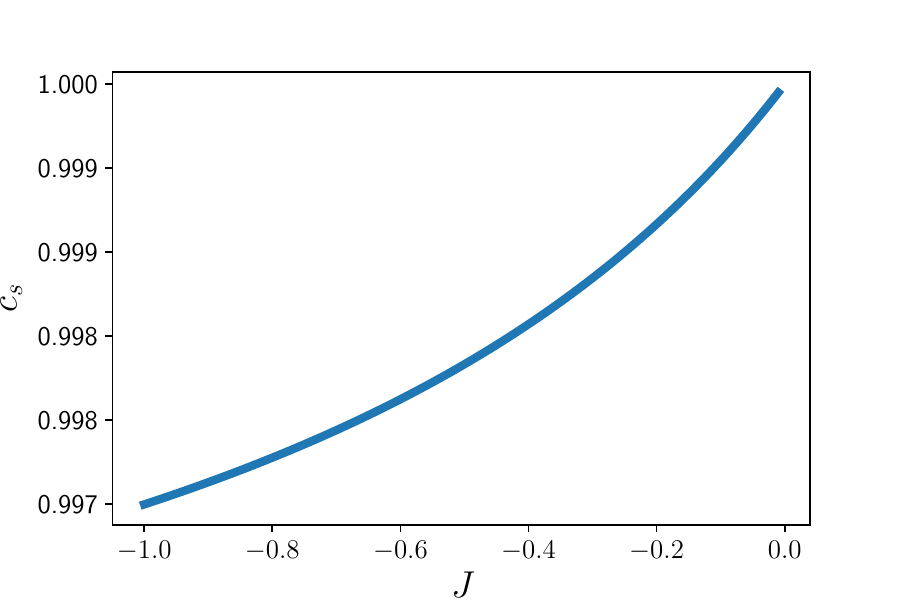}%}%
%\qquad
%\subfigure[]{%
\label{fig:cslt1}%
\includegraphics[height=2in]{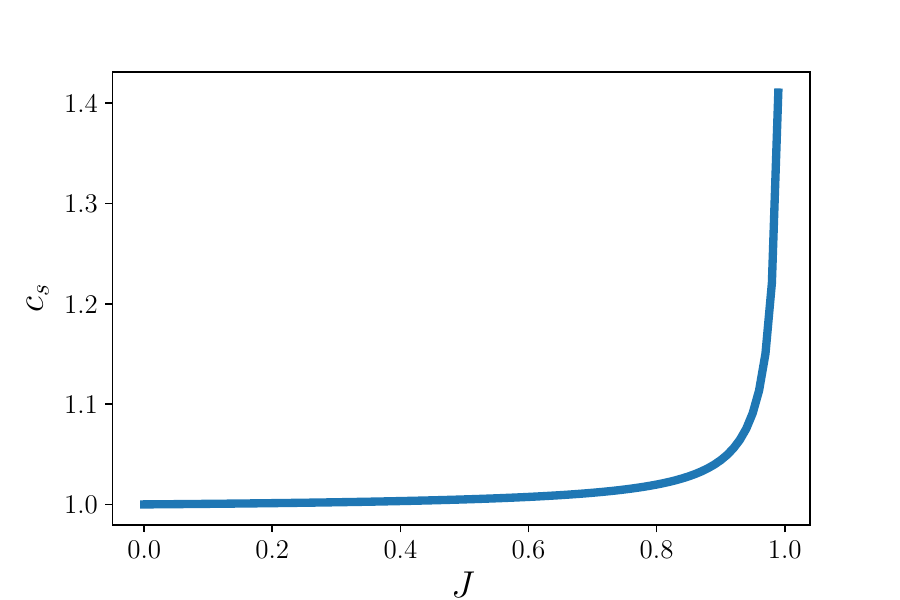}%}%
\caption{Plot showing the behaviour of $c_s$ with respect to parameter $J$.}
\label{fig:cs-plot}
\end{figure}

The vector Galileon action in FRW background is~\cite{2017-Debottam.Shankaranarayanan-JCAP}:
\begin{eqnarray}
\label{eq:TheModelFRW}
\mathcal{S}_{VEC} =  2D \, \int\,d^4x  \Big[- \,\frac{a^\prime{}^2}{a^4}\,A_i^\prime{}^2 +   \frac{a^{\prime\prime}}{a^3}\, \left(\partial_i A_j\right)^2  - \frac{{a^{\prime}}^2}{a^4}\,\left(\partial_i A_j\right)^2\Big] \, .
\end{eqnarray} 
where $A_{\mu}$ satisfies Coulomb gauge condition \eqref{def:CoulombG}. $D$ is the coupling constant with the dimensions of the square of the inverse of energy, i. e. $D \equiv 1/\Lambda^2$. We define the following dimensionless parameter $J$:
\begin{align}\label{eq:J-def}
J = 4 D H^2 = 4 \left(\frac{H}{\Lambda} \right)^2 \,  . 
\end{align}
Like the EFT action \eqref{EFT:L}, the 
above action (\ref{eq:TheModelFRW}) is not in canonical form; hence, we need to rewrite the action in canonical form. To do that, we define the canonical vector field $\mathbb{A}_i$ as:
\begin{align}\label{eq:def-canonicalA}
\mathbb{A}_i = \left( 1 - J \right)^{1/2} \, A_i \qquad \implies \qquad 
\left( 1 - J \right)^{1/2} \, A_i^{\prime} = \mathbb{A}_i^{\prime} + \frac{J ( {\mathcal{H}}^{\prime} - \mathcal{H}^2 )  }{ \mathcal{H} (1 - J ) } \mathbb{A}_i  \, .
\end{align}
Substituting Eqs. (\ref{eq:def-canonicalA}, \ref{eq:J-def}) in Eq.~\eqref{def:VECaction} and setting $N(\eta) = a(\eta)$ leads to: 
\begin{align}\label{eq:tot_action-canonical}
\mathcal{S}_{\rm{tot}} &=  \frac{1}{2} \, \int\,d^4x  \left[ \,\,\mathbb{A}_i^{\prime}{}^2  - \left[  \left(  \frac{\mathcal{H}^{\prime} - \mathcal{H}^2  }{\mathcal{H}}  \right)^2 \frac{J}{ (1 - J)^2 }   +   \frac{  J \left( \mathcal{H}^{\prime\prime} - 3 \mathcal{H}^{\prime} \mathcal{H} + \mathcal{H}^3 \right)  }{(1-J) \, \mathcal{H}}   \right]  \,\mathbb{A}_i^2 \right.
\nonumber \\
&{}{} \qquad  \left.  -  \left( 1 -  \frac{   J( \mathcal{H}^{\prime} - \mathcal{H}^2 ) }{(1-J) \mathcal{H}^2}  \right) \,   \left(\partial_i \mathbb{A}_j\right)^2 \right] \, ,
\end{align}
which is in the canonical form. The equation of motion for the canonical vector field can be obtained by varying the action (\ref{eq:tot_action-canonical}) with respect to $\mathcal{A}_i$:
{\small
\begin{align}\label{eq:eom-canonicalA}
\mathbb{A}_i^{\prime\prime} -  c_s^2 \, \nabla^2 \mathbb{A}_i  + \left[  \left(  \frac{\mathcal{H}^{\prime} - \mathcal{H}^2  }{\mathcal{H}}  \right)^2 \frac{J}{ (1 - J)^2 }   +   \frac{  J \left( \mathcal{H}^{\prime\prime} - 3 \mathcal{H}^{\prime} \mathcal{H} + \mathcal{H}^3 \right)  }{(1-J) \, \mathcal{H}}   \right] \,\mathbb{A}_i   = 0 \, 
\end{align}
}
where $c_s$ is the propagation speed of electromagnetic fluctuations and is given by:
\begin{align}\label{eq:def-cs}
c_s  = \sqrt{1 -  \frac{   J( \mathcal{H}^{\prime} - \mathcal{H}^2 ) }{(1-J) \mathcal{H}^2}} 
= \sqrt{\frac{1 - J (1 - \epsilon_1)}{ 1 - J} }  \, .
\end{align}
In obtaining the last expression, we have used the definition \eqref{def:Slowroll}. Substituting the Fourier decomposing of the canonical vector field
$\mathbb{A}_i$ (\ref{eq:fourierd}) in Eq.~\eqref{eq:eom-canonicalA}, we get:
{\small
\begin{align}\label{eq:eom-canonicalA-fourier}
\mathbb{A}_k^{\prime\prime} + \left[
c_s^2 \, k^2  + \left( \,\,  \left(  \frac{\mathcal{H}^{\prime} - \mathcal{H}^2  }{\mathcal{H}}  \right)^2 \frac{J}{ (1 - J)^2 }   +   \frac{  J \left( \mathcal{H}^{\prime\prime} - 3 \mathcal{H}^{\prime} \mathcal{H} + \mathcal{H}^3 \right)  }{(1-J) \, \mathcal{H}}   \right) \,\,  \right] \mathbb{A}_k = 0 \, .
\end{align}
}
In terms of the slow-roll parameters, the above expression reduces to: 
\begin{align}\label{eq:eom-SR-canonicalA}
\mathbb{A}_k^{\prime\prime} +  \left[ c_s^2 \, k^2 -    \frac{\mathcal{H}^2 J \epsilon_1}{(1 - J )^2} \, 
\left( [1 - J] (1  - 3\epsilon_1 + \epsilon_2)
- J\epsilon_1   \right) \right] \,\mathbb{A}_k = 0  \,  ,
\end{align}
where $\epsilon_1$ and $\epsilon_2$ are the slow-roll parameters defined in Eq.\eqref{def:Slowrollpara}. (See Appendix \ref{app:VG-slow-roll}.)
This is a key expression regarding which we would like to discuss the following: 
First, the above expression is exact and valid for any cosmological scenario. Although we have expressed the equation in terms of slow-roll parameters, we have not set $\epsilon_1 << 1$. Second, positive permittivity provides a condition on the value of $J$ to be $J < 1$. Hence, the transformation 
(\ref{eq:def-canonicalA}) is well-defined for all values of $J < 1$.
Third, in the limit of $J \to 0$, $c_s = 1$ and matches with standard electrodynamics. However, for any non-zero value of $J$, the sound speed is not unity; depending on the value of $J$, it can be greater or less than $1$. In the case of inflation, $\epsilon_1$ is positive, and hence, for the positive values of $J$, $c_s > 1$. However, in the case of super-inflation ($\dot{H}>0$)~\cite{2001-Gunzig.etal-PRD,2014-Basak-Shanki-JCAP}, $ \epsilon_1<0$ leading to the speed of sound less than 1~\cite{2014-Basak-Shanki-JCAP}. In the case of negative values of $J$, the sound speed is always less than unity. \ref{fig:cs-plot} contains the plot of $c_s$ as a function of $J$. It shows that $c_s > 1$ for positive $J$ during inflation.

Our next step is to solve Eq.~\eqref{eq:eom-SR-canonicalA}.
It is not possible to solve the above equation exactly for arbitrary $a(\eta)$. Hence, we consider two --- sub-horizon and super-horizon ---  limits of the above differential equation and match the mode functions at the sound horizon.
 
In the sub-horizon limit ($c_s  k | \eta| \gg 1$), we assume that the modes satisfy the Bunch-Davies vacuum, i. e., the mode approaches the Minkowski space behavior in the asymptotic past:
\begin{align}\label{eq:BunchDavies}
\lim_{k\eta \to -\infty}  \mathbb{A}_k (\eta) =  \frac{1}{\sqrt{2c_s k}}e^{-ic_s k\eta}  \,\,   .
\end{align}
For the physical gauge field ($A_{k}(\eta)$), using the relation (\ref{eq:def-canonicalA}), the above condition translates to: 
\begin{align}\label{eq:BunchDavis-ActualVec}
\lim_{k\eta \to -\infty} A_k (\eta) =  \frac{1}{\sqrt{2 k c_s \, (1 - J)}}  
%\, \left( \frac{1 - J ( 1 - \epsilon_{1})}{ 1 - J}  \right)^{-1/4}  \,
e^{-ik c_s \eta} 
%\, \sqrt{ \frac{1 - J + J \epsilon_{1}}{ 1 - J} } }
\end{align}
As we will see below, the above relation would be convenient to match the solution at the sound-horizon crossing to fix the arbitrary coefficients. In the super-horizon limit ($ c_s k | \eta | << 1$), we have 
\begin{align}\label{eq:superhor}
A_k (\eta) \simeq  C_1 +  C_2 \,\, \frac{ (1 - J)^{1 + \frac{2 J \epsilon_1 }{ 1 - J } } }{ 1 - J + 2 J \epsilon_1} \left( - k \eta \right)^{1 + \frac{2 J \epsilon_1 }{ 1 - J } }   \,\,  ,
\end{align}
where $C_1$ and $C_2$ are arbitrary coefficients. Since ${2 J \epsilon_1 }/(1 - J)$ is positive, and hence at the end of inflation, i.e., $\eta \rightarrow 0$, the second term in the above expression is negligible. In other words, the vector field freezes at the end of inflation and is approximately constant $C_1$.

As mentioned earlier, we demand that the sub-horizon modes \eqref{eq:BunchDavis-ActualVec} and super-horizon modes \eqref{eq:superhor} are continuous and differentiable at the sound-horizon. From Eq.(\ref{eq:eom-SR-canonicalA}), we obtain 
the horizon crossing time $\eta_*$ as
\begin{align}\label{eq:eta-cross}
\eta_*  = -  \frac{1}{k}\left[  \, \frac{ J_* \epsilon_1^* ( 1-J_* - \epsilon_1^*(3 - 2 J_*) + \epsilon_2^* (1-J_*) )   }{ (1-J_*) (1-J_* + J_* \epsilon_1^*) }   \, \right]^{1/2} \, ,
\end{align}
where $"*"$ refers to the quantities evaluated at the horizon crossing time $\eta_*$. Matching the mode functions at $\eta_*$ leads to: 
\begin{align}\label{eq:C1}
C_1 &= \frac{  ( 1 - J_* + 2 J_* \epsilon_1^* ) - i \sqrt{  J_* \epsilon_1^*\left( 1 - J_* - \epsilon_1^* (3 - 2 J_*) + \epsilon_2^*(1-J_*) \right)  } }{  \sqrt{2 c_s^* k (1 - J_*) }  \, (1 - J_* + 2 \epsilon_1^* J_*) }    e^{i \sqrt{ \frac{ J_* \epsilon_1^*\left( 1 - J_* - \epsilon_1^* (3 - 2 J_*) + \epsilon_2^*(1-J_*) \right) }{ (1-J_*)^2 } } }
   \\
\label{eq:C2}
C_2 &=   \frac{i \, {c_s^*}^{\frac{1}{2}+\frac{2J_* \epsilon_1^*}{1 - J_*}}}{\sqrt{2k(1-J_*)}} \left(  J_* \epsilon_1^*\left( 1 - J_* - \epsilon_1^* (3 - 2 J_*) + \epsilon_2^*(1-J_*) \right)  \right)^{\frac{ 2 J_*\epsilon_1^*}{J_* - 1}}  e^{i \sqrt{ \frac{ J_* \epsilon_1^*\left( 1 - J_* - \epsilon_1^* (3 - 2 J_*) + \epsilon_2^*(1-J_*) \right) }{ (1-J_*)^2 } } }
\end{align}

\subsection{Power spectrum and estimation of magnetic field strength}
\label{sec:power_spectrum}

To study the observable effects, we evaluate the energy
density of the electromagnetic fields~\cite{2013-Durrer.Neronov-Arxiv}. The $0-0$th component of the energy-momentum tensor $T_{\mu\nu}$ in the FRW background (\ref{def:FRW}) is
\begin{align}\label{eq:T00-def}
 T_{0 0} = - \frac{N^2}{a^3} \frac{\delta \mathcal{L}}{\delta N} \, .
\end{align}
Substituting the action \eqref{def:VECaction}
in the above expression and setting 
$N(\eta) = a(\eta)$, we get:
\begin{align}\label{eq:rho}
\rho 
= \frac{1}{2 a^4 } (1 - 3 J) \, \delta^{i j} A_i^\prime A_j^\prime 
+  \frac{1}{2 a^4 } ( 1 +  2 J ) \,  \delta^{i k} \delta^{j l}\, \partial_i A_j \,\partial_k A_l  + \frac{J}{H \,a^5}\, \delta^{i j}\, A_i{}^\prime\, \nabla^2 A_j \, .
\end{align}

The first term is the energy density of the Electric field $(\rho_E)$. 
The second and the third terms are the energy densities of the
magnetic field $(\rho_B)$ and $(\rho_{B.B^\prime})$, respectively.
Note that the first two terms decay as $a^{-4}$ while the last term decays as $a^{-5}$. Hence, one can ignore the last term in evaluating the energy density of electromagnetic fluctuations. As we will see below, imposing the condition that the energy density is always positive provides a condition on $J$.

Using the decomposition (\ref{eq:fourierd}), the electric and magnetic
part of the perturbation spectrum per logarithmic interval can be
written as:
\begin{eqnarray}\label{eq:SpectraB}
&& \mathcal{P}_{B}(k) \equiv  \frac{\mbox{d}}{\mbox{dln}k} \langle 0|\hat{\rho}_{B^2}|0 \rangle = \frac{ (1 + 2 J ) }{2 \pi^2} \frac{k^5}{a^4} \left|A_k\right|^2 \\
 \label{eq:SpectraE}
&&  \mathcal{P}_{E}(k) \equiv  \frac{\mbox{d}}{\mbox{dln}k} \langle0|\hat{\rho}_{E^2}|0\rangle =  \frac{ (1 - 3 J) }{ 2 \pi^2 } \frac{k^3}{a^4} \left|A_k^\prime\right|^2 \\
  \label{eq:SpectraBB}
&& \mathcal{P}_{_{B.B^\prime}}(k) \equiv  \frac{\mbox{d}}{\mbox{dln}k} \langle0|\hat{\rho}_{_{B.B^\prime}}|0\rangle = -\frac{J }{ 4 \pi^2 } \frac{k^5}{a^4 \mathcal{H}} ( A_k^\prime A_k^* + {A_k^\prime}^* A_k ) ~~~~.
 \end{eqnarray}
During most of the Universe's history,  the electrical conductivity of the Universe is high. Hence, the electric fields decay and do not contribute to the energy density. Hence, like in Sec. \eqref{sec:EFT-Spectrum}, we will concentrate on the magnetic field component\footnote{In some models, the electric power spectrum dominates over the magnetic part~\cite{2009-Demozzi.etal-JCAP}. In our case, the electric power spectrum does not contribute.}. 
Imposing the condition that energy density is non-negative implies that $J \geq -1/2$. Combining the earlier constraint, this implies $J$ in the range $[-0.5, 1)$ is well-defined.

Substituting Eq.~(\ref{eq:superhor}) in Eq.~(\ref{eq:SpectraB}) leads to:
\begin{eqnarray}\label{eq:PowerSpectraB}
&& \mathcal{P}_{B}(k) = \frac{H^4 (- k \, \eta_f)^4 }{4 \pi^2 } \frac{ (1+2J_f) ( 1 - J_* (2 - 5\epsilon_1^*) + J^2_* (1 - 5\epsilon_1^*)  )  }{c_A^* (1-J_*) (1-J_*+2J_*\epsilon_1^*)^2}
 \end{eqnarray}
where $\eta_f$ refers to the end of inflation and $J_f = 4 D H_f^2$.  

We now calculate the energy density of the generated electromagnetic fields by integrating the power spectrum over the Fourier modes $k_i < k < k_f$ where $k_i$ and $k_f$ refer to the initial and final modes leaving the horizon during inflation, i. e.:
\begin{align}\label{eq:rho_B}
    \rho_{B} = \int_{k_i}^{k_f} \mathcal{P}_{B}(k) \, \mbox{dln}k
    = \frac{H_f^4 (- k_f \, \eta_f)^4 }{16 \pi^2 } \frac{ (1+2J_f) ( 1- J_*(2-5\epsilon_1^*) + J^2_* (1 - 5\epsilon_1^*) )  }{c_A^* (1-J_*) (1-J_*+2J_*\epsilon_1^*)^2} \, .
\end{align}
We want to stress that the above expression is for a generic inflation model and depends on the value of $J_*$. 
In the rest of the section, we determine the range of $J_*$ for which the 
generated magnetic fields are sufficient for the observable large-scale magnetic fields such that the generated magnetic fields do not affect the background FRW metric. To avoid the back-reaction, the energy density of the generated fields must be less than the total background energy density during inflation. Assuming slow-roll inflation $\epsilon_1 << 1$, 
the back-reaction parameter ($\xi$) for the Galileon model is~\cite{2009-Demozzi.etal-JCAP,2016-Subramanian-Arxiv,2020-Talebian.etal-arXiv}:
\begin{align}\label{eq:back-reaction}
\xi \equiv \frac{\rho_{ \rm{B} } }{ \rho_{\rm{Inf}} } \approx \frac{  H_f^2  }{ 48 \pi^2 \, M_{ \rm{pl} }^2 } \, \frac{1}{(1 - \, J_* )}   <  1
\end{align}
where $\rho_{\rm{Inf}} = 3 H_f^2 M_{ \rm{pl} }^2$ is the background energy density during inflation and $\rho_B$ is given by Eq.(\ref{eq:rho_B}). As mentioned above, we assumed $\epsilon_1 << 1$ and set $-k_f \eta_f \sim 1$. From the above condition, we get the following constraint:
\begin{align}\label{eq:Jc_constraint}
\frac{1}{1 - J_*} \,   <    \, 2.25 \times 10^{13}
\end{align}
Combined with the earlier constraint of $J$, we see that $J_*$ close to $1$ can lead to large amplification. Let us now consider two cases:
\begin{enumerate}
\item[${\bf J_* > 0}$:] When $J_*$ is very close to 1, we can have large amplification in the magnetic field strength.  
Let us now estimate the magnetic field strength corresponding to $(1 - J_*)^{-1}  \approx  2.25 \times 10^{13}$ at the comoving wavenumber $k = 1~{\rm Mpc}^{-1}$.  The magnetic field for a mode with wave vector $k$ at the end of inflation is given by:
\begin{align}\label{eq:Bf-exp}
B_f \approx \sqrt{ \frac{ H_f^4 k^4 \eta_f^4 }{4 \pi^2 c_s^* (1-J_*) } }  \approx 1.58 \times 10^{13}  \, \rm{G} \, ,
\end{align}
where we have set $| \eta_f | = 10^{-20} \rm{Mpc}$~\cite{2014-Martin.etal-PhyDarkUniv} and since $J_* \to 1$ we have approximated $c_s^* \simeq 1$. 

In the radiation-dominated (or matter-dominated) epoch, the magnetic field decays adiabatically as $B \propto a^{-2}$. We then have the following relation between the magnetic field strength at present and the end of inflation:
\begin{align}\label{eq:B0-Bf-rel}
B_0 = B_f  \left( \frac{a_f}{a_0}  \right)^2 .
\end{align}
Assuming the instantaneous reheating and the entropy conservation, i.e., $g T^3 a^3 = \rm{constant}$ during its evolution, where $T$ is the temperature of the relativistic fluid, and $g$ is the number of effective relativistic degrees of freedom~\cite{2016-Subramanian-Arxiv} gives
\begin{align}\label{eq:a0/af-rel}
\frac{a_0}{a_f} =  \left( \frac{90}{8\pi^3}  \right)^{1/4}  \frac{g_f^{1/12} }{g_0^{1/3} } \frac{\sqrt{ H_f \, M_{ \rm{pl} } } }{T_0} 
\approx 0.9 \times 10^{29} \, \left(  \frac{H_f}{10^{-5}  \, M_{ \rm{pl} }  }  \right)^{1/2}
\end{align}
where we have taken $g_f \approx 100$ and $g_0 = 2.64$. Substituting Eq.~(\ref{eq:a0/af-rel}) in Eq.~(\ref{eq:B0-Bf-rel}),the present day magnetic field strength at $\rm{Mpc}^{-1}$ scale is:
\begin{align}\label{eq:B0estimate}
B_0 = 1.5 \times 10^{-45} \, \rm{G}
\end{align}
\item[${\bf J_* < 0}$:] In this case, $1/(1 - J_*)  \approx 1$. The magnetic field for a mode with wave vector $k$ at the end of inflation is given by:
\begin{align}\label{eq:Bf-exp2}
B_f \approx \sqrt{ \frac{ H_f^4 k^4 \eta_f^4 }{4 \pi^2 c_s^* } }  \approx 1.58 \times 10^{7}  \, \rm{G} \, ,
\end{align}
where, here again, we have set $| \eta_f | = 10^{-20} \rm{Mpc}$~\cite{2014-Martin.etal-PhyDarkUniv} and $c_s^* \simeq 1$. Following the same analysis, the present day magnetic field strength at $\rm{Mpc}^{-1}$ scale is:
\begin{align}\label{eq:B0estimate2}
B_0 = 1.5 \times 10^{-51} \, \rm{G}
\end{align}
\end{enumerate}
In estimating the magnetic field strength, we 
have used the standard values of inflation, however, if inflation occurs over the energy scales such that $10^{-10} \lesssim H_f / M_{ \rm{pl} } \lesssim 10^{-5} $~\cite{2021-Tripathy.etal-arXiv}, then the adiabatic expansion rate will be much less, i. e.,
\begin{align}
10^{26} \lesssim \frac{a_0}{a_f} \lesssim 10^{29}~~.
\end{align}
In that case, we gain six orders of magnitude in the magnetic field strength. 

Our analysis shows that appreciable magnetic field strength can be achieved when $J_* \to 1$. However, as can be see in \eqref{fig:cs-plot}, $J_* > 0$ leads to super-luminal electromagnetic perturbations. Let us now compare the above result with the EFT analysis. Following the discussion in Appendix \eqref{app:VGModel}, we 
find that $s_2$ and $e_1$ are the non-zero parameters that describe vector Galileon model. Substituting these in Eq.~\eqref{eq:cAslowroll}, we have:
\begin{equation}
  c_A = 1 - \frac{1}{2}\left[s_2 - e_1 (1-\epsilon_1)\right] \left(\frac{H}{\Lambda}\right)^2.
\end{equation}
Using the relation \eqref{eq:VG-EFTComp}, we have:
\begin{equation}
c_A = 1 + \frac{J \epsilon_1}{8} \,  .
\end{equation}
The EFT analysis shows that $J < 0$ avoids super-luminal propagation. Hence, we need another physical condition --- the modes should be sub-luminal. This conclusion is identical to the one we obtained from the EFT analysis. Thus, our analysis shows that conformal invariance breaking is necessary but not sufficient condition to generate sufficient magnetic fields in the early Universe.

Another interesting thing to note is that by setting the above parameters in the EFT power spectrum \eqref{PS-sr}, we can see that the vector Galilean model will never lead to a scale-invariant power spectrum and is consistent with the analysis in this section. 

\section{Conclusions and Discussions}
\label{sec:discussion}

The origin of primordial magnetic fields is still unresolved and requires physics beyond the standard models of cosmology and particle physics. Although inflation provides a causal mechanism for the generation of primordial density perturbations, it can not generate the appreciable primordial magnetic field in the early Universe. It has been argued that conformal invariance breaking is a sufficient condition to generate primordial magnetic fields during inflation. In this chapter, using EFT based on expansion about the Hubble parameter $(H)$ and its derivatives, we show that the generation of primordial magnetic fields requires two
necessary conditions --- conformal invariance breaking and causal propagation. We have also shown that a broad class of magnetogenesis models can be reproduced from the EFT that is a sum of series in $H/\Lambda$, time-derivatives of $H$ --- ($H'/\Lambda^2$), ($H''/\Lambda^3$), $\cdots$ --- and their products, for instance, $H H'/\Lambda^3$, $H H''/\Lambda^4$, $\cdots$.

Like the EFT of inflation, the EFT
of magnetogenesis requires the inclusion of fluctuations in the matter and metric degrees
of freedom. As shown in Appendix A, the gauge field is the relevant gauge-invariant variable for the EFT. Demanding that the EFT of magnetogenesis breaks conformal invariance, however, satisfies local Lorenz invariance and gauge invariance, we obtained the general second-order EFT action. This action depends on two expansion scalar functions --- $f_1(H, a, \Lambda)$ and $f_2(H, a, \Lambda)$. We then showed that the EFT action \eqref{EFT:L} could reproduce all the known magnetogenesis models. 
To make the computation of the power-spectrum tractable and to highlight the importance of speed of perturbations, we truncate the series \eqref{eft:A} up to second order. However, the truncation of the series to compute the power spectrum has no bearing on the EFT expansion \eqref{eft:A}.

By truncating the expansion scalar functions to $\Lambda^{-2}$, we derived the EOM of the gauge field and obtained the power spectrum in the slow-roll inflation scenario. From Eq.~(\ref{PS-srFin}), we see that the amplification in the magnetic power spectrum --- evaluated at the horizon crossing $(-k \eta c_A = 1$) ---  is possible whenever $3 s_1 - 5 d_1 > 0$. Specifically, we see for sub-luminal ($d_1 < s_1$) or super-luminal ($s_1 < 0$) modes, the leading order correction term in the power-spectrum ${3 s_1-5d_1} > 0$.
In other words, we can have large amplifications even for super-luminal fluctuations. To avoid EFTs with superluminal fluctuations~\cite{2006-Adams.etal-JHEP}, we need another physical condition --- the modes should be sub-luminal.

We then considered a specific model of inflationary magnetogensis where the vector Galileon breaks the conformal invariance. We extensively studied the magnetic field generation during inflation, considering the total action. Due to the absence of coupling between scalar field and electromagnetic field, the model does not lead to a strong coupling problem~\cite{2017-Debottam.Shankaranarayanan-JCAP}; hence, the model does not require fine-tuning, which is an interesting feature of our model.
Furthermore, the evolution of vector modes is frozen on the super-horizon scale. We showed that the model predicts the present-day magnetic field of strength $B_0 \sim 10^{-45} {\rm G}$ on the cosmological (Mpc) scale which is several orders higher compared to the standard electromagnetic action, and there is no back-reaction problem. For low-scale inflationary models i.e., $10^3 - 10^4 \, {\rm GeV}$, the model generates $B_0 \sim 10^{-33} \rm{G}$. However, this comes with the price --- super-luminal electromagnetic perturbations. In other words, the model generates large magnetic fields and does not violate any other conditions; it leads to super-luminal perturbations. 

Modifications of general relativity provide an alternative explanation for cosmological inflation~\cite{2022-Shankaranarayanan-Joseph-GRG}. 
Modified gravity theories have extra degrees of freedom that might have interesting physical consequences in the early Universe. For instance, Stelle gravity~\cite{1978-Stelle-GRG} contains massive tensor modes, and these modes carry more energy than the scalar modes in $f(R)$ gravity models~\cite{Chowdhury:2022ktf}. While the analysis presented in this chapter can be extended to $f(R)$ gravity models, it is not straightforward to extend to Stelle gravity. Investigating the EFT of inflation and magnetogenesis in these modified gravity models is interesting. Also, it will be interesting to extend the EFT of magnetogenesis to bouncing models. The analysis presented in this chapter can be extended to helical fields.

\chapter{Gertsenshtein-Zel$'$dovich effect: A plausible explanation for fast radio bursts?}
\label{ch:GZeffect}

Technological advancement has fuelled research in high-energy astrophysical phenomena at larger redshift ranges, and we are in a position to address some unresolved questions starting from pulsar emission mechanism~\cite{2021MNRAS_pulsar} to short bursts such as Gamma-ray bursts (GRBs)~\cite{2016-Levan.etal-SSR}, Fast radio bursts (FRBs)~\cite{2008-Lorimer-LivRevRel,2019-Cordes.Chatterjee-AnnRevAA,2019-Platts.etal-PhyRept}. To date, more than 600 FRBs have been reported in various catalogues~\cite{Petroff:2016tcr,2019-Platts.etal-PhyRept, Pastor-Marazuela:2020tii,2021-Rafiei.etal-APJ}. $99\%$ of these FRBs have the following three characteristic features: observed peak flux ($S_{\nu}$) 
varies in the range $0.1~{\rm Jy} < S_{\nu} < 700~{\rm Jy}$, coherent radiation and the pulse width is less than a second \cite{2021-Rafiei.etal-APJ,Petroff:2016tcr}. These observations have posed the following questions: What causes these extreme high-energy transient radio-bursts from distant galaxies, lasting only a few milliseconds each~\cite{2008-Lorimer-LivRevRel,2019-Cordes.Chatterjee-AnnRevAA,2019-Platts.etal-PhyRept}? Why do some FRBs repeat at unpredictable intervals, but most do not~\cite{2019-Platts.etal-PhyRept}?
Does strong gravity provide an active role? 

%\textcolor{red}{This work is an attempt to address origin of the FRBs and its coherent nature. More specifically, what is the physical mechanism that results in large amount of coherent radiation in a short time}~\cite{2008-Lorimer-LivRevRel,2019-Cordes.Chatterjee-AnnRevAA,Petroff:2016tcr,2019-Platts.etal-PhyRept, Pastor-Marazuela:2020tii,2021-Rafiei.etal-APJ}. Since the time scale of these events is less than a second and the emission is coherent, the astrophysical processes that explain these events \emph{cannot} be thermal~\cite{2008-Lorimer-LivRevRel}. 
In the previous chapters, we have looked at the generation and effects of large scale magnetic fields. In this chapter, we will discuss the interesting consequences of small-scale strong magnetic fields, for example, magnetar (see Fig.~\ref{fig:magneticuniverse} in chapter~\ref{ch:intro}). As we have briefly discussed in chapter~\ref{ch:intro}, the origin of FRBs is one of the open problems in modern cosmology and astrophysics. In this chapter, we will attempt to address the origin of the FRBs and their coherent nature of radiation. More specifically, what is the physical mechanism that results in large amount of coherent radiation in a short time
~\cite{2008-Lorimer-LivRevRel,2019-Cordes.Chatterjee-AnnRevAA,Petroff:2016tcr,2019-Platts.etal-PhyRept, Pastor-Marazuela:2020tii,2021-Rafiei.etal-APJ}.

Naturally, many models have been proposed to explain the origin of FRBs. All these models try to provide a physical mechanism that results in large amount of coherent radiation in a short time~\cite{2008-Lorimer-LivRevRel,2019-Cordes.Chatterjee-AnnRevAA,Petroff:2016tcr,2019-Platts.etal-PhyRept, Pastor-Marazuela:2020tii,2021-Rafiei.etal-APJ}. Since the time scale of these events is less than a second, and the emission is coherent, the astrophysical processes that explain these events \emph{cannot} be thermal~\cite{2008-Lorimer-LivRevRel}.

Broadly, these models can be classified into two categories~\cite{2022-Zhang-arXiv}: FRBs created by interaction of an object with a pulsar/magnetar and FRBs created from the magnetar/pulsar itself~\cite{2008-Lorimer-LivRevRel,2019-Cordes.Chatterjee-AnnRevAA,Petroff:2016tcr,2019-Platts.etal-PhyRept, Pastor-Marazuela:2020tii,2021-Rafiei.etal-APJ}. The first category can further be classified into two broad classes. In the first class, the energy powering FRBs comes from
the neutron star magnetosphere/wind themselves, and an orbiting object converts this energy into
radiation. In the second class, the object falls onto the neutron star, and its gravitational energy partly gets converted to FRBs.
In the second class, many models involving non-thermal processes such as Synchrotron radiation~\cite{Book-Lorimer.Kramer-PulsarAstronomy}, black hole super-radiance~\cite{2018-Conlon.Herdeiro-PLB}, evaporating primordial black hole \cite{1977-Rees-Nature,2020-Carr.Kuhnel}, spark from cosmic strings~\cite{2008-Vachaspati-PRL}, Quark Novae~\cite{2015-Shand.etal-RAA},  {synchrotron maser shock model~\cite{2020ApJ...900L..26W},  radiation from reconnecting current
sheets in the far magnetosphere~\cite{2020ApJ...897....1L},  curvature emission from charge bunches~\cite{2022ApJ...927..105W} } have been proposed. Several classes of FRB models predict prompt multiwavelength counterparts and
specify the ratio between the energy emitted by the counterpart
and by the FRB~\cite{2017-Zhang-ApJL,2019-Metzger.etal-MNRAS}.

However, despite the use of exotic new physics, no single model has provided a universal explanation for the enormous energy released in these events. It is important to note that all these mechanisms require  
electromagnetic interaction to generate FRBs.
Due to the nature of electromagnetic interaction, small-scale emission mechanisms usually predominate over large-scale coherent electromagnetic processes (like astrophysical masers and pulsar radio emission). In this chapter, we provide an alternative framework that overcomes this and can explain the observed coherence in FRBs.

%Many models involving non-thermal processes such as Synchrotron radiation~\cite{Book-Lorimer.Kramer-PulsarAstronomy}, black hole super-radiance~\cite{2018-Conlon.Herdeiro-PLB}, evaporating primordial black hole \cite{1977-Rees-Nature,2020-Carr.Kuhnel}, spark from cosmic strings~\cite{2008-Vachaspati-PRL}, Quark Novae~\cite{2015-Shand.etal-RAA},  {synchrotron maser shock model~\cite{2020ApJ...900L..26W},  radiation from reconnecting current sheets in the far magnetosphere~\cite{2020ApJ...897....1L},  curvature emission from charge bunches~\cite{2022ApJ...927..105W} } have been proposed. However, despite the use of exotic new physics, no single model has been able to provide a universal explanation for the enormous energy released in these events. 

As shown below, one key missing ingredient is the \emph{dynamics of strong-gravity}. The Spatio-temporal changes in the strong-gravity regime --- oscillons,  phase transitions, plasma instability, primordial black holes, reheating --- generate gravitational waves (GWs) in a broad range of frequencies ($10^{-15} - 10^{15}$ Hz) \cite{1974-Hawking.Carr-MNRAS,2009PhRvL.103k1303A,2015-Kuroda.etal-IJMPD,2019-Ejlli.etal-EPJC,2020-Aggarwal.etal-arXiv,2020-Chen.Rajendran.etal-arXiv,2021-Pustovoit.etal-JOP}. Like EM waves, GWs are generated by the time-varying quadrupole moment~\cite{2009-Sathyaprakash.Schutz-LivRevRel,2000-Schutz-arXiv}. Since all masses have the same gravitational sign and tend to clump together, they produce large coherent bulk motions that generate \emph{energetic, coherent GWs}~\cite{2007-Hendry.Woan-AG}. Thus, if a mechanism that converts incoming coherent GWs to EM waves exists, we can explain the extremely energetic, coherent nature of FRBs~\cite{2018-Popov.etal-Usp,2022-Lieu.etal-CQG}. In this chapter, we construct a model that uses this feature. 

Since FRBs are highly energetic, an attentive reader might wonder do incoming GWs carry such large energies. GWs carry an enormous amount of energy. For example, typical GWs from a compact binary collapse with amplitude $h \sim 10^{-22}$ carry the energy of the order of $10^{20}~\rm{Jy}$ \cite{2009-Sathyaprakash.Schutz-LivRevRel,2000-Schutz-arXiv}. If the GWs indeed carry a lot of energy, can this energy transform into other observable forms of energy?  

Currently, there is evidence of GWs in the frequency range $10^{-9} - 10^{4}~{\rm Hz}$ from LIGO-VIRGO-KAGRA and PTA observations~\cite{2016-Abtott.etal_LIGOScientific-PRL,2023-Agazie.etal-AAS}.
While most of the current effort has focused on these frequency ranges, there is a surge in activity for the possibility of detecting GWs in the MHz-GHz frequency range~\cite{2020-Aggarwal.etal-arXiv}. New physics beyond the standard model of particle physics, like an exotic compact object, can produce observable GW signals in this frequency ranges~\cite {2020-Aggarwal.etal-arXiv,2020-Chen.Rajendran.etal-arXiv,2021-Pustovoit.etal-JOP}.

\begin{figure*}[!ht]
\centering
\includegraphics[height=1.6in]{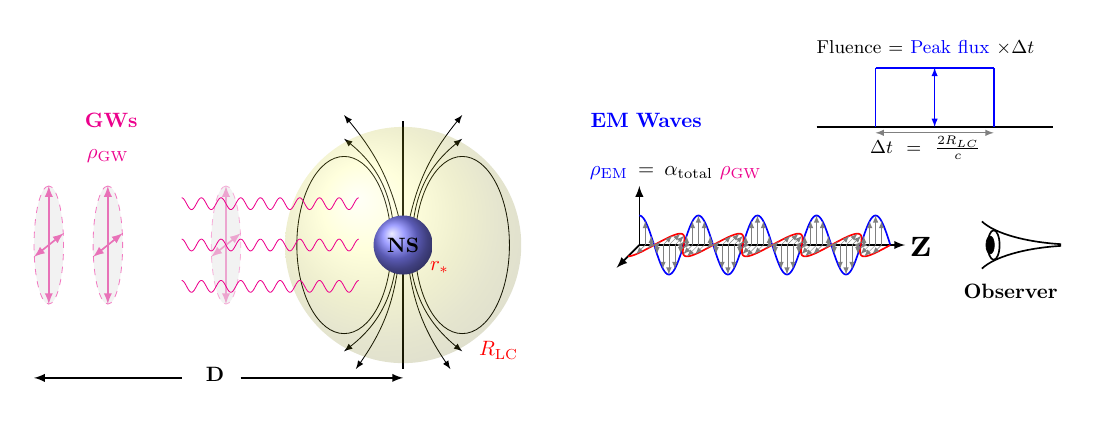}%
\caption{Schematic depiction of GZ effect. The externally generated GWs at distance $D$ from the NS is converted to EM Waves in the magnetosphere.The blue region corresponds to NS with radius $r_*$ and the yellow region around NS corresponds to magnetosphere of radius $R_{\rm LC}$. The black curves correspond to the magnetic field lines.}
\label{fig:setupFigure}
\end{figure*}

A physics maxim is that energy can be transformed between different forms. Although the total energy is conserved, the efficiency of the transformation depends on the energy scale, background dynamics, and external conditions (parameters). Energy transformation is one way to probe strong gravity regions like the early universe, black-holes, and NS. In this chapter, we propose a novel approach that uses the energy conversion from incoming, coherent GWs to electromagnetic (EM) waves that can \emph{explain milli-second bursts like FRBs}.

GWs get converted to EM waves in the presence of strong transverse magnetic fields --- Gertsenshtein-Zel'dovich (GZ) effect~\cite{1962-Gertsenshtein-JETP,1974-Zeldovich-SJETP,2018-Zheng.Wei.Li-PRD,2021-Domcke.Garcia-Cely-PRL}. 
To understand the GZ effect, consider coherent GWs with a frequency $\omega_g$ passing through a region with a high transverse static magnetic field ${\bf B}$. The propagation of GWs leads to compression and stretching of the magnetic field proportional to $h {\bf B}$ ($h$ is the amplitude of GWs), which acts as a source leading to the generation of EM waves ~\cite{1962-Gertsenshtein-JETP,1974-Zeldovich-SJETP}. The induced (resultant) EM waves generated will have maximum amplitude at resonance, i.e., the frequency of EM waves is identical to $\omega_g$. (For details, see appendix~\ref{appsec:E&B-solution}) 

In quantum mechanical language, the GZ effect is analogous to the mixing of neutrino flavors --- the external field \emph{catalyzes} a resonant mixture of photon and graviton states~\cite{2023-Palessandro.Rothman-PDU}. The external magnetic field provides the extra angular momentum necessary for the spin-1 (photon) field to mix with the spin-2 (graviton) field. Thus, the GZ mechanism involves the transfer of energy from the incoming, coherent GWs to emitted, coherent EM radiation in the presence of the background magnetic field. 
The maximum efficiency of this conversion can be achieved if the background field is strong at the resonance frequency (when the frequency of the emitted EM radiation is the same as the incoming GWs).
Hence, the background magnetic field acts as a catalyst in this mechanism.

We show that the \emph{GZ effect} may provide an interpretation for a class of non-repeating FRBs. 
To explain the energy bursts in FRBs, we propose a model with the following two realistic assumptions: 
(i) the astrophysical object is compact and has a strong gravity environment, and 
(ii) the object possesses a small time-dependent magnetic field on top of the large, effective static, transverse magnetic field. 
These two assumptions principally lead us to stellar remnants, such as NS and magnetars with magnetic field strength ranging from $10^{8} - 10^{15}~\rm{G}$~\cite{2020-Andersen.etal.CHIME-FRB-Nature,2015-Belvedere.etal-ApJ,2019-Cordes.Chatterjee-AnnRevAA,2008-Lorimer-LivRevRel}. 
The small time-dependent magnetic fields arise due to the rotation of the NS about its axis with frequency $\omega_B$~\cite{2015-Belvedere.etal-ApJ,2008-Melikidze.gil-ProcIAU,2019-Platts.etal-PhyRept,2019-Pons.Vigan-arXiv,2012-Pons.etal-AandA}. We consider $\omega_B$ in the range $[1, 10^3]~{\rm Hz}$~\cite{Book-Lorimer.Kramer-PulsarAstronomy}. 
%Since the time-dependence is linked to the rotation, $\omega_B$ is related to the rotation period of the NS in the range $1 < \omega_B (s^{-1}) < 10^3$. 
As a result, the effective magnetic field at a given point in the NS magnetosphere is ${\bf B}(t) = {\bf B}^{(0)} + \delta {\bf B} \sin(\omega_B t)$. It has been noted that $|\delta {\bf B}/{\bf B}^{(0)}|$ can be as large as $0.1$~\cite{2012-Pons.etal-AandA}. Here, we take $|\delta {\bf B}/{\bf B}^{(0)}| \sim  10^{-2}$.

\section{Model}

\ref{fig:setupFigure} gives the schematic depiction of the physical model to explain the energy burst in FRBs. Consider GWs generated due to exotic compact objects (such as Boson stars, Oscillons, gravastars)~\cite{2020-Aggarwal.etal-arXiv,2020-Chen.Rajendran.etal-arXiv,2021-Pustovoit.etal-JOP} passing through the magnetosphere of NS 
at a distance $D$. In the figure, the magnetosphere is depicted as a cylinder. GZ effect converts GWs to EM waves as they pass through the magnetosphere~\cite{1962-Gertsenshtein-JETP,1974-Zeldovich-SJETP}. This conversion occurs at all points in the magnetosphere. 
Therefore, a faraway observer will see the integrated effect happening in the entire magnetosphere in this short duration.
For example, the light cylinder radius ($R_{\rm LC}$) for a typical NS is $\sim 10^{7}-10^{9}~{\rm cm}$, implying that the GWs take less than one second to cover the entire magnetosphere. This is one of the primary ingredients supporting our analysis for the FRB observations. 

To compute the GZ effect at a point in the magnetosphere, we consider source-free Maxwell's equations on the background space-time with GW fluctuations (see details in Appendix~\eqref{appsec:E&B-solution}). The linearized Einstein's equations, up to first order in the space-time 
perturbations are highly accurate. In this approximation, the effects of GWs on the stress-tensor (Riemann tensor) are negligible. Hence, we consider background space-time to be Minkowski (in cartesian coordinates)~\cite{Book-Gravitation_MTW}). 
The two polarizations of GW (with frequency $\omega_g$ and wave-vector $k_g$) propagating along the z-direction are:
\begin{align}\label{eq:h-Expression}
h_+  = %h_{xx} = - h_{yy} =
A_+ \, e^{i \left( k_g z - \omega_g t \right) }, 
h_{\times}  = %h_{xy} =  h_{yx} = 
i A_{\times} \, e^{i \left( k_g z - \omega_g t \right) } ,
\end{align}
where $A_+$ and $A_{\times}$ are the constant amplitudes of the GWs. We assume that both the modes of GWs are generated with an equal amount of energy, i.e., $|A_{+}| = |A_{\times}| $  --- the isospectrality condition in general relativity~\cite{Chandrasekhar_BlackHoles-Book}.
Taking the distance between GW source  { (possibly an exotic compact object)} and NS to be $D = 1 \,  \rm{kpc}$, gives $h \simeq A_{+} = 10^{-20}$ at $1 \, \rm{GHz}$  { (see Appendix \ref{appsec:ECO})}. In this chapter, we have assumed $h$ to be 
three orders smaller ($\sim 10^{-23}$) near the NS.

As mentioned above, the key requirement of the GZ-effect is the presence of the transverse magnetic field to the direction of propagation of coherent GWs. Besides, the time taken by the GWs to pass through the entire magnetosphere is much smaller than the rotation period of the millisecond pulsar/magnetar ($\omega_B^{-1}$). Given the direction of propagation of GWs along the z-axis, the effective time-dependent transverse magnetic field is taken to be
$\textbf{B}(t) = \left( 0, B^{(0)}_y + \delta B_y \sin (\omega_B t), 0  \right)$
~\cite{2019-Pons.Vigan-arXiv,2012-Pons.etal-AandA}. 

Although the effective magnetic field depends on the distance from the surface of the object~\cite{2008-Lorimer-LivRevRel,2015-Belvedere.etal-ApJ,2019-Cordes.Chatterjee-AnnRevAA,2020-Andersen.etal.CHIME-FRB-Nature}, we assume that ${\bf B}(t)$ is independent of the distance from the surface up to 
$R_{\rm LC }$~\cite{Book-Lorimer.Kramer-PulsarAstronomy}.
In appendix~(\ref{appsec:Integration-2}), we explicitly show that the above assumption that the background magnetic field can be treated as a constant in the entire magnetosphere gives \emph{identical results} to that of the background field decreasing radially, i. e.,
\begin{align}
\left(  B_r , B_{\theta} , B_{\phi}  \right) = B_{*} \, \left( \frac{r_* }{r} \right)^3 \, \left(  2 \cos\theta, \sin\theta , 0  \right)
\end{align}
where $B_*$ is the magnetic field on the NS surface. More specifically, assuming that the NS magnetic field is dipolar, we show in appendix~(\ref{appsec:Integration-2}) that we can approximate the average magnetic field at any point in the magnetosphere to be constant. In other words, the total conversion factor we obtain using the above assumption mimics the realistic NS regions.

Given the above setup, we now evaluate the GZ-effect in the magnetosphere of the NS and compare it with observational quantities in two steps: 
\begin{enumerate}
    \item  First step involves evaluating 
    the conversion from coherent GWs to EM waves at a typical point inside the magnetosphere. This is referred to as \emph{conversion factor} ($\alpha$). We then obtain the total conversion factor ($\alpha_{\rm tot}$) inside the entire magnetosphere \emph{at resonance} (the frequency of EM waves is identical to $\omega_g$).  {This conversion factor includes only those contributions that are along the direction of line-of-sight and coinciding with the incoming GWs. Also, only the transverse magnetic field component to the direction of propagation at each point of the magnetosphere will contribute to the emitted EM waves. This can potentially explain the coherent nature of FRBs~\cite{2013-Katz-PRD,2017-Kumar.etal-MNRAS}.} 
    \item For a given conversion factor, we obtain the Poynting vector of the resultant EM waves along the direction of propagation ($S_z$). Then, we compare the theoretically derived Poynting vector with the observation of peak flux with the reported FRBs.
\end{enumerate}
The Poynting vector is a well-defined quantity for photons that travel from the source to the observer without any hindrance~\cite{Book-Carroll.Ostlie-CUP,Book-Condon.Ransom-PUP,Book-Zhang-GRB-CUP}. More specifically, assuming there is no absorption/emission of the photons during the entire journey, the Poynting vector (of the EM waves) $S_z$ is conserved (independent of the distance between the magnetosphere and observer) and is valid for emissions from compact sources. Hence, the Poynting vector estimated at a small angle (along the direction of the incoming gravitational waves) remains the same at the source and the detector. Furthermore, as shown in \ref{fig:setupFigure}, the incoming, coherent GW is along the $z-$axis in the entire magnetosphere. Therefore, the cumulative effect of the emitted EM waves is in the same direction.

To evaluate the conversion factor, solving the linearized {covariant Maxwell's} equations leads to the following electric and magnetic fields induced due to GWs, i.e., $\tilde{E}_x$ and $\tilde{B}_y$ as\footnote{We thank Archana Pai for pointing out a mistake in $\tilde{E}_x$ where $1/2$ factor was missing in the first term.}: 
\begin{align}\label{eq:E_x-Final}
\tilde{E}_x &\simeq  - \frac{A_{+} }{4} \, B^{(0)}_y \left( 1   -  \xi \,  \omega_B t  \,\, \right) \,\, e^{i \left(k_g z - \omega_g t\right) } 
\\
\label{eq:B_y-Final}
\tilde{B}_y &\simeq  - \frac{A_{+} }{4} \, B^{(0)}_y \left(1  +  2 \xi \,  \omega_g t  \,\,\right) e^{i \left(k_g z - \omega_g t\right) } \, ,
\end{align}
where $\xi \equiv \delta B_y/B_y^{(0)}$. Note that 
the amplitude of $\tilde{B}_y$ has a dependence on $\omega_g$, while the amplitude of $\tilde{E}_x$ has $\omega_B$ dependence on $\omega_g$. This is because the induced electric field arises due to the time-varying magnetic field (see appendix~\ref{appsec:E&B-solution}).

The conversion factor ($\alpha$) --- ratio of the energy density of EM wave and GWs --- gives the efficiency of the process at resonance. $\alpha$ for this process is 
\begin{align}\label{eq:alpha-at-point}
\!\!\! \alpha \equiv \frac{ \rho_{\rm EM } }{\rho_{\rm GW }} 
\simeq \frac{G |B^{(0)}_y|^2 }{4 c^2 } 
\left[2   \left( \frac{\xi z}{c}\right)^2 + 2 \frac{\xi}{\omega_g} \frac{z}{c} 
+ \frac{1 }{\omega_g^2} \right] 
\end{align}
where $z$ refers to the radial distance in the magnetosphere. For details, see appendix~\ref{appsec:Integration}.

\begin{figure}[ht]
\centering
%\subfigure[]{%
\label{fig:alphaTot_magnt}%
\includegraphics[height=3.2in]{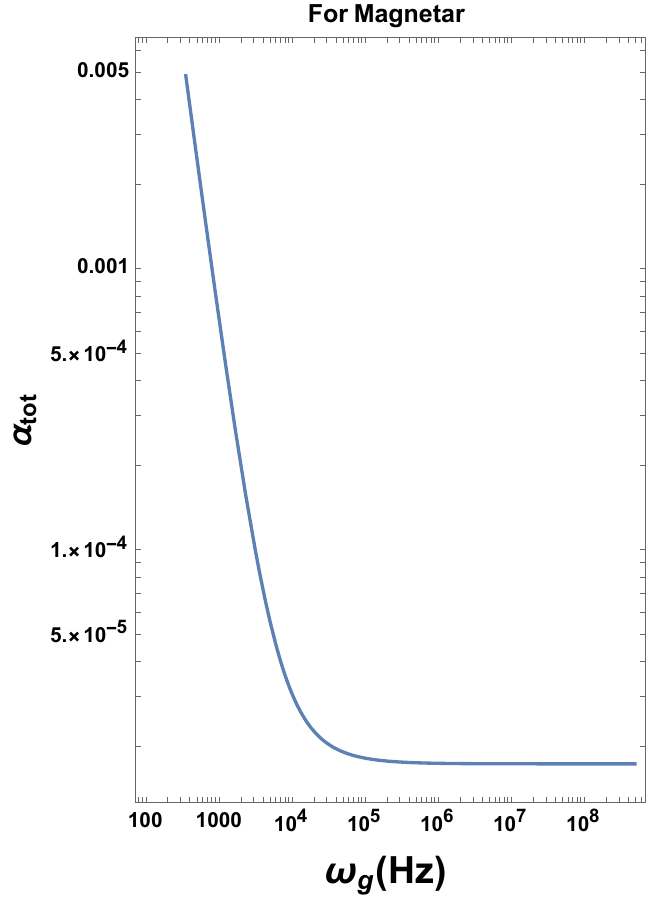} %}%
\qquad
%\subfigure[]{%
\label{fig:alphaTot_ns}%
\includegraphics[height=3.2in]{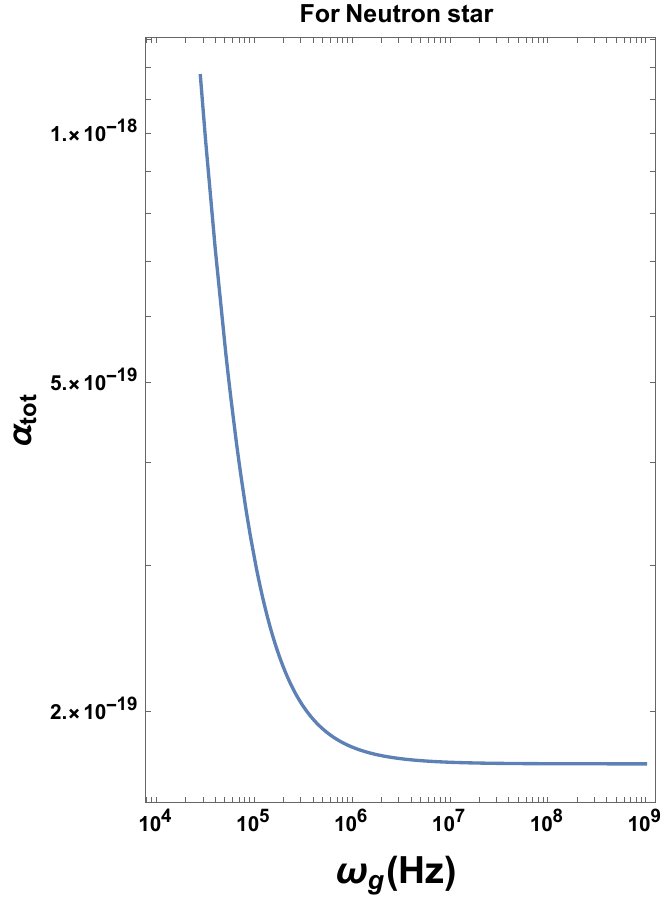} %}%
\caption{Log-Log plot of $\alpha_{\rm tot}$ versus $\omega_g$ for typical magnetar (left plot) and typical NS (right plot). For magnetar, we set $B^{(0)}_y = 10^{15} \, {\rm G}, R_{\rm LC} = 10^9~ \, {\rm cm},  \omega_B = 1 \, {\rm Hz}$. For NS, we set $B^{(0)}_y = 10^{10}  {\rm G} , R_{\rm LC} = 10^7 {\rm cm}  , \omega_B = 1  {\rm kHz}$.}
\label{fig:Plot-convfact}
\end{figure}

The above expression is the conversion factor at a single point on the magnetosphere. Assuming that there is no cross-correlation of GZ-effect at two distinct points in the magnetosphere, we obtain the total conversion by integrating over the entire magnetosphere (from the surface of the compact object to the light cylinder $R_{\rm LC}$). (The cross-correlation corresponds to the induced EM waves at two distinct points affecting each other. This physically corresponds to higher-order effects 
($A_{+} \partial_{z} \tilde{B}_y$) which are neglected.). Thus, the total conversion factor is 

\begin{align}\label{eq:alpha_tot}
\alpha_{\rm{tot} } \simeq
\frac{\pi G |B^{(0)}_y|^2 }{c^2} \left[  \frac{2}{3} \left[\frac{ \xi R_{\rm LC}}{c} \right]^2 \!\! + \frac{ \xi R_{\rm LC}}{ \omega_g c} + \frac{1}{\omega_g^2} \right].
\end{align}
See appendix~\ref{appsec:Integration} for details. It is important to note that $\alpha_{\rm tot}$ is \emph{independent of the amplitude} of GWs. 
To understand the variation of $\alpha_{\rm tot}$ with $\omega_g$, in \ref{fig:Plot-convfact}, we have plotted the conversion factor for %two compact objects--- 
magnetar % {should we avoid discussing about the magnetar?}
%(for which $B^{(0)}_y = 10^{15} \, {\rm G} \, \, , R_{\rm LC} = 10^9~ \, {\rm cm} \,\,  ,  \omega_B = 1 \, {\rm Hz}$) 
and NS/milli-second pulsar.
%(for which $B^{(0)}_y = 10^{10} \, {\rm G} \, \, , R_{\rm LC} = 10^7 {\rm cm} \,\,  , \omega_B = 1 \, {\rm kHz}$)
%~\cite{2019-Cordes.Chatterjee-AnnRevAA,2015-Belvedere.etal-ApJ,2021-White.etal-arXiv,2008-Melikidze.gil-ProcIAU}. 

\begin{figure}[ht]
\centering
%\subfigure[]{%
\label{fig:Term-magn}%
\includegraphics[height=2in]{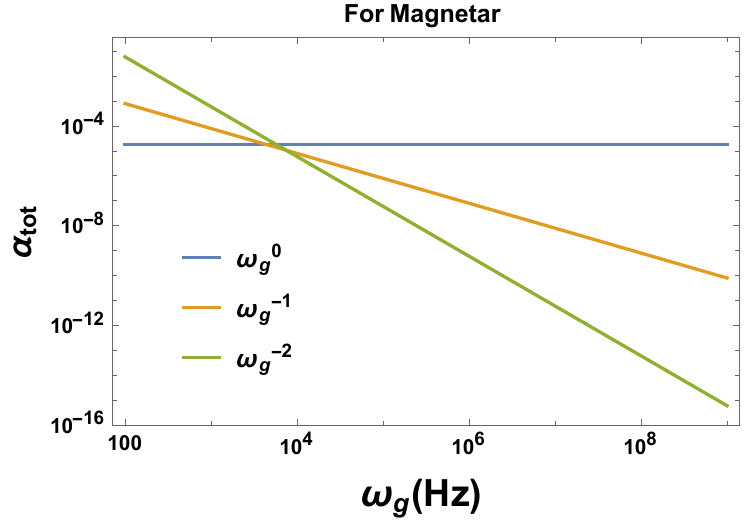} %}%
\quad
%\subfigure[]{%
\label{fig:Terms_ns}%
\includegraphics[height=2in]{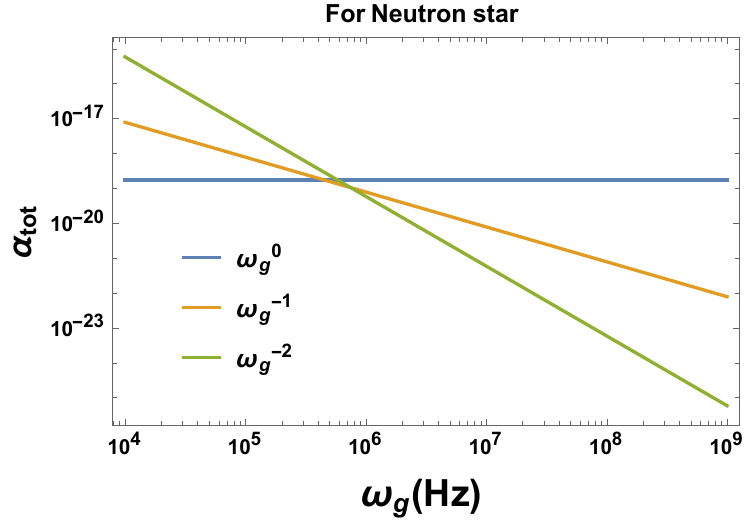} %}%
\caption{Log-Log plot of the three terms in the RHS of Eq.~(\ref{eq:alpha_tot}) versus $\omega_g$ for typical magnetar (top plot) and typical NS (bottom plot). For magnetar, we have set $B^{(0)}_y = 10^{15} {\rm G}, R_{\rm LC} = 10^9~ {\rm cm},  \omega_B = 1 {\rm Hz}$.
For NS/milli-second pulsar, we have set $B^{(0)}_y = 10^{10} {\rm G}, R_{\rm LC} = 10^7 {\rm cm}, \omega_B = 1 {\rm kHz}$.}
\label{fig:Plot-TermsCompare}
\end{figure}

From the plots, we infer that the total conversion factor is almost insensitive at higher frequencies ($> 1~{\rm MHz}$) for both types of compact objects. 
This is because the second and third terms in RHS of Eq.~\eqref{eq:alpha_tot} are inversely proportional to $\omega_g$ and, hence, 
the contribution to $\alpha_{\rm tot}$ is only from the first term. To elaborate, \ref{fig:Plot-TermsCompare} plots each of these terms in Eq.~\eqref{eq:alpha_tot} which 
shows a clear distinction between milli-second Pulsar and magnetar. In both cases, the cross-over occurs below 
$1~{\rm MHz}$. Since we are interested in radio frequency in the GHz range, the total conversion factor \eqref{eq:alpha_tot} is independent of the incoming, coherent GW frequency.

This leads to the important question: What is the efficiency of the GZ-effect near magnetar and NS? Table \eqref{table1} lists the total conversion factor (4th column) for magnetar and NS for different frequencies. We want to emphasize the following points: First, as mentioned earlier, we have assumed the amplitude of the GWs, i.e., $A_+ = A_\times =  1.4\times 10^{-23}$~\cite{2015-Kuroda.etal-IJMPD,2020-Aggarwal.etal-arXiv}. However, as we show below, even with this conservative value, the model explains the peak flux of FRBs. Second, $\alpha_{\rm tot}$ is very high near the magnetar for low-frequency GWs. Specifically, the conversion is $0.06 \% $ and $0.25 \%$ at 1000 Hz and 500 Hz, respectively (for the magnetar $B_y^{(0)} = 10^{15} \, {\rm G}$). For high-frequency GWs, as we show below, even a total conversion factor $(\alpha_{\rm tot})$ of $10^{-19}$ can lead to appreciable energy in the radio frequency (GHz range). 

Further, to compare the generated EM energy density in the entire magnetosphere with the observations, we compute the flux of the induced EM waves (\ref{eq:E_x-Final}, \ref{eq:B_y-Final}) by calculating the Poynting vector~\cite{1979-Rybicki.Lightman-Book}. 
Rewriting Eq.~(\ref{eq:alpha_tot}) as a quadratic equation in $\xi R_{\rm LC}/c$ provides the functional dependence in-terms of $\alpha_{\rm tot}$. This leads to:
%Solving Eq. (\ref{eq:alpha_tot}), we obtain $\xi R_{\rm LC}/c$ in-terms of $\alpha_{\rm tot}$ that leads to:
%

\begin{align}\label{eq:PoyntingVec-in-alpha}
    S_z \simeq \frac{A_+^2 |B_y^{(0)}|^2 c }{256 \pi} 
    \left[ \sqrt{\frac{24  c^2  \omega_g^2 \alpha_{\rm tot}}{ \pi G | B^{(0)}_y |^2 } - 15} - \frac{12 c^2 \omega_g  \omega_B \alpha_{\rm tot}}{\pi G | B^{(0)}_y |^2 } - 1\right] .
\end{align}

\begin{figure}[ht]
\centering
%\subfigure[]{%
\label{fig:MagnetarPV-Jy}%
\includegraphics[height=2in]{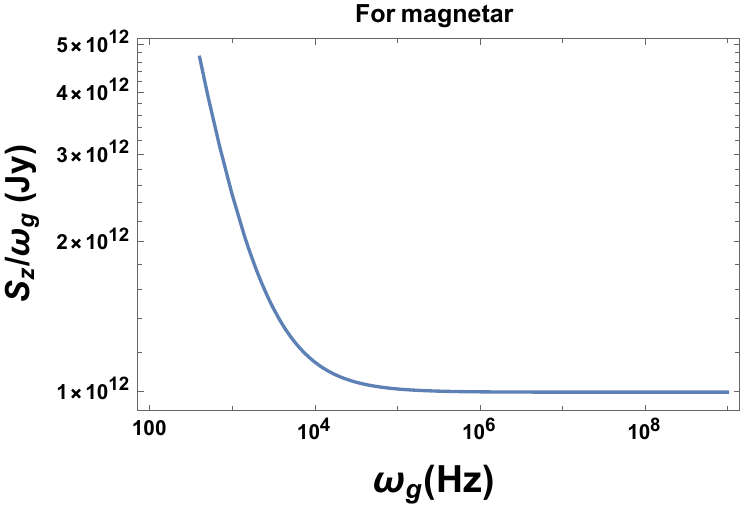} %}%
\quad
%\subfigure[]{%
\label{fig:NeutronStarPV-Jy}%
\includegraphics[height=2in]{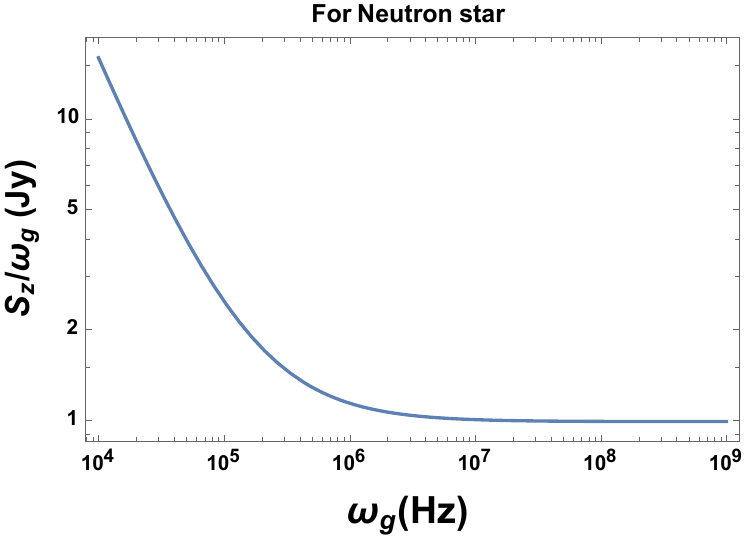} %}%
\caption{
Log-Log plot of $S_z/\omega_g$ versus $\omega_g$ for typical magnetar (top plot) and typical NS (bottom plot). For magnetar, we have set $B^{(0)}_y = 10^{15}  {\rm G}, R_{\rm{LC}} = 10^9~ \, {\rm cm},  \omega_B = 1  \rm{Hz}$. 
For NS/milli-second pulsar, we have set $B^{(0)}_y = 10^{10} \, {\rm G}  , R_{\rm{LC}} = 10^7 {\rm cm} , \omega_B = 1 \, \rm{kHz}$. For both plots we have set  $A_{+} = 1.4\times 10^{-23}$ corresponding to a typical GW source~\cite{2020-Aggarwal.etal-arXiv}. 
%We have used $1 \, \rm{erg \,cm^{-3}} = 10^{23} \rm{Jy \, cm^{-1}  s  Hz }$
}
\label{fig:PoyntVec-Jy}
\end{figure}

Note that $\alpha_{\rm{tot}}$ in RHS of the above expression is a function of $\omega_g$ and the parameters of magnetar/NS. See appendix~\ref{appsec:poyntingVec} for details.
Thus, the Poynting vector of the induced EM waves in the vicinity of magnetar/NS can be obtained by substituting the 
the parameters of magnetar/NS, $\omega_g$ and $A_+$. 
To compare with the peak flux of FRBs, \ref{fig:PoyntVec-Jy} contains the plot of the Poynting vector per unit frequency $(S_z/\omega_g)$ as a function of $\omega_g$ for magnetar and NS. 

%s

The last column in Table \eqref{table1} contains the spectral flux density (Poynting vector per unit frequency) for generic parameter ranges of $R_{\rm LC}$ and $B_{y}^{(0)}$. Given the parameters listed in the table \eqref{table1}, our model predicts a range of spectral flux density that can be as small as $0.1~{\rm Jy}$ (milli-second Pulsar) and can be as large as $10^{11}~{\rm Jy}$ (Magnetar).  We are now in a position to compare the results of the model with radio observations in GHz frequency. 

\section{A plausible explanation for FRBs}

As mentioned earlier, more than 600 (non-repeating) FRBs are reported in various catalogues~\cite{Petroff:2016tcr,2019-Platts.etal-PhyRept, Pastor-Marazuela:2020tii,2021-Rafiei.etal-APJ}
$99\%$ of these FRBs were found to have the following three characteristic features: peak flux ($S_{\nu}$) varies in the range $0.1~{\rm Jy} < S_{\nu} < 700~{\rm Jy}$, the pulse width is less than one second and coherent radiation~\cite{2021-Rafiei.etal-APJ,Petroff:2016tcr}. In Ref.~\cite{2016-Petroff.etal-PASA}, the authors classified the quantities associated with FRBs into two types --- observed and derived. Peak flux and width are observed quantities, while Luminosity and Luminosity distance are derived quantities. The uncertainty in the Dispersion Measure versus redshift relation can have a cascading effect on the derived quantities of the FRBs, leading to uncertainty in the origin of these events~\cite{2019-Pol.etal-arXiv}. Hence, to reduce the systematic bias, we focus on observed Peak flux and estimate the same using the Poynting vector of the emitted EM waves ($S_{z}/\omega_{g}$). 
%{in the units of Jy}. 

%Hence in this work to rewe have focused on the observed the derived quantities, such as the object's Luminosity distance will be directly affected.
%
%It can also be understood that to estimate the redshift of any given FRB, most of the models such as FRUitBat, Pol etal~\cite{2019-Pol.etal-arXiv} use the $DM_{\rm lss}$ ($DM_{\rm lss} = DM_{\rm obs} - DM_{\rm host} - DM_{\rm MW}$) along with the host galaxy identification techniques. These methods have an uncertainty cascading effect on the distance estimation of the FRBs. Consequently, the derived quantities, such as the object's Luminosity {distance} will be directly %\textbf{affect this uncertainty} {affected}.}

\begin{table}[h]
\centering
%\scriptsize
\begin{tabular}{|c|c|c|c|c|c|}
	\hline
	    $R_{\rm LC }$   &  $B_y^{(0)}$    &   $\omega_g $    &  $\alpha_{\rm tot}$  & $\rho_{\rm EM} \,  $ & $\frac{S_z}{\omega_g}$ \\ 
	   (cm)  &  (Gauss)  &  (MHz) &  &  ($\rm{Jy \, cm^{-1} \, s \, Hz }$)  &  (Jy)  \\
	\hline
   $10^9$ &  $10^{15}$  & $1$  & $1.74 %246 
   \times 10^{-5}$ & $9.3 \times 10^{10} $ & $9.95\times 10^{11}$  \\ 
 \hline
   $10^9$  &  $10^{12}$  & $500$  &  $ 1.72 \times 10^{-11}$  & $2.3 \times 10^{10}  $ & $9.94\times 10^{5}$  \\ 
 \hline
     $10^8$  &  $10^{11}$  & $1400$  &  $ 1.72 \times 10^{-15}$  & $1.8 \times 10^7  $ & $ 961.57 $  \\ 
 \hline
   $10^7$  &  $10^{10}$  & $1400$  &  $1.72 \times 10^{-19}$  & $1.8 \times 10^{3} $ & $0.99$  \\ 
 \hline
   $10^8$  &  $10^{9}$  & $1400$  &  $1.72 \times 10^{-19}$ &  $1.8 \times 10^{3} $ & $0.09$  \\ 
 \hline
\end{tabular}
\caption{The table contains numerical values of the total conversion factor ($\alpha_{\rm tot}$), energy density of EM waves ($\rho_{\rm EM}$), and spectral flux density (Poynting vector per unit frequency). The first two rows are for a typical Magnetar and the last three rows are for a typical NS.  We have set  
%$G = 6.67 \times 10^{-8}~\rm{dyne \, cm^2 gm^{-2} }, c = 3 \times 10^{10} \rm{cm \,\, s^{-1}}$, \, 
$A_{+} = 1.4\times 10^{-23}$ corresponding to a typical GW source~\cite{2020-Aggarwal.etal-arXiv}. 
%We have used $1 \, \rm{erg \,cm^{-3}} = 10^{23} \rm{Jy \, cm^{-1} \, s \, Hz }$.
}
\label{table1}
\end{table}

As mentioned earlier, $R_{\rm LC}$ of a 
typical NS is 
$\sim 10^{7}-10^{9}~\rm{cm}$ and it takes less than one second for the GWs to pass through the entire magnetosphere. This directly implies that the induced EM waves due to GZ-effect will appear as a burst lasting for less than one second. 
Thus, our model provides a natural explanation for FRBs lasting less than a second. 
Further, we see from the last column of Table \eqref{table1} that our model predicts the burst of EM wave with the flux $< 1000~\rm{Jy}$. Thus, our model naturally explains the observed peak flux and pulse width of $99\%$ of the reported FRBs.

Having estimated the peak flux and pulse width for our model, we can now calculate other quantities associated with FRBs. In particular, we calculate two such quantities --- Fluence~(${\cal F})$~\cite{2015-Keane.Petroff-MNRAS,2018-Macquart.Ekers-MNRAS} and Isotropic Equivalent Luminosity (IEL)~\cite{2019-Lu.Kumar-MNRAS}. The Fluence is defined as the product of the burst width $\Delta t = (2R_{LC})/c$ and the peak flux ($S_z/\omega_g$)~\cite{2015-Keane.Petroff-MNRAS,2018-Macquart.Ekers-MNRAS}:
\[
{\cal F} = (S_z/\omega_g) \times \Delta t .
\]
Using $S_z$ from the table \ref{table1}, we see that our model predicts the Fluence in the range $1 < {\cal F} ({\rm Jy~ms}) < 10^7$.
We now rewrite the energy density $(\rho_{\rm EM})$ of emitted EM waves in terms of the isotropic equivalent luminosity ($L_{\rm IEL}$). Assuming that the progenitor is at a distance $d$ from the observer, $L_{\rm IEL}$ is given by~\cite{2019-Lu.Kumar-MNRAS}: 
\begin{align}\label{eq:luminosity}
    \rho_{\rm EM} = \frac{L_{\rm IEL}}{4\pi d^2 c}.
\end{align}
The above expression assumes that $d \gg$ {size of the progenitor} is justified for the reported FRBs. Due to the lack of confirmed detection in other wavelengths, the distance of the FRBs is not well constrained~\cite{2022-Zhang-arXiv}. The dispersion measure of the FRB gives one indirect estimation \cite{2013-Thornton.etal-Science,2016-Ravi.etal-Science}. It is expected that the distance of the FRBs is in the range $[10~{\rm kpc} - 1~{\rm Gpc}]$.  
%${\rm IEL} [10^{39}- ]$

From the FRB catalogs~\cite{Petroff:2016tcr,2019-Platts.etal-PhyRept, Pastor-Marazuela:2020tii,2021-Rafiei.etal-APJ}, we see that FRBs have a peak flux of around $100 ~{\rm Jy}$. For our model, this corresponds to the energy density of the emitted EM waves to be $\sim 9.2 \times 10^5~{\rm Jy \, cm}^{-1}{\rm s\, Hz}$. Taking the above distance range ($[10~{\rm kpc} - 1~{\rm Gpc}]$), the above energy density translates to $L_{\rm IEL}$ in the range 
$[3.1 \times 10^{39} - 10^{49}]~{\rm erg/s}$. Thus, our model can explain the inferred luminosity of FRBs~\cite{2017-Law.etal-ApJ}.

Given this, we can now identify the possible progenitors of FRBs. To identify this, we consider two FRBs --- FRB120127~\cite{FRB120127} and FRB011025~\cite{FRB011025}. These two sources represent a typical FRB in the catalogs. The observations of these two FRBs at $1.5 ~\rm{and} ~1.3~\rm{GHz}$, show a typical peak flux of $0.62^{+0.35}_{-0.10}~\rm {Jy}$ and $0.54^{+0.11}_{-0.07}~\rm {Jy}$, respectively.
From the last row of the Table \eqref{table1}, we see that our model predicts the progenitor should be a millisecond pulsar with an effective magnetic field strength of $10^{10}~{\rm G}$ and $R_{\rm{LC}} \sim 10^{7}~\rm {cm}$. Fluence for these two FRBs is around $0.66~\rm{Jy \, ms}$ and isotropic equivalent luminosity $(L_{\rm IEL})$ at a distance ${d = 10~{\rm Mpc}}$ is $3\times 10^{42} \rm{erg \, s^{-1}}$.

FRB 200428 is confirmed to be a galactic FRB~\cite{Bochenek2020Natur}. Recent observations indicate that the magnetar SGR 1935+2154 residing in the Milky Way is associated with the FRB 200428 with a fluence of $>1.5 \times 10^6~{\rm Jy ~ms}$ in the $1.28 - 1.4$ GHz band detected by the STARE2 radio array~\cite{Bochenek2020Natur,2020Natur.587...54C}. It is reported that this magnetar has a surface dipole magnetic field of $\sim 2.2 \times 10^{14}$ G, which can be deduced from the period slow-down rate of $3.24~{\rm s}$. Substituting the constant magnetic field approximation $B_y(0) \sim 5 \times 10^{11}~{\rm G}$ with $R_{\rm LC} \sim 10^9~{\rm cm}$ and $A_+ \sim 10^{-26}$ in Eq.\eqref{eq:PoyntingVec-in-alpha}, our model predicts peak flux $\sim 4.8 \times 10^4~{\rm Jy}$ and Fluence to be $1.65 \times 10^7 {\rm Jy~ms}$. Thus our model predicts isotropic equivalent luminosity at $~10 \ {\rm kpc}$ to be $L_{\rm IEL} \sim 1.4\times 10^{43} \rm{erg \, s^{-1}}$.
Consequently, our model has the potential to explain these observations and play a crucial role in any future FRB and progenitor association (if any).

We can do a similar analysis for 
all the FRBs in the catalog with a pulse width of less than one second~\cite{2021-Rafiei.etal-APJ,Petroff:2016tcr}. Our model predicts that the progenitor can be a NS with an effective magnetic field strength in the range $10^{9} - 10^{11}~{\rm G}$ and rotation frequency $1 < \omega_B < 1000$. Our model can provide an explanation for the observed peak flux for a class of \emph{non-repeating} FRBs and predicts FRBs with extremely low and high peak flux. Note that our model is not sensitive to the galactic environment. Such future detections will further strengthen the model. 

As mentioned earlier, the model uses the GZ mechanism to explain the origin of FRBs. GZ mechanism involves the transfer of energy from the incoming GWs to emitted EM radiation in the presence of the background magnetic field. Although our model falls in the first category where FRBs are generated from the magnetar/NS, electromagnetic radiation is generated when GWs pass through the magnetosphere. In other words, the background magnetic field acts as a catalyst in this mechanism. This can be verified by evaluating the ratio of the energy density of the outgoing EM waves ($\rho_{\rm EM}$) with the energy density of the background magnetic field ($\rho_{B}$). For a magnetic field strength of $10^{10} {\rm G}$, this ratio is minuscule:
\[
\frac{\rho_{\rm EM}}{\rho_B} \sim 2.2 \times 10^{-39} .
\]
Thus, the proposed mechanism is not expensive  compared to the magnetic energy present in the magnetosphere.

\section{Discussions}

Our mechanism requires that the emitted GWs pass through the NS, resulting in non-repeating FRBs along the line of sight of the observer. In other words, we have assumed that the EM emission due to the GZ-mechanism in the pulsar magnetosphere is directional dependent. The intervening medium does not impact the radiation from this mechanism. More importantly, the mechanism requires that the emitted GWs pass through the NS, resulting in FRBs along the line of sight of the observer. The probability of this event is a product of the probability that the GW passes through the NS and the probability that the emitted EM is along the line of sight of the observer. Hence, the probability of seeing such an event is not very high. More importantly, any existing NS/Magnetar in any galaxy can produce FRB.

%Our mechanism requires that the emitted GWs pass through the NS, resulting in \textcolor{blue}{non-repeating} FRBs along the line of sight of the observer. The probability of this event is a product of the probability that the GW passes through the NS and the probability that the emitted EM is along the line of sight of the observer. Hence, the probability of seeing such an event is not very high.

Further, it is estimated that around $10^8 - 10^9$ NSs are present in the Milky Way galaxy, roughly $1\%$ of the total number of stars in the galaxy~\cite{2006-Diehl.etal-Nature,2010-Sartore.etal-AandA}. Also, it is estimated that the magnetar formation rate is approximately $1 - 10$ percent of all pulsars~\cite{2015-Gullon.etal-MNRAS,2019-Beniamini.etal-MNRAS}.   {
One of the basic assumption of our model is that, given GW signal in GHz frequency, all NSs can act as source of FRBs at all times. This assumption translates to the fact that the maximum FRB events per day can be $10^8 - 10^9$.  However, the observed FRB rate is $10^{3}$ for the entire sky per day. This can be attributed to the fact that the probability of this event is a product of the probability that the GW passes through the NS times the probability that the emitted EM is along the line of sight of the observer.}
%Given the observed FRBs rate of $10^{3}$ for the entire sky per day, the number of NSs in our galaxy is six orders higher than the FRB rate. 
%
 {Consequently, our model predicts the observed FRB event rate and the coherent nature of FRBs~\cite{2013-Katz-PRD}. Since, GWs can be generated up to 14~GHz~\cite{2022-Aggarwal.etal-PRL,2020-Ito.etal-EPJC}, our model naturally does not have high-energy counterpart.} 

As mentioned in the Introduction, the co-planar property of the EM wave emitted due to the GZ mechanism is due to the coherent nature of the incoming GWs~\cite{2007-Hendry.Woan-AG}. 
Due to the co-planar property, it can maintain the flux~\cite{2022-Lieu.etal-CQG}. Hence, the three key features of FRBs are naturally explained in our model.

A variety of processes generate GWs in a broad range of frequencies~\cite{1974-Hawking.Carr-MNRAS,2009PhRvL.103k1303A,2015-Kuroda.etal-IJMPD,2019-Ejlli.etal-EPJC,2020-Aggarwal.etal-arXiv,2021-Pustovoit.etal-JOP}. However, it is possible to detect these waves only in a limited range. The proposed model provides an indirect mechanism to detect GWs at high frequencies. Interestingly, our model also provides a way to test modified theories of gravity. In this chapter, we have focused on GWs in general relativity. Certain modified theories like Chern-Simons gravity lead to birefringence~\cite{2006-Alexander.Peskin.Jabbari-PRL} which can explain the polarized nature of FRBs. However, exact rotating solutions in these theories are unknown and require sophisticated numerics. Thus, the high-frequency GWs will provide a unique view of the Universe --- it's birth and evolution.

\chapter{Conclusions and future outlook}
\label{ch:conclusion}

Due to technological advancements and the availability of vast amounts of data from various observations, we are currently in an era of precision cosmology. Although precision cosmology enhances our understanding of the Universe, it also challenges established theoretical ideas, giving rise to intriguing new questions. To seek answers to these questions, new ideas are introduced, often requiring going beyond the standard model of cosmology. These issues are commonly referred to as \emph{open problems} (see sec~\ref{sec:open_problem}). Finding solutions to these problems requires either extending the standard model or considering the fine-tuning of model parameters. In this thesis, we looked at three problems
\begin{enumerate}
    \item What is the origin of large-scale magnetic fields? 
    \item  Why there is more matter than antimatter and how to create the observed amount of matter-antimatter asymmetry in the Universe? 
    \item  What is the origin of Fast radio bursts in the sky?
\end{enumerate}
and proposed novel mechanisms to resolve them. 
 %The primary objective is to comprehensively examine these problems and propose novel mechanisms to resolve them. 
%The thesis also gives a novel picture of how we can use magnetic fields as a tool to probe the Universe. 

%In this thesis, we discussed the magnetic fields, present in the Universe at all length scales, starting from small-scale but very strong magnetic fields of magnetars to large-scale weak magnetic fields in galaxies. Interestingly, they provide novel probes to the Universe at relevant scales.
Chapter~\ref{ch:intro} discussed these open problems in detail. The common theme of this thesis is the generation of cosmological magnetic fields and using these fields as a tool to probe the Universe at very large (scales of galaxies) and small scales (scales of compact objects). 
%This has been possible due to the omnipresence of the magnetic fields in the Universe.
As discussed in chapter~\ref{ch:intro}, the magnetic fields are present at all length scales, from small-scale but very strong magnetic fields of magnetars to large-scale weak magnetic fields in galaxies. We also saw that understanding the origin and evolution of the magnetic fields on different epochs provides novel probes to the Universe at relevant scales. 
Unfortunately, the origin of magnetic fields on large scales is still unresolved because it is difficult to generate sufficiently strong primordial seed magnetic fields that are coherent over large scales. This requires physics beyond the standard models of cosmology. Although inflation provides a causal mechanism for generating primordial density perturbations, it can not generate the appreciable primordial magnetic field in the early Universe. It has been argued that conformal invariance breaking is necessary to generate sufficiently strong primordial magnetic fields during inflation. Various mechanisms and models have been proposed in the literature to break the conformal invariance of the electromagnetic fields; however, they mostly involve fine-tuning the coupling parameters. In this thesis, we have proposed novel mechanisms to generate the primordial magnetic field during inflation and looked at the effect on matter-antimatter asymmetry. We have also constructed a generic framework---effective field theory of magnetogenesis which can reproduce a large class of magnetogenesis models proposed in the literature.

In chapter~\ref{ch:GalScalElect}, we constructed Galileon scalar electrodynamics action, which preserves Galilean symmetry in field space and local gauge invariance. Due to the complex scalar field, the number of Galilean symmetry invariant terms is reduced. In flat space-time, we have explicitly shown that the equations of motion are second order. In curved space-time, due to the non-commutative nature of the covariant derivatives, the minimal coupling of the matter and gravity term leads to higher derivatives in the equation of motion. We introduced non-minimal coupling terms to the Galileon field that makes the equations of motion second-order in an arbitrary curved space-time. 
As an application of the model, we considered the case when the Galileon scalar electrodynamics dominated the early Universe. We have shown that the model leads to inflation for a range of parameters. Since the model breaks conformal invariance, one possible application of the model is to study the phenomenological consequences of generating primordial magnetic fields during inflation. 
%The non-minimal coupling term is different compared that the one used in the literature for the real scalar field~\cite{2009-Deffayet.etal-PRD}. However, in the real scalar field limit, the complex scalar Galileon action is identical to the real scalar Galileon action obtained by Deffayet et al~\cite{2009-Deffayet.etal-PRD}. 

In chapter~\ref{ch:helical-PMF}, we have proposed a viable scenario for generating helical magnetic fields during inflation, which does not require coupling the electromagnetic field to the scalar field. We have shown that the model leads to the generation of the helical fields due to the coupling of the electromagnetic fields with the dual of Riemann tensor. To our knowledge, Riemann tensor coupling has not been discussed in the literature to generate primordial helical magnetic fields. We have shown the key features of the model: First, it does not require the coupling of the electromagnetic field with the scalar field. Hence, there are no extra degrees of freedom, and it will not lead to a strong-coupling problem. Second, the conformal invariance is broken due to the coupling to the Riemann tensor. Since the curvature is large in the early Universe, the coupling term will introduce non-trivial corrections to the electromagnetic action. However, the modification term will not contribute at late-times, and the theory is identical to standard electrodynamics. Third, the power spectrum of the helical magnetic fields generated has a slight red-tilt for slow-roll inflation. This differs from the scalar field coupled models where the power-spectrum has a blue-tilt. We have also identified the reason for this difference. Fourth, our model is free from backreaction for a range of scale-factor during inflation.

After discussing the generation of the helical magnetic field, in chapter~\ref{ch:PMF_Baryo}, we have studied a viable baryogenesis scenario in the early Universe that does not require any extension to the Standard Model of particle physics. The crucial ingredient is the generation of primordial helical magnetic fields due to Riemann coupling proposed in chapter~\ref{ch:helical-PMF}. The advantage of the primordial helical magnetic fields is that the non-zero helicity suggests a non-zero contribution in the CP violation term. An interesting feature of the model is the stretching of the primordial helical magnetic fields to super-horizon scales during inflation --- the same mechanism that leads to primordial density perturbations. While the helical modes generated around 40 - 60 e-foldings before the end of inflation lead to the observed large-scale magnetic fields, the helical modes that renter the horizon very early (at the beginning of the  radiation-dominated epoch) lead to the baryon asymmetry. Also, we have explicitly shown that Davidson's conditions (see Ref.~\cite{1996-Davidson-PLB}) are necessary \emph{but not} sufficient in order to map the symmetries of the universe in the presence of the magnetic field with Sakharov's conditions for baryogenesis. The key missing ingredient is the requirement of \emph{primordial helical magnetic fields}.
We have shown that the BAU parameter predicted by our model is independent of any specific inflation model and reheating dynamics; however, it depends on the scale at which inflation ends and reheating temperature. 
%The BAU parameter \eqref{eq:baryon_Asym-M} obtained in our model is inversely proportional to reheating temperature. 
Assuming the exit of inflation at $10^{14}$~GeV, for the observed amount of baryon asymmetry $\eta_B \sim 10^{-10}$, we obtained  that the reheating temperature should be in the range $10^{12} - 10^{14}$~GeV, which is consistent with the constraints on the reheating temperature \cite{2006-Bassett.Tsujikawa.Wands-RevModPhys,1984-Ellis.Kim.Nanopoulos-PLB,1999-Benakli.Davidson-PRD}. 
%This means that our model \emph{does not} prefer a very low-energy reheating temperature~\cite{1999-Benakli.Davidson-PRD}.  

In chapter~\ref{ch:EFTmagnetogenesis}, using effective field theory, we showed that the generation of primordial magnetic fields requires two
necessary conditions --- conformal invariance breaking and causal propagation. We explicitly showed that the metric vector perturbations do not have temporal evolution and do not affect the dynamics of the gauge field. Hence, the gauge field is the relevant gauge-invariant variable for the EFT. We obtained the general second-order EFT action by demanding that the EFT of magnetogenesis breaks conformal invariance; however, it satisfies local Lorentz invariance and gauge invariance. This action depends on two expansion scalar functions --- $f_1(H, a, \Lambda)$ and $f_2(H, a, \Lambda)$. We then showed that the EFT action could reproduce a large class of the known magnetogenesis models. For most models, it is sufficient to consider up to the second order in the expansion parameters. We also compared our approach to previous approaches and showed the advantages of our approach. 
We considered a specific model of inflationary magnetogensis where the vector Galileon breaks the conformal invariance. We extensively studied the magnetic field generation during inflation, considering the total action, i.e., standard electrodynamics action with vector Galileon modification terms. Due to the absence of coupling between the scalar and electromagnetic fields, the model does not lead to a strong coupling problem~\cite{2017-Debottam.Shankaranarayanan-JCAP}; hence, the model does not require fine-tuning, which is an interesting feature of our model.
Furthermore, the evolution of vector modes is frozen on the super-horizon scale. We showed that the model predicts the present-day magnetic field of strength $B_0 \sim 10^{-45} {\rm G}$ on the cosmological (Mpc) scale, which is several orders higher compared to the standard electromagnetic action, and there is no back-reaction problem. However, this comes with the price --- super-luminal electromagnetic perturbations. In other words, the model generates large magnetic fields and does not violate any other conditions; it leads to super-luminal perturbations. Thus, our EFT framework suggests that to construct a field theoretical consistent model of inflationary magnetogenesis, we need two necessary conditions--- breaking of conformal invariance and causal propagation of perturbations. 
%By truncating the expansion scalar functions to $\Lambda^{-2}$, we derived the EOM of the gauge field and obtained the power spectrum in the slow-roll inflation scenario. Moreover, we explicitly showed that the magnetic power spectrum could be large, irrespective of the value of the sound speed of the perturbations ($c_A$). In other words, we can have large implications even for super-luminal fluctuations. Hence, our analysis shows that conformally breaking is necessary but insufficient for magnetic field generation in the early Universe. We thus need another physical condition --- the modes should be sub-luminal.

In chapter~\ref{ch:GZeffect}, we proposed a novel mechanism that explains the Fast radio bursts that utilize the strong magnetic field of a compact object. The energy conversion from GWs to electromagnetic waves occurs when GWs pass through the magnetosphere of compact objects (neutron stars or magnetars) due to the Gertsenshtein-Zel$'$dovich effect.
This conversion produces bursts of electromagnetic waves in the GHz range, leading to FRBs. The mechanism requires that the incoming GWs pass through the neutron star with a magnetic field transverse to the direction of propagation of GWs, resulting in FRBs along the line of sight of the observer. The probability of this event is a product of the probability that the GW passes through the NS and the probability that the emitted EM is along the line of sight of the observer. Hence, the probability of seeing such an event is not very high. 
We showed that our model explains why the FRB rates are low and why certain FRBs are random. Furthermore, our model predicts that the progenitor of these events is NS in any galaxy. Moreover, importantly, our model predicts that FRBs can also be extragalactic in origin. We have shown that the mechanism presented in chapter~\ref{ch:GZeffect} can explain three key features of FRBs---peak flux, pulse width, and coherent nature of radiation. We have also discussed the implications of the mechanism in providing the indirect detection of high-frequency gravitational waves.
%Further, it is estimated that around $10^8 - 10^9$ NSs are present in the Milky Way galaxy, roughly $1\%$ of the total number of stars in the galaxy~\cite{2006-Diehl.etal-Nature,2010-Sartore.etal-AandA}. Also, it is estimated that the magnetar formation rate is approximately $1 - 10$ percent of all pulsars~\cite{2015-Gullon.etal-MNRAS,2019-Beniamini.etal-MNRAS}.  Given the observed FRBs rate of $10^{3}$ for the entire sky per day, the number of NSs in our galaxy is six orders higher than the FRB rate. 
%Also, there is a huge international effort with existing telescopes (uGMRT, VLA, ATCA) and upcoming facilities like SKA to search for NS/pulsars outside the Milky Way. The detection of extragalactic milli-second Pulsar will validate our model.

\section{Future outlook}

As discussed in chapters~\ref{ch:helical-PMF} and \ref{ch:PMF_Baryo}, the Riemann coupling to the electromagnetic fields generates the primordial helical magnetic fields and has interesting consequences for understanding the creation of baryon asymmetry in the universe. We saw that the perturbations equation~\eqref{eq:equation_of_motion} contains second-order derivatives of $\mathcal{H}$. Since ${\cal H}$, and ${\cal H}''$ are different for inflation and bounce models~\cite{2020-Nandi-PLB}, the helical modes may provide signatures to distinguish the two paradigms. Also, in chapter~\ref{ch:PMF_Baryo}, we have seen that small-scale modes of the helical fields which exit the horizon near the end of inflation lead to baryogenesis. However, %we have only focused on the electromagnetic fields and the effects of the helical fields on baryogenesis. 
it will be interesting to extend the analysis to Gluons and study the effects on the asymmetry generated in quarks and the baryons. This could be done by considering the coupling the similar mass dimension 6 operator (Riemann coupling) to the non-abelian SU(2) gauge fields. Another interestingly aspect of the small scales modes discussed in chapter~\ref{ch:PMF_Baryo} would be to see if these modes have sufficiently large power, because they could also provide a mechanism for the formation of primordial black holes in the early Universe. 

Modifications of general relativity provide an alternative explanation for cosmological inflation~\cite{2022-Shankaranarayanan-Joseph-GRG}. 
Modified gravity theories have extra degrees of freedom that might have interesting physical consequences in the early Universe. For instance, Stelle gravity~\cite{1978-Stelle-GRG} contains massive tensor modes, and these modes carry more energy than the scalar modes in $f(R)$ gravity models~\cite{Chowdhury:2022ktf}. While the analysis in chapter~\ref{ch:EFTmagnetogenesis} can be extended to $f(R)$ gravity models, it is not straightforward to extend to Stelle gravity. Investigating the EFT of inflation and magnetogenesis in these modified gravity models would be interesting to explore. Also, it will be interesting to extend the EFT of magnetogenesis to bouncing models. The EFT action constructed in chapter~\ref{ch:EFTmagnetogenesis} can also be extended to include the parity-breaking term, leading to the helical magnetic field. In chapter~\ref{ch:EFTmagnetogenesis}, we have seen that causal propagation must be considered while constructing the inflationary magnetogenesis models; exploring this issue in other magnetogenesis scenarios would be interesting. In chapter~\ref{ch:GZeffect}, we have explained the properties of the non-repeating FRBs and focussed on the transient incoming GWs. It would be interesting to see if the Gertsenshtein-Zel'dovich mechanism can also explain the repeating nature of FRBs and also if we can incorporate the incoming continuous GWs to explain the FRBs.

 %Use FileName if the file is stored in the same folder as Thesis.tex. 

%=====================================================================
% APPENDIX
%  Appendices, if any, must precede the cited literatures.
%  Appendices shall be numbered in Roman Capitals (e.g. Appendix IV)

\begin{appendices}
\chapter{Galileon scalar electrodynamics: Details} 
\label{ap:GalScED}

In this Appendix, we provide more details on the results from chapter \ref{ch:GalScalElect}.

\section{Conformal transformation in 4-D curved space-time}\label{CT}
	Conformal transformation is a local change of scale, which is given by scaling the metric by a smooth non-vanishing function of space-time $\Omega(x^{\mu})$~\cite{Book-General-Relativity-Wald,Book-General-Relativity-Carroll}:
	\begin{equation}\label{CT1}
	\tilde{g}_{\mu\nu} = \Omega^2(x)g_{\mu\nu}
	\end{equation}
	Let us see how this transformation affects the electromagnetic field tensor $F^{\mu\nu}$ in arbitrary curved space-time.
	\begin{equation}
	F_{\mu\nu} = \nabla_{\mu}A_{\nu} - \nabla_{\nu}A_{\mu} = \partial_{\mu}A_{\nu} -  \Gamma^{\lambda}_{\mu\nu}A_{\lambda} - \partial_{\nu}A_{\mu} +  \Gamma^{\lambda}_{\nu\mu}A_{\lambda} = \partial_{\mu}A_{\nu} - \partial_{\nu}A_{\mu}
	\end{equation}
	Therefore, $F_{\mu\nu}$ is invariant under the conformal transformation (\ref{CT1}), and is not affected by geometry of the space. \\
	Now we will see how other geometrical quantities which are derived from the metric, change under this transformation. The Christoffel symbol changes as 
	\begin{equation}
	\tilde{\Gamma}^{\lambda}_{\mu\nu} = \Gamma^{\lambda}_{\mu\nu} + C^{\lambda}_{\mu\nu}
	\end{equation}
	where $C^{\lambda}_{\mu\nu} = \Omega^{-1}\left( \delta^{\lambda}_{\mu}\nabla_{\nu}\Omega + \delta^{\lambda}_{\nu}\nabla_{\mu}\Omega - g_{\mu\nu}g^{\rho\lambda}\nabla_{\rho}\Omega \right)$\\
	%\vspace{0.5}
Putting this value of Christoffel connetion in the Riemann tensor in the conformal frame (Jordan frame) we get,
	\begin{equation}
	\tilde{R}^{\lambda}_{\sigma\mu\nu} = R^{\lambda}_{\sigma\mu\nu} + \nabla_{\mu}C^{\lambda}_{\nu\sigma} - \nabla_{\nu}C^{\lambda}_{\mu\sigma} + C^{\lambda}_{\mu\rho}C^{\rho}_{\nu\sigma} - C^{\lambda}_{\nu\rho}C^{\rho}_{\mu\sigma}
	\end{equation}
Therefore, Riemann tensor changes under the conformal transformaton, contracting with the metric gives the Ricci tensor
	\begin{equation}
	\tilde{R}_{\mu\nu} = R_{\mu\nu} - [2\delta^{\alpha}_{\mu}\delta^{\beta}_{\nu} + g_{\mu\nu}g^{\alpha\beta}]\Omega^{-1}(\nabla_{\alpha}\nabla_{\beta}\Omega)+[4\delta^{\alpha}_{\mu}\delta^{\beta}_{\nu} - g_{\mu\nu}g^{\alpha\beta}]\Omega^{-2}(\nabla_{\alpha}\Omega)(\nabla_{\beta}\Omega)
	\end{equation}
	and Ricci curvature scalar as
	\begin{equation}
	\tilde{R} = \Omega^{-2}R - -6g^{\alpha\beta}\Omega^{-3}(\nabla_{\alpha}\nabla_{\beta}\Omega) 
	\end{equation}
	Hence, we can see that Christoffel connections, Riemann tensor, Ricci tensor and Ricci scalar transform according to the relation given above. So, introducing these quantities in the action break the conformal invariance of the action. It is important to note that for higher derivative field theories ( e.g., Galileon fields) the conformal invariance is explicitly broken as 
	\begin{equation}
	\tilde{\nabla}_{\mu}\phi = \nabla_{\mu}\phi = \partial_{\mu}\phi
	\end{equation}
	\begin{align}
	\tilde{\nabla}_{\mu}\tilde{\nabla}_{\nu}\phi = \nabla_{\mu}\nabla_{\nu}\phi - (\;\delta^{\alpha}_{\mu}\;\delta^{\beta}_{\nu}\; + \;\delta^{\beta}_{\mu}\;\delta^{\alpha}_{\nu}\;)\;\Omega^{-1}\;(\nabla_{\alpha}\Omega)(\nabla_{\beta}\Omega)
		\end{align}
		where $\phi$ is any real scalar field.
\section{Fixing the coefficients of $\mathcal{L}_4$}\label{Generic}

In this Appendix, we fix all the coefficients in the Lagrangian \eqref{L4}. Demanding that the Lagrangian is invariant under the Galileon transformation \eqref{eq:GalileanShift} leads to the 
following constraint:
\begin{align}\label{LCGGT}
2A_1 \,\partial_{\mu} \partial^{\mu}\pi^* \partial_{\nu} \partial^{\nu}\pi^* \partial_{\alpha}\pi a^{\alpha} 
+ A_2 \,\partial_{\mu}\partial^{\mu}\pi^* \partial_{\nu} \partial^{\nu}\pi \partial_{\alpha}\pi^* a^{\alpha} 
\nonumber\\
+  B_1\,(\partial_{\mu}\partial^{\mu}\pi^* \partial^{\nu} \partial^{\alpha}\pi \partial_{\nu}\pi^* a_{\alpha} + \partial_{\mu}\partial^{\mu}\pi^* \partial^{\nu} \partial^{\alpha}\pi \partial_{\alpha}\pi {a_{\nu}}^*) 
+ 2B_2\,\partial_{\mu}\partial^{\mu}\pi^* \partial^{\nu} \partial^{\alpha}\pi^* \partial_{\nu}\pi a_{\alpha} 
\nonumber\\
+ 2 C_1\,\partial_{\mu}\partial_{\nu} \pi^*\partial^{\mu}\partial^{\nu}\pi^* \partial^{\alpha}\pi a_{\alpha} 
+ C_2\, \partial_{\mu}\partial_{\nu}\pi^* \partial^{\mu}\partial^{\nu}\pi \partial^{\alpha}\pi^* a_{\alpha} \nonumber\\
+ 2 D_1\, \partial^{\mu}\partial^{\nu}\pi^* \partial_{\nu}\partial_{\alpha}\pi^* \partial_{\mu}\pi a^{\alpha}
+ D_2 \,\partial^{\mu}\partial^{\nu}\pi^* \partial_{\nu}\partial_{\alpha}\pi \partial_{\mu}\pi^* a^{\alpha}
+ D_3\,\partial^{\mu}\partial^{\nu}\pi \partial_{\nu}\partial_{\alpha}\pi^* \partial_{\mu}\pi^* a^{\alpha} +c.c. = 0 \, .
\end{align}    
Here, c.c. denotes the complex conjugate, we have considered only first order terms in $a^{\mu}$ and have ignored the quadratic terms of the infinitesimal shift parameters. All the first order terms in Eq.~(\ref{LCGGT}) either do not contribute or contribute total derivative terms. After integrating by parts (and ignoring the total derivative terms as they do not contribute to the action) and collecting similar terms, we get
\begin{align}\label{a1}
\left[(2A_1 - B_1)\,(\Box \pi^*)^2\partial_{\alpha}\pi + (A_2 -2 B_2 - B_1)\,\Box \pi^*\Box \pi \partial_{\alpha}\pi^*
- ( B_1 +2C_1)\,\partial^{\nu}\Box\pi^*\partial_{\nu}\pi^* \partial_{\alpha}\pi\right. \nonumber\\
\left. - (B_1 + D_3)\,\partial^{\mu}\Box \pi \partial_{\mu}\pi^*\partial_{\alpha}\pi^*+ 
(C_2 -2 D_1 -D_3)\, \partial_{\mu}\partial_{\nu}\pi^* \partial^{\mu}\partial^{\nu}\pi \partial_{\alpha}\pi^* \right. \nonumber\\
\left.
- (2D_1 + 2B_2)\,\partial^{\mu}\Box \pi^*\partial_{\mu}\pi \partial_{\alpha}\pi^* - (2C_1 - D_2)\,\partial_{\mu}\partial_{\nu}\pi^* \partial^{\nu}\partial_{\alpha}\pi \partial^{\mu}\pi^*\right]a^{\alpha} = 0
\end{align}
Only way to satisfy the above equation for an arbitrary $\pi$ and $\pi^*$ is that all the coefficients vanish. This leads to the following relations:
\begin{align}\label{Coeff}
B_1 &= 2 A_1;~C_1 = - A_1 \nonumber\\
B_2 &= -A_1 + \frac{A_2}{2} = -E\nonumber\\
C_2 &= - A_2;~D_2 = -2 A_1 \nonumber\\
D_1 &= A_1 - \frac{A_2}{2} = E\nonumber\\
D_3 &= -2 A_1 \, .
\end{align}
It is straightforward to see from the above expressions that all 
the parameters can be written in terms of two  parameters $A_1$ and $A_2$. Substituting the coefficients (\ref{Coeff}) in Eq.~(\ref{L4}) leads to the following generic complex scalar Lagrangian $\mathcal{L}_{4}$;
 \begin{align}\label{LCGG}
{\mathcal{L}}_{4}&{} = A_1 [\,\, (\,\,(\Box \pi)^2 \partial_{\alpha}\pi^* \partial^{\alpha}\pi^* +  2 \Box \pi\partial_{\nu}\pi\partial^{\nu} \partial^{\alpha}\pi^* \partial_{\alpha}\pi^* - \partial_{\mu}\partial_{\nu}\pi \partial^{\mu}\partial^{\nu}\pi \partial^{\alpha}\pi^* \partial_{\alpha}\pi^* + \rm{c.c.}\,\,) \nonumber \\&{}-2 \partial_{\mu}\pi^*\partial^{\mu}\partial^{\nu}\pi \partial_{\nu}\partial_{\alpha}\pi^* \partial^{\alpha}\pi ]
 + A_2 [\Box \pi^* \Box \pi \partial_{\alpha}\pi^* \partial^{\alpha}\pi  - \partial_{\mu}\partial_{\nu}\pi^*\partial^{\mu}\partial^{\nu}\pi \partial^{\alpha}\pi^* \partial_{\alpha}\pi ] \nonumber \\&{}
+ E [\partial_{\mu}\pi^*\partial^{\mu}\partial^{\nu}\pi \partial_{\nu}\partial_{\alpha}\pi \partial^{\alpha}\pi^* -\Box \pi\partial_{\nu}\pi^* \partial^{\nu} \partial^{\alpha}\pi \partial_{\alpha}\pi^* + \rm{c.c.}  ]
\end{align}    
Now, there are only two free parameters $A_1$ and $A_2$ in the Lagrangain.     
The action for the above generic Lagrangian (\ref{LCGG}) is given by,
\begin{equation}\label{S_GCC}
{S}_{4} = \int d^4 x~\mathcal{L}_{4}
\end{equation}

We can identify the generic action (\ref{S_GCC}) as the summation of the following actions:
\begin{equation}
\begin{split}\label{S_i}
S^{(1)}_{4} &= A_1 \int d^4 x \; (\, \partial_{\mu} \partial^{\mu}\pi \partial_{\nu} \partial^{\nu}\pi \partial_{\alpha}\pi^* \partial^{\alpha}\pi^* + \rm{c.c.} \,\,)\\
{S}^{(2)}_{4} &= A_2 \int d^4 x \; \partial_{\mu} \partial^{\mu}\pi^* \partial_{\nu} \partial^{\nu}\pi \partial_{\alpha}\pi^* \partial^{\alpha}\pi\\
{S}^{(3)}_{4} &= 2 A_1 \int d^4 x \; (\, \partial_{\mu}\partial^{\mu}\pi\partial_{\nu}\pi \partial^{\nu} \partial^{\alpha}\pi^* \partial_{\alpha}\pi^* + \rm{c.c.}\,\,)\\
{S}^{(4)}_{4} &= -E \int d^4 x \; (\,\partial_{\mu}\partial^{\mu}\pi\partial_{\nu}\pi^* \partial^{\nu} \partial^{\alpha}\pi \partial_{\alpha}\pi^* + \rm{c.c.} \,\,)\\
{S}^{(5)}_{4} &= -A_1 \int d^4 x \; (\, \partial_{\mu}\partial_{\nu}\pi \partial^{\mu}\partial^{\nu}\pi \partial^{\alpha}\pi^* \partial_{\alpha}\pi^* + \rm{c.c.}\,\,)\\
{S}^{(6)}_{4} &= -A_2 \int d^4 x \; \partial_{\mu}\partial_{\nu}\pi^*\partial^{\mu}\partial^{\nu}\pi \partial^{\alpha}\pi^* \partial_{\alpha}\pi\\
{S}^{(7)}_{4} &= E \int d^4 x \; (\, \partial_{\mu}\pi\partial^{\mu}\partial^{\nu}\pi^* \partial_{\nu}\partial_{\alpha}\pi^* \partial^{\alpha}\pi + \rm{c.c.}\,\,)\\
{S}^{(8)}_{4} &=  -2 A_1 \int d^4 x \; \partial_{\mu}\pi^*\partial^{\mu}\partial^{\nu}\pi^* \partial_{\nu}\partial_{\alpha}\pi \partial^{\alpha}\pi\\
{S}^{(9)}_{4} &= -2 A_1\int d^4 x \; \partial_{\mu}\pi^*\partial^{\mu}\partial^{\nu}\pi \partial_{\nu}\partial_{\alpha}\pi^* \partial^{\alpha}\pi
\end{split}
\end{equation}
where
\begin{equation}\label{S4}
{S}_{4} = \sum_{i=1}^{9} \; {S}^{(i)}_{4}
\end{equation}
Having obtained the generic 4th order Galileon action  (\ref{S_i}), our next step is to calculate the equations of motion of $\pi$ ( for $\pi^*$ is straight forward). We define the quantity $\mathcal{E}_i$ as the variation of the ${S^{(i)}}_{CGG}$ with respect to $\pi$ (with respect to $\pi^*$ for ${\mathcal{E}_i}^*$) as, 
\begin{equation}\label{E_i}
\mathcal{E}^i_4 = \frac{\delta {S}^{(i)}_{4}}{\delta \pi}
\end{equation}
Hence, equation of motion for the generic action (\ref{S4}) is given by,
\begin{align}\label{E_CGG}
\mathcal{E}_{4} = \sum_{i=1}^{9} \; \mathcal{E}_4^i
\end{align}
Using Eq.~(\ref{E_i}), the equations of motion corresponding to nine actions are
\begin{align}
\label{eq:e1}
\mathcal{E}^{(1)}_4 &= 2A_1\,[\partial_{\mu}\partial^{\mu}\Box \pi\, \partial_{\alpha}\pi^*\partial^{\alpha}\pi^* + 4\,\partial_{\mu}\Box \pi\, \partial_{\alpha}\pi^*\partial^{\mu}\partial^{\alpha}\pi^*  + 2\, \Box \pi\, \partial_{\alpha} \Box \pi^*\,\partial^{\alpha}\pi^*+ 2 \,\Box \pi\, \partial_{\mu}\partial_{\alpha}\pi^*\,\partial^{\mu}\partial^{\alpha}\pi^* \nonumber\\ &\hspace{1cm}- 2\,\partial^{\alpha}\Box \pi^*\, \Box \pi^*\, \partial_{\alpha}\pi - (\Box \pi^*)^2 \,\Box \pi]
\\
\vspace{1cm}\nonumber\\
\label{eq:e2}
\mathcal{E}^{(2)}_4 &= A_2\,[\partial_{\mu}\partial^{\mu}\Box \pi^*\, \partial_{\alpha}\pi\partial^{\alpha}\pi^* + 2\,\partial_{\mu}\Box \pi^*\, \partial^{\mu}\partial^{\alpha}\pi \partial_{\alpha}\pi^*  + 2\, \partial_{\nu}\Box \pi^*\, \partial_{\alpha}\pi \partial^{\nu}\partial^{\alpha}\pi^* +  \Box \pi^*\, \partial_{\alpha} \Box \pi \,\partial^{\alpha}\pi^* \nonumber \\&{}
+ \Box \pi^*\,\partial^{\alpha}\Box \pi^*\, \partial_{\alpha} \pi + 2\, \Box \pi^*\, \partial_{\nu}\partial_{\alpha}\pi \partial^{\nu}\partial^{\alpha}\pi^* - \partial_{\alpha}\Box \pi^*\,\Box \pi \,\partial^{\alpha}\pi^* - \Box\pi^*\, \partial_{\alpha} \Box \pi\, \partial^{\alpha}\pi^* - (\Box \pi^*)^2\, \Box \pi]
\\
\vspace{1cm}\nonumber\\
\label{eq:e3}
\mathcal{E}^{(3)}_4 &= 2A_1\,[\,2\,\partial^{\alpha}\partial^{\nu}\Box \pi^*\, \partial_{\nu}\pi^*\,\partial_{\alpha}\pi + 2\,\partial^{\nu}\Box \pi^*\, \partial^{\alpha}\pi\,\partial_{\nu}\partial_{\alpha}\pi^* + 2\partial^{\alpha}\Box \pi^* \Box \pi^*\,\partial_{\alpha}\pi + (\Box \pi^*)^2\, \Box \pi \nonumber\\&{}+ 2\,\partial^{\mu}\partial^{\nu}\partial^{\alpha}\pi^*\,\partial_{\mu}\partial_{\nu}\pi\, \partial_{\alpha}\pi^*
+ 2\, \partial^{\mu}\partial^{\nu}\partial^{\alpha}\pi^*\, \partial_{\mu}\partial_{\alpha}\pi^*\, \partial_{\nu}\pi +  2\,\partial^{\nu}\partial^{\alpha}\pi^*\,\partial_{\mu}\partial_{\nu}\pi\, \partial^{\mu}\partial_{\alpha}\pi^* 
\nonumber\\&{} - \Box \pi \,\partial^{\nu}\partial^{\alpha}\pi^*\, \partial_{\nu}\partial_{\alpha}\pi^*]
\\
\vspace{1cm}\nonumber\\
\label{eq:e4}
\mathcal{E}^{(4)}_4 &= -E\,[\,2\,\partial^{\nu}\Box \pi^*\, \partial_{\nu}\partial_{\alpha}\pi\,\partial^{\alpha} \pi^* + 4\,\partial^{\mu}\partial^{\nu}\partial^{\alpha}\pi\, \partial_{\mu}\partial_{\nu}\pi^*\,\partial_{\alpha}\pi^* +   2\,\partial^{\mu}\partial^{\nu}\pi^*\,\partial_{\nu}\partial_{\alpha}\pi\, \partial_{\mu}\partial^{\alpha}\pi^* \nonumber\\&{}
+ 2\,\partial_{\alpha}\partial_{\nu}\Box \pi \,\partial^{\nu}\pi^*\, \partial^{\alpha}\pi^*\, 
 + 2\,\partial^{\nu}\Box \pi\, \partial_{\alpha}\partial_{\nu}\pi^*\, \partial^{\alpha}\pi^* + 2\,\partial^{\nu}\Box \pi\, \Box \pi^*\, \partial_{\nu}\pi^* +
2\,\partial^{\alpha}\Box \pi^*\,\Box \pi \,\partial_{\alpha}\pi^*  
\nonumber\\&{}
+ \Box \pi\,\partial_{\nu}\partial_{\alpha}\pi^* \,\partial^{\nu}\partial^{\alpha}\pi^* + (\Box \pi^*)^2\,\Box \pi 
- \partial_{\nu}\Box \pi^*\, \partial^{\nu}\partial^{\alpha}\pi^*\, \partial_{\alpha}\pi - 2\,\Box \pi^*\, \partial^{\alpha}\Box \pi^* \,\partial_{\alpha}\pi  
\nonumber\\&{}
- 2 \,\Box \pi^*\,\partial^{\nu}\partial^{\alpha}\pi^*\,\partial_{\nu}\partial_{\alpha}\pi - \partial_{\alpha} \Box \pi^* \,\partial^{\nu}\partial^{\alpha}\pi^*\,\partial_{\nu}\pi]
\end{align}
\begin{align}
\vspace{1cm}\nonumber\\
\label{eq:e5}
\mathcal{E}^{(5)}_4 &= -2A_1\,[\,\partial_{\nu}\partial^{\nu} \Box\pi\, \partial_{\alpha}\pi^*\,\partial^{\alpha}\pi^* + 2\,\partial^{\nu}\Box \pi\, \partial^{\alpha}\pi^*\,\partial_{\nu}\partial_{\alpha}\pi^* + 2\,\partial^{\mu}\Box \pi\, \partial^{\alpha}\pi^*\,\partial_{\mu}\partial_{\alpha}\pi^* 
\nonumber\\&{} + 2\,\partial^{\mu}\partial^{\nu} \pi\, \partial_{\nu}\partial_{\alpha}\pi^*\,\partial_{\mu}\partial^{\alpha}\pi^* 
+ 2\,\partial^{\mu}\partial^{\nu} \pi\, \partial^{\alpha}\pi^*\,\partial_{\nu}\partial_{\mu}\partial_{\alpha}\pi^*\,  -2\,\partial_{\nu}\partial_{\mu}\partial_{\alpha}\pi^*\, \partial^{\mu}\partial^{\nu} \pi^*\, \partial^{\alpha}\pi  
\nonumber\\&{} - \Box \pi\,\partial_{\mu}\partial_{\nu}\pi^*\, \partial^{\mu}\partial^{\nu}\pi^* ]
\\
\vspace{1cm}\nonumber\\
\label{eq:e6}
\mathcal{E}^{(6)}_4 &= -A_2\,[\,\partial_{\nu}\partial^{\nu}\Box \pi^*\, \partial_{\alpha}\pi\,\partial^{\alpha}\pi^* + \partial_{\nu}\Box \pi^*\, \partial^{\nu}\partial^{\alpha}\pi\, \partial_{\alpha}\pi^* + \partial_{\nu}\Box \pi^*\, \partial_{\alpha} \pi\,\partial^{\nu}\partial^{\alpha}\pi^* + \partial_{\mu}\Box \pi^*\, \partial^{\mu}\partial^{\alpha}\pi \,\partial_{\alpha}\pi^* \nonumber\\&{}+ \partial_{\mu}\partial_{\nu}\pi^*\, \partial^{\mu}\partial^{\alpha}\pi\,\partial^{\nu} \partial_{\alpha}\pi^* + \partial_{\mu}\Box \pi^* \,\partial_{\alpha}\pi\, \partial^{\mu}\partial^{\alpha}\pi^* 
+ \partial_{\mu}\partial_{\nu}\pi^* \,\partial^{\nu}\partial^{\alpha}\pi\, \partial^{\mu}\partial_{\alpha}\pi^* \nonumber\\&{} +  \partial_{\mu}\partial_{\nu}\pi^*\,\partial_{\alpha}\pi\, \partial^{\mu}\partial^{\nu}\partial^{\alpha}\pi^* - \partial^{\mu}\partial^{\nu}\pi \,\partial^{\alpha}\pi^*\, \partial_{\alpha}\partial_{\mu}\partial_{\nu}\pi^*- 
 \Box \pi^*\, \partial_{\mu}\partial_{\nu}\pi^* \,\partial^{\mu}\partial^{\nu}\pi ]
\\
\vspace{1cm}\nonumber
\label{eq:e7}
\mathcal{E}^{(7)}_4 &= E\,[\,2\,\partial^{\mu}\partial^{\alpha}\Box \pi\, \partial_{\mu}\pi^*\,\partial_{\alpha}\pi^* + 4\,\partial^{\mu}\partial^{\nu}\partial^{\alpha}\pi\, \partial_{\nu}\partial_{\mu}\pi^*\,\partial_{\alpha}\pi^* + 2\,\partial_{\mu}\Box \pi\, \Box \pi^*\,\partial^{\mu}\pi^* 
\nonumber\\&{} + 2\,\partial_{\alpha}\Box \pi\, \partial_{\mu}\pi^*\, \partial^{\mu}\partial^{\alpha}\pi^*  + 2\,\Box\pi^* \,\partial_{\nu}\partial_{\alpha}\pi\, \partial^{\nu}\partial^{\alpha}\pi^* + 2\,\partial^{\nu}\Box\pi^*\, \partial_{\nu}\partial_{\alpha}\pi\, \partial^{\alpha}\pi^*  
\nonumber\\&{} 
+ 2\,\partial_{\nu}\partial_{\alpha} \pi \,\partial^{\nu}\partial^{\mu} \pi^*\, \partial_{\mu}\partial^{\alpha} \pi^* 
+ 2\,\partial_{\nu}\partial_{\alpha} \pi\,\partial_{\mu} \pi^* \,\partial^{\nu}\partial^{\mu}\partial^{\alpha} \pi^* 
- 2\,\partial^{\nu}\Box \pi^* \,\partial_{\nu}\partial_{\alpha} \pi^*\, \partial^{\alpha} \pi 
\nonumber\\&{} -2\,\partial^{\mu}\partial^{\nu} \pi^*\, \partial_{\mu}\partial_{\nu}\partial_{\alpha} \pi^*\, \partial^{\alpha} \pi - 2\,\partial^{\mu}\partial^{\nu} \pi^* \,\partial_{\nu}\partial_{\alpha} \pi^*\, \partial_{\mu}\partial^{\alpha} \pi] \\
%%%%
\vspace{1cm}\nonumber\\
\label{eq:e8}
\mathcal{E}^{(8)}_4 &= -2\,A_1\,[\,\partial^{\mu}\partial^{\alpha}\Box \pi^*\, \partial_{\mu}\pi^*\,\partial_{\alpha}\pi + \partial^{\mu}\Box\pi^*\, \partial_{\alpha}\partial_{\mu}\pi^*\,\partial^{\alpha}\pi + \partial^{\mu}\Box \pi^* \,\Box \pi\,\partial_{\mu}\pi^* 
\nonumber \\&{}\hspace{1cm} 
+2\,\partial_{\alpha}\partial_{\mu}\partial_{\nu}\pi^*\, 
\partial^{\nu}\partial^{\mu}\pi^*\, \partial^{\alpha}\pi 
+ \partial_{\mu}\partial_{\nu}\pi^* \,
\partial^{\nu}\partial^{\mu}\pi^*\, \Box\pi
+ \partial^{\alpha}\partial^{\mu}\partial^{\nu}\pi^* \,\partial_{\mu}\pi^*\, \partial_{\nu}\partial_{\alpha}\pi 
\nonumber \\&{}\hspace{1cm} + \partial^{\mu}\partial^{\nu}\pi^*\, \partial_{\mu}\partial_{\alpha}\pi^* \,\partial_{\nu}\partial^{\alpha}\pi + \partial^{\mu}\partial^{\nu}\pi^*\,\partial_{\mu}\pi^*\,\partial_{\nu}\Box \pi   - \partial^{\mu}\partial^{\nu}\partial^{\alpha}\pi^*\, \partial_{\nu}\partial_{\alpha} \pi\, \partial_{\mu}\pi^*
\nonumber\\&{}\hspace{1cm}
-\partial^{\mu}\partial^{\nu} \pi^* \partial_{\nu}\Box \pi \partial_{\mu}\pi^* 
 - \partial^{\mu}\partial^{\nu}\pi^* \partial_{\nu}\partial_{\alpha} \pi \partial^{\alpha}\partial_{\mu} \pi^*]
\\
\vspace{1cm}\nonumber\\
\label{eq:e9}
\mathcal{E}^{(9)}_4 &= -2\,A_1\,[\,\partial^{\mu}\partial^{\alpha}\Box \pi^*\, \partial_{\mu}\pi^*\,\partial_{\alpha}\pi + \partial^{\mu}\partial^{\nu}\partial^{\alpha}\pi^*\,\partial_{\nu}\partial_{\mu}\pi^*\, \partial_{\alpha}\pi + \partial^{\mu}\partial^{\nu}\partial^{\alpha} \pi^*\, \partial_{\mu} \pi^*\,\partial^{\nu}\partial^{\alpha}\pi \nonumber\\&{}\hspace{1cm} + \partial_{\alpha}\Box\pi^*\,\Box\pi^* \,
\partial^{\alpha}\pi  + \partial_{\alpha}\partial_{\nu}\pi^*\, 
\partial^{\nu}\Box \pi^*\, \partial^{\alpha}\pi+ \partial_{\nu}\partial_{\alpha}\pi^*\, 
\Box \pi^*\, \partial^{\nu}\partial^{\alpha}\pi 
\nonumber\\&{}\hspace{1cm} + \partial_{\alpha}\Box \pi^*\, \partial_{\mu}\pi^*\,\partial^{\mu}\partial^{\alpha}\pi + 
-\partial^{\mu}\partial^{\nu} \pi\, \partial_{\nu}\Box \pi^*\, \partial_{\mu}\pi^* 
]
\end{align}
Thus, the equation of motion for the generic 4th order action \eqref{S4} is:
\begin{align}\label{eq:Ecgg}
\mathcal{E}_{4} &= \left(A_1 + \frac{A_2}{2}\right)[ \,\,\Box \pi\, \partial_{\mu}\partial_{\nu}\pi^*\, \partial^{\mu} \partial^{ \nu}\pi^* + 2\,\Box \pi^*\, \partial_{\mu}\partial_{\nu}\pi \,\partial^{\mu} \partial^{ \nu}\pi^* 
\nonumber\\&{}\hspace{1cm} -2\,\partial_{\mu}\partial_{\nu}\pi^*\, \partial^{\nu}\partial^{\alpha}\pi\,\partial^{\mu}\partial_{\alpha}\pi^*  -(\,\Box \pi^*)^2\,\Box \pi\,]
\end{align}
To compare with the earlier results~\cite{2009-Nicolis-PRD}, we discuss special cases by taking some specific values of the coefficients $A_1$ and $A_2$.\\
\\
\textbf{Case 1: $A_1 = 0$}
Equations of motion for $\pi$ is given by
\begin{align}\label{eq:case1}
\mathcal{E}_4^{A_1 = 0} &= \frac{A_2}{2}[ \,\,\Box \pi\, \partial_{\mu}\partial_{\nu}\pi^*\, \partial^{\mu} \partial^{ \nu}\pi^* + 2\,\Box \pi^*\, \partial_{\mu}\partial_{\nu}\pi \,\partial^{\mu} \partial^{ \nu}\pi^*  -2\,\partial_{\mu}\partial_{\nu}\pi^*\, \partial^{\nu}\partial^{\alpha}\pi\,\partial^{\mu}\partial_{\alpha}\pi^*  -(\,\Box \pi^*)^2\,\Box \pi\,]
\end{align}
which in the limit $\pi = \pi^*$ gives
\begin{align}\label{eq:case1_scalarlimit}
\mathcal{E}_4^{A_1 = 0} &= \frac{A_2}{2}\,[\, 3\, \Box \pi \,\partial_{\mu}\partial_{\nu}\pi\, \partial^{\mu} \partial^{ \nu}\pi - 2\,\partial_{\mu}\partial_{\nu}\pi\, \partial^{\nu}\partial^{\alpha}\pi\,\partial^{\mu}\partial_{\alpha}\pi -(\Box \pi)^3\,\,]
\end{align}
\textbf{Case 2: $A_2 = 0$}
Equations of motion for $\pi$ is given by
\begin{align}\label{eq:case2}
\mathcal{E}_4^{A_2 = 0} &= A_1[ \,\,\Box \pi\, \partial_{\mu}\partial_{\nu}\pi^*\, \partial^{\mu} \partial^{ \nu}\pi^* + 2\,\Box \pi^*\, \partial_{\mu}\partial_{\nu}\pi \,\partial^{\mu} \partial^{ \nu}\pi^*  -2\,\partial_{\mu}\partial_{\nu}\pi^*\, \partial^{\nu}\partial^{\alpha}\pi\,\partial^{\mu}\partial_{\alpha}\pi^*  -(\,\Box \pi^*)^2\,\Box \pi\,]
\end{align}
which in the limit $\pi = \pi^*$ gives
\begin{align}\label{eq:case2_scalarlimit}
\mathcal{E}_4^{A_2 = 0} &= A_1\,\left[\, 3\, \Box \pi \,\partial_{\mu}\partial_{\nu}\pi\, \partial^{\mu} \partial^{ \nu}\pi - 2\,\partial_{\mu}\partial_{\nu}\pi\, \partial^{\nu}\partial^{\alpha}\pi\,\partial^{\mu}\partial_{\alpha}\pi -(\Box \pi)^3\,\,\right]
\end{align}
\textbf{Case 3: $A_1 = A_2$}
Equations of motion for $\pi$ is given by
\begin{align}\label{eq:case3}
\mathcal{E}_4^{A_1 \neq A_2} &= A_1\,\left[\, \frac{3}{2}\,\Box \pi\, \partial_{\mu}\partial_{\nu}\pi^* \,\partial^{\mu} \partial^{ \nu}\pi^* + 3\,\Box \pi^*\, \partial_{\mu}\partial_{\nu}\pi \,\partial^{\mu} \partial^{ \nu}\pi^*
\right. \nonumber\\&{}\hspace{1cm} \left. -3\,\partial_{\mu}\partial_{\nu}\pi^*\, \partial^{\nu}\partial^{\alpha}\pi\,\partial^{\mu}\partial_{\alpha}\pi^*  -\frac{3}{2}\,(\Box \pi^*)^2\,\Box \pi\,\,\right]
\end{align}
which in the limit $\pi = \pi^*$ gives
\begin{align}\label{eq:case3_scalarlimit}
\mathcal{E}_4^{A_1 = A_2} &= \frac{3 A_1}{2}\,\,[\, 3\, \Box \pi\, \partial_{\mu}\partial_{\nu}\pi \,\partial^{\mu} \partial^{ \nu}\pi - 2\,\partial_{\mu}\partial_{\nu}\pi\, \partial^{\nu}\partial^{\alpha}\pi\,\partial^{\mu}\partial_{\alpha}\pi -(\Box \pi)^3\,\,]
\end{align}
\textbf{Case 4: $A_1 \neq A_2$}
Equations of motion for $\pi$ is given by
\begin{align}\label{eq:case4}
\mathcal{E}_4^{A_1 \neq A_2} &= \left(A_1 + \frac{A_2}{2}\right)[ \,\,\Box \pi\, \partial_{\mu}\partial_{\nu}\pi^*\, \partial^{\mu} \partial^{ \nu}\pi^* + 2\,\Box \pi^*\, \partial_{\mu}\partial_{\nu}\pi \,\partial^{\mu} \partial^{ \nu}\pi^*  -2\,\partial_{\mu}\partial_{\nu}\pi^*\, \partial^{\nu}\partial^{\alpha}\pi\,\partial^{\mu}\partial_{\alpha}\pi^* \nonumber\\&{}
 -(\,\Box \pi^*)^2\,\Box \pi\,]
\end{align}
which in the limit $\pi = \pi^*$ gives
\begin{align}\label{eq:case4_scalarlimit}
\mathcal{E}_4^{A_1 \neq A_2} &= \left(A_1 + \frac{A_2}{2}\right)[\, 3\, \Box \pi\, \partial_{\mu}\partial_{\nu}\pi \,\partial^{\mu} \partial^{ \nu}\pi - 2\,\partial_{\mu}\partial_{\nu}\pi\, \partial^{\nu}\partial^{\alpha}\pi\,\partial^{\mu}\partial_{\alpha}\pi -(\Box \pi)^3\,\,]
\end{align}
From the above cases, it is interesting to see that equations of motion are independent of the values of $A_1$ and $A_2$ and the equations of motion of all the four cases are identical to Ref.~\cite{2009-Nicolis-PRD} in flat space-time. Hence, for simplicity, we set $A_1 = 0$ in the Lagrangian for the fourth order complex scalar Galileon Lagrangian.
%%%%%
\section{Gauge fixing}\label{GaugeAppend}

In this Appendix, we show that the Galileon complex scalar action \eqref{action_full} is indeed invariant under the Gauge transformation.
Replacing, $\partial_{\mu} \rightarrow {D}_{\mu} = \partial_{\mu} + \alpha A_{\mu} $ (where we fix $\alpha$ at the end of the calculation) implies 
\begin{align}\label{G1}
\partial_{\mu}\partial^{\nu}\pi \rightarrow {D}_{\mu}{D}^{\nu}\pi= (\partial_{\mu} + \alpha A_{\mu} )(\partial^{\nu} + \alpha A^{\nu})\pi
={\partial_{\mu}\,^{\nu}{\pi}+ \alpha  (\partial_{\mu}{A^{\nu}} \pi+ A^{\nu}    \partial_{\mu}{\pi}+ A_{\mu}    \partial^{\nu}{\pi})+ \alpha\alpha A_{\mu} A^{\nu}   \pi}
\end{align}
Under the local U(1) gauge transformation: 
$\pi \rightarrow \pi e^{-i e \theta(x)}$, above expression becomes
\begin{align}\label{G2}
{D}_{\mu}{D}^{\nu}(\pi e^{-i e \theta}) &=e^{-ie\theta}(\partial_{\mu}\,^{\nu}{\pi}+\alpha \partial_{\mu}{A^{\nu}} \pi+A^{\nu} \alpha \partial_{\mu}{\pi}+A_{\mu} \alpha \partial^{\nu}{\pi}+A_{\mu} A^{\nu} \alpha \alpha \pi\nonumber \\
&{}-ie (\partial^{\nu}{\pi} \partial_{\mu}{\theta}+\partial_{\mu}{\pi} \partial^{\nu}{\theta}+\partial_{\mu}\,^{\nu}{\theta} \pi+A^{\nu} \alpha \partial_{\mu}{\theta} \pi+A_{\mu} \alpha \partial^{\nu}{\theta} \pi)+ie ie \partial_{\mu}{\theta} \partial^{\nu}{\theta} \pi
\end{align}
Under the gauge transformation $A_{\mu} \rightarrow A_{\mu} + \partial_{\mu}\theta$, Eq.~(\ref{G2}) becomes,
\begin{align}\label{G3}
{D}_{\mu}{D}^{\nu}(\pi e^{-i e \theta})&\rightarrow e^{-ie\theta}[\partial_{\mu}\,^{\nu}{\pi}+\alpha( \partial_{\mu}{A^{\nu}} \pi + \partial_{\mu}{\partial^{\nu}\theta} \pi+ A^{\nu}  \partial_{\mu}{\pi}+ \partial^{\nu}\theta  \partial_{\mu}{\pi} +A_{\mu}  \partial^{\nu}{\pi}+   \partial_{\mu}\theta  \partial^{\nu}{\pi}\nonumber \\&{}+ \alpha  A_{\mu} A^{\nu}\pi + \alpha A_{\mu} \partial^{\nu}\theta \pi + \alpha A^{\nu} \partial_{\mu}\theta \pi + \alpha \partial_{\mu}\theta \partial^{\nu}\theta \pi)
-ie (\partial^{\nu}{\pi} \partial_{\mu}{\theta}+\partial_{\mu}{\pi} \partial^{\nu}{\theta}+\partial_{\mu}\,^{\nu}{\theta} \pi\nonumber \\&{}+\alpha A^{\nu} \partial_{\mu}{\theta} \pi +\alpha \partial^{\nu}\theta  \partial_{\mu}{\theta} \pi +\alpha A_{\mu} \partial^{\nu}{\theta} \pi + \alpha \partial_{\mu} \partial^{\nu}{\theta} \pi )+ie ie \partial_{\mu}{\theta} \partial^{\nu}{\theta} \pi]
\end{align}
For the ${D}_{\mu}{D}^{\nu}\pi$ to be invariant under the local gauge transformations, we set  $\alpha = ie$ in Eq.~(\ref{G3}). Hence, the action (\ref{action_CSG}) is invariant under the following 
simultaneous transformations:
\begin{align}\label{eq:gauge_trans}
A_{\mu} \rightarrow A_{\mu} + \partial_{\mu}\theta\,\,;\hspace{0.5cm}
\pi \rightarrow \pi e^{-ie\theta(x)} 
\end{align}
and the quantity ${D}_{\mu}{D}^{\nu}\pi$ will transform as
\begin{align}\label{eq:termTransform}
{D}_{\mu}{D}^{\nu}(\pi e^{-ie\theta})= (\partial_{\mu}\,^{\nu}{\pi}+\alpha \partial_{\mu}{A^{\nu}} \pi+A^{\nu} \alpha \partial_{\mu}{\pi}+A_{\mu} \alpha \partial^{\nu}{\pi}+A_{\mu} A^{\nu} \alpha \alpha \pi)e^{-ie\theta}
\end{align}
Similar analysis can be done for $\pi^* \rightarrow \pi^* e^{ie\theta(x)}$ also.
%%%%%
\section{Galileon scalar electrodynamics in curved space-time}
\label{app:C}

In this Appendix, we show that the action \eqref{action_m} leads to 
higher derivative equations of motion. We also show addition of the 
non-minimal term leads to second order equations of motion. The variation of the action (\ref{action_m}) with respect to $\pi$ yield the EOM:
\begin{small}
\begin{align}
\mathcal{E}^{\rm{min}}_4 &= \frac{\omega}{2 \lambda_{\pi}^6}[\,\,2\,\nabla_{\mu}\nabla^{\mu}\pi^*\, \nabla_{\nu}\nabla_{\alpha}\pi\,\nabla^{\nu}\nabla^{\alpha}\pi^* -\nabla_{\mu}\nabla^{\mu}\pi^*\, \nabla_{\nu}\nabla^{\nu}\pi^*\, \nabla_{\alpha} \nabla^{\alpha}\pi - 2\,\nabla_{\mu}\nabla_{\nu}\pi^*\, \nabla^{\mu}\nabla^{\alpha}\pi^*\,\nabla^{\nu}\nabla_{\alpha}\pi\nonumber\\ &{}+\nabla_{\mu}\nabla^{\mu}\pi \,\nabla^{\mu}\nabla^{\nu}\pi^*\, \nabla_{\mu}\nabla_{\nu}\pi^* + 2\,\nabla_{\alpha}\pi\, \nabla^{\alpha}\pi^*\,(\,\nabla^{\nu}\nabla_{\nu}\nabla_{\mu}\nabla^{\mu}\pi^*  - \nabla^{\nu}\nabla^{\mu}\nabla_{\mu}\nabla_{\nu}\pi^*\, )\nonumber\\&{} + \nabla^{\alpha}\pi^*\, \nabla_{\nu}\pi^*\,(\,\nabla^{\mu}\nabla_{\mu}\nabla^{\nu}\nabla_{\alpha}\pi + \nabla^{\alpha}\nabla^{\nu}\nabla_{\mu}\nabla^{\mu}\pi\, )  - \nabla^{\alpha}\pi^*\, \nabla_{\mu}\pi^*\,(\,\nabla^{\nu}\nabla^{\mu}\nabla_{\nu} \nabla_{\alpha}\pi + \nabla_{\alpha}\nabla_{\nu}\nabla^{\mu}\nabla^{\nu}\pi\, )\nonumber\\&{} +3\,\nabla^{\nu}\nabla^{\alpha}\pi\, \nabla_{\alpha}\pi^*\,(\,\nabla_{\nu}\nabla_{\mu}\nabla^{\mu}\pi^* - \nabla_{\mu}\nabla^{\mu}\nabla_{\nu}\pi^*\,) + 2\,\nabla^{\nu}\nabla^{\alpha}\pi^*\, \nabla_{\alpha}\pi\,(\,\nabla_{\nu}\nabla_{\mu}\nabla^{\mu}\pi^* - \nabla_{\mu}\nabla^{\mu}\nabla_{\nu}\pi^*\,)\nonumber\\&{}
+ \nabla^{\mu}\nabla_{\mu}\pi\, \nabla^{\alpha}\pi^*\,(\,\nabla_{\nu}\nabla^{\nu}\nabla_{\alpha}\pi^* - \nabla_{\alpha}\nabla^{\nu}\nabla_{\nu}\pi^*\,)+ 2\,\nabla^{\mu}\nabla^{\nu}\pi^*\, \nabla^{\alpha}\pi^*\,(\nabla_{\mu}\nabla_{\nu}\nabla_{\alpha}\pi - \nabla_{\nu}\nabla_{\mu}\nabla_{\alpha}\pi\,)\nonumber\\&{}
+\nabla^{\nu}\nabla^{\alpha}\pi^*\, \nabla_{\alpha}\pi^*\,(\,\nabla_{\nu}\nabla_{\mu}\nabla^{\mu}\pi - \nabla_{\mu}\nabla^{\mu}\nabla_{\nu}\pi\,) + 2\,\nabla^{\mu}\nabla^{\nu}\pi\, \nabla^{\alpha}\pi^*\,(\,\nabla_{\alpha}\nabla_{\mu}\nabla_{\nu}\pi^* - \nabla_{\nu}\nabla_{\alpha}\nabla_{\mu}\pi^*\,)\nonumber\\&{}
+2\,\nabla^{\mu}\nabla^{\nu}\pi^*\, \nabla^{\alpha}\pi\,(\,\nabla_{\mu}\nabla_{\alpha}\nabla_{\nu}\pi^* - \nabla_{\nu}\nabla_{\mu}\nabla_{\alpha}\pi^*\,)\,\,]
\end{align} 
\end{small}
Using the following commutation properties of covariant derivatives:
\begin{align}\label{eq:commutation}
[\nabla_{\mu},\nabla_{\nu}]\nabla^{\alpha}\pi &= R^{\alpha}_{\rho\mu\nu}\,\nabla^{\rho}\pi; \;
\nabla_{\nu}\nabla_{\mu}\nabla_{\alpha}\pi - \nabla_{\alpha}\nabla_{\mu}\nabla_{\nu}\pi = R^{\rho}_{\mu\alpha\nu}\nabla_{\rho}\pi; \; 
\nonumber\\&{}
\nabla_{\alpha}\nabla_{\nu}\nabla^{\nu}\pi - \nabla_{\nu}\nabla^{\nu}\nabla_{\alpha}\pi = -R_{\nu\alpha}\nabla^{\nu}\pi,
\end{align}
we get
\begin{small}
\begin{align}
\mathcal{E}_4^{\rm{min}} = \frac{\omega}{2 \lambda_{\pi}^6}[\,-2\,\nabla_{\alpha}\pi\, \nabla^{\alpha}\pi^*\,\nabla^{\nu}\nabla^{\mu}\pi^*\,R_{\mu\nu} -\nabla_{\alpha}\pi\, \nabla^{\alpha}\pi^*\,\nabla^{\mu}\pi^*\,\nabla_{\mu}R 
+2\,\nabla^{\mu}\pi^*\,\nabla^{\alpha}\pi^*\,\nabla^{\nu}\nabla^{\rho}\pi\, R_{\rho\mu\alpha\nu} \nonumber\\ - \nabla^{\mu}\pi^*\,\nabla^{\alpha}\pi^*\,\nabla_{\mu}\nabla^{\nu}\pi\, R_{\nu\alpha} - \nabla^{\mu}\pi^*\,\nabla^{\alpha}\pi^*\,\nabla^{\rho}\pi\,\nabla_{\rho}R_{\alpha\mu} -3\,\nabla^{\alpha}\nabla^{\nu}\pi\, \nabla_{\alpha}\pi^* \,\nabla^{\mu}\pi^*\, R_{\mu\nu}\nonumber\\ -2\,\nabla^{\mu}\pi^*\,\nabla_{\alpha}\pi\,\nabla^{\nu}\nabla^{\alpha}\pi^*\,R_{\mu\nu} + \nabla^{\nu}\pi^*\,\nabla^{\alpha}\pi^*\,\nabla^{\mu}\nabla_{\mu}\pi\, R_{\nu\alpha} -2\,\nabla^{\nu}\nabla^{\alpha}\pi^*\,\nabla_{\alpha}\pi^*\,\nabla^{\mu}\pi\, R_{\mu\nu}\nonumber\\ +2\,\nabla_{\mu}\nabla^{\mu}\pi^*\, \nabla_{\nu}\nabla_{\alpha}\pi\,\nabla^{\nu}\nabla^{\alpha}\pi^* -\nabla_{\mu}\nabla^{\mu}\pi^*\, \nabla_{\nu}\nabla^{\nu}\pi^*\, \nabla_{\alpha} \nabla^{\alpha}\pi - 2\,\nabla_{\mu}\nabla_{\nu}\pi^*\, \nabla^{\mu}\nabla^{\alpha}\pi^*\nabla^{\nu}\nabla_{\alpha}\pi\nonumber\\ +\nabla_{\mu}\nabla^{\mu}\pi\, \nabla^{\mu}\nabla^{\nu}\pi^* \,\nabla_{\mu}\nabla_{\nu}\pi^*\,\,]
\end{align}
\end{small}
Note that the second and fifth terms in the above equation are higher derivative (of metric) terms. In order to remove these terms we need to add some non-minimal terms in the action in such a way that on varying the total action, all the derivatives of the Ricci tensor and Ricci scalar vanish. The equation of motion for the non-minimal action (\ref{non-min}) is given by:
\begin{align}
\mathcal{E}^{\rm{nm}}_4 &= \frac{\omega}{4 \lambda_{\pi}^6}[\,2\,\nabla_{\mu}\nabla^{\mu}\pi^*\, \nabla_{\nu}\pi^*\,\nabla^{\nu}\pi\, R 
+ 2\,\nabla_{\mu}\nabla_{\nu}\pi\, \nabla^{\mu}\pi^*\,\nabla^{\nu}\pi^*\, R  + 2\, \nabla_{\mu}\pi^* \,\nabla^{\mu}\pi\,\nabla^{\alpha}\pi^* \,\nabla_{\alpha}R\nonumber\\&{}
-\nabla_{\mu}\nabla^{\mu}\pi\, \nabla_{\nu}\pi^*\,\nabla^{\nu}\pi^*\, R+ 2\,\nabla_{\mu}\nabla^{\mu}\pi\, \nabla^{\nu}\pi^*\,\nabla^{\alpha}\pi^*\, R_{\nu\alpha} + 4\, \nabla^{\alpha}\nabla^{\mu}\pi^*\,\nabla_{\alpha}\pi\,\nabla^{\nu}\pi^*\, R_{\mu\nu}\nonumber\\&{} + 2\,\nabla^{\alpha}\pi\, \nabla^{\mu}\pi^*\, \nabla^{\nu}\pi^*\, \nabla_{\alpha} R_{\mu\nu} + 4\,\nabla^{\mu}\nabla_{\alpha}\pi^*\,\nabla^{\alpha}\pi^*\, \nabla^{\nu}\pi  R_{\mu\nu} + + 2\nabla_{\alpha}\pi^*\nabla^{\alpha}\pi^* \nabla^{\mu}\nabla^{\nu}\pi\, R_{\mu\nu}\,\,]
\end{align}
If we add the above non minimal action in the action (\ref{action_m}) then it will cancel all the higher order derivative terms in the EOM. So varying the action
${S}_4^{\rm{min}} + {S}^{\rm{nm}}_4$ with respect to $\pi$, we get 
\begin{align}\label{eq:S4eomCurved}
\mathcal{E}^{\prime} &= \frac{\omega}{2 \lambda_{\pi}^6}\,\,[\,\,2\,\Box\pi^*\, \nabla_{\nu}\nabla_{\alpha}\pi\,\nabla^{\nu}\nabla^{\alpha}\pi^* -(\,\Box\pi^*)^2\, \Box\pi +\Box\pi\, \nabla^{\mu}\nabla^{\nu}\pi^*\, \nabla_{\mu}\nabla_{\nu}\pi^* 
\nonumber\\ &{} 
- 2\,\nabla_{\mu}\nabla_{\nu}\pi^*\, \nabla^{\mu}\nabla^{\alpha}\pi^*\,\nabla^{\nu}\nabla_{\alpha}\pi + \Box\pi^*\,\nabla_{\mu}\pi^*\,\nabla^{\mu}\pi\, R -\frac{1}{2}\Box\pi\,\nabla_{\alpha}\pi^*\,\nabla^{\alpha}\pi^* \,R \nonumber\\ &{} + \nabla_{\alpha}\pi^*\,\nabla_{\mu}\pi^*\,\nabla^{\mu}\nabla^{\alpha}\pi \,R 
 +2 \,\Box\pi\, \nabla^{\nu}\pi^*\,\nabla^{\mu}\pi^* \,R_{\nu\mu} +\nabla_{\alpha}\pi^*\, \nabla^{\alpha}\pi^*\,\nabla^{\nu}\nabla^{\mu}\pi \,R_{\mu\nu} \nonumber\\ &{}  - 2\,\nabla_{\alpha}\pi\, \nabla^{\alpha}\pi^*\,\nabla^{\nu}\nabla^{\mu}\pi^*\,R_{\mu\nu} 
 - \nabla^{\mu}\pi^*\,\nabla^{\alpha}\pi^*\,\nabla_{\mu}\nabla^{\nu}\pi\, R_{\nu\alpha} \nonumber\\ &{}  + 2\, \nabla^{\mu}\pi^*\,\nabla^{\alpha}\pi^*\,\nabla^{\nu}\nabla^{\rho}\pi\, R_{\rho\mu\alpha\nu} -3\,\nabla^{\alpha}\nabla^{\nu}\pi\, \nabla_{\alpha}\pi^*\, \nabla^{\mu}\pi^*\,R_{\mu\nu}  ]
\end{align}
\section{Consistency check with real scalar galileon }\label{sec:consistencyRealGal}
In this section, we show systematically that our fourth order action ($S^{\rm{min}}_4 + S^{\rm{nm}}_4$ ) exactly matches with the results of Deffayet et al~\cite{Deffayet2009} in the $\pi = \pi^*$ limit.  Eq.~ (\ref{eq:S4pi}) can be written as,
\begin{align} \label{eq:Smin+Snm}
\left. {S}_4^{\rm{min}} + {S}_4^{\rm{nm}} \right|_{\pi = \pi^*}  &= A_2 \, \int d^4x\,\,\sqrt{-g}\,\, \left[\, (\Box\pi)^2 \,\nabla_{\alpha}\pi\, \nabla^{\alpha}\pi + \Box \pi\, \nabla_{\nu}\pi \, \nabla^{\nu}\nabla^{\alpha}\pi \,\nabla_{\alpha}\pi  \right.
\nonumber\\
& ~~ - \left. \,\nabla_{\mu}\nabla_{\nu}\pi\, \nabla^{\mu}\nabla^{\nu}\pi\, \nabla_{\alpha} \pi\, \nabla^{\alpha}\pi 
 - \nabla^{\mu}\nabla^{\nu}\pi\, \nabla_{\nu}\nabla_{\alpha}\pi\,\nabla_{\mu}\pi\, \nabla^{\alpha}\pi\, \right]  \\
&{} 
 + \frac{A_2}{4} \int d^4x\,\,\sqrt{-g}\,\,\nabla_{\alpha}\pi \nabla^{\alpha}\pi \nabla^{\mu}\pi\nabla^{\nu}\pi\,\,G_{\mu\nu} 
 \nonumber\\&{}
  - \frac{3A_2}{4} \int d^4x\,\,\sqrt{-g}\,\, \nabla_{\alpha}\pi\nabla^{\alpha}\pi \nabla^{\mu}\pi \nabla^{\nu}\pi R_{\mu\nu} \nonumber
\end{align}
where $G_{\mu\nu} = R_{\mu\nu} - \frac{1}{2} \, g_{\mu\nu} \, R$. 
Concentrating on the last term in the above equation (\ref{eq:Smin+Snm}) and using the relation $(\nabla_{\mu}\nabla_{\nu}\nabla^{\nu}\pi - \nabla_{\nu}\nabla^{\nu}\nabla_{\mu}\pi)\,\nabla^{\mu}\pi = -R_{\mu\nu}\nabla^{\mu}\pi \nabla^{\nu}\pi$, ignoring the total derivative terms, we get:
\begin{small}
\begin{align}\label{eq:residualTerm}
  -\frac{3A_2}{4} \nabla_{\alpha}\pi\nabla^{\alpha}\pi \nabla^{\mu}\pi \nabla^{\nu}\pi R_{\mu\nu} &= -\frac{3A_2}{4} \nabla_{\alpha}\pi\nabla^{\alpha}\pi \nabla_{\mu}\nabla^{\mu}\nabla_{\nu}\pi \nabla^{\nu}\pi + \frac{3A_2}{4} \nabla_{\alpha}\pi\nabla^{\alpha}\pi \nabla_{\nu}\nabla^{\mu}\nabla_{\mu}\pi \nabla^{\nu}\pi 
   \nonumber \\
  &{} = \frac{3A_2}{2} \nabla_{\mu}\nabla_{\alpha}\pi \nabla^{\mu}\nabla^{\nu}\pi \nabla_{\nu}\pi \nabla^{\alpha}\pi + \frac{3A_2}{4} \nabla_{\mu}\nabla_{\nu}\pi \nabla^{\mu}\nabla^{\mu}\pi \nabla^{\nu}\pi \nabla_{\alpha}\pi\nabla^{\alpha}\pi   \nonumber \\
  &{}\hspace{1cm} -\frac{3A_2}{4} (\Box \pi )^2\nabla_{\alpha}\pi\nabla^{\alpha}\pi - \frac{3A_2}{2} \Box \pi \nabla^{\mu}\nabla^{\nu}\pi \nabla_{\mu}\pi \nabla_{\nu}\pi \, .
\end{align}
\end{small}
Using Eqs.~(\ref{eq:Smin+Snm}) and (\ref{eq:residualTerm}), we get:
\begin{align} \label{eq:CovariantGal_action}
\left. {S}_4^{\rm{min}} + {S}_4^{\rm{nm}} \right|_{\pi = \pi^*}  &= \frac{A_2}{4} \, \int d^4x\,\,\sqrt{-g}\,\, \left[\, (\Box\pi)^2 \,\nabla_{\alpha}\pi\, \nabla^{\alpha}\pi -2 \Box \pi\, \nabla_{\nu}\pi \, \nabla^{\nu}\nabla^{\alpha}\pi \,\nabla_{\alpha}\pi  \right.
\nonumber\\
&{} \hspace{1.5cm} - \left. \,\nabla_{\mu}\nabla_{\nu}\pi\, \nabla^{\mu}\nabla^{\nu}\pi\, \nabla_{\alpha} \pi\, \nabla^{\alpha}\pi 
 + 2 \nabla^{\mu}\nabla^{\nu}\pi\, \nabla_{\nu}\nabla_{\alpha}\pi\,\nabla_{\mu}\pi\, \nabla^{\alpha}\pi\, \right] \nonumber \\
&{} 
 + \frac{A_2}{4} \int d^4x\,\,\sqrt{-g}\,\,\nabla_{\alpha}\pi \nabla^{\alpha}\pi \nabla^{\mu}\pi\nabla^{\nu}\pi\,\,G_{\mu\nu} 
\end{align}
Thus, the above action (\ref{eq:CovariantGal_action}) matches with the action in Ref.~\cite{Deffayet2009}.
It is important to note that there is an overall sign difference between our result and obtained by Deffayet et al~\cite{Deffayet2009}. This is due to the signature convention of the metric and hence our definition of energy-momentum tensor i.e.  
$T_{\mu\nu} = \frac{2}{\sqrt{-g}}\frac{\delta S}{\delta g^{\mu\nu}}$, differs by an overall negative sing compared to Deffayet et al~\cite{Deffayet2009}.

\subsection{Galileon scalar electrodynamics in curved space time for $A_2 = 0$}\label{sec:A1_terms}
Before discussing the other special case $A_2 = 0$, we obtain the non-minimal coupling terms for the full action.
Consider the action corresponding to the full Lagrangian (\ref{LCGG}) and varying the action with respect to $\pi$ gives the equation of motion for the generic action in curved spacetime. As our main focus is to obtain the non-minimal action, we collect only the fourth order derivative terms in the field as we have done in Appendix (\ref{app:C}).
After collecting the fourth order terms we get,
\begin{align}\label{eq:app_LCGGminimal}
\mathcal{E}^{\rm{min}}_4 &= \frac{A_2}{2} \left[\,\,2\nabla_{\alpha}\pi\, \nabla^{\alpha}\pi^*\,(\,\nabla^{\nu}\nabla_{\nu}\nabla_{\mu}\nabla^{\mu}\pi^*  - \nabla^{\nu}\nabla^{\mu}\nabla_{\mu}\nabla_{\nu}\pi^*\, ) \right. \nonumber\\&{} \left. + \nabla_{\alpha}\pi^*\, \nabla^{\mu}\pi^*\,(\,\nabla_{\nu}\nabla^{\nu}\nabla^{\alpha}\nabla_{\mu}\pi + \nabla_{\mu}\nabla^{\alpha}\nabla_{\nu}\nabla^{\nu}\pi  - \nabla_{\nu}\nabla^{\alpha}\nabla^{\nu} \nabla_{\mu}\pi - \nabla_{\mu}\nabla_{\nu}\nabla^{\alpha}\nabla^{\nu}\pi\, )\right] \nonumber \\&{} 
+ A_1 \left[\,\,2 \nabla_{\alpha}\pi^*\, \nabla^{\alpha}\pi^*\,(\,\nabla^{\nu}\nabla_{\nu}\nabla_{\mu}\nabla^{\mu}\pi  - \nabla^{\nu}\nabla^{\mu}\nabla_{\mu}\nabla_{\nu}\pi\, ) \right. \\
%%%%%
&{} \left. + \nabla^{\alpha}\pi^*\, \nabla^{\mu}\pi^*\,(\,\nabla_{\nu}\nabla^{\alpha}\nabla^{\nu}\nabla_{\mu}\pi + \nabla_{\mu}\nabla_{\nu}\nabla^{\alpha}\nabla^{\nu}\pi  - \nabla^{\nu}\nabla_{\nu}\nabla^{\alpha} \nabla_{\mu}\pi - \nabla_{\mu}\nabla^{\alpha}\nabla_{\nu}\nabla^{\nu}\pi\, ) \right. \nonumber\\&{} \left. + 2 \nabla_{\alpha}\pi^*\, \nabla_{\nu}\pi\,(\,\nabla^{\mu}\nabla_{\mu}\nabla^{\nu}\nabla^{\alpha}\pi^* + \nabla^{\nu}\nabla^{\alpha}\nabla^{\mu}\nabla_{\mu}\pi^*  - \nabla^{\nu}\nabla^{\mu}\nabla^{\alpha} \nabla_{\mu}\pi^* - \nabla_{\mu}\nabla_{\alpha}\nabla^{\mu}\nabla_{\nu}\pi^*\, ) \right]
\nonumber 
\end{align} 
using the commutation properties of the covariant derivatives and using (\ref{eq:commutation}) we get
\begin{align} \label{eq:higherordertermsLCGG}
\mathcal{E}^{\rm{min}}_4 &= -\frac{A_2}{2} \nabla_{\alpha}\pi^* \nabla^{\alpha}\pi \nabla^{\mu}\pi^* \nabla_{\mu}R - A_1\nabla_{\alpha}\pi^* \nabla^{\alpha}\pi^* \nabla^{\mu}\pi \nabla_{\mu}R  -\frac{A_2}{2} \nabla^{\alpha}\pi^* \nabla^{\mu}\pi^* \nabla^{\nu}\pi \nabla_{\nu}R_{\alpha\mu} \nonumber\\ &{} + A_1 \nabla^{\alpha}\pi^* \nabla^{\mu}\pi^* \nabla^{\nu}\pi \nabla_{\nu}R_{\alpha\mu} - 2A_1 \nabla^{\alpha}\pi^* \nabla^{\mu}\pi^* \nabla^{\nu}\pi \nabla_{\mu}R_{\alpha\nu} \, .
\end{align} 
We now consider the following general non-minimal action
\begin{align}\label{eq:general_nm_action}
    S^{\rm{nm}}_4 = \int d^4x \sqrt{-g} \,\, \mathcal{L}^{\rm{nm}}_4
\end{align}
where 
\begin{align}\label{eq:general_nm_lagrangian}
 \mathcal{L}^{\rm{nm}}_4 &= - \left[ \nabla_{\alpha}\pi^* \nabla^{\alpha}\pi^* \nabla^{\mu}\pi \nabla^{\nu}\pi + \nabla_{\alpha}\pi \nabla^{\alpha}\pi \nabla^{\mu}\pi^* \nabla^{\nu}\pi^* \right]\left(g_1 \, R_{\mu\nu} + g_2 \, g_{\mu\nu} R \right) \nonumber\\&{} -\nabla_{\alpha}\pi^* \nabla^{\alpha}\pi \nabla^{\mu}\pi^* \nabla^{\nu}\pi ]\left(h_1 \, R_{\mu\nu} + h_2 \, g_{\mu\nu} R \right)
\end{align}
where $g_1, g_2, h_1$ and $h_2$ are arbitray constants which we will fix later.
Variation of the action (\ref{eq:general_nm_action}) with respect to $\pi$ and collecting only derivatives of the Ricci tensor and Ricci scalar terms because we are interested in fixing the coefficients  $g_1, g_2, h_1,h_2$, we obtain
\begin{align}\label{eq:higherorderterms_nm}
\mathcal{E}^{\rm{nm}}_4 &= g_1 \nabla_{\alpha}\pi^* \nabla^{\alpha}\pi^* \nabla^{\mu}\pi \nabla_{\mu}R + 4g_2 \nabla_{\alpha}\pi^* \nabla^{\alpha}\pi^* \nabla^{\mu}\pi \nabla_{\mu}R + \frac{h_1}{2} \nabla_{\alpha}\pi^* \nabla^{\alpha}\pi \nabla^{\mu}\pi^* \nabla_{\mu}R \\ &{} + 2h_2 \nabla_{\alpha}\pi^* \nabla^{\alpha}\pi \nabla^{\mu}\pi^* \nabla_{\mu}R + 2g_1 \nabla^{\mu}\pi^* \nabla^{\nu}\pi^* \nabla^{\alpha}\pi \nabla_{\alpha}R_{\mu\nu} + h_1 \nabla^{\mu}\pi^* \nabla^{\nu}\pi  \nabla^{\alpha}\pi^* \nabla_{\alpha}R_{\mu\nu} \, .
\nonumber
\end{align} 
Adding Eqs.~(\ref{eq:higherordertermsLCGG}) and (\ref{eq:higherorderterms_nm}) such that all the derivative terms of Ricci tensor and scalar vanish, we obtain the following relations
:
\begin{align}
    -A_1 + g_1 + 4g_2 &= 0 \\
    -\frac{A_2}{2} + A_1 + 2g_1 &= 0\\
    -\frac{A_2}{2} + 2h_2 + \frac{h_1}{2} &= 0\\
    -2A_1 + h_1 &= 0
\end{align}
which implies 
\begin{align}
    g_1 &=  \frac{A_2}{4}  -\frac{A_1}{2} \\
    g_2 &= \frac{3A_1}{8} -\frac{A_2}{16}\\
    h_1 &= 2A_1\\
    h_2 &= -\frac{A_1}{2} + \frac{A_2}{4} \, .
\end{align}
Substituting the above values in Eq.~(\ref{eq:general_nm_lagrangian}), we get
\begin{align}\label{eq:general_nm_lagrangianA1A2}
 \mathcal{L}^{\rm{nm}}_4 &= \frac{A_1}{2}\left[ \nabla_{\alpha}\pi^* \nabla^{\alpha}\pi^* \nabla^{\mu}\pi \nabla^{\nu}\pi + \nabla_{\alpha}\pi \nabla^{\alpha}\pi \nabla^{\mu}\pi^* \nabla^{\nu}\pi^* \right]\left( R_{\mu\nu} - \frac{3}{4} \, g_{\mu\nu} R \right) \nonumber\\&{}
- \frac{A_2}{4}\left[ \nabla_{\alpha}\pi^* \nabla^{\alpha}\pi^* \nabla^{\mu}\pi \nabla^{\nu}\pi + \nabla_{\alpha}\pi \nabla^{\alpha}\pi \nabla^{\mu}\pi^* \nabla^{\nu}\pi^* \right]\left( R_{\mu\nu} - \frac{1}{4} \, g_{\mu\nu} R \right) \nonumber\\&{} -2A_1\nabla_{\alpha}\pi^* \nabla^{\alpha}\pi \nabla^{\mu}\pi^* \nabla^{\nu}\pi ]\left( R_{\mu\nu} - \frac{1}{4}\, g_{\mu\nu} R \right) - \frac{A_2}{4}\nabla_{\alpha}\pi^* \nabla^{\alpha}\pi \nabla^{\mu}\pi^* \nabla^{\nu}\pi \, g_{\mu\nu} R
\end{align}
We note that the above non-minimal action identical with $A_1 = 0$ case considered in Appendix~(\ref{app:C}). In $\pi = \pi^*$ limit, this case also leads to identical equations of motion as in Ref.~\cite{Deffayet2009}.

\section{Vector Galileon action $S_{VEG}$}
\label{app:D}

In Ref.~\cite{2017-Nandi.Shankaranarayanan-JCAP}, the authors obtained a vector Galileon model that leads to second-order equations. 
For completeness, in this appendix, we have listed below the vector Galileon action. The complete vector Galileon action can be written as
\begin{equation}\label{eq:VEG}
{S}_{VEC} = {S}_{\rm VG} + \lambda_{\rm VG} \sum_{i=1}^{12} {S}_{Vi}.
\end{equation}
where, $\lambda_{VG}$ is coupling constant 
\begin{align}
\label{eq:VecGalAction}
{S}_{VG} &= \lambda_{\rm VG}\,\int \,d^4 x\, \sqrt{-g}\,
\epsilon^{\alpha \gamma \nu} \epsilon^{\mu \eta \beta}\,
\nabla_{\alpha \beta}A_\gamma \, \nabla_{\mu \nu}A_\eta \\
{S}_{V1} &= E_1\,\int d^4x\, \sqrt{-g}\, {g}^{\mu \nu}
{g}^{\alpha \beta} {g}^{\gamma \delta}\, R_{\mu \nu}
\,{\nabla}_{\alpha}{A}_{\gamma}\, {\nabla}_{\beta}{A}_{\delta}
\\ {S}_{V2} &= E_2\,\int d^4x\, \sqrt{-g}\,{g}^{\mu \alpha} {g}^{\nu \beta} {g}^{\gamma \delta}\,{R}_{\mu
	\nu}\, {\nabla}_{\alpha}{A}_{\gamma}\,
{\nabla}_{\beta}{A}_{\delta} \\
{S}_{V3} &= E_3\,\int d^4x\, \sqrt{-g}\,{g}^{\mu \nu} {g}^{\alpha \beta}
{g}^{\gamma \delta} \,{R}_{\mu \nu}
\,{\nabla}_{\alpha}{A}_{\beta}
\,{\nabla}_{\gamma}{A}_{\delta} \\ 
{S}_{V4} &=E_4\,\int d^4x\, \sqrt{-g}\,{g}^{\mu \nu} {g}^{\alpha
	\delta} {g}^{\gamma \beta}\, {R}_{\mu \nu}
\,{\nabla}_{\alpha}{A}_{\beta}\,
{\nabla}_{\gamma}{A}_{\delta} \\ 
{S}_{V5} &=E_5\,\int d^4x\, \sqrt{-g}\,{g}^{\mu \gamma} {g}^{\alpha
	\beta} {g}^{\nu \delta}\, {R}_{\mu \nu}\,
{\nabla}_{\alpha}{A}_{\beta}\,
{\nabla}_{\gamma}{A}_{\delta} \\ 
{S}_{V6} &=E_{6}\,\int d^4x\, \sqrt{-g}\,{g}^{\mu \alpha} {g}^{\nu
	\delta} {g}^{\gamma \beta}\, {R}_{\mu \nu}\,
{\nabla}_{\alpha}{A}_{\beta}\,
{\nabla}_{\gamma}{A}_{\delta} \\
{S}_{V7} &=E_{7}\,\int d^4x\, \sqrt{-g}\,{g}^{\mu \alpha} {g}^{\nu
	\beta} {g}^{\gamma \zeta} {g}^{\delta \eta}
\,{R}_{\alpha \beta \gamma \delta}\,
{\nabla}_{\mu}{A}_{\nu}\, {\nabla}_{\zeta}{A}_{\eta}\\
{S}_{V8}&= E_{8}\,\int d^4x\,
\sqrt{-g}\,{g}^{\mu \alpha} {g}^{\eta \beta}
{g}^{\gamma \zeta} {g}^{\delta \nu} \,{R}_{\alpha
	\beta \gamma \delta}\, {\nabla}_{\mu}{A}_{\nu}
\,{\nabla}_{\zeta}{A}_{\eta}~~~~~~
\\ 
{S}_{V9} &= E_{9}\,\int d^4x\,\sqrt{-g}\,{g}^{\alpha \beta} {g}^{\gamma \delta}
{g}^{\mu \nu}\, {R}_{\alpha \beta}\,{R}_{\gamma \delta}\, {A}_{\mu}{A}_{\nu}\\ 
{S}_{V10} &= E_{10}\,\int d^4x\, \sqrt{-g}\,{g}^{\alpha \beta}{g}^{\gamma \mu} {g}^{\delta \nu}\,
{R}_{\alpha \beta}\, {R}_{\gamma \delta}\,{A}_{\mu} {A}_{\nu} \\ 
{S}_{V11}&=
E_{11}\,\int d^4x\, \sqrt{-g}\,{g}^{\alpha\gamma} {g}^{\beta \delta} {g}^{\mu \nu}\,
{R}_{\alpha \beta}\, {R}_{\gamma \delta}\,{A}_{\mu} {A}_{\nu}~~~~~~
\\ {S}_{V12} &= E_{12}\,\int d^4x\,\sqrt{-g}\,{g}^{\alpha \gamma} {g}^{\beta
	\mu} {g}^{\delta \nu}\, {R}_{\alpha\beta}\, {R}_{\gamma \delta}\, {A}_{\mu}{A}_{\nu}
\end{align} 
where $E_i$'s are dimensionless coefficients. Note that $E_i$ are related to $\lambda_{\rm VG}$. 	
\chapter{Helical magnetic field from Riemann coupling: Details} 
\label{ap:helical_pmf}

In this Appendix, we provide more details on the results from chapter \ref{ch:helical-PMF}.

\section{Series expansion of the Bessel function}\label{app:Series-exp}

For completeness, in this appendix, we obtain the series expansion of the Bessel functions to the leading order terms~\cite{2010-Olver.etal-Book}. It is useful to define the following quantity:
\begin{align}\label{eq:floor-functions}
F(\tau)  = \exp\left[\frac{i \pi}{\alpha} \left\lfloor \frac{\pi -\arg (\tau)-\arg
   \left(\frac{ \sqrt{k} \, \varsigma}{\alpha }\right) }{2 \pi }\right\rfloor \right]
\end{align}
where $\left\lfloor \cdots \right\rfloor$ represents the floor function. Up to leading order, the series expansion for the Bessel functions are:
\begin{align}
& J_{  \frac{1}{2 \alpha}} \left( \frac{\varsigma \, \sqrt{k} }{\alpha} \tau \, \right) =  \frac{F(\tau) }{\Gamma \left( 1+ \frac{1}{2\alpha}  \right) } \left( \frac{\varsigma \, \sqrt{k} }{2\alpha} \right)^{\frac{1}{2\alpha}}
   \tau^{ \frac{1}{2\alpha} }\left( 1 - \frac{\varsigma^2 k \tau^2}{2 \alpha (1+2\alpha) }+O\left(\tau^3\right)\right), \\
& Y_{  \frac{1}{2 \alpha}} \left( \frac{\varsigma \, \sqrt{k}  }{\alpha} \tau \, \right) = \frac{F(\tau) }{\pi} \left( \frac{\varsigma \, \sqrt{k} }{2\alpha} \right)^{\frac{1}{2\alpha}}
  \Gamma \left( - \frac{1}{2\alpha}  \right) \cos\left( \frac{\pi}{2\alpha} \right) \left(  - \tau^{\frac{1}{2\alpha}} +  \frac{ k \, \varsigma^2 \, \tau^{2 + \frac{1}{2\alpha}}}{2 \alpha (1 + 2\alpha) }+O\left(\tau^3\right)\right)
  \nonumber \\
  &~~~~~~~~~~~~~~~~~~~~~+ \frac{F(\tau)^{-1} }{\pi} \left( \frac{2\alpha}{ \varsigma \, \sqrt{k} } \right)^{\frac{1}{2\alpha}}
  \Gamma \left( \frac{1}{2\alpha}  \right)  \left( -  \tau^{ - \frac{1}{2\alpha} } + \frac{ k \, \varsigma^2 \,\tau^{2 - \frac{1}{2\alpha}}}{2 \alpha (-1+2\alpha) }+O\left(\tau^3\right)\right), \\
 %%% 
& J_{  \frac{1}{2 \alpha}} \left( -\frac{i \varsigma \, \sqrt{k} }{\alpha} \tau \, \right) 
=  \frac{\tilde{F}(\tau) }{\Gamma \left( 1+ \frac{1}{2\alpha}  \right) } \left( - \frac{ i \,\varsigma \, \sqrt{k} }{2\alpha} \right)^{\frac{1}{2\alpha}}
  \tau^{ \frac{1}{2\alpha} } \left(  1 + \frac{\varsigma^2 k \tau^2}{2 \alpha (1+2\alpha) }+O\left(\tau^3\right)\right) \\
& Y_{  \frac{1}{2 \alpha}} \left(-i \frac{\varsigma \, \sqrt{k}  }{\alpha} \tau \, \right) = -\frac{\tilde{F}(\tau) }{\pi} \left( -\frac{ i \, \varsigma \, \sqrt{k} }{2\alpha} \right)^{\frac{1}{2\alpha}}
  \Gamma \left( - \frac{1}{2\alpha}  \right) \cos\left( \frac{\pi}{2\alpha} \right) \tau^{\frac{1}{2\alpha}} \left(   1 +  \frac{ k \varsigma^2\tau^2 }{2 \alpha (1 + 2\alpha) }+O\left(\tau^3\right)\right) \nonumber
\end{align}
\begin{align}  
&~~~~~~~~~~~~~~{} - \frac{\tilde{F}(\tau)^{-1} }{\pi} \left( -\frac{2\alpha}{ \, i \, \varsigma \, \sqrt{k} } \right)^{\frac{1}{2\alpha}}
  \Gamma \left( \frac{1}{2\alpha}  \right)  \tau^{ - \frac{1}{2\alpha} }  \left( 1 + \frac{ k \, \varsigma^2 \,\tau^2}{2 \alpha (-1+2\alpha) }+O\left(\tau^3\right)\right).
\end{align}
At leading order, the helicity modes are:
\begin{align}\label{Appeq:A+}
A_+(\tau,k) &= F(\tau) \, \left( \frac{\varsigma \, \sqrt{k} }{2\alpha} \right)^{\frac{1}{2\alpha}} \left[  \frac{C_1 }{\Gamma \left( 1+ \frac{1}{2\alpha}  \right) } 
   -  \frac{C_2 }{\pi}  \Gamma \left( - \frac{1}{2\alpha}  \right) \cos\left( \frac{\pi}{2\alpha} \right) \right] \nonumber \\
 &-~~ C_2 \frac{F(\tau)^{-1} }{\pi} \left( \frac{2\alpha}{\varsigma \, \sqrt{k} } \right)^{\frac{1}{2\alpha}}
  \Gamma \left( \frac{1}{2\alpha}  \right)  \tau^{ - \frac{1}{\alpha} }   \\
% \end{align} 
%  
%\begin{align}
\label{Appeq:A-}
A_-(\tau,k) &=\tilde{F}(\tau) \left( -i \frac{\varsigma \, \sqrt{k} }{2\alpha} \right)^{\frac{1}{2\alpha}}  \left[ \,  \frac{C_3 }{\Gamma \left( 1+ \frac{1}{2\alpha}  \right) }  -  \frac{C_4}{\pi}
\Gamma \left( - \frac{1}{2\alpha}  \right) \cos\left( \frac{\pi}{2\alpha} \right)  \right] \nonumber\\
&{}-~~C_4 \frac{\tilde{F}(\tau)^{-1} }{\pi} \left( -\frac{2\alpha}{ i \varsigma \, \sqrt{k} } \right)^{\frac{1}{2\alpha}}
  \Gamma \left( \frac{1}{2\alpha}  \right)  \tau^{ -\frac{1}{\alpha} } 
\end{align}
\begin{figure}[!hbt]
	\centering
	\includegraphics[width=0.65\textwidth]{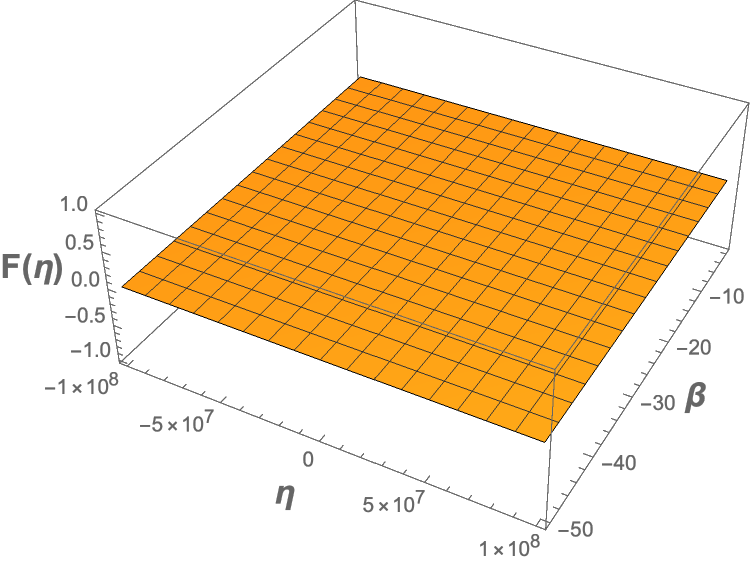}
	\caption{Plot of the Floor function for the range of $-10^8 < \eta < 10^8$ and $-50 < \beta < -2$.}
	\label{fig:Plot}
\end{figure}

It is convenient to define the following quantities:
\begin{align}\label{eq:quantities-F}
\mathcal{F}(\tau) &= F(\tau) \, \left( \frac{\varsigma }{2\alpha} \right)^{\frac{1}{2\alpha}}, \\
\label{eq:quantities-Ftilde}
\tilde{\mathcal{F}}(\tau) &= \tilde{F}(\tau) \, \left( -i\frac{\varsigma }{2\alpha} \right)^{\frac{1}{2\alpha}}, \\
\label{eq:quantities-C}
C(\tau) &= F(\tau) \, \left( \frac{\varsigma }{2\alpha} \right)^{\frac{1}{2\alpha}} \left[  \frac{C_1 }{\Gamma \left( 1+ \frac{1}{2\alpha}  \right) } 
   -  \frac{C_2 }{\pi}  \Gamma \left( - \frac{1}{2\alpha}  \right) \cos \left( \frac{\pi}{2\alpha} \right) \right], 
 \end{align} 
\begin{align}  
   \label{eq:quantities-Ctilde}
\tilde{C}(\tau) & = \tilde{F}(\tau) \left( -i \frac{\varsigma}{2\alpha} \right)^{\frac{1}{2\alpha}}  \left[ \,  \frac{C_3 }{\Gamma \left( 1+ \frac{1}{2\alpha}  \right) } 
-\frac{C_4}{\pi}
\Gamma \left( - \frac{1}{2\alpha}  \right) \cos\left( \frac{\pi}{2\alpha} \right)  \right]
\end{align}
\ref{fig:Plot} is the 3-D plot of the floor function for the range of values of $\eta$ and $\beta$ that are consistent with generic models of inflation. As can be seen from the plot, the floor function is zero in the range of interest. Due to this reason, we have suppressed the time dependence in above quantities i.e., $\mathcal{F}(\tau) = \mathcal{F},\tilde{\mathcal{F}}(\tau) = \tilde{\mathcal{F}}$ and $C(\tau) = C, \tilde{C}(\tau) = \tilde{C}$. We have used this in computing the energy densities in Sec. \eqref{sec:Helical}. 

%
%%%%%%%%  S E C T I O N %%%%%%%%%%
%
%
%%%%%%%%%%%%%   Case M = H %%%%%%%%%%%%%%%%%
\section{Power spectrum for slow-roll inflation}
\label{app:slow-roll-case}

In Sec. \eqref{sec:Helical}, we obtained the power-spectrum for the helical fields in the power-law inflation. In this section, we obtain the power-spectrum for slow-roll inflation. 

To do that, we first obtain the power-spectrum for the power-law inflation by 
assuming that $M$ is slowly varying and is related to the Hubble parameter 
$H$, i. e., $M \sim H \sim  \mathcal{H}/{a}$.

\subsection{Power law inflation}
The equation of motion (\ref{eq:eom_helicity}) for power inflation (\ref{eq:powerLaw}) is: 
\begin{align}\label{eq:eom_helicity-PL-beta}
A_h^{\prime\prime} + \left[  k^2 + \frac{8hk\xi}{\eta} \right] \, A_h= 0.
\end{align}
where $\xi = \frac{\beta (\beta + 2)}{(\beta + 1) }$, solution of the above equation for super horizon modes are given by
\begin{align}\label{eq:+sup-mod-M=H}
A_+(\eta, k) &= 2 \sqrt{2 k \xi \eta} \left[ \, D_1 J_1\left(4 \sqrt{ 2 k \xi \eta} \right)+2 i D_2 Y_1\left(4 \sqrt{2 k \xi \eta} \, \right)\right]\\
\label{eq:-sup-mod-M=H}
A_-(\eta, k) &= -2 \sqrt{2 k \xi \eta} \left[  \, D_3 I_1\left(4 \sqrt{2 k \xi \eta} \right)+ 2 D_4 \,  K_1\left(4 \sqrt{2 k \xi \eta} \right)\right] \, ,
\end{align}
{   where $D_1,D_2(D_3,D_4)$ are the arbitrary constants, we assume these are of the same order of $C_1,C_2(C_3,C_4)$ respectively.} As was shown earlier, we can set $\left| A_-(\eta, k)   \right| = 0$, and positive mode can be approximated by power series at leading order as 
%\begin{align}\label{eq:+sup-mod-M=H-series}
%A_+(\eta, k) &= 8D_1 \xi k \eta - D_2\,  \left[\,\, 64 F(\eta,k) \xi k \eta - \frac{2i}{\pi} \,\, \right].
%\end{align}
%
\begin{align}\label{eq:+sup-mod-M=H-series}
A_+(\eta, k) &= 8D_1 \xi k \eta + D_2\, \frac{2i}{\pi} \, .
\end{align}
where we have used $F(\eta,k) = 0$ in the above eq. (\ref{eq:+sup-mod-M=H-series}), therefore spectral magnetic energy density at horizon exit is given by
%
%
%\begin{align}\label{eq:spectral_rhoB-case-M=H}
%\left.  \frac{d\rho_B}{d\rm{ln}k} \right|_{k_* \eta_* \sim 1} &=  \frac{( - \eta_0)^{4\beta + 4}}{\left( 2\pi \right)^2} k_*^{4\beta + 9}  \left( \,\, \left| 8D_1 \xi - D_2 \left( 64 F(\eta_*,k_*) \xi - \frac{2i}{\pi} \right)  \right|^2 \right.
%  \nonumber\\
%  &{}
% \left.+ \left| 4D_3 \xi  + D_4 \left( 32 \pi i F(\eta_*,k_*) \xi + 1   \right) \, \, \right|^2  \,\,\,\, \right).
%\end{align}
%
%
\begin{align}\label{eq:PowSpec_B_M=H0}
\left.  \frac{d\rho_B}{d\rm{ln}k} \right|_{k_* \eta_* \sim 1} =  \frac{( - \eta_0)^{4\beta + 4}}{\left( 2\pi \right)^2} k_*^{4\beta + 9}  \left( \,\, \left| 8D_1 \xi + D_2 \frac{2i}{\pi}  \right|^2  \,\,\, \right)
\end{align}
Unlike the exact calculation in Sec. \eqref{sec:Helical}, this approximation leads to scale invariant spectrum for $\beta = - \frac{9}{4}$. Substituting the values of the coefficients $D_1$ and $D_2$ from eq.(\ref{eq:Coeff-value}), we obtain:
 \begin{align}\label{eq:PowSpec_B_M=H}
\left.  \frac{d\rho_B}{d\rm{ln}k} \right|_{k_* \eta_* \sim 1} \approx 10^{59} \rm{GeV}^4= 10^3 \mathcal{H}^4
\end{align}
Thus, $R \sim 10^{-9}$. This implies that the above approximation leads to a reduction in the helical field power-spectrum. We will now use this procedure to evaluate the power-spectrum in slow-roll inflation. 
 %
%
%%%%%%%%%%  S E C T I O N %%%%%%%%%%%%%%%
%
%
\subsection{ Slow roll inflation}
The slow-roll parameter $\epsilon$ in terms of $\mathcal{H}$ is defined by 
\begin{align}
\epsilon = 1 - \frac{\mathcal{H}^{\prime}}{ \mathcal{H}^2 }
\end{align}
A necessary condition for inflation is  $\epsilon< 1$. For the leading order slow-roll, we have:
\begin{equation}
\mathcal{H} \approx - \frac{ (1+\epsilon) }{\eta};~\mathcal{H}^{\prime\prime} \approx -\frac{2(1+\epsilon)}{\eta^3};
\end{equation}
Substituting these in Eq.~(\ref{eq:eom_helicity}), we have:
\begin{align}\label{eq:eom_helicity_sr-epsilon}
A_h^{\prime\prime} +\left[  k^2 - \frac{16kh}{\eta} \frac{\epsilon}{1+\epsilon}\,  \right] \, A_h \approx 0.
\end{align}
Note that the above equation (\ref{eq:eom_helicity_sr-epsilon}) can also be obtained by substituting $\beta = -2-\epsilon$ in Eq.~(\ref{eq:eom_helicity-PL-beta}) where the expression of $\xi$ will be $\xi = -\frac{2\epsilon}{1+\epsilon}$. Hence, the super-horizon mode solution can be obtained by substituting $\xi = -\frac{2\epsilon}{1+\epsilon}$ in eq.(\ref{eq:+sup-mod-M=H}) and (\ref{eq:-sup-mod-M=H}). Thus after setting $ \left| A_-(\eta,k) \right|= 0$, spectral magnetic energy density for slow roll case will have the form
\begin{align}\label{eq:PowSpec_B-M=H-slowroll}
  \left.  \frac{d\rho_B}{d\rm{ln}k} \right|_{k_* \eta_* \sim 1}  &= {  \frac{1}{\left( 2\pi \right)^2} \frac{k_*^{1 - 4\epsilon}}{  (-\eta_0)^{4\epsilon + 4} } \left( \,\, \left| -\frac{16\epsilon }{1+\epsilon}D_1  +  \frac{2i}{\pi} D_2  \right|^2  \,\,\, \right)  }
\end{align}
Unlike the exact calculation in Sec. \eqref{sec:Helical}, this approximation leads to blue-tilt spectrum. Thus, this is not a  approximation to obtain power-spectrum for helical magnetic fields.

\chapter{Helical magnetic field from Riemann coupling lead to baryogenesis: Details} 
\label{apdetails:PMF_baryo}

In this Appendix, we provide more details on the results from chapter \ref{ch:PMF_Baryo}.

\section{Quantization in the Helicity basis}
\label{app:helicity_basis}
In this section, we briefly discuss the evolution of the quantum fluctuations of the electromagnetic field in the helicity basis~~\cite{2018-Sharma.Subramanian.Seshadri.PRD}.  Decomposition of the vector potential in Fourier domain leads to:
\begin{align}\label{eq:FourierT}
A^{i}(\vec{x}, \eta) =  \int \frac{d^3 k}{(2\pi)^3} \sum_{\lambda = 1,2} \varepsilon^i_{\lambda} \left[ A_{\lambda}(k,\eta) b_{\lambda}(\vec{k}) e^{ik\cdot x}  
+ A^*_{\lambda}(k,\eta)  b^{\dagger}_{\lambda}(\vec{k}) e^{- ik\cdot x} \right]
\end{align}
where $b(\textbf{k})$ and $b^{\dagger}(\textbf{k})$ are the annihilation and creation operators respectively for a given comoving mode $\textbf{k}$, and $\varepsilon_{\lambda}^i$ is the orthogonal basis vector which in right-handed coordinate system~\cite{2018-Sharma.Subramanian.Seshadri.PRD} is given by
\begin{align}\label{eq:basisVector}
\varepsilon^{\mu} = \left( \frac{1}{a}, \textbf{0} \right), \,\,\,\, \varepsilon^{\mu} = \left( 0, \frac{ \hat{\varepsilon}^i_{\lambda} }{a} \right), \,\,\,\, \varepsilon^{\mu}_3 = \left(  0, \frac{\hat{\textbf{k}}}{a} \right) \quad  \text{for} \quad \lambda = 1, 2 \, ,
\end{align}
3-vectors $\hat{\varepsilon}^i_{\lambda}$ are unit vectors orthogonal to $\hat{\textbf{k}}$ and to each other. Substituting Eq.~(\ref{eq:basisVector}) in 
Eq.~(\ref{eq:FourierT} ) and defining the new variable 
$\bar{A}_{\lambda} = a(\eta) \,  A_{\lambda}(k,\eta)$, we have:
\begin{align}\label{eq:Fdecomposition}
A_{i}(\textbf{x}, \eta) = \int \frac{d^3 k}{(2\pi)^3} \sum_{\lambda = 1,2} \,\hat{\varepsilon}_{i \lambda} \left[ \bar{A}_{\lambda} b_{\lambda}(\textbf{k}) e^{ i \textbf{k} \cdot \textbf{x} }  
+ \bar{A}^*_{\lambda}  b^{\dagger}_{\lambda}(\vec{k}) e^{- i \textbf{k} \cdot \textbf{x} } \right] \, .
\end{align}
Substituting  Eq.~\eqref{eq:Fdecomposition} in Eq.~\eqref{eq:equation_of_motion}, we get:
\begin{align}\label{eq:EOM_fourier_space}
\sum_{\lambda = 1,2}b_{\lambda} \left[  \hat{\varepsilon}_{i \lambda}  \bar{A}_{\lambda}^{\prime\prime} + \frac{4i}{M^2} \epsilon_{i j l} k_j \hat{\varepsilon}_{l \, \lambda} \bar{A}_{\lambda} \, \left( \frac{a^{\prime\prime\prime} }{a^3} - 3\frac{ a^{\prime\prime}  a^{\prime}  }{a^4} \right) +  k^2 \hat{\varepsilon}_{i \lambda} \bar{A}_{\lambda}\right] = 0 
\end{align}
where we have used $\partial_j \partial_j = -k^2$. 

Since the action \eqref{eq:action} contains parity breaking term (helicity term), it is useful to work in the helicity basis. The helicity basis vectors $\varepsilon_+$ and $\varepsilon_-$ corresponding to $h = +1$ and $h = -1$ are defined as
\begin{align}\label{eq5:helicity_basis}
\varepsilon_{\pm} = \frac{1}{\sqrt{2}} \left(   \hat{\varepsilon}_1 \pm i  \hat{\varepsilon}_2  \right).
\end{align}
Assuming that the wave propagates in the $z-$direction, the vector potential in the helicity basis is given by:
%the case where the wave is propagating along the $\varepsilon_3$ direction, in %coulomb gauge (radiation gauge), the vector potential takes the form
%
\begin{align}\label{eq5:decomp_A_helicity}
\bar{\textbf{A}} = \bar{A}_1 \hat{\varepsilon}_1 + \bar{A}_2  \hat{\varepsilon}_2 = A_+ \varepsilon_+ + A_- \varepsilon_-
\end{align}
where $A_+$($A_-$) refer to the vector potential with positive (negative) helicity. 
The ground state in the helicity basis is defined as
\begin{align}\label{eq:GS_helicity}
b_h(\textbf{k}) | 0 \rangle = 0 
\end{align}
and satisfy the following commutation relations:
\begin{align}\label{eq:comm-b_h}
\left[ b_h(\textbf{k}), b^{\dagger}_{h^{\prime}}(\textbf{q})  \right] &= \left( 2\pi \right)^3 \, \delta^3(\textbf{k} - \textbf{q}) \, \delta_{h h^{\prime}} \\
\left[ b_h(\textbf{k}), b_{h^{\prime}}(\textbf{q})  \right] &= 0 = \left[ b^{\dagger}_h(\textbf{k}), b^{\dagger}_{h^{\prime}}(\textbf{q})  \right] \, .
\end{align}

Rewriting \eqref{eq:EOM_fourier_space} in the Helicity basis and replacing
$\epsilon_{i j l} \partial_j A_l \longrightarrow  -k \sum_{h = \pm 1} h A_h \varepsilon_{h}$, we have:
\begin{align}\label{App_eq:eom_helicity}
A_h^{\prime\prime} + \left[  k^2 - \frac{4kh}{M^2} \, 
\Gamma(\eta)  \right] A_h= 0 \, ,
\end{align}
where,
\begin{equation}
\label{def:Gamma}
  \Gamma(\eta) = \frac{a^{\prime\prime\prime}}{a^3} - 3\frac{a^{\prime\prime} a^{\prime} }{a^4} = \frac{1}{a^2} \left(\mathcal{H}'' - 2 \mathcal{H}^3\right) \, .  
\end{equation}
%
%
%and the ground state helicity density as
%\begin{align}\label{eq:rhoh}
%\rho_h \left(\eta, k \right) \equiv -\langle 0 | A_i B^i | 0 \rangle = \int \frac{dk}{k} \, \frac{d\rho_h}{d\rm{ln}k} = \int \frac{dk}{k} \frac{1}{2\pi^2} \frac{k^4}{ a^3} \left( \,\, \left| A_+\left(\eta, k \right) \right|^2 - \left| A_-\left(\eta, k \right) \right|^2 \,\,  \right).
%\end{align}
%
%
%where $d\rho_{\Upsilon}/d (\ln k)$ for $\Upsilon \in \{ E,B,h\}$ is the spectral energy contained in logarithmic interval in $k-$space. Note that the helicity density is the difference between the two helicity spectrums. Hence, the magnetic helicity is maximum if one mode is enhanced and the other mode is suppressed~\cite{2013-Durrer.Neronov-Arxiv}. 
%

\section{Generation and evolution of helical modes }
\label{app:Helical}

Substituting the power-law inflation scale factor (\ref{eq:powerLaw}) in 
Eq.~(\ref{eq:eom_helicity}), we have:
\begin{align}\label{eq:arbitrayBeta_eta}
{A_h^{\prime\prime} + \left[ k^2 - \frac{8kh}{M^2} 
\frac{ \beta (\beta+1) ( \beta + 2)}{\eta_0^3} 
\left( \frac{-\eta_0}{\eta} \right)^{(2 \beta+5)} \right]  \, A_h = 0 } \, .
\end{align}
Helicity term  vanishes for de-sitter case ($\beta = -2$), which is consistent with the fact that the de Sitter symmetry will not be preserved in the presence of helicity terms. However, it will be non-zero for the approximately de Sitter universe i.e., $\beta = -2-\epsilon$. 
%For the power-law inflation model, the scalar and tensor perturbations can be evaluated exactly. However, as can be seen, it is not possible to obtain an exact expression. To obtain the solution, we consider two regions. 
%In sub-horizon limit, the wavelength of the mode is smaller than the Hubble radius, i. e., $H \ll k$. 
In sub horizon limit ($\left| - k \eta \right| \gg 1$), Eq.~(\ref{eq:arbitrayBeta_eta}) simplifies to:
\begin{align}\label{eq:sub-horizon}
A_h^{\prime\prime} + k^2 A_h \approx 0 
\end{align}
and assuming that the quantum field is in the vacuum state at asymptotic past 
(Bunch-Davies vacuum state), we have:
\begin{align}
A_h = \frac{1}{\sqrt{k}} e^{-ik\eta}.
\end{align}
On super-Horizon scales ($\left| - k \eta \right| << 1$), Eq.~(\ref{eq:arbitrayBeta_eta}) becomes:
\begin{align}\label{eq:sup_mode_alpha_tau-varsigma}
{   \alpha^2 \frac{d^2 A_h}{d\tau^2} + \frac{\alpha(\alpha+1)}{\tau} \frac{d A_h}{d \tau} + h \, k \, \varsigma^2 A_h  = 0     }
\end{align}
where  
\begin{equation}
\label{def:varsigma}
{\varsigma^2 \equiv -\frac{1}{M^2 \, \eta_0} (2\alpha - 3)(2\alpha - 1)(2\alpha + 1)\, ,
\tau = \left( -\frac{\eta_0}{ \eta} \right)^{\alpha}  \, ,
\alpha = \beta + \frac{3}{2} }
\end{equation}
Note that $\tau$ (dimensionless variable $0 < \tau < \infty$) and $\eta$ (negative during inflation) are linearly related. [At the start of inflation, $\tau$ is large and vanishes 
at the end of inflation.] Note that $\alpha = -\frac{1}{2}$ corresponds to de-sitter and $ \alpha \leq - \frac{1}{2} $.
The solutions for the above equation (\ref{eq:sup_mode_alpha_tau-varsigma}) are:
\begin{subequations}
\begin{align}\label{eq:sup_mode_h+}
A_{+}(\tau,k) &= \tau^{- \frac{1}{2\alpha} } \, J_{  \frac{1}{2 \alpha}} \left( \frac{\varsigma \sqrt{k} }{\alpha} \tau \, \right)  C_1+ \tau^{- \frac{1}{2\alpha} } \, Y_{ \frac{1}{2 \alpha} }  \left( \frac{ \varsigma \sqrt{k} }{\alpha}\tau \right)  C_2
\\
\label{eq:sup_mode_h-}
A_{-}(\tau,k) &=  \tau^{- \frac{1}{2\alpha} } \, J_{  \frac{1}{2 \alpha}} \left(  -i \frac{ \varsigma \, \sqrt{k} }{\alpha} \tau  \right)  C_3+ \tau^{- \frac{1}{2\alpha} }  \, Y_{  \frac{1}{2 \alpha} } \left(  - i \, \frac{ \varsigma\, \sqrt{k} }{\alpha} \tau  \right)  C_4 \, ,
\end{align}
\end{subequations}
where $C_1, C_2, C_3, C_4$ are arbitrary constants of dimension $L^{1/2}$. 
For the two helicity modes, we fix the constants $C_1, C_2$ ($C_3, C_4)$ by matching $A_h$ and $A_h'$ at the transition time of sub-horizon and super-horizon modes at $k_* \sim \eta_*^{-1}$ where $*$ refers to the 
quantities evaluated at the horizon-exit.

Although the analysis can be done for any general value of $\alpha$, to keep the calculations tractable, we obtain the constants for $\alpha = -1$.  There are two reasons for this choice: First, in this special case, $\tau \propto \eta$ and the super-horizon modes can be written in terms of $\eta$ using the linear relation. 
Second, the constants $C_1, C_2, C_3, C_4$ have a weak dependence of $\alpha$ 
and, hence, finding the value for a given value of $\alpha$ will be accurate within an order~\cite{2020-Kushwaha.Shankaranarayanan-PRD}. Thus, matching the solutions and the derivatives at the horizon-exit, we get:
\begin{align}\label{eq:Coefficients}
C_1 &= -e^i \,  \sqrt{ \frac{\pi \eta_0}{ 2} } \left( \frac{1}{\sqrt{\Theta}} \rm{sin}\Theta 
 + i \sqrt{ \Theta }  \,  \rm{cos} \Theta   \right), \,\,\,\,
C_2 = -i \, e^i \,  \sqrt{ \frac{\pi \eta_0}{ 2} } \left( \frac{1}{\sqrt{\Theta}} \rm{cos}\Theta 
 - i \sqrt{ \Theta }  \,  \rm{sin} \Theta   \right) \\
C_3 &= e^i \,  \sqrt{ \frac{\pi \eta_0}{ 2 } } \left( \frac{1}{\sqrt{i \Theta}} \rm{sinh}\Theta 
 +  \sqrt{ i \Theta }  \,  \rm{cosh} \Theta   \right), \,\,\,\,
C_4 = -i \, e^i \,  \sqrt{ \frac{\pi \eta_0}{ 2 } } \left( \frac{1}{\sqrt{i \Theta}} \rm{cosh} \Theta
 +  \sqrt{ i \Theta }  \,  \rm{sinh} \Theta   \right). \nonumber
\end{align}
where $\Theta = \sqrt{ \frac{15 \eta_*}{M^2 \eta_0^3} }$ is the dimensionless constant.

In Ref.~\cite{2020-Kushwaha.Shankaranarayanan-PRD}, 
the current authors derived the magnetic field spectral energy density and is given by
\begin{align}\label{eq:powerSpetrum}
\frac{d\rho_B}{d\rm{ln}k}   =  \left| {\cal C}(k_*, \alpha) \right|^2 \, \left( \frac{ k }{k_*} \right)^{ 2 - 4\alpha }   k^{3 + 4\alpha +\frac{1}{2\alpha}}  
 +  \left| {\cal C}_2(k_*,\alpha)  \right|^2 \,
   \left( \frac{ k }{k_*} \right)^{ 4 - 4\alpha }   k^{1 + 4\alpha - \frac{1}{2\alpha}}  
\end{align}  
\begin{figure}[!hbt]
%\centering
%\subfigure[]{%
\label{fig:Powerfirst}%
\includegraphics[height=2in]{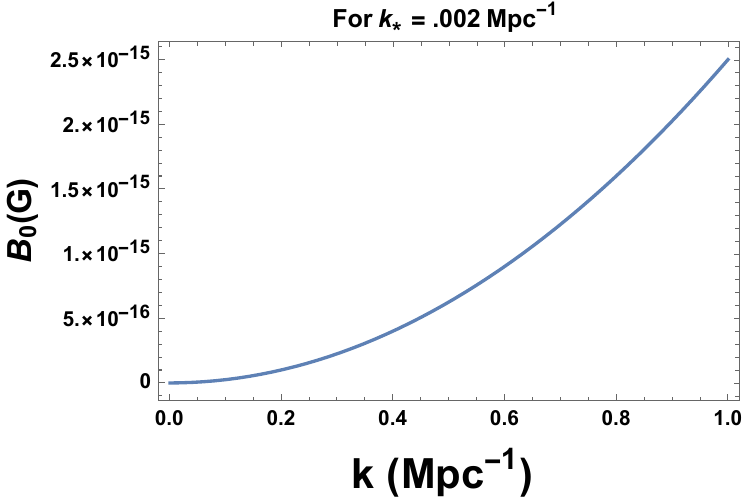}%
\quad
%\subfigure[]{%
\label{fig:Powertsecond}%
\includegraphics[height=2in]{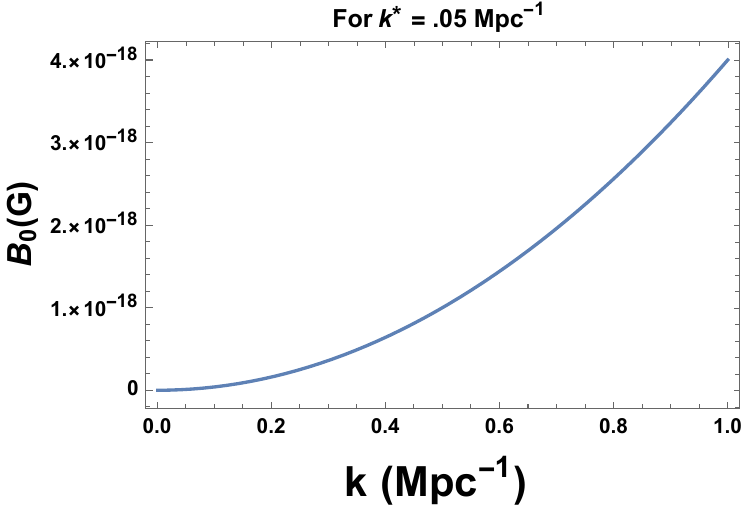}%
\caption{ Plot showing the strength of the present day primordial helical magnetic field at different length scales (from Mpc to Gpc), for two different pivot scales.}
\label{fig:PowerSpectrum}
\end{figure}
where $k_*$ is the pivot scale, and 
${\cal C}(k_*, \alpha)$ and ${\cal C}_2(k_*, \alpha)$ are constants that depend on the inflationary energy scale (See Eq. 44 in Ref. \cite{2020-Kushwaha.Shankaranarayanan-PRD}). For de Sitter inflation ($\alpha = - 1/2$), 
the present day magnetic field as a function of $k$ is given by:
\begin{equation}
B_0(k) \sim 10^{-20} \left( \frac{k}{k_*} \right)^2 \, \rm{G} 
\end{equation}
where we have included only the leading order contribution and have discarded the subleading contribution.

In \ref{fig:PowerSpectrum} we have plotted the power spectrum of the present day primordial helical magnetic field at different length scales for two pivot scales $k_* = 0.002 \, Mpc^{-1}$~\cite{Liddle.Lyth-Book} and $k_* = 0.05 \, Mpc^{-1}$~\cite{2018-Planck}. One can see from figure \ref{fig:PowerSpectrum} (left plot) that for around Mpc scale, the value of present day magnetic field is $10^{-15} G$. This is consistent with the current observations~\cite{2010-Kahniashvili.Ratra.etal-PRD,2013-Durrer.Neronov-Arxiv,2016-Subramanian-Arxiv,2004-Giovannini-IJMPD,2020-Vachaspati-RepProgPhys}.

Using the fact that modes exit the horizon around 5 e-foldings, 
\begin{align}\label{eq:eta_*}
\eta_* = \eta_{end} \cdot 10^2
\end{align} 
and $\mathcal{H} \sim {\eta_0}^{-1} \sim 10^{14} \rm{GeV} $, for $M \sim 10^{14} -10^{17} \rm{GeV}$~\cite{2020-Kushwaha.Shankaranarayanan-PRD,2004-Shankaranarayanan.Sriramkumar-PRD}, we obtain 
\[
\Theta  \approx  \sqrt{ \frac{\eta_{end} \cdot 10^{45} GeV^3  }{M^2 }} \, , 
\]
which is very small value.
Note also that 
\[
\eta_{end} = -\frac{1}{a(\eta_{end}) H} = -\frac{e^{-N_{\rm Inf}}}{H_{\rm Inf}} \approx 10^{-41} \, {\rm GeV}^{-1} \, .
\]
Using the fact that $\Theta$ is very small, we get
 \begin{align}\label{eq:ApproxCoefficients}
|C_1| \approx | C_3| \approx \sqrt{\Theta \, \eta_0} \,\, , \qquad \text{and} \qquad 
|C_2| \approx | C_4| \approx \sqrt{ \frac{\eta_0}{\Theta}}.
\end{align}
Hence, we obtain the following relations among the coefficients $|C_1| \approx |C_3| << |C_2| \approx |C_4|$.

\section{BAU parameter for arbitrary values of $\Lambda$ and $\mu$}
\label{app:Calculations}

\begin{figure}[!hbt]
%\centering
%\subfigure[]{%
%\label{fig:first}%
\includegraphics[height=2in]{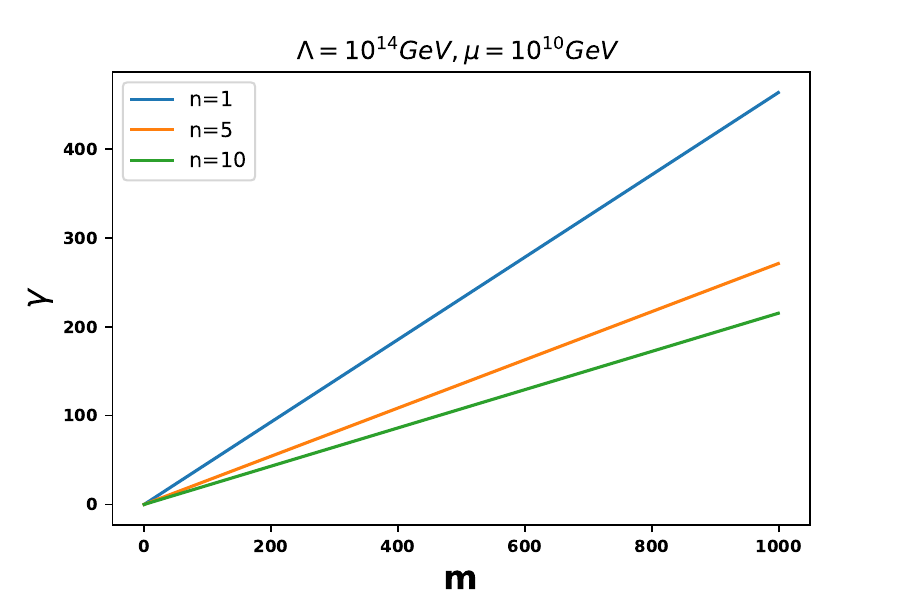}%
\quad
%\subfigure[]{%
%\label{fig:third}%
\includegraphics[height=2in]{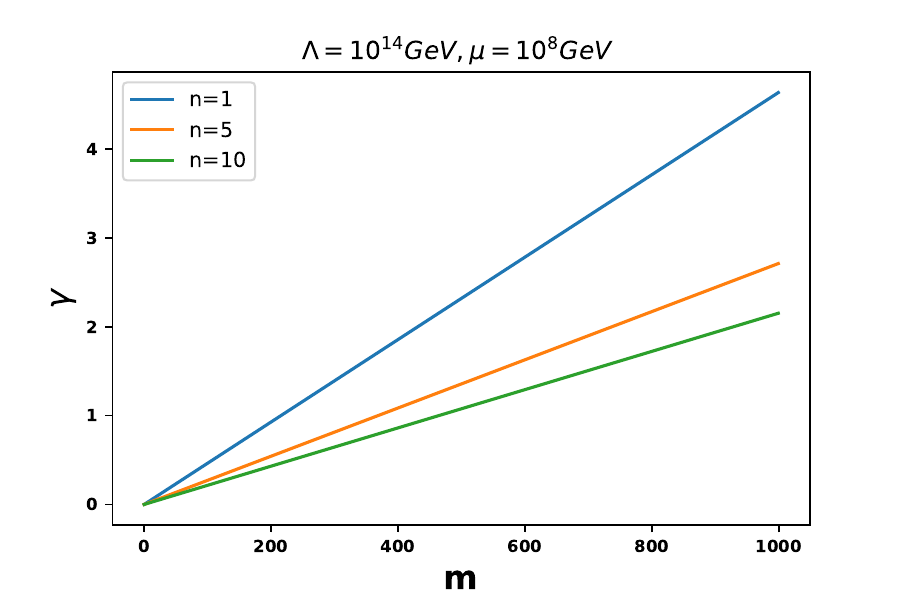}%
\quad
%\subfigure[]{%
%\label{fig:third}%
\includegraphics[height=2in]{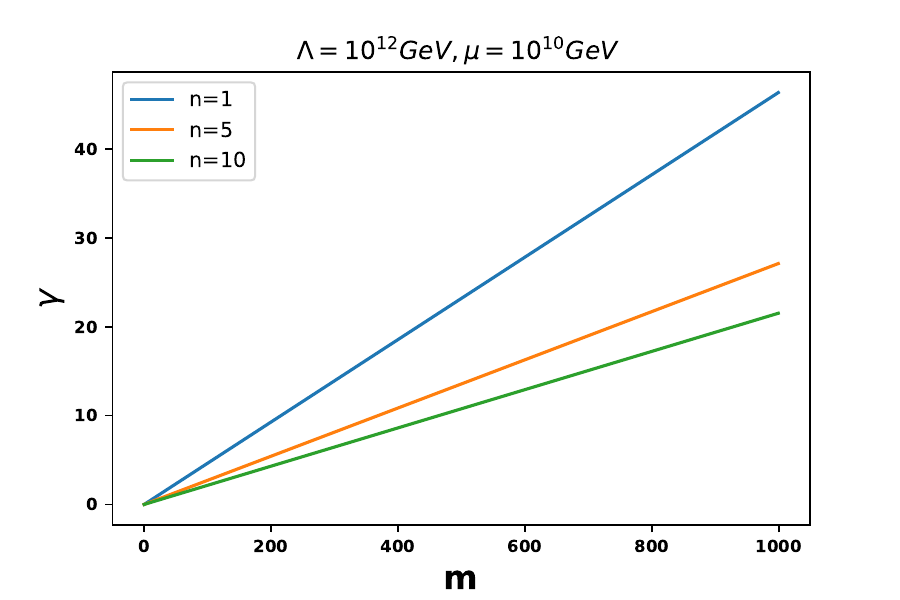}%
\quad
%\subfigure[]{%
%\label{fig:fourth}%
\includegraphics[height=2in]{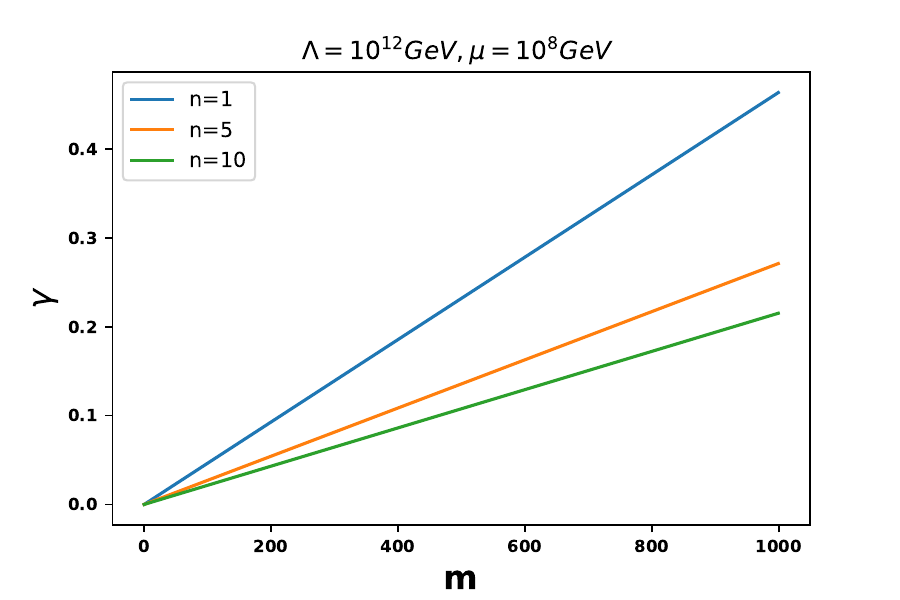}%
\caption{ Plot showing the behaviour of reheating temperature $T_{RH}$ (vertical axis) with parameter $M$ (horizontal axis), for different ranges of $\Lambda,\mu$. In upper panel ( and lower panel) $\Lambda$ is fixed at $10^{14} GeV$ ( and $10^{12} GeV$) and lower energy scale is varied.}
\label{fig:reheatingGeneric}
\end{figure}
Following Ref.\cite{2016-Subramanian-Arxiv}, we first evaluate the contribution of dilution factor for arbitrary values of $\Lambda$ and $\mu$. Assuming instantaneous reheating, the universe transited to radiation domination after inflation.  Using entropy conservation i.e. $g a^3T^3 = \mbox{constant}$ during its evolution, where $g$ is effective relativistic degrees of freedom, we have: 
\begin{equation}
\frac{a_{\mu}}{a_{\Lambda}} =\left( \frac{g_{\Lambda}}{g_{\mu}}\right)^{1/3} \frac{T_{\Lambda}}{T_\mu}
\label{eq:AppC01}
\end{equation}
where $a_{\Lambda}, a_{\mu}$ are the scale factors at which the helical modes with energy $\Lambda$ and $\mu$ reentered the radiation dominated Universe, $g_{\Lambda}, g_{\mu}$ are the effective relativistic degrees of freedom at which the helical modes with energy $\Lambda$ and $\mu$ reentered the radiation dominated Universe, and $T_{\Lambda}, T_{\mu}$ are the Universe temperatures at which the helical modes with energy $\Lambda$ and $\mu$ reentered the radiation dominated Universe.  The Friedmann equation is: 
\begin{align}
\label{eq:AppC02}
H^2 = \frac{1}{3 M_{\rm P}^2} \left[ g_{\Lambda} \left(\frac{\pi^2}{30} \right) T_{\Lambda}^4 \right] \, .
\end{align}
The baryogenesis occurs as soon as the helical modes reenter the Hubble radius during the radiation-dominated epoch. For simplicity, we assume that baryogenesis occurs at the start of radiation-dominated epoch, hence we take energy scale of $\Lambda$ to be of order $H$. Substituting $T_{\Lambda}$ from Eq. \eqref{eq:AppC02} in Eq. \eqref{eq:AppC01}, we get,
  \begin{align}
 \frac{a_{\mu}}{a_{\Lambda}} = \frac{g_\Lambda^{1/12}}{g_\mu^{1/3}} \left( \frac{90}{\pi^2} \right)^{1/4} \frac{M_{\rm P}^{1/2} \Lambda^{1/2}}{T_\mu}  \approx \frac{ 10^{9} \sqrt{\Lambda}}{\mu},
 \end{align}
where $g_\Lambda \sim 100$ (during reheating) and $g_\mu$ is of the order 10. Physically, we see that the Universe has expanded by a factor of $\frac{10^9 \, \sqrt{\Lambda}}{\mu}$ when the helical modes of energy $\Lambda$ and $\mu$ reenter the radiation dominated epoch and hence, 
in Eq.~(\ref{eq:n_B}), the inverse of this factor will act as the dilution factor. 

Setting $\Lambda = 10^{14}$~GeV, and 
$\mu = 10^{10}$~ GeV, in the above expression, we have: 
\[
\frac{a_{\mu}}{a_{\Lambda}} = \frac{g_f^{1/12}}{g_1^{1/3}} \left( \frac{90}{\pi^2} \right)^{1/4} \frac{\left( M_{\rm P} \Lambda\right)^{1/2}}{\mu}  \approx 10^{6} \, .
\]
For the modes that exit the horizon during inflation at $N=5$, we have: 
\[ 
\frac{1}{a^3} \sim 10^{-33} \frac{\mu^3}{\Lambda^{3/2}} .
\]
Therefore, for generic scales of baryogenesis, the BAU parameter \eqref{eq:baryon_Asym} is given by
\begin{align}\label{eq:baryon_Asym-M-mu}
\eta_B  \approx  \frac{ 10^{-38} \cdot \eta_0^2 }{ \sqrt{\eta_{end} \cdot 10^{45} GeV^3} } \frac{M^3  \left( \Lambda^3 \mu^3 - \mu^6 \right)  }{ \Lambda^{3/2} \,  T^3_{\rm{RH}}  } \quad \approx \quad 
10^{-11} \left(  \frac{M}{M_P} \right)^3  \frac{ \left( \Lambda^3 \mu^3 - \mu^6 \right) }{ \Lambda^{3/2} \,  T^3_{\rm{RH}}  }.
\end{align}
Using the parametrization in Eq. \eqref{eq:parametrization}, we have
\begin{align}
\frac{m^3 \left( \delta^3 \Delta^3 \, 10^{12} - \Delta^6  \right) }{ \delta^{3/2} \gamma^3} \approx n \, 10^{22}
\end{align}
where $\mu = \Delta 10^8$~GeV, $\Delta \in \{ 1, 100  \}$ and $ \gamma \in \{ 10^{-2}, 1000  \} $. \ref{fig:reheatingGeneric} shows the behaviour of the reheating temperature as a function of $M$, for different ranges of $\Lambda, \mu$. From the plots we infer that the results obtained in Sec. \eqref{sec:baryon_asymm} by neglecting $\mu$ are consistent with the results 
in this Appendix.
\section{Effect of the Riemann coupling near Schwarzschild black hole }
\label{app:blackhole}

In this appendix, we will show that $S_{\rm CB}$ coupling is tiny near the
solar mass Schwarzschild black-holes. Specifically, we evaluate at the 
Schwarzschild radius of a non-rotating spherically symmetric black hole of mass $\mu$. Since the calculation is order of magnitude, we calculate the Kretschmann scalar $K = R_{\mu\nu\alpha\beta} R^{\mu\nu\alpha\beta}$ for the Schwarzschild black-hole. For this case, the Kretschmann scalar $K$ at a radial distance $r$ from the black-hole center is given by:
\begin{align}\label{Kretschmann-scalar}
K = \frac{48 G^2 \mu^2}{c^4 r^6}
\end{align}
which implies that the Riemann tensor $\sim \sqrt{K} $. The coupling term
\begin{align}
\frac{\sqrt{K}}{M^2} \sim \frac{\sqrt{48} G \mu}{c^2 r^3} \frac{1}{M^2}.
\end{align}
For Schwarzschild radius $r_h = {2G \mu}/{c^2}$, the coupling term becomes 
\begin{align}\label{app:Kretschmann-scalar-coupling}
\frac{\sqrt{K}}{M^2} \sim   \sqrt{\frac{3}{4}}  \frac{c^4}{ G^2 \, \mu^2 \, M^2 } \, .
\end{align}
We set $\mu = 1 M_\odot $ where $ M_\odot \approx 10^{30} \rm{Kg}$ is the solar mass. Since the result might be interesting to various astrophysical and cosmological phenomenon, we will calculate the value of the coupling $\sqrt{K}/M^2$ (\ref{app:Kretschmann-scalar-coupling}) in units which are preferred in early universe cosmology (natural units) and gravity (geometrized units).

Natural units are preferred in the early Universe Physics, and all scales are rewritten in terms of $GeV$. For simplicity, we write the following quantities in terms of $m_P$ as,
\begin{align}\label{app:NaturalUnits}
G \approx M_P^{-2}, \qquad c = 1, \qquad M \approx 10^{-2} M_P, \qquad \mu = 10^{30} \rm{Kg} \approx 10^{38} M_P 
\end{align}
where we have used $M_P \approx 10^{19} \rm{GeV} \approx 10^{-8} \rm{Kg}$ and $M = 10^{17} \rm{GeV} \approx 10^{-2} M_P$. Using Eq.(\ref{app:NaturalUnits}) in Eq.(\ref{app:Kretschmann-scalar-coupling}), we get 
\begin{align}\label{app:KS-coupling-NaturalUnits}
\frac{\sqrt{K}}{M^2} \approx 10^{-72}.
\end{align}
To understand the effect in astrophysical phenomenon, we calculate the value of the coupling (\ref{app:Kretschmann-scalar-coupling}) in geometrized units ($c = G = 1$). Since our model parameter $M \approx 10^{17} \rm{GeV}$ has unit of energy, we need to substitute the conversion $M \rightarrow \frac{M G}{c^4}$ in
Eq.~(\ref{app:Kretschmann-scalar-coupling}), which gives 
\begin{align}\label{app:KS-coupling-SI}
\frac{\sqrt{K}}{M^2} \approx   \sqrt{\frac{3}{4}}  \frac{c^{12} }{ G^4 \, \mu^2 \, M^2 }.
\end{align}
Using the following values,
\begin{align}
c = G = 1,  \qquad 
M = 10^{-10} \rm{Kg},  \qquad 
\mu \approx 10^{30} \rm{Kg}
\end{align}
we obtain
\begin{align}\label{app:KS-coupling-SI_Units}
\frac{\sqrt{K}}{M^2} \approx 10^{-40} \, \rm{Kg}^{-4}.
\end{align}
Hence, from the above analysis, it is clear that the effect near the solar mass size black-hole is negligible. Note that since Riemann tensor is coupled with gauge kinetic terms $F_{\mu\nu}F^{\mu\nu}$, the coupling term contains the frequency of the electromagnetic waves, i.e., $\omega^2$, hence the coupling will have effects at very high frequency. 
The coupling constant \eqref{app:Kretschmann-scalar-coupling} 
is of the order of one or greater for black-holes of mass $\mu \sim 100 M_P$, i.e.,

\begin{align}
 \frac{ \sqrt{K} }{M^2} \sim 1, \qquad \text{for} \quad \mu \sim 100 M_{P}  
\end{align}
Such size primordial black-holes form in the very early Universe. For such black-holes, Hawking temperature is~\cite{2008-Das.Shanki.Sur-Review} 
\begin{align}
T_H = \left(  \frac{\hbar c^3}{G k_B}  \right) \frac{1}{8\pi \mu} \quad \approx  \quad 10^{-3} M_P 
\end{align}
Since the Hawking radiation is thermal, we can obtain the peak wavelength of the black-body spectrum from Wien's displacement law $\lambda_{\rm{max} } T = \rm{constant}$. The wavelength $\lambda_{\rm{max} }$ corresponding to $100 M_P$ mass black holes is given by: 
\begin{align}
 \lambda_{\rm{max} } \sim T_H^{-1} \approx 10^3 M_P^{-1} 
\end{align}
Using the conversion $1~\rm{GeV} = 5.06 \times 10^{15} m^{-1}$ which gives $M_P \approx 10^{34} m^{-1}$, we get  
\begin{align}
 \lambda_{\rm{max} } \approx 10^3 M_P^{-1} \approx 10^{-31} m, \qquad \implies \nu = \frac{c}{\lambda_{\rm{max}} } = 10^{39} s^{-1}. 
\end{align}
Assuming these PBHs are produced just after bigbang, the above frequency will be redshifted by a factor $10^{20}$. Thus, the redshifted frequency is $\sim 10^{19} Hz$ (Gamma-ray) and can have potential signatures. However, it is important to note that at that scale, we need to include higher-order corrections. 
As mentioned earlier, in this analysis, we have ignored the mass dimension 8 operators~\cite{2019-Ruhdorfer.etal-JHEP}. At the Planck scale, we need to include dimension 8 and beyond. Hence, to understand these effects of PBH of Planck mass size, we need to include mass-dimension 8 operators and beyond.
\chapter{Effective field theory of magnetogenesis: Details} %Effective field theory of magnetogenesis identify necessary and sufficient conditions
\label{apdetail:EFTmagnetogenesis}

In this Appendix, we provide more details on the results from chapter~\ref{ch:EFTmagnetogenesis}.

\section{Decoupling of vector perturbations and gauge fields}
\label{sec:Decoupling}

As mentioned in the Introduction, we aim to construct an EFT of magnetogenesis. Like any other effective theory, the EFT of magnetogenesis includes two components: Symmetries and degrees of freedom~\cite{2007-Burgess-ARNPS,2020-Penco-Arxiv}. Thus, like any perturbation theory in gravity, the EFT of magnetogenesis requires the inclusion of fluctuations in the matter and metric degrees of freedom. As mentioned above, in the EFT of inflation, one makes a specific gauge choice in which the inflaton fluctuations vanish~\cite{2007-Cheung.etal-JHEP}. To consistently investigate magnetogenesis, we need to include the vector perturbations in the metric~\cite{2004-Battefeld.Brandenberger-PRD,2004-Lewis-PRD,2014-Basak-Shanki-JCAP}. In this appendix, we show that the vector perturbations do not have temporal evolution; hence, only the gauge field is the relevant gauge-invariant variable for the EFT.

We consider vector perturbations about a spatially flat $(3+1)-$dimensional FRW line element:
\begin{equation}
ds^2 = \overline{g}_{\mu\nu} dx^{\mu} dx^{\nu} = 
N^2(t) dt^2 - a^2(t) \delta_{ij} dx^i dx^j = 
N^2(\eta) \, d \eta^2 - a(\eta)^2\delta_{ij} dx^i dx^j \, .
\label{def:FRW}
\end{equation}
The only degree of freedom in the metric is the usual expansion parameter $a(t)$, a function only of time by virtue of the homogeneity of the three-space. The lapse function $N(t)$ simply represents the time reparametrization invariance.

The conformal time ($\eta$), 
is related to the cosmic time by the relation $t =\int d\eta a(\eta)$, $a(\eta)$ is the scale factor and $N(\eta) = a(\eta)$. $H$ is the
Hubble parameter given by $H \equiv \dot{a}/{a}$ while $\mathcal{H} \equiv a'/a$ is related to the Hubble parameter by the relation $\mathcal{H} = H\,a$.

At the linear level, for a single scalar field inflation with gauge-field $(A_{\mu})$, the metric perturbations ($\delta g_{\mu\nu}$) can be categorized into three distinct types — scalar, vector and tensor perturbations~\cite{1984-Kodama.Sasaki-PTPS,1992-Mukhanov.etal-Phy.Rep.}. As mentioned above, we will consider the gauge field perturbations for magnetogenesis. Hence, the perturbed FRW line-element corresponding to vector perturbations can be written as~\cite{1984-Kodama.Sasaki-PTPS,1992-Mukhanov.etal-Phy.Rep.,2004-Battefeld.Brandenberger-PRD},
\begin{equation}
\label{eq:VectPert}
ds^2 
= a^2(\eta)\left[ d\eta^2 + 2 S_i d \eta dx^i - \left(\delta_{ij}+ 2 F_{(i,j)} \right) \, dx^i dx^j \right] \, ,
\end{equation}
where $S_i$ and $F_i$ are pure vector contributions and are divergenceless~\cite{1984-Kodama.Sasaki-PTPS,1992-Mukhanov.etal-Phy.Rep.}:
\begin{equation}
    \partial_i S^i = 0; ~~
    \partial_i F^i = 0;~~F_{(i,j)}=\frac{F_{i,j}+F_{j,i}}{2} \, .
\end{equation}
The above-perturbed metric has four unknown vector-mode functions. Since there are only two physically relevant vector modes, we must make certain choices on $S_i$ and $F_i$. We choose~\cite{2004-Battefeld.Brandenberger-PRD}:
\begin{equation}
    F_i = 0.
\end{equation}

To keep calculations transparent, let us consider the following action~\cite{2021-Giovannini-PLB}: 
\begin{equation}\label{test-L}
\mathcal{S}= S_{\rm EH} + S_{\rm Inflaton} + S_{\rm EM} + S_{\rm EM-NM}
\end{equation}
where,
\begin{eqnarray}
S_{\rm EH} &=& - \int d^4x \sqrt{-g} \frac{M_{\rm Pl}^2}{2} R \, , \nonumber \\
%%%%
S_{\rm Inflaton} &=& \int d^4x \sqrt{-g} \left[ \frac{1}{2}\partial_{\mu} \phi \partial ^{\mu} \phi - V(\phi)  \right], \nonumber \\
\label{def:EMaction}
S_{\rm EM} &=& - \frac{1}{4} 
\int d^4x \sqrt{-g} \,  
F_{\mu \nu} F^{\mu \nu} 
\, , \\
%%%
\label{def:EM-Nonminimal}
S_{\rm EM-NM} &=& \frac{1}{M^2}\int d^4x \sqrt{-g}\left[\Lambda_1(\varphi)R F_{\mu \nu} F^{\mu \nu} +
\Lambda_2(\varphi) R^{\nu}_{\mu}F_{\nu \alpha} F^{\mu \alpha}+ \Lambda_3(\varphi)R_{\mu\nu\alpha\beta}F^{\mu\nu}F^{\alpha\beta}\right. \nonumber \\ 
& & \left. 
+ \Lambda_4(\varphi)C_{\mu\nu\alpha\beta}F^{\mu\nu}F^{\alpha\beta}+ 
 \Lambda_5(\varphi)
\Box \varphi F_{\mu \nu} F^{\mu \nu} +
\Lambda_6(\varphi)\partial_{\mu}\varphi
\partial^{\nu}\varphi F^{\mu \alpha}F_{\nu \alpha}
\right.  \nonumber \\ 
& & \left. + \Lambda_7(\varphi)
\nabla_{\mu}\nabla^{\nu}\varphi F_{\nu \alpha} F^{\mu \alpha}\right] \nonumber
\end{eqnarray}

$\phi$ is the canonical scalar (inflaton) field, $V(\phi)$ is the self-interacting potential, $\varphi = \phi/M$ with $M$ being some characteristic scale,
$F_{\mu \nu} \equiv \partial_{\mu} A_{\nu} - \partial_{\nu} A_{\mu}$ is the electromagnetic tensor. Eq.~\eqref{def:EMaction} is the standard gauge-field action while action \eqref{def:EM-Nonminimal} corresponds to all possible non-minimal coupling terms of the gauge field with gravity and the inflaton~\cite{2021-Giovannini-PLB}. Owing to the fact that the background space-time is FRW, the natural choice for the \emph{classical} gauge field in the background space-time is $\bar{A}_{\mu} = 0$. Splitting the gauge-field in the background and perturbations \eqref{eq:VectPert}, we have:
\begin{equation}
    A_{\mu} = \bar{A}_{\mu} + \delta A_{\mu} = \delta A_{\mu}.
\end{equation}
Like scalar perturbations, gauge field perturbations are quantum fluctuations. 
In terms of the Stewart-Walker lemma, the quantities which vanish in the FRW universe are gauge invariant
perturbations~\cite{1974-Stewart-Walker-PRSLA}. 
In this case, it is easy to check that $\delta A_{\mu}$ is gauge invariant. As mentioned above, there are only two physically relevant vector modes on the metric side. This implies that the gauge field should have only two physically relevant modes. One of the simplest and most well-known choice is the Coulomb gauge, i. e., 
\begin{equation}
\delta A_0 = 0, \partial_i \delta A^{i}=0 \, .
\label{def:CoulombG}
\end{equation}

With the above choices for the metric and gauge field, expanding action \eqref{test-L} to second order in $S_i$ and $A_{\mu}$, we have:
\begin{equation}\label{vec-L}
    \mathcal{S}^{(2)} = \frac{M_{\rm Pl}^2}{4}
    \int d^4x a^2  \left(\partial_j S^i\right)^2 +
    \frac{1}{2} \int d^4x \, \left[g_1(a,H,M){\delta A_i^{\prime}}^2 - \partial_j (g_2(a,H,M)\delta A_i)^2  \right] \, ,
\end{equation}

where $g_1(a,H,t,M)$ and $g_2(a,H,t,M)$ are functions of scale factor 
$a$, Hubble parameter $H$. In the case of minimal coupling, $g_1, g_2$ become unity.
This is a crucial result, and the EFT we develop in this work depends on this, regarding which we would like to discuss the following points: First, 
the vector perturbation $S_i$ does not have any temporal evolution, while the gauge fields have temporal evolution even if one considers non-minimal couplings of the  gauge field with gravity and inflaton field as in
\eqref{def:EM-Nonminimal}. Since $S_i$ does not evolve in time, it does not affect the dynamics of the gauge field. Thus, $S_i$ can be considered an irrelevant degree of freedom. Second, 
the above action does not contain any coupling between $\delta A_i$ and $S_i$. Hence, the vector perturbation $S_i$ does not affect the dynamical evolution of the gauge field $\delta A_{i}$. This is even more interesting considering that the analysis is also valid for non-minimal coupling. Even though initially, the energy density of the gauge field may be negligible compared to the inflaton energy, due to non-minimal coupling, the energy density of the gauge field will evolve non-trivially during
the inflationary phase. 
Consequently, non-minimal coupling terms may cause isocurvature density perturbations~\cite{2005-Shankaranarayanan.Lubo-PRD}, and 
the gauge field can not be treated as a spectator field. However, from Eq. \eqref{vec-L}, we see that the vector modes do not evolve in time, and the only relevant degree of freedom is the gauge field.  See Ref.~\cite{Kubota:2022pit} for a similar study in the context of inflationary paradigm with a spectator field with non-minimal coupling. Third, the above analysis can be extended to the case where the 3-space is not flat. In that case, also $S_i$ does not evolve in time. Lastly, we can extend the computation for $f(R)$ theories as in second order in
perturbation $S_i$ will not have any time evolution for this kind of theory.
This conclusion can also be drawn for 
Horava-Lifshitz gravity \cite{2010-Gong.etal-PRD}. 
%Fifth, the above analysis can be extended to the bouncing scenario~\cite{2020-Nandi-PLB}, implying that the effective degrees of freedom for the vector perturbations is the gauge field $A_i$.

Lastly, constructing EFT magnetogenesis requires consistently counting the number of degrees of freedom to write down the EFT Lagrangian. Hence, we need to find any possible interaction between the vector modes of perturbation $S_i$ and $\delta A_i$. From \eqref{test-L}, one can see that the lowest order
interaction between $S_i$ and $\delta A_i$ would come from the standard electromagnetic term
$F_{\mu\nu}F^{\mu\nu}$ and is:
\begin{equation}
    F_{\mu \nu}F^{\mu \nu} = 2 S^i g^{jk} \, \partial_0 \delta A_j \, 
    \left(\partial_i \delta A_k - \partial_k \delta A_j\right) \, .
\end{equation}
As can be seen, this interaction term is third-order. Similarly, the non-minimal terms in Eq. \eqref{def:EM-Nonminimal} can also produce higher order interaction between gauge field
and vector modes.
While these are important for higher-order computations where one also has to include scalar and tensor perturbations~\cite{2010-Christopherson.etal-PRD}; for the current work, these terms are irrelevant. Hence, the interaction between $S_i$ and $\delta A_i$ can be ignored for the EFT of magnetogenesis containing quadratic terms, and the only relevant degrees of freedom will be the gauge field $A_{\mu}$.

\section{Correspondence between EFT parameters and magnetogenesis models}\label{app:coup-details}

In this appendix, we provide detailed calculations that show a one-to-one mapping between the magnetogenesis model and EFT parameters defined in Eq.~\eqref{eft:A}. Specifically, we consider 
Ratra model~\cite{1991-Ratra-Apj.Lett}, gravitational coupling of the electromagnetic field~\cite{1988-Turner.Widrow-PRD,2020-Kushwaha.Shankaranarayanan-PRD}, higher-order gravitational coupling~\cite{2022-Bertolami.etal-arXiv} and Galileon vector fields~\cite{2017-Debottam.Shankaranarayanan-JCAP}.

\subsection{Class of Ratra Model with integer exponent}

The Lagrangian corresponding to the conformal invariance breaking term in the Ratra model is~\cite{1991-Ratra-Apj.Lett}:
\begin{equation}
\mathcal{S}_{\rm Ratra} = \int d^4x \sqrt{-g} \, f^2(\phi)F_{\mu\nu}F^{\mu\nu} 
\end{equation}
where $f(\phi)$ is an arbitrary function of the inflaton field. 
While there have been many simple choices, here, we consider one choice~\cite{2012-Barnaby.Namba.Peloso-PRD}:
\begin{equation}
\label{def:fphi}
f(\phi) = f_0 \exp{\left(-\frac{C \phi^2}{\Lambda^2}\right)},
\end{equation}
where $f_0, C$ are positive constants. Note that in the limit of $\Lambda \to \infty$, $f(\phi) \to f_0$ and is consistent with the series expansion~\eqref{eft:A}.

For the inflationary potential~\cite{Chervon:2017kgn}
\begin{equation}
V(\phi) = -\frac{8}{3} M^2_{\rm Pl} \lambda_2 \phi^2 + \lambda_2 \phi^4 \, ,
\end{equation}
the scalar field $\phi$ evolves as:
\begin{equation}
    \phi^2 = \sqrt{3} M_{\rm Pl} H
\end{equation}
Substituting the above expression in Eq. \eqref{def:fphi}, we have:
\begin{equation}
    f(\phi) = f_0 \exp{\left(-\sqrt{3} \tilde{C} \frac{H}{\Lambda}\right)} \qquad \tilde{C} = C \frac{M_{\rm Pl}}{\Lambda}
\end{equation}
Expanding the above form of $f(\phi)$ implies that the series will only consist of integer powers of $H$ and can be easily mapped to the EFT action \eqref{EFT:L}.
%Thus, the Ratra model for the above form of $f(\phi)$ correspond to the terms 
%$\left(\frac{\mathcal{H}}{\Lambda}\right)^{C \sqrt{\frac{p}{2}}}$ term  in the EFT action \eqref{EFT:L}. Similarly, for other forms of $f(\phi)$ one can obtain a one-to-one mapping with the EFT parameters. 
Recently, in Ref.~\cite{2022-Durrer.etal-arXiv}, authors have considered magnetogenesis in the Higgs-Starobinsky model.  This leads to a generalized form of $f(\phi)$ in the Einstein frame.  By repeating the 
same procedure, we can obtain a one-to-one mapping with the EFT parameters $s_n, d_n$.

%\subsubsection{Slow-roll Inflation}
%\textcolor{blue}{
%For slow roll inflation the inflaton field value can be estimated as,
%\begin{equation}
%    V(\phi) \sim \frac{1}{8 \pi G} (3 H^2 + \dot{H})
%\end{equation}
%So for potentials of the type of $\lambda_n \phi^n$ (where $\lambda_n $ is some
%coupling constant)we can have,%
%\begin{equation}
%    \phi \sim \frac{1}{8 \pi G \lambda_n} (3 H^2 + \dot{H})^{\frac{1}{n}}
%\end{equation}
%Now if we consider a conformal symmetry breaking coupling function for magnetogenesis %as,
%\begin{equation}
%    f(\phi) \sim (\phi / \phi_0)^m, 
%\end{equation}
%then we can see that for a choice of $(m/n) \rightarrow$ integer, the model can be 
%mapped to our EFT expansion. 
%Other choices of potential can also be mapped to our EFT expansion for appropriate 
%choices of parameters.}

\subsection{Gravitational Coupling}

The Lagrangian corresponding to the non-minimal coupling of the electromagnetic field tensor with curvature terms are~\cite{1988-Turner.Widrow-PRD}:
\begin{equation}\label{gr-coup}
\mathcal{S}_{\rm NC} =\frac{1}{4}\int d^4 x \left(- \Gamma_1 R F_{\mu\nu}F^{\mu\nu}
    - \Gamma_2 R_{\mu\nu}g_{\alpha \beta}F^{\mu\alpha}F^{\nu\beta}
    - \Gamma_3 R_{\mu\nu\alpha\beta} F^{\mu\nu}F^{\alpha\beta}\right)
\end{equation}
where $\Gamma_1, \Gamma_2$ and $\Gamma_3$ are coupling constants. For the FRW background \eqref{def:FRW}, the Ricci scalar ($R$) is: 
\begin{equation}
\label{FRW:Ricciscalar}
    R = -6 \left( \frac{\mathcal{H}^2+\mathcal{H}^{\prime}}{a^2}\right).
\end{equation}
Substituting in the term in Eq. \eqref{gr-coup}, we can see that the conformal invariance breaking terms correspond to 
$s_2 =b_1 = d_2 =  e_1 = -3$ in Eq.~\eqref{eft:A}. Let us now focus on the 
second term ($R_{\mu\nu}g_{\alpha \beta}F^{\mu\alpha}F^{\nu\beta}$). For the FRW background, we have: 
\begin{equation}
   R_{\mu\nu}g_{\alpha \beta}F^{\mu\alpha}F^{\nu\beta}=
   -\frac{1}{a^2}\left(\frac{\mathcal{H}^2}{2}+\mathcal{H}^{\prime} \right)
   (A_i^{\prime})^2+ 
   \frac{1}{a^2} \left(\frac{\mathcal{H}^{\prime}}{2}+\mathcal{H}^2\right)
    (\partial_i A_j)^2
\end{equation}
Comparing the above expression with Eq.~\eqref{eft:A}, we see that the conformal invariance breaking terms correspond to $s_2=-\frac{1}{2}, b_1 =-1$ 
and $d_2 =-1, e_1=-\frac{1}{2}$. Finally, expanding $R_{\mu\nu\alpha\beta} F^{\mu\nu}F^{\alpha\beta}$ term for the FRW background, we have: 
\begin{equation}
    R_{\mu\nu\alpha\beta} F^{\mu\nu}F^{\alpha\beta} = 
    -\frac{\mathcal{H}^{\prime}}{a^2}(A_i^{\prime})^2 + \frac{\mathcal{H}^2}{a^2} (\partial_i A_j)^2.
\end{equation}
Here again, choosing $b_1=d_1=-1$ in \eqref{eft:A} corresponds to the Riemann coupling terms in the above action 
\eqref{gr-coup}.

\subsection{Higher Order Gravitational Coupling}

The Lagrangian corresponding to the non-minimal coupling of the electromagnetic field tensor with higher-order Ricci scalar is~\cite{2022-Bertolami.etal-arXiv}:
\begin{equation}
\mathcal{S}_{\rm Higher} 
=\int d^4x \sqrt{-g} \, R^3 F_{\mu\nu}F^{\mu\nu}
\end{equation}
Substituting the Ricci scalar \eqref{FRW:Ricciscalar} in the above expression, it can be shown that the above 
action has a one-to-one correspondence with EFT action \eqref{eft:A}, for the following choice of parameters: 
\[
s_6 = -6, b_3 = -6, g_{41} = -18, g_{22} = 
-18, d_6 = -6, e_3 = -6, h_{41} = -18, h_{22} = -18 \, .
\]

\subsection{Vector Galileon Model}
\label{app:VGModel}

Let us now consider the conformal invariance breaking term in the vector Galileon model~\cite{2017-Debottam.Shankaranarayanan-JCAP}:
 \begin{equation}
\mathcal{S}_{\rm VG} =  2 \, D 
\int d^4 x \left[-\frac{a'^2}{N^3 a}(A_i')^2 + \left(\frac{a''}{N a^2} - \frac{a' N'}{N^2 a^2}\right)(\partial_i A_j)^2 \right] 
\label{Galileon}
\end{equation}
where for the FRW metric \eqref{def:FRW} the lapse function $N = a(\eta)$. Comparing 
\eqref{Galileon} and \eqref{eft:A}, 
we see that all parameters except $s_2$  and $e_1$ vanish in \eqref{eft:A}. We thus get: 
\begin{equation}
\label{eq:VG-EFTComp}
s_2 = e_1, ~{\rm and}~D = - \frac{s_2}{\Lambda^2} \, .
\end{equation}
%========  Energy Density for EFT ================

%
\section{EFT action with different coupling functions for the components of the gauge field}\label{general-propagation}
In general, we can have different coefficients in front of components of the gauge field $(A_1, A_2, A_3)$ in the EFT action~(\ref{EFT:L}). However, as shown in this appendix, this will lead to different propagation speeds for these three components. To see this, we consider the following action:
\begin{equation}\label{general-coeff}
\mathcal{S} = \int d^4 x \left\lbrace\left( f_{1} \, A_1^{\prime 2} + f_{1,2} \, A_2^{\prime 2} + f_{1,3} \, A_3^{\prime 2}\right) - \left( f_{2} \, (\partial_j A_1)^2 + f_{2,2} \,(\partial_j A_2)^2 + f_{2,3} \, (\partial_j A_3)^2\right)\right\rbrace.
\end{equation}
First, we can introduce the following field redefinitions of the gauge field components $(A_2, A_3)$:
\begin{equation}\label{redef}
\tilde{A}_2^{\prime} = \left(\frac{f_{1,2}}{f_1}\right)^{1/2} A_2^{\prime} ; \quad 
\tilde{A}_3^{\prime} = \left(\frac{f_{1,3}}{f_1}\right)^{1/2} A_3^{\prime}  \, .
\end{equation}
Thus, \eqref{general-coeff} becomes:  
\begin{equation}\label{general-coeff2}
\mathcal{S} = \int d^4 x \left\lbrace f_{1} \left( A_1^{\prime 2} +  \tilde{A}_2^{\prime 2} + \tilde{A}_3^{\prime 2}\right) - \left( f_{2} \, (\partial_j A_1)^2 + f_{2,2} \,(\partial_j A_2)^2 + f_{2,3} \, (\partial_j A_3)^2\right)\right\rbrace.
\end{equation}
Here we can notice that the first three terms have the same coupling functions as $f_1$, but the action is written in terms of $A_1$, $A_2$, $A_3$ and $\Tilde{A}_2$ and $\Tilde{A}_3$. To write down the action in terms of $A_1$, $\Tilde{A}_2$ and $\Tilde{A}_3$ we next focus on the single component $A_2$ in \eqref{general-coeff2} which can be written in terms of $\Tilde{A}_2$ as,
\begin{equation}
    A_2 = \left(\frac{f_{1}}{f_{1,2}}\right)^{1/2} \tilde{A}_2 - \int d\eta \, \frac{d}{d \eta}\left(\frac{f_{1}}{f_{1,2}}\right)^{1/2} \tilde{A}_2(\eta)
\end{equation}
then $f_{2,2} \,(\partial_j A_2)^2$ term in \eqref{general-coeff2} becomes:
\begin{align}
    f_{2,2} \,(\partial_j A_2)^2 &= f_{2,2} \, \left(\frac{f_{1}}{f_{1,2}}\right) (\partial_j \tilde{A}_2)^2 + f_{2,2}\int d\eta \, d\eta^{\prime} \, \frac{d}{d \eta}\left(\frac{f_{1}}{f_{1,2}}\right)^{1/2} \partial_j \tilde{A}_2(\eta) \frac{d}{d \eta^{\prime}}\left(\frac{f_{1}}{f_{1,2}}\right)^{1/2} \partial_j \tilde{A}_2(\eta^{\prime}) \nonumber \\ 
    &{} - 2 f_{2,2} \, \left(\frac{f_{1}}{f_{1,2}}\right)^{1/2} \partial_j \tilde{A}_2 \int d\eta \, \frac{d}{d \eta}\left(\frac{f_{1}}{f_{1,2}}\right)^{1/2} \partial_j \tilde{A}_2(\eta)\\
    &= F_{2,2} (\partial_j \tilde{A}_2)^2
\end{align}
where, $F_{2,2}$ is a function of $f_1$, $f_{1,2}$, $f_{2,2}$ and $\eta$. A similar exercise can be done for the $A_3$ component to express it in terms of $\Tilde{A}_3$, and we can introduce a new coupling function as $F_{2,3}$. Hence, after the field redefinition \eqref{general-coeff2} can be written as,
\begin{equation}\label{general-coeff3}
\mathcal{S} = \int d^4 x \left\lbrace f_{1} \left( A_1^{\prime 2} +  A_2^{\prime 2} + A_3^{\prime 2}\right) - \left( f_{2} \, (\partial_j A_1)^2 + F_{2,2} \,(\partial_j A_2)^2 + F_{2,3} \, (\partial_j A_3)^2\right)\right\rbrace.
\end{equation}
Here we have omitted the "tilde" from the fields. We can also notice that even if the temporal derivatives of the field components have the same coupling function, the spatial derivative part of the field components have different coupling functions. Hence, the propagation speeds of the three components are:
\begin{equation}
c_{A_1} = \left(\frac{f_{2}}{f_{1}}\right)^{1/2} ; \quad     \tilde{c}_{A_2} = \frac{F_{2,2}}{f_1} ; \quad
     \tilde{c}_{A_3} = \frac{F_{2,3}}{f_1}
\end{equation}
Hence, the propagation speed of the different components is different. The analysis in this work can be extended for this case.
%
%\begin{equation}
%    c_{A_1} = \left(\frac{f_{2}}{f_{1}}\right)^{1/2} ; \quad
 %   c_{A_2} = \left(\frac{f_{2,2}}{f_{1,2}}\right)^{1/2} ; \quad
 %   c_{A_3} = \left(\frac{f_{2,3}}{f_{1,3}}\right)^{1/2}
%\end{equation}

\section{Energy density for EFT action in terms of Lapse function}
\label{app:rhoEFT}
Vector perturbations can affect the background metric and matter configuration in which the perturbations propagate. This has been extensively studied for density perturbations during inflation and is also operational for other perturbations like vector perturbations. In order to quantify the effect of these EFT perturbations on the FRW background, we need to obtain the energy density ($\rho_{\rm EFT}$) corresponding to the EFT action \eqref{EFT:L}. 
%========

 To obtain the expression for the energy density, it is convenient to use the ADM formalism which simplifies the calculations significantly. The metric in ADM formalism is given by~\cite{1992-Mukhanov.etal-Phy.Rep.}:
 \begin{align}\label{appeq:ADM}
     ds^2 = N^2 d\eta^2 - \gamma_{ij} (N^i d\eta + dx^i) (N^j d\eta + dx^j)
 \end{align}
 where $N$ is lapse function, $N^i$ is shift vector and $\gamma_{ij}$ is the metric on constant $\eta-$hypersurface. Note that with $N^i =0$ and $\gamma_{ij} = a^2(\eta) \delta_{ij}$, we obtain flat FRW metric (\ref{def:FRW}). 
In the flat 
FRW background (\ref{def:FRW}), for arbitrary $N(\eta)$, the standard electromagnetic action becomes
\begin{align}\label{appeq:EMaction-FRW}
 S_{EM} = \frac{1}{2} \int d^4x \, \left[ \frac{a}{ N} {A_i^{\prime}}^2 - \frac{N}{a}(\partial_i A_j)^2  \right] \, ,
\end{align}
and the vector Galileon action (ignoring the total derivative term) is ~\cite{2017-Debottam.Shankaranarayanan-JCAP}:
\begin{eqnarray}
\label{appeq:TheModelFRW-simplify}
 \mathcal{S}_{VEC} =  2D \, \int\,d^4x  \Big[- \,\frac{a^\prime{}^2}{N^3\, a}\,A_i^\prime{}^2 +   \frac{2 {a^{\prime}}^2}{N\,a^3}\, \left(\partial_i A_j\right)^2  - \frac{2 a^{\prime}}{N\,a^2}\,\partial_i A_j \, \partial_i A_j^{\prime}\Big] \, .
\end{eqnarray} 
The lapse function does not have a kinetic term, and the variation of the total (gravity and matter) action w.r.t $N$ will lead to a constraint. Rewriting the total action 
in the form 
\begin{equation}
S_{\rm Grav-Mat} = S_{\rm EH} + S_{\rm tot} = \int d^3x a^3 \int dt \left[ p_a \dot{a} + p_{A_i} \dot{A}_i - N \, H_{A_i} \right] \, ,
\end{equation}
where $p_a$ is the canonical momentum corresponding to $a$, $p_{A_i}$ is the canonical momentum corresponding to $A_i$ and $H_{A_i}$ is the Hamiltonian corresponding to the EM field, including the Vector Galileon term.

We can now extend the above analysis for the EFT action \eqref{EFT:L}. In other words,  the EFT action can be written as:
\begin{equation}
\mathcal{S}_{\rm EFT} = \int d^3x a^3 \int dt \left[  p_{A_i} \dot{A}_i - N \, H_{\rm EFT} \right] \, ,
\end{equation}
where $H_{\rm EFT} = \int d^3x a^3 \rho_{\rm EFT}$ is the Hamiltonian 
corresponding to the EM field with EFT expansion. The corresponding energy density is:
\begin{align}\label{appeq:rho-def}
\rho_{\rm EFT} = T^0_0 = g^{00} T_{00} = - \frac{1}{a^3} \frac{\delta \mathcal{S}_{\rm EFT} }{\delta N} \, .
\end{align}
Note that in order to use the above method,
it is convenient to rewrite the expansion parameter in terms of the lapse function in the FRW line element (\ref{def:FRW}).
However, when we wrote the expansion parameters in Eq.~(\ref{eft:A}) we have set $N(\eta) = a(\eta)$. Hence, we first need to reinstate the lapse function $N(\eta)$ in Eq.~(\ref{eft:A}). This can be done in two steps: 
\begin{enumerate}
   \item[Step 1:] We first write the EFT parameters $f_i(\mathcal{H}, a, \eta)$, defined in Eq.~(\ref{eft:A}) (for $i = 1,2$), in terms of the cosmic time:  
   \begin{align}
    f_i(\mathcal{H}, a, \eta)  
    \xrightarrow{a\, d\eta \, \rightarrow \, dt}
    f_i(H, a, t)   
    \end{align}
\item[Step 2:] Rewrite the EFT parameters in cosmic time $f_i(H,a,t)$ in terms of conformal time and arbitrary lapse function $N(\eta)$:  
    \begin{align}
    f_i(H, a, t)   
    \xrightarrow{dt \, \rightarrow \, N d\eta}
    f_i(\mathcal{H}, a, N, \eta) 
    \end{align}
\end{enumerate}    
We need to be careful to obtain $f_i(\mathcal{H}, a, N, \eta)$,  as the zeroth order terms (i.e., $s_0, d_0$) in Eq.~(\ref{eft:A}) are independent of $N(\eta)$. 
%a(\eta)$ in the setting $N(\eta) = a(\eta)$(which is used in Eq.(\ref{eft:A}) ). 
We fix these zeroth-order terms in the EFT expansion parameters with respect to the standard electromagnetic action for the metric (\ref{def:FRW}): 
    \begin{align}
        S_{\rm EM} = -\frac{1}{4} \int d^4x \sqrt{-g} F_{\mu\nu} F^{\mu\nu} = \int d^4x \left[ \frac{a(\eta)}{2N(\eta)} (A_i^{\prime})^2 - \frac{N(\eta)}{2a(\eta)} (\partial_i A_j)^2 \right]
    \end{align}
Thus, by fixing the zeroth-order EFT parameters with $S_{\rm EM}$ as $a(\eta) s_0/2N(\eta)$ and $N(\eta) d_0/2a(\eta)$, we will obtain the expansion parameters $f_1$ and $f_2$. Note that in the setting $N(\eta) = a(\eta)$ we get the zeroth order terms as $s_0, d_0$ (cf. Eq.(\ref{eft:truncated})), which are time-independent and can be set to unity. 

Following these two steps, the expansion parameters ($f_1, f_2$), in terms of lapse function, take the following forms:
\begin{equation}
\label{appeq:EP-lapse}
\begin{split}
 &  f_1(\mathcal{H}, a, N, \eta) = 
    \frac{a}{N} s_0 + \sum_{n=1}^{\infty} \frac{s_n}{N^n}  \left(\frac{\mathcal{H}}{\Lambda}\right)^n
    + \sum_{n=1}^{\infty} \frac{b_n}{a} \frac{1}{N^n}
    \left(\frac{1}{\Lambda^2} \frac{d}{d\eta} \left( \frac{a}{N} \mathcal{H} \right) \right)^n+...
\\
& f_2(\mathcal{H}, a, N, \eta) = 
    \frac{N}{a} d_0 + \sum_{n=1}^{\infty} \frac{d_n}{N^n}  \left(\frac{\mathcal{H}}{\Lambda}\right)^n
    + \sum_{n=1}^{\infty} \frac{e_n}{a} \frac{1}{N^n}
    \left(\frac{1}{\Lambda^2} \frac{d}{d\eta} \left( \frac{a}{N} \mathcal{H} \right) \right)^n+...
\end{split}
\end{equation}
As mentioned above, we have separated the zeroth-order terms in the expansion parameters and are fixed with respect to the standard electrodynamics action. 
Setting $N(\eta) = a(\eta)$ in the above expression matches with Eq.(\ref{eft:A}). 
Using the expansion parameters (\ref{appeq:EP-lapse}) in terms of lapse function and varying the action (\ref{eft:A}) with respect to the lapse function $N \rightarrow N+\delta N$ gives:
\begin{align}\label{appeq:deltaS1}
    \delta \mathcal{S}_{\rm EFT} &= - \int d^4x \, \left[ \left( \frac{a}{N^2} s_0 + \sum \frac{n}{N^{n+1}} s_n \left( \frac{\mathcal{H}}{\Lambda} \right)^n + \sum \frac{n}{a N^{n+1} \Lambda^{2n}} b_n \left( \frac{d}{d\eta}\left( \frac{a \mathcal{H}}{N} \right) \right)^n\right) \delta N \, (A_i^{\prime})^2 
    \right. \nonumber\\
    &{} \left. 
    - (A_i^{\prime})^2 \sum \frac{n}{a N^n \Lambda^{2n}} b_n \left( \frac{d}{d\eta}\left( \frac{a \mathcal{H}}{N} \right) \right)^{n-1} \frac{d}{d\eta}\left( \frac{a \mathcal{H}}{N^2} \delta N \right) + .....
    \right. \nonumber\\
    &{} \left.
    + \left( \frac{1}{a} d_0 - \sum \frac{n}{N^{n+1}} d_n \left( \frac{\mathcal{H}}{\Lambda} \right)^n - \sum \frac{n}{a N^{n+1} \Lambda^{2n}} e_n \left( \frac{d}{d\eta}\left( \frac{a \mathcal{H}}{N} \right) \right)^n\right) \delta N \, (\partial_i A_j)^2  \right. \nonumber\\
    &{} \left. 
    + (\partial_i A_j)^2 \sum \frac{n}{a N^n \Lambda^{2n}} e_n \left( \frac{d}{d\eta}\left( \frac{a \mathcal{H}}{N} \right) \right)^{n-1} \frac{d}{d\eta}\left( \frac{a \mathcal{H}}{N^2} \delta N \right) + .....
    \right].
\end{align}
Doing integration by parts in the second and fourth line in the equation (\ref{appeq:deltaS1})
and ignoring the total derivative terms gives:
\begin{align}\label{appeq:deltaS}
    \delta \mathcal{S}_{\rm EFT} &= - \int d^4x \, \left[ \left( \frac{a}{N^2} s_0 + \sum \frac{n}{N^{n+1}} s_n \left( \frac{\mathcal{H}}{\Lambda} \right)^n + \sum \frac{n}{a N^{n+1} \Lambda^{2n}} b_n \left( \frac{d}{d\eta}\left( \frac{a \mathcal{H}}{N} \right) \right)^n\right) \, (A_i^{\prime})^2 
    \right. \nonumber\\
    &{} \left. 
    +  \frac{a \mathcal{H}}{N^2} \frac{d}{d\eta} \left( (A_i^{\prime})^2 \sum \frac{n}{a N^n \Lambda^{2n}} b_n \left( \frac{d}{d\eta}\left( \frac{a \mathcal{H}}{N} \right) \right)^{n-1}  \right) + .....
    \right. \nonumber\\
    &{} \left.
    + \left( \frac{1}{a} d_0 - \sum \frac{n}{N^{n+1}} d_n \left( \frac{\mathcal{H}}{\Lambda} \right)^n - \sum \frac{n}{a N^{n+1} \Lambda^{2n}} e_n \left( \frac{d}{d\eta}\left( \frac{a \mathcal{H}}{N} \right) \right)^n\right) \, (\partial_i A_j)^2  \right. \nonumber\\
    &{} \left. 
    - \frac{a \mathcal{H}}{N^2} \frac{d}{d\eta} \left( (\partial_i A_j)^2 \sum \frac{n}{a N^n \Lambda^{2n}} e_n \left( \frac{d}{d\eta}\left( \frac{a \mathcal{H}}{N} \right) \right)^{n-1}  \right) + .....
    \right]\delta N .
\end{align}
Now using Eq.~(\ref{appeq:rho-def}) and Eq.~(\ref{appeq:deltaS}), the energy density corresponding to the EFT action is:
\begin{align}\label{appeq:rhoEFT}
    \rho &= \frac{1}{a^4}\left[ \left( \frac{a^2}{N^2} s_0 + \sum \frac{n a}{N^{n+1}} s_n \left( \frac{\mathcal{H}}{\Lambda} \right)^n + \sum \frac{n}{ N^{n+1} \Lambda^{2n}} b_n \left( \frac{d}{d\eta}\left( \frac{a \mathcal{H}}{N} \right) \right)^n\right) \, (A_i^{\prime})^2 
    \right. \nonumber\\
    &{} \left. 
    +  \frac{a^2 \mathcal{H}}{N^2} \frac{d}{d\eta} \left( (A_i^{\prime})^2 \sum \frac{n}{a N^n \Lambda^{2n}} b_n \left( \frac{d}{d\eta}\left( \frac{a \mathcal{H}}{N} \right) \right)^{n-1}  \right) + .....
    \right. \nonumber\\
    &{} \left.
    + \left( d_0 - \sum \frac{n a}{N^{n+1}} d_n \left( \frac{\mathcal{H}}{\Lambda} \right)^n - \sum \frac{n}{ N^{n+1} \Lambda^{2n}} e_n \left( \frac{d}{d\eta}\left( \frac{a \mathcal{H}}{N} \right) \right)^n\right) \, (\partial_i A_j)^2  \right. \nonumber\\
    &{} \left. 
    - \frac{a^2 \mathcal{H}}{N^2} \frac{d}{d\eta} \left( (\partial_i A_j)^2 \sum \frac{n}{a N^n \Lambda^{2n}} e_n \left( \frac{d}{d\eta}\left( \frac{a \mathcal{H}}{N} \right) \right)^{n-1}  \right) + .....
    \right].
\end{align}

Setting $N(\eta) = a(\eta)$ in the above expression and using the truncated series expansion (\ref{eft:truncated}), we obtain the energy density for the EFT action:
\begin{align}
   \rho_{\rm EFT} = \rho_{\rm E} + \rho_{\rm B} + \rho_{\rm mixing}  
\end{align}
where 
\begin{subequations}
\label{eq:EFT-rho}
\begin{align}
    \rho_{\rm E} &= \frac{1}{a^4} \left( s_0 + \frac{s_1}{a} \frac{\mathcal{H}}{\Lambda} + \frac{1}{a^2} \left( \frac{\mathcal{H}}{\Lambda} \right)^2 
    \left[2 s_2 - b_1 (1 +  \epsilon_1) \right] \right) {A_i^{\prime}}^2 \\
    \rho_{\rm B} &= \frac{1}{a^4} \left( d_0 - \frac{d_1}{a} \frac{\mathcal{H}}{\Lambda} - 
    \frac{1}{a^2} \left( \frac{\mathcal{H}}{\Lambda} \right)^2 
    \left[2 d_2 - e_1 (1 + \epsilon_1) \right] \right) (\partial_i A_j)^2
    \\
    \rho_{\rm mixing} &= \frac{2}{a^6} \frac{\mathcal{H}}{\Lambda^2} \left(   b_1 A_i^{\prime} A_i^{\prime\prime} + e_1 \partial_i A_j \, \partial_i A_j^{\prime} \right)
\end{align}
\end{subequations}
where in the last expression $A_i^{\prime\prime}$ can be substituted by using the equations of motion (\ref{eq:GFieldEOM}) in the real space.

\iffalse
\section{EOM of $\delta A_{\mu}$}
\label{app:EOM}

Considering different scenarios the general equation of motion of $\delta A_{\mu}$ can be written as,

\begin{equation}
     \mathcal{A}_{i,k}^{\prime \prime} + \left(c_{\mathcal{A}}(t) k^2  + m^2(t) \right) \mathcal{A}_{i,k} = 0
\end{equation}

Here $\mathcal{A}_i$ is canonically normalised field and the normalisation factor will depend on the particular scenario we are considering. The sound speed $c_{\mathcal{A}}$ can appear from two effects: $1.$ from the coupling between scalar and $\mathcal{A}_i$ as discussed earlier, $2.$ the Gallileon scenario can also affect the sound speed.

Here, one important thing to note is that as $\pi A_i A_i$ coupling affects the sound speed, this put an automatic constraint on the mode function $\pi$. Because, $0< c_{\mathcal{A}}<1$, the mode functions $\pi$ has to obey $\pi < 1$.
\fi

\section{EFT magnetogenesis power-spectrum for power-law inflation}
\label{app:power-law}

In Sec. \eqref{sec:EFT-Spectrum}, we obtained the power-spectrum during slow-roll inflation. In this Appendix, we obtain the power spectrum for a generic power-law and de Sitter inflation. 

The procedure we follow to obtain the magnetic power-spectrum is the same as in the case of slow-roll. In the case of de Sitter, the power spectrum is exact while in the case of power-law inflation, we use WKB approximation. 
The scale factor during the power-law inflation is
\begin{equation}
    a(\eta) = (-H_0 \eta)^{\beta+1}.
\end{equation}
where $\beta \leq 2$ for power-law inflation, $\beta=-2$ corresponds to 
exact de Sitter and $H_0$ denotes the characteristic energy scale associated with inflation.

Like in Sec.~\eqref{sec:EFT-Spectrum}, 
to solve Eq. \eqref{eq:GFieldEOM}, we introduce new variables such that all the time dependence can be converted into momentum
dependence~\cite{2004-Shankaranarayanan.Sriramkumar-PRD}:
\begin{equation}\label{new-var}
x = \ln{\frac{\beta + 1}{k \eta}}, 
\quad \mathcal{A}_k = e^{-\frac{x}{2}} u_k,
\end{equation}
In terms of these new variables, $\mathcal{H}$ and its derivatives are given by:
\begin{eqnarray}\label{H-deriv}
 \mathcal{H} = \frac{a^{\prime}}{a} = e^x k, & & \mathcal{H}^{\prime} = \frac{a^{\prime\prime}}{a} - \left(\frac{a^{\prime}}{a}\right)^2 = - \frac{1}{(\beta+1)} e^{2x}k^2,\\
\mathcal{H}^{\prime\prime} = \frac{2}{(\beta+1)^2} e^{3x} k^3, & & 
\mathcal{H}^{\prime\prime\prime} = \frac{a^{\prime\prime\prime\prime}}{a}- 4 \frac{a^{\prime} a^{\prime\prime\prime}}{a^2}+12 \left(\frac{a^{\prime}}{a}\right)^2 \frac{a^{\prime\prime}}{a} - 6 \left(\frac{a^{\prime}}{a}\right)^4
= - \frac{6}{(\beta+1)^3} e^{4x} k^4 \nonumber.
\end{eqnarray}

In the redefined variables the sound speed \eqref{def:CA} and $Z''/Z$ become:
\begin{eqnarray}
\label{eq:cA-Powerlaw}
c_{A}^2 &=& 1+ (d_1-s_1) \frac{e^x}{a}  \frac{k}{\Lambda}+ 
\left[ (s_1^2-s_2-s_1 d_1+d_2) 
- \frac{(e_1-b_1)}{(\beta+1)} e^x \right]
\frac{e^x}{a^2}  
    \left(\frac{k}{\Lambda}\right)^2, 
    \\
\label{eq:Z-powerlaw}
\frac{Z^{\prime\prime}}{Z} &=& \frac{s_1}{2}\left\{1+\frac{3}{\beta+1}+\frac{2}{(\beta+1)^2}\right\} \frac{e^{3x}}{a}
    \frac{k^3}{\Lambda} \nonumber \\
& +& \left\{\frac{s_1^2}{4}+2s_2-\frac{s_1^2+5s_2-2b_1}{\beta+1}
    -\frac{\frac{s_1^2}{4}+2b_1}{(\beta+1)^2}-
    \frac{3b_1}{(\beta+1)^3}
    \right\} \frac{e^{4x}}{a^2}
    \frac{k^4}{\Lambda^2}.
\end{eqnarray}
where the scale factor $a(\eta)$ in 
the new variables is:
\begin{equation}\label{scale-factorx}
    a(x)=\left(H_0 e^{-x}\frac{\beta+1}{k}\right)^{\beta+1}.
\end{equation}
Substituting Eqs.~(\ref{eq:cA-Powerlaw}, \ref{eq:Z-powerlaw}) 
in Eq. \eqref{eq:GFieldEOM}, we have:
{\small
\begin{equation}\label{u-eom}
\frac{d^2 u_k}{d x^2} - 
\left[\frac{1}{4} 
- e^{-2x}(\beta+1)^2 
- \left\{M_1 e^{\beta x}+M_3 e^{(\beta+2)x}
+\left(M_2 
+M_4e^{2x}\right)e^{(2\beta+2)x}\frac{k^{\beta+2}}{\Lambda}\right\}
\frac{k^{\beta+2}}{\Lambda} \right]
u_k=0.
\end{equation}
}
where,
{\small
\begin{eqnarray}
M_1 &=& \frac{(\beta+1)^2}{\left(H_0 (\beta+1)\right)^{(\beta+1)}}(d_1-s_1),\\
M_2 &=& \frac{(\beta+1)^2}{\left(H_0 (\beta+1)\right)^{2(\beta+1)}}\left(s_1^2-s_2+s_1 d_1 +d_2-\frac{e_1-b_1}{\beta+1}\right),\\
M_3 &=& \frac{s_1}{2} \frac{(\beta+1)^2}{\left(H_0 (\beta+1)\right)^{(\beta+1)}} \left(1+\frac{3}{\beta+1}+\frac{2}{(\beta+1)^2}\right),\\
M_4 &=& \frac{(\beta+1)^2}{\left(H_0 (\beta+1)\right)^{2(\beta+1)}} 
      \left\{\frac{s_1^2+8 s_2}{4}-\frac{s_1^2-5s_2+2b_1}{\beta+1}
   -\frac{3s_1^2+4s_2+20 b_1}{4(\beta+1)^2}-
    \frac{3b_1}{(\beta+1)^3}\right\} \, .
\end{eqnarray}
}
Eq.~\eqref{u-eom} is the equation of motion for generic power-law inflation. 
It is not possible to obtain an exact analytical solution for a generic power law. Hence, in the rest of this appendix, we obtain the exact expression for the de Sitter case ($\beta = - 2$) and obtain the WKB solution for an arbitrary value of $\beta$.

\subsection{Power spectrum for de Sitter}

Setting $\beta=-2$ in \eqref{u-eom}, we have: 
\begin{equation}\label{eom-redefined}
\frac{d^2 u_k}{d x^2}+c_A^2 e^{-2x} u_k -m_2^2 u_k = 0 \, ,
\end{equation}
where,
\begin{eqnarray}
c_A^2 = 1+C_1 \frac{H_0}{\Lambda}+C_2 \frac{H_0^2}{\Lambda^2}\, , &~~~& 
m_2^2 = \frac{1}{4} -C_4 \frac{H_0^2}{\Lambda^2} \\
C_1 = s_1-d_1 \, , ~~C_4 = \frac{s_1^2}{2}-4 s_2 \, &~~~&
C_2 = s_1^2-s_2+s_1 d_1 +d_2 +e_1-b_1 \, .
\end{eqnarray}
As in the slow-roll case, the solution to Eq.~\eqref{eom-redefined} are Hankel functions~\cite{abramowitz+stegun}:
\begin{equation}
    u_k(x) = \alpha H_{m_2}^{(1)}(e^{-x}m_1) + \beta H_{m_2}^{(2)}(e^{-x}m_1),
\end{equation}
Following the steps from Eq.~\eqref{eq:EFT-Modefunction} 
to Eq.~\eqref{eq:BunchDaviesEFT}, 
the gauge-field mode functions ($A_k$) in the super-Hubble scales reduce to: 
\begin{equation}
A_k(\eta) = \sqrt{\frac{\pi}{4}}\left(\frac{-i}{\pi}\right) \Gamma(m_2) \, \frac{c_A}{Z}(-\eta)^{\frac{1}{2}}
\left(\frac{- c_A k\eta }{2}\right)^{-m_2},
\end{equation}
where,
\begin{equation}
   Z = 1 + d_1 \frac{H_0}{\Lambda}+(d_2+e_1) \frac{H_0^2}{\Lambda^2}
\end{equation}
The magnetic field power spectrum is:
\begin{equation}
\label{de-sitter-ps}
    \mathcal{P}_B = \frac{H^4}{8 \pi^3 Z^2 c_A^3} \Gamma^2(m_2) (-c_A k \eta)^{4+C_4\frac{H_0^2}{\Lambda^2}}
\end{equation}
This matches with the power-spectrum \eqref{PS-srFin} for slow-roll inflation by setting $\epsilon_1 = 0$.

\iffalse
\textcolor{red}{\textbf{Note}: Here one important thing to note is that in this power 
spectrum if both $C_3$ and $C_4$ are negligible with respect to the first term i,e 4 
in the exponent of $(c_A k \tau)$ then the power spectrum will be proportional to 
$\eta^4$ which corresponds to the fact that the power spectrum falls as
$\frac{1}{a^4}$ which is similar to the standard electromagnetic case. This is
actually consistent with the fact that in the EFT expansion the leading order term
i,e $s_0$ and $d_0$ are dominant over their EFT correction and ultimately the
Lagrangian can be approximated as the standard EM case unless we tune the parameters
$C_3$ or $C_4$ such that they are comparable to the leading order term.}
\fi

\subsection{Power spectrum for generic power-law inflation}

Unlike de Sitter, we can not obtain an exact analytical expression for generic power-law inflation. Instead, we use the WKB approximation to obtain the spectrum~\cite{2004-Shankaranarayanan.Sriramkumar-PRD}. Rewriting \eqref{u-eom} as
\begin{equation}
\frac{d^2 {u}_{k}}{dx^2}  + 
\omega^2(x)\, {u}_{k}(x) 
= 0,\label{eq:demukcux} 
\end{equation}
where $\omega^2(x)$ is 
\begin{equation}\label{gen-beta}
\omega^2(x)=e^{-2x}(\beta+1)^2+M_1 e^{\beta x}\frac{k^{\beta+2}}{\Lambda}
+M_2 e^{(2\beta+2)x}\frac{k^{2\beta+4}}{\Lambda^2}
+M_3 e^{(\beta+2)x}\frac{k^{\beta+2}}{\Lambda}
+M_4e^{(2\beta+4)x}\frac{k^{2\beta+4}}{\Lambda^2}-\frac{1}{4}    
\end{equation}
In the sub-horizon limit ($x\rightarrow -\infty$), the third term in RHS of the above expression dominates for $\beta<-2$. Hence, in the sub-horizon region ($I$), we have:
\begin{equation}
u^{\rm I}_{k}(x) = \frac{1}{\sqrt{\omega_I(x)}} \,
\left[ A_k \, \exp \, i\int^{x}dy 
\, \omega_I(y)
+ B_k \, \exp \, - i\int^{x}dy\, \omega_I(y) \right]
,\label{eq:wkb}
\end{equation}
where  ${\bar A}(k)$ and ${\bar B}(k)$ are $k$-dependent constants that
are to be fixed by the initial conditions, and
\begin{equation}
\omega_I^2(x) =  \frac{M_2}{\Lambda^2} \, e^{(2\beta+2)x} 
\, k^{2\beta+4} \, .
\end{equation}

As in the slow-roll and power-law cases, we shall assume that the gauge field is in the Bunch-Davies vacuum \eqref{eq:BunchDaviesEFT} on the sub-Hubble scales. This leads to:
\begin{eqnarray}
A_k =  -\frac{i (\beta +1)}{ \sqrt{32} \, \omega_I(x_i)} \, , 
&~~~&
B_k =\frac{1}{\sqrt{32} \, \omega_I(x_i)} \left(i \beta +4 \omega_I(x_i)+ i\right) \, . 
\end{eqnarray}
In the super-horizon limit ($x\rightarrow \infty$), the first term in the RHS of \eqref{gen-beta} dominates for $\beta < -2$. Hence, 
the super-horizon region (II), we have:
\begin{equation}\label{region-IIWKB}
    u^{\rm II}_k(x) = \frac{C_k}{\sqrt{\omega_{\rm II}(x)}} \exp{\left( \int_x^{x_f} \vert\omega_{\rm II}(y)\vert dy\right)} + \frac{D_k}{\sqrt{\omega_{\rm II}(x)}} \exp{\left( \int_x^{x_f} -\vert\omega_{\rm II}(y)\vert dy\right)}
\end{equation}
Near the horizon crossing ($x=0$), the WKB approximation breaks down and in this region, we can approximate $\omega(x) = \alpha \, x$, 
where $\alpha$ depends on $M_1, M_2, M_3, \Lambda$ and $\beta$. 
Matching the modes at the horizon exit, we have the following relation between $C_k, D_k$ with $A_k, B_k$:
\begin{eqnarray}
C_k = \frac{A_k+B_k}{\sqrt{8}} e^{-\Psi};~~D_k = \frac{B_k-A_k}{\sqrt{2} i} e^{\Psi} \,
\end{eqnarray}
where,
\begin{equation}
    \Psi = \int_{x_*}^{x_f}\alpha \, y \, dy
\end{equation}
Taking into account only the growing mode, the modulus squared of the mode function is:
\begin{equation}\label{powerlaw-spec}
\left\vert\mathcal{A}_k\right\vert^2 = \frac{\left\vert B_k-A_k\right\vert^2}{2} e^{-x}\frac{e^{2\Psi}}{\sqrt{\omega(x)}}.
\end{equation}
Substituting this in Eq.~\eqref{eq:pMS}, the magnetic power spectrum at the super-horizon scales is:
\begin{equation}
    \mathcal{P}_B (k) = \frac{k^5}{2 \pi^2 a^4} \frac{\left\vert B_k-A_k\right\vert^2}{2 Z} e^{-x}\frac{e^{2\Psi}}{\sqrt{\omega(x)}}
\end{equation}
%
%
%============ Vector Galileon Appendix =========
%
\section{Vector Galileon: Equation of motion in terms of Slow-roll parameters}
\label{app:VG-slow-roll}

In case of slow-roll inflation, the slow-roll parameters are defined
as~\cite{2014-Martin.etal-PhyDarkUniv}:
\begin{eqnarray}
\epsilon_1 = -\frac{H^{\prime}}{ a H^2} = 1 - \frac{\mathcal{H}^{\prime}}{\mathcal{H}^2};~~~
&&\epsilon_{2} = \frac{\epsilon^{\prime}_1}{a H \epsilon_1} 
= \frac{ \left( 2{\mathcal{H}^{\prime}}^2 - \mathcal{H}\mathcal{H}^{\prime\prime} \right) }{\mathcal{H}^2 \left(\mathcal{H}^2 - \mathcal{H}^{\prime} \right) } ~~.
\label{def:Slowrollpara}
\end{eqnarray}
%
%which gives, 
%\begin{eqnarray}
%a^{\prime} = a^2 H \\
%a^{\prime\prime} = - \frac{(\epsilon_1 - 2) a^{\prime}{}^2 }{a} \\
%\end{eqnarray}
%
The equation of motion (\ref{eq:eom-canonicalA}) for the canonical vector field $\mathbb{A}_k$  in terms of slow-roll parameters can be obtained as
\begin{align}\label{eqapp:eom-canonical-SR}
\mathbb{A}_k^{\prime\prime} +  \left[ c_s^2 \, k^2 -    \frac{ {\mathcal{H}}^2 J \epsilon_1}{(1 - J )^2} \, \left(1 - J - 3\epsilon_1 + 2 J\epsilon_1 + \epsilon_2 (1 - J)  \right) \right] \,\mathbb{A}_k = 0 \, ,
\end{align}
where $c_s$ is defined in Eq. \eqref{eq:def-cs}.
Substituting Eq.(\ref{eq:def-canonicalA}) in Eq.~(\ref{eqapp:eom-canonical-SR}), the equation of motion of the physical vector field $A_{k}$ is:
%for the action (\ref{eq:eom-canonicalA-fourier})
%
\begin{align}
A_k^{\prime\prime} + \frac{2 \mathcal{H} J \epsilon_1 }{ (1 - J) } A_k^{\prime} 
+ \frac{ k^2 (1 - J + J \epsilon_1 ) }{ (1 - J) } A_k = 0 \, .
\end{align}

\chapter{Details on GZ effect: A plausible explanation for fast radio bursts?} 
\label{ap:chapGZeffect}

In this Appendix, we provide more details on the results from chapter~\ref{ch:GZeffect}.

\section{Gertsenshtein-Zel$'$dovich effect and induced electromagnetic waves}
\label{appsec:E&B-solution}
%
%\textcolor{red}{After the historical detection of GWs in the frequency range $100 - 10^4~{\rm Hz}$, there is a surge in activity for the possibility of detection of GWs in the MHz-GHz frequency range. (For details, see Aggarwal et al~\cite{2020-Aggarwal.etal-arXiv}.)}

Consider a background space-time with GWs~\cite{Book-Gravitation_MTW}:
\begin{align}\label{eq:Metric}
g_{\alpha\beta} = \eta_{\alpha\beta} + h_{\alpha\beta} \, , \qquad 
g^{\alpha\beta} = \eta^{\alpha\beta} - h^{\alpha\beta}
\end{align}
where $h_{\alpha\beta} << 1$ is the GW fluctuation, and $\eta_{\alpha\beta} = \rm{diag}(1,-1,-1,-1)$ is Minkowski space-time. The background space-time is generally curved; however, the results derived for Minkowski space-time carry through for the conversion factor computations. For the generic curved background, the Riemann corrections contribution is tiny~\cite{Book-Gravitation_MTW}. Hence, we only report the results for the Minkowski space-time. 

For the GWs propagating in the $z-$axis, we have $h_{xx} = - h_{yy} = h_{+}$, $h_{xy} = h_{\times}$. 
%The metric describing the gravitational fluctuations and the background is given by:
%
%\begin{align}\label{eq:Metric}
%g_{\alpha\beta} = \eta_{\alpha\beta} + h_{\alpha\beta} \quad = \quad  \begin{pmatrix}
%1 & 0 & 0 & 0 \\
%0 & -1+h_+ & h_{\times} & 0\\
%0 & h_{\times} & -1-h_+ & 0\\
%0 & 0 & 0 & -1
%\end{pmatrix} ,
%\hspace{1cm} 
%g^{\alpha\beta} = \begin{pmatrix}
%1 & 0 & 0 & 0 \\
%0 & -1-h_+ & -h_{\times} & 0\\
%0 & -h_{\times} & -1+h_+ & 0\\
%0 & 0 & 0 & -1
%\end{pmatrix}.
%\end{align}
%
%where $h_{\alpha\beta} << 1$ is the gravitational wave fluctuation, and $\eta_{\alpha\beta} = \rm{diag}(1,-1,-1,-1)$ is Minkowski spacetime. The inverse of the metric can be obtained as $g^{\alpha\beta} = \eta^{\alpha\beta} - h^{\alpha\beta}$ up to linear order.
%\textcolor{red}{Since the equations are linear, we consider a monochromatic circularly polarized gravitational wave propagating along the z-direction.}
Eq. \eqref{eq:h-Expression} corresponds to a monochromatic circularly polarized GW~\cite{Book-Gravitation_MTW,2018-Zheng.Wei.Li-PRD}. As mentioned earlier, the presence of magnetic field transverse to the direction of propagation of GWs acts as a catalyst for the conversion process \cite{1974-Zeldovich-SJETP}. As discussed, we consider the total magnetic field in the magnetosphere to be%~\cite{2018-Zheng.Wei.Li-PRD}
\begin{align}\label{eq:bg-magnetic_field}
\textbf{B}(t) = \left( 0, B^{(0)}_y + \delta B_y \sin (\omega_B t), 0  \right)
\end{align}
where $B^{(0)}_y$ is static magnetic field and $ B_y \sin (\omega_B t)$ is the alternating (time-varying) magnetic field with frequency $\omega_B $ and $| \delta B_y| < |B^{(0)}_y| $.
Also, we assume that the amplitude of the alternating magnetic field is two orders lower than the static magnetic field, i.e., 
$| {B_y}/{B^{(0)}_y}| \approx 10^{-2}$, so that time-varying magnetic field has significant effects on the conversion~(\cite{2012-Pons.etal-AandA,2019-Pons.Vigan-arXiv}). The induced electric field due to the time-varying magnetic field is~[\cite{Book-Jackson-Classical_Electrodynamics}]
\begin{align}\label{eq:InducedElectricField}
\textbf{E}(z,t) = \left( - \frac{z\, \delta \omega_B B_y}{c} \cos(\omega_B t)  , 0 , 0 \right) \, .
\end{align}
%
%\textcolor{red}{In the absence of GWs, the non-zero  components of the background electromagnetic (EM) field tensor $F^{(0)}_{\alpha\beta}$  is:
%
%\begin{align}\label{eq:FT-component}
%&{} F^{(0)}_{0 1} = E_x = - F^{(0)}_{1 0}   = - \frac{ z \, \delta B_y \omega_B}{c} \cos(\omega_B t); \nonumber\\ 
%&{} F^{(0)}_{1 3} = B_z = - F^{(0)}_{3 1}  =  B^{(0)}_y + \delta B_y \sin (\omega_B t) \,.
%\end{align}
%}
In the presence of GWs, the EM field tensor is:
\begin{align}\label{eq:EMField-tensor-dd_matrix}
F_{\alpha\beta} = F^{(0)}_{\alpha\beta} + F^{(1)}_{\alpha\beta} 
= \begin{pmatrix}
0 & \mathcal{E}_x & \tilde{E}_y & \tilde{E}_z \\
-\mathcal{E}_x & 0 & -\tilde{B}_z & \mathcal{B}_y \\
-\tilde{E}_y & \tilde{B}_z & 0 & -\tilde{B}_x\\
-\tilde{E}_z & - \mathcal{B}_y & \tilde{B}_x & 0  
\end{pmatrix}  ,
\end{align}
where $F^{(1)}_{\alpha\beta}$ is the field tensor induced due to the GWs that needs to be determined. Similarly, the induced electric 
$[\tilde{ \textbf{E} } = \left(  \tilde{E}_x, \tilde{E}_y, \tilde{E}_z \right)]$ 
and magnetic field vectors  $[\tilde{\textbf{B} }= \left( \tilde{B}_x, \tilde{B}_y, \tilde{B}_z \right)$] due to GWs are to be determined.
Note that $\mathcal{B}_y = B^{(0)}_y + \delta B_y \sin (\omega_B t) + \tilde{B}_y $ and $ \mathcal{E}_x = - (z \, \delta B_y \omega_B/{c}) \cos(\omega_B t) + \tilde{E}_x$. 
The covariant Maxwell's equations (in the source-free region) are:
%
%\begin{subequations}
\begin{align}
\label{eq:MaxwellEq12}
\partial_{\mu} \left( \sqrt{-g} F^{\mu\nu} \right) = 0;~~
%\label{eq:MaxwellEq34}
 \partial_{\mu} \left( \sqrt{-g} \tilde{F}^{\mu\nu} \right) = 0
 %\textcolor{red}{ \partial_{\mu} \left( \sqrt{-g} \eta^{\mu\nu\alpha\beta} F_{\alpha\beta} \right) =} 0 
\end{align}
%\end{subequations}
where $\tilde{F}^{\mu\nu} =  \epsilon^{\mu\nu\alpha\beta} F_{\alpha\beta}/2$ is the dual of EM field tensor, $\epsilon^{\mu\nu\alpha\beta}  = \eta^{\mu\nu\alpha\beta}  /{ \sqrt{-g} }$ and $\eta^{\mu\nu\alpha\beta}$ is defined as $\eta^{0123} = 1 = - \eta_{0123} $ is an antisymmetric tensor.
Substituting Eq.~(\ref{eq:EMField-tensor-dd_matrix}) in Eq.~\eqref{eq:MaxwellEq12}, and treating $F^{(1)}_{\alpha\beta} $ and $h_{\alpha\beta}$ as first-order perturbations, we have
%\textcolor{red}{
%
%\begin{subequations}
%\begin{align}
%\label{eq:MaxwellEq1-1stOrder}
%\partial_{\mu} \left[ \left(  \eta^{\mu\alpha} h^{\nu\beta} +  h^{\mu\alpha} \eta^{\nu\beta} \right)  F^{(0)}_{\alpha\beta} -  \eta^{\mu\alpha} \eta^{\nu\beta}  F^{(1)}_{\alpha\beta}   \right] = 0 \\
%\label{eq:MaxwellEq2-1stOrder}
% \partial_{\mu} \left(  \eta^{\mu\nu\alpha\beta} F^{(1)}_{\alpha\beta}  \right) = 0 \,\, .
%\end{align}
%\end{subequations}
%}
\begin{subequations}
\begin{align}
\label{eq:dtEx}
&{} \frac{1}{c} \partial_t \tilde{E}_x - \partial_y \tilde{B}_z + \partial_z \tilde{B}_y 
+ \left( B^{(0)} + \delta B_y \sin(\omega_B t) \right) \, \partial_z h_{+}  
\nonumber\\ 
&{}- \frac{ z \, \delta B_y \omega_B}{c^2} \frac{\partial }{\partial t} \left( h_{+} \, \cos(\omega_B t) \right) = 0 \\
\label{eq:dtBy}
&{}\frac{1}{c} \partial_t \tilde{B}_y - \partial_x \tilde{E}_z + \partial_z \tilde{E}_x  = 0 \, ,
\end{align}
\end{subequations}
where we have expressed in terms of electric and magnetic fields. Since GWs (and EM waves) propagate along the z-direction, we have $\tilde{E}_z = \tilde{B}_z = 0$. Substituting Eq.~(\ref{eq:h-Expression}) in  the above wave equations lead to the following wave equations for $\tilde{E}_x, \tilde{B}_y$:
\begin{subequations}
\begin{align}\label{eq:WaveEquation-Ex-FF}
\frac{1}{c^2} \frac{\partial^2  \tilde{E}_x }{\partial t^2} - \partial_z^2 \tilde{E}_x  
&= - f_{E}(z^{\prime},t^{\prime}) \\
\label{eq:WaveEquation-By-FF}
\frac{1}{c^2} \frac{\partial^2  \tilde{B}_y }{\partial t^2} - \partial_z^2 \tilde{B}_y 
&= - f_{B}(z^{\prime},t^{\prime})
\end{align}
\end{subequations}
where $f_{E/B}(z^{\prime},t^{\prime})$ are the forcing functions and are given by: 
\begin{subequations}
\begin{align}\label{eq:ForcingFunction-E}
&f_{E}(z^{\prime},t^{\prime}) =  
\frac{ i A_{+} \delta B_y k_g }{ 2 c}  
\left[ \omega_{+}  e^{i ( k_g z^{\prime} - \omega_{+} t^{\prime}) } - \omega_{-} 
e^{i  ( k_g z^{\prime} - \omega_{-} t^{\prime})} \right]  
\nonumber\\
& + \frac{ z^{\prime} A_{+} \delta B_y \omega_B }{ 2 c^3}  \left[\omega_{+}^2 e^{i ( k_g z^{\prime} - \omega_{+} t^{\prime} ) } + \omega_{-}^2 e^{i ( k_g z^{\prime} - \omega_{-} t^{\prime}) }  \right]  \nonumber \\
& + \frac{A_{+} B^{(0)} k_g \omega_g  }{c}  
e^{i ( k_g z^{\prime} - \omega_g t^{\prime}) } \\
\label{eq:ForcingFunction-B}
& f_{B}(z^{\prime},t^{\prime}) = 
\frac{ i A_{+} \delta B_y k_g^2  }{ 2 } \left[ e^{i \left( k_g z^{\prime} - \omega_{+} t^{\prime} \right) }  - e^{i \left( k_g z^{\prime} - \omega_{-} t^{\prime} \right) } 
\right]
\nonumber\\
& + \frac{ i A_{+} \delta B_y \omega_B   }{ 2 c^2}    \left[ \omega_{+}  e^{i \left( k_g z^{\prime} - \omega_{+} t^{\prime} \right) }  -  \omega_{-}  e^{i \left( k_g z^{\prime} - \omega_{-} t^{\prime} \right) } 
\right]
\nonumber\\
& -  \frac{ z^{\prime} A_{+} \delta B_y \omega_B k_g   }{ 2 c^2}  \left[ \omega_{+}  e^{i( k_g z^{\prime} - \omega_{+} t^{\prime}) }   - \omega_{-}  e^{i( k_g z^{\prime} - \omega_{-} t^{\prime}) }  \right] \nonumber \\
& + A_{+} B^{(0)} k_g^2   e^{i \left( k_g z^{\prime} - \omega_g t^{\prime} \right) } 
\end{align}
\end{subequations}
The solutions to the wave equations (\ref{eq:WaveEquation-Ex-FF}) and (\ref{eq:WaveEquation-By-FF}) are given by: 
\begin{align}\label{eq:ResponseToForcingFunction}
F_{E/B}(z,t) = \int \int dz^{\prime} dt^{\prime}  G( t,t^{\prime} ; z,z^{\prime} )  f_{E/B}( t^{\prime},z^{\prime} ) 
\end{align}
where $F_{E/B}(z,t)$ denotes the corresponding solution of the forcing function $f_{E/B}(z^{\prime},t^{\prime})$ and $G( t,t^{\prime} ; z,z^{\prime})$ is the retarded Green's function corresponding to the wave equation and is given by:
\begin{align}\label{eq:GreenF}
G( t,t^{\prime} ; z,z^{\prime} ) = \frac{c }{2} \Theta \left( \,\, c(t-t^{\prime}) - |z-z^{\prime}| \,\, \right) 
\end{align}
where $\Theta-$function is non-zero only for $t \geq  t^{\prime} + \frac{|z-z^{\prime}|}{c}$. Substituting the forcing functions for the electric (\ref{eq:ForcingFunction-E}) and the magnetic (\ref{eq:ForcingFunction-B}) fields, in the integral equation (\ref{eq:ResponseToForcingFunction}), leads to:
\begin{align}\label{eq:E_x-Expression}
\tilde{E}_x &\simeq  -\frac{B^{(0)}_y A_{+} }{4} \,\, e^{i \left(k_g z - \omega_g t\right) }  
\\
& +  \frac{ \delta B_y A_{+}  \omega_B t }{2} \,\, e^{i \left(k_g z - \omega_g t\right) } 
+  \frac{ \delta B_y A_{+} }{4 i } \, \left( \frac{\omega_B}{\omega_g} \right) \,\, e^{i \left(k_g z - \omega_g t\right) }  
\nonumber \\
\label{eq:B_y-Expression}
\tilde{B}_y &\simeq  -\frac{B^{(0)}_y A_{+} }{4} \,\, e^{i \left(k_g z - \omega_g t\right) }  \\
& -  \frac{ \delta B_y A_{+}  \omega_g t }{2} \,\, e^{i \left(k_g z - \omega_g t\right) } 
 +  \frac{ \delta B_y A_{+} }{4 i } \, \left( \frac{\omega_B }{\omega_g} \right)^2 \,\, e^{i \left(k_g z - \omega_g t\right) }  
 \,\,  .
 \nonumber 
\end{align}
We want to mention the following points regarding the above expressions: First, in obtaining the above expressions, we have assumed that $ \omega_B << {\omega_g}$. This is valid in our case because the conversion from GWs to EM waves occurs in the 
sub-GHz frequency range. This approximation leads to $\omega^n_{\pm} \simeq \omega^n_g \left( 1 \pm n \frac{\omega_B}{\omega_g} \right)$ and allows us to approximate $\omega_{\pm} \approx \omega_g$ in the exponentials. Note that we have also ignored the terms with $e^{ i\omega_g t}$ since they will lead to wave propagating along negative z-direction i.e., $e^{i \left(k_g z + \omega_g t\right) }$, and hence are not relevant to our analysis.
Second, the last term in both expressions is tiny and can be neglected leading to Eqs.~(\ref{eq:E_x-Final}, \ref{eq:B_y-Final}). 
%\textcolor{red}{Thus, we have
%
%\begin{align}\label{eq:E_x-Expression-Final-1-app}
%\tilde{E}_x &\simeq  -\frac{B^{(0)}_y A_{+} }{2} \,\, e^{i \left(k_g z - \omega_g t\right) }   +  \frac{ \delta B_y A_{+}  \omega_B t }{2} \,\, e^{i \left(k_g z - \omega_g t\right) }  \\
%\label{eq:B_y-Expression-Final-1-app}
%\tilde{B}_y &\simeq  -\frac{B^{(0)}_y A_{+} }{4} \,\, e^{i \left(k_g z - \omega_g t\right) }   -  \frac{\delta B_y A_{+}  \omega_g t }{2} \,\, e^{i \left(k_g z - \omega_g t\right) } 
%\end{align}
%
%In the next section, we will use the above expressions to obtain the conversion factor for the entire magnetosphere.}
Lastly, GZ effect is a pure gravitational effect (due to incoming GWs) and the generation of EM waves does not require any source term (plasma or charged particles). Hence, if the incoming GWs are coherent, the emitted EM waves are coherent at resonance.

\section{Evaluating the radius of the magnetosphere}
\label{appsec:Integration-2}

We show that the assumption that the background magnetic field can be treated as a constant in the entire magnetosphere gives identical results to that of the background field decreasing radially. 

In evaluating $\alpha_{\rm tot}, $ we have assumed that the background magnetic field is a constant until light cylinder radius $R_{\rm LC}$. In this section, we show that this assumption is physically consistent. 

To do that, we consider the magnetic field of the pulsar magnetosphere in vacuum to be dipolar. 
We evaluate the average dipolar magnetic field in the magnetosphere. In spherical polar coordinates, the dipolar magnetic field is given by~\cite{2016-Cerutti.Beloborodov-SpaceSciRev,2016-Petri-JPlasmaPhy}:
%We would like to mention that the strength of the magnetic field decreases as we go away from the surface of NS, which is maximum near the surface; therefore, it is important to know the optimum distance (or radius of the magnetosphere) up to which the magnetic field strength can be taken approximately same. To understand this, let us consider the dipole magnetic field of the pulsar magnetosphere in a vacuum, which is spherical polar coordinates is given by~\cite{2016-Cerutti.Beloborodov-SpaceSciRev,2016-Petri-JPlasmaPhy}
%
\begin{align}\label{eq:B-stellar_equator}
\left(  B_r , B_{\theta} , B_{\phi}  \right) = B_{*} \, \left( \frac{r_* }{r} \right)^3 \, \left(  2 \cos\theta, \sin\theta , 0  \right)
\end{align}
where $B_*$ is the magnetic field on the NS surface. 

We now compute the average magnetic field on the volume between the radius of the NS ($r_*$) and 
$R_{ \rm{LC}}$:
\begin{equation}\label{eq:Vol_avg_magnetic}
\!\! \bar{ \textbf{B} }(r,\theta,\phi) = 
\frac{1 }{V}
\int_{r_{*} }^{ R_{ \rm{LC} } }  \!\!\!\!\!\! dr  \int_0^{\pi}  \!\!\!\! d\theta \int_0^{2 \pi } \!\!\!\! d\phi \, r^2 \sin\theta \, \textbf{B} (r, \theta, \phi) 
\end{equation}
where $V$ is the volume enclosed by both the surfaces:
\begin{align}\label{Volume}
V = \frac{4 \pi}{3} \left(   R_{ \rm{LC} }^3  - r_{*}^3  \right).
\end{align}
From Eqs.~(\ref{eq:Vol_avg_magnetic}, \ref{Volume}), we define the integral $\mathcal{I}$ as: 
{\small
\begin{align}\label{eq:Integration-I-Def}
\!\!\!\! \mathcal{I} \equiv V \, \bar{ \textbf{B} }(r,\theta,\phi)  =  
\int_{r_{*} }^{ R_{ \rm{LC} } } \!\!\!\! dr \int_0^{\pi} \!\!\!\! 
d\theta \int_0^{2 \pi } \!\!\!\! d\phi \, r^2 \sin\theta 
\textbf{B} (r, \theta, \phi) .
\end{align}
}
Note that the volume average of the radial component of the magnetic field vanishes.
The non-zero contribution comes from the angular variation of the magnetic field, i.e.,
\begin{align}\label{eq:Vol_avg_magnetic-theta}
\bar{\textbf{B}}_{\theta}  &= \frac{3 \pi B_* r_{*}^3 }{4 \left(  R_{ \rm{LC} }^3 - r_{*}^3  \right) } \ln \left| \frac{ R_{ \rm{LC} } }{ r_* }  \right| .
\end{align}
From Eqs.~(\ref{eq:Integration-I-Def}, \ref{eq:Vol_avg_magnetic-theta}), we obtain
\begin{align}\label{eq:I-eval-1}
\mathcal{I} &= 
 \pi^2 B_* r_{*}^3  \ln \left| \frac{ R_{ \rm{LC} } }{ r_* }  \right|
\end{align}
We now compare this with the assumption that the 
magnetic field is approximately constant until $R_{\rm LC}$. To do this, we substitute the 
average magnetic field obtained in Eq.~(\ref{eq:Vol_avg_magnetic-theta}) inside the integral in  Eq.(\ref{eq:Integration-I-Def}). 
Doing the radial integral between $r_*$ to  $\mathcal{R}$ leads to:
\begin{align}\label{eq:I-eval-2}
\mathcal{I} &=  
\pi^2 B_* r_{*}^3 \frac{ ( \mathcal{R}^3 - r_*^3 ) }{ \left(  R_{ \rm{LC} }^3 - r_{*}^3  \right) }  \,   \ln \left| \frac{ R_{ \rm{LC} } }{ r_* }  \right|.
\end{align}

Setting $\mathcal{R} = R_{ \rm{LC} } $ in the above expression, we see that the two expressions (\ref{eq:I-eval-1}, \ref{eq:I-eval-2}) are approximately the same. Thus, even if there is a radial and angular variation in the magnetic field in the magnetosphere, we can approximate the background field to be approximately constant in the region $ r_* \leq r \leq R_{ \rm{LC} }$. Thus, our expression for the total conversion factor mimics the realistic NS regions. 

\section{Conversion factor from entire magnetosphere}
\label{appsec:Integration}
In this appendix, we estimate the total conversion factor from the entire magnetosphere of the compact object. 
The energy density carried by these induced EM waves is~\cite{Book-Jackson-Classical_Electrodynamics}
\begin{align}\label{eq:rho_EM}
\rho_{\rm{EM} }  
%= \frac{ | \tilde{E}_x |^2 + | \tilde{B}_y |^2 }{8 \pi} 
\simeq \frac{ |A_{+} |^2 |B^{(0)}_y|^2 }{64 \pi}  \left[ 1   + 2 \, \xi^2 \omega_g^2 \, t^2 + 2 \, \xi \omega_g \, t  \,\,  \right] \, .
\end{align}
where $\xi = \delta B_y/B_y^{(0)}$. As mentioned earlier, the conversion is maximum when EM waves are approximately the same as the incoming waves. 
Having obtained the energy carried by the induced EM waves, we need to obtain what fraction of wave energy is converted to EM waves~\cite{1974-Zeldovich-SJETP}? To do this, we calculate the energy density carried by the GWs~\cite{Book-Gravitation_MTW}, i. e.:
\begin{align}\label{eq:rho_GW}
\rho_{\rm{GW} } = \frac{c^2 \omega_g^2}{32 \pi G } \left(  |A_{+}|^2 + |A_{\times}|^2 \right) = \frac{c^2 \omega_g^2}{16 \pi G } \,\,  |A_{+}|^2 \, ,
\end{align}
wherein second equality, we have used the fact that both the modes of GWs are generated with an equal amount of energy, i.e., $|A_{+}| = |A_{\times}| $ also referred to as isospectrality relation~\cite{Chandrasekhar_BlackHoles-Book}. From Eqs.~(\ref{eq:rho_EM}, \ref{eq:rho_GW}), we obtain Eq.~\eqref{eq:alpha-at-point}. 

To obtain the conversion factor in the entire magnetosphere, we define the following dimensionless parameter:
\begin{align}\label{eq:define-X}
X = \frac{r}{R_{ \rm{LC} }}
\end{align}
where $r$ is the distance from the surface of neutron star to a point in the magnetosphere, i.e., $r_* \lesssim \,  r  \lesssim  \, R_{\rm{LC}}$, and $r_* (= 10^6 \rm{cm})$ is the radius of the Neutron star (NS). In the case of Magnetar $R_{\rm LC} = 10^9~{\rm cm}$ and for NS $R_{\rm LC} = 10^7~{\rm cm}$, hence, $X_* = \frac{r_*}{R_{\rm{LC}} } < 1$. In other words, the range of $X$ is $ 0.001 \lesssim \,   X   \lesssim \,  1 $. 

Since we are interested in the EM waves reaching the observer, we are interested in evaluating the conversion factor along the observer's line of sight. Thus, we have 
\begin{align}\label{eq:z=XR}
z = r = XR_{\rm{LC}}.
\end{align}
where $\theta = 0$ corresponds to the direction along the line-of-sight of the observer.  Thus, the total conversion factor is given by the integral
\begin{align}\label{eq:alpha_tot-def}
\alpha_{\rm{tot} } = \Omega \int_{X_*}^{1} \alpha \,\, dX  ,
\end{align}
where $\Omega$ is the total solid angle which is an overall constant factor because Eq.(\ref{eq:alpha_tot-def}) is independent of the angular coordinates.

\section{Poynting vector}
\label{appsec:poyntingVec}

In astrophysical observations of compact objects, a quantity of interest is the energy flux density which is the Poynting vector. This section evaluates the Poynting vector for magnetar and neutron star in SI unit and Jansky Hz, a widely used unit for spectral flux density in radio observations.

The Poynting vector of the induced electric field (\ref{eq:E_x-Final}) and induced magnetic field (\ref{eq:B_y-Final}) is~\cite{Book-Jackson-Classical_Electrodynamics}:

\begin{align}\label{eq:poyntingVector}
    S_z &= \frac{c}{8 \pi} \tilde{E}_x \times \tilde{B}^*_y \\ 
    & \simeq \frac{A_+^2 \, |B_y^{(0)}|^2  \, c }{128 \pi} \left[ 1 + 2 \omega_g \xi \frac{R_{\rm{LC}}}{c} - 2 \omega_g \omega_B \xi^2 \left( \frac{R_{\rm{LC}}}{c}\right)^2  \right] \nonumber 
\end{align}
where $\tilde{B}^*_y$ is the complex conjugate of the induced magnetic field $\tilde{B}_y$.  As mentioned above, the frequency of the alternating magnetic field is $1 \rm{Hz}$ for magnetar and $10^3 \, \rm{Hz}$ for the millisecond pulsar. Here again, we have assumed that $\omega_B << \omega_g$. Rewriting the above Poynting vector in terms of $\alpha_{\rm{tot} }$, we get Eq.~\eqref{eq:PoyntingVec-in-alpha}. 

\section{Current status of high-frequency GWs}\label{appsec:ECO}
Primordial black-holes, Exotic compact objects and early Universe can generate high-frequency GW (HFGW) in MHz to GHz~\cite{2008-Akutsu.etal-PRL,2008-Nishizawa.etal-PRD,2016-Holometer-PRL,2017-Chou.etal-PRD,2020-Ito.etal-EPJC,2021-Domenech-EPJC,2021-Goryachev.etal-PRL,2022-Aggarwal.etal-PRL}. These can generate GWs up to 14~GHz~\cite{2020-Ito.etal-EPJC}. Over the last decade, many HFGW detectors are proposed, and some of them are operational. For instance, the Japanese 100 MHz detector with a $0.75~{\rm m}$ armlength interferometer has been operational for a decade~\cite{2008-Akutsu.etal-PRL,2008-Nishizawa.etal-PRD}, Holometer detector has put some limit on GWs at MHz~\cite{2016-Holometer-PRL,2017-Chou.etal-PRD}. The Bulk Acoustic GW detector experiment recently reported two MHz events after 153 days of operation~\cite{2021-Goryachev.etal-PRL}. A GHz GW detector is also proposed~\cite{2020-Ito.etal-EPJC}. These detectors are ideally suited for searching for physics beyond the standard model (SM), like primordial black-holes, exotic compact objects and early Universe. For instance, exotic compact objects with characteristic strain $h$~\cite{2020-Aggarwal.etal-arXiv}:
\[
h \lesssim  10^{-19} C^{5/2} \left( \frac{\rm MHz}{f} \right) \left( \frac{\rm Mpc}{D} \right)
\]
lead to the following GW amplitudes: 
 \[
 h_{\rm 1.4 GHz, 10kpc} \lesssim 10^{-21} \, ,
  h_{\rm 1.4 GHz, 1Mpc} \lesssim 10^{-23} \ .
 \]

After 153 days of operation, the Bulk Acoustic GW detector experiment recently reported two MHz events~\cite{2021-Goryachev.etal-PRL}. According to Goryachev et al. ~\cite{2021-Goryachev.etal-PRL}, the data corresponding to two MHz events fits best a single energy depositing event. The authors mention that the strongest observed signal results in a required characteristic strain amplitude of $h_c \approx 2.5 \times 10^{-16}$, corresponding to a PBH merger of $m_{\mathrm{PBH}}<4 \times 10^{-4} M_{\odot}$ (which gives a maximum frequency at inspiral of $5.5 \ \mathrm{MHz}$), at a distance of $D \approx 0.01 \mathrm{pc}$. It is important to note that these are not conclusive enough. However, these detections also imply that these events are not rare.

As mentioned above, many HFGW detectors are proposed, and better estimates will be available as more and more detectors will be operational in the coming decades. This can provide better estimate of these events in the next decade.

\section{Calculating isotropic Equivalent Luminosity and Fluence}
In this appendix, we calculate the fluence and Isotropic Equivalent Luminosity (IEL). The observed fluence is defined as the product of the burst width $\Delta t = (2R_{LC})/c$ and the peak flux ($S_z/\omega_g$)~\cite{2015-Keane.Petroff-MNRAS,2018-Macquart.Ekers-MNRAS}. %The last to one column in table~\ref{table2} shows the observed fluence.
The energy density of the FRB electromagnetic (EM) waves
at a distance $d$ from the source (NS/magnetar) in the limit $d \gg \text{source size} (r_*)$ is given by~\cite{2019-Lu.Kumar-MNRAS}: 
\begin{align}\label{eq:luminosity}
    \rho_{\rm EM} = \frac{L_{\rm IEL}}{4\pi d^2 c},
\end{align}
where $L$ is the isotropic equivalent luminosity. 
%In the ~\ref{table2} we estimate the $L_{\rm IEL}$ for a range of distance $d$ from pc to Mpc.

\end{appendices} 
%=====================================================================
% BIBLIOGRAPHY
%   This should follow the appendices, if any, otherwise summary and
%   conclusions chapter.
% Choose your bibliography style
% plain is the basic style, others include ieeetr, siam, asm, etc
%\bibliographystyle{acl}
% Add the bib file
%\bibliography{ref}
\bibliographystyle{utphys.bst}
\bibliography{Thesis.bbl}
%\bibliography{ThesisReferences}
%\addcontentsline{toc}{chapter}{References}
%=====================================================================
% PUBLICATIONS
%  publications if any may be listed after the literature cited.

%=====================================================================
% ACKNOWLEDGMENTS
%   This is the last item in the thesis. It should be signed by
%   author, with date.
%\include{Acknowledgements}

\end{document}